\documentclass[11pt,a4paper]{article}
\pdfoutput=1
\usepackage{jheppub}

\usepackage[top=4.5cm, bottom=0.cm, outer=-0.25cm, inner=4.75cm]{geometry}
\usepackage{placeins}
\usepackage{latexsym}
\usepackage[bbgreekl]{mathbbol}
\usepackage{mathtools}
\usepackage{amsmath,amssymb,amsthm,amsbsy,amsfonts,mathrsfs}
\usepackage{multirow}
\usepackage{slashed}
\usepackage{float}
\usepackage{esint}
\usepackage{epsfig}
\usepackage{lscape}
\usepackage{graphicx}
\usepackage[utf8]{inputenc}
\usepackage{subfigure}
\usepackage{nicefrac}
\usepackage{rotating}
\usepackage{afterpage}
\usepackage{bbm}
\usepackage{color,xcolor}
\usepackage{cancel}
\usepackage{stackrel}
\usepackage[framemethod=tikz]{mdframed}
\usepackage{empheq}
\usepackage[many]{tcolorbox}
\usepackage{simplewick}
\usepackage[font=footnotesize,labelfont=bf,justification=centerlast,width=.94\textwidth]{caption}
\usepackage{tikz}
\usepackage{tikz-feynman}
\usepackage{tcolorbox}
\tcbuselibrary{theorems}

\newcommand*{\orangebox}{%
  \tcboxmath[colback=red!10!white, colframe=white, size=fbox, arc=10pt, boxrule=0.8pt]%
}

\newcommand*{\cyanbox}{%
  \tcboxmath[colback=cyan!10!white, colframe=white, size=fbox, arc=2pt, boxrule=0.8pt]%
}

\newmdenv[%
middlelinecolor=blue!20!,
middlelinewidth=1pt,
backgroundcolor=blue!10!,
roundcorner=3pt
]{identity}

\newmdenv[%
middlelinecolor=black,
middlelinewidth=1pt,
backgroundcolor=blue!20!,
roundcorner=3pt
]{titlebox}

\newmdenv[%
middlelinecolor=red!20!white,
middlelinewidth=1pt,
backgroundcolor=red!10!white,
roundcorner=10pt,
subtitlebelowline=true,
frametitle={Calculation},
frametitlefont={\normalfont\bfseries\sffamily\color{red!40!white}},
]{calculation}

\def\){\right)}
\def\({\left( }
\def\]{\right] }
\def\[{\left[ }

\def\NO{\nonumber}

\newcommand{\be}{\begin{equation}}
\newcommand{\ee}{\end{equation}}

\def\bea{\begin{eqnarray}}
\def\eea{\end{eqnarray}}

\def\bal#1\eal{\begin{align}#1\end{align}}

\def\bald{\begin{aligned}}
\def\eald{\end{aligned}}

\def\bsub{\begin{subequations}}
\def\esub{\end{subequations}}

\def\beqx{\begin{displaymath}}
\def\eeqx{\end{displaymath}}

\newcommand{\bmat}{\left(\begin{array}}
\newcommand{\emat}{\end{array}\right)}

\def\a{\alpha}
\def\b{\beta}
\def\c{\chi}
\def\d{\delta}
\def\e{\epsilon}
\def\f{\phi}
\def\g{\gamma}
\def\h{\eta}
\def\j{\psi}
\def\k{\kappa}
\def\l{\lambda}
\def\m{\mu}
\def\n{\nu}
\def\o{\omega}
    
\def\p{\pi}

    \def\th{\theta}
\def\r{\rho}
\def\s{\sigma}
\def\t{\tau}
\def\x{\xi}

\def\D{\Delta}
\def\F{\Phi}
\def\G{\Gamma}

\def\O{\Omega}
    \def\Om{\Omega}

    \def\Th{\Theta}

\def\ve{\varepsilon}

    \def\vth{\vartheta}

\def\vf{\varphi}

\def\ca{{\cal A}}
\def\cb{{\cal B}}

\def\ci{{\cal I}}
\def\cj{{\cal J}}

\def\cn{{\cal N}}
\def\co{{\cal O}}
\def\cp{{\cal P}}
\def\cq{{\cal Q}}

\def\ct{{\cal T}}

\def\bo{{\raise-.3ex\hbox{\large$\Box$}}}               
\def\pa{\partial}                                       
\def\face{{\raise.2ex\hbox{$\displaystyle \bigodot$}\mskip-2.2mu \llap {$\ddot
        \smile$}}}                                   
\def\>{\rangle}                                      
\def\<{\langle}                                      

\def\tx#1{\text{#1}}
\def\wt#1{\widetilde{#1}}                            
\def\lbar#1{\ensuremath{\overline{#1}}}              
\def\leftrightarrowfill{$\mathsurround=0pt \mathord\leftarrow \mkern-6mu
        \cleaders\hbox{$\mkern-2mu \mathord- \mkern-2mu$}\hfill
        \mkern-6mu \mathord\rightarrow$}        
\def\dvec#1{\vbox{\ialign{##\crcr
        \leftrightarrowfill\crcr\noalign{\kern-1pt\nointerlineskip}
        $\hfil\displaystyle{#1}\hfil$\crcr}}}           
\def\tr{{\rm tr \,}}                                    
\def\diag{{\rm diag \,}}                                
\def\Re{{\rm Re\,}}                                     
\def\Im{{\rm Im\,}}                                     

\def\-{\hphantom{-}}

\def\dbar#1{\raisebox{0.pt}{\ensuremath{\stackrel{\rule{6pt}{1pt}}{#1}\hskip-.5pt}}}       

\def\bbx#1\ebx{\begin{empheq}[box={\tcbhighmath[colframe=blue!20!white,colback=blue!10!white]}]{align} #1 \end{empheq}}

\def\bbxd#1\ebxd{\begin{identity} \vskip -.4cm #1 \end{identity}\vskip-.2cm}

\def\btbox#1\etbox{\begin{titlebox} \vskip -.0cm #1 \end{titlebox}\vskip-.0cm}

\def\lb{\ensuremath{<\hskip-.1cm}}
\def\rb{\ensuremath{\hskip-.1cm >}}

\title{Supersymmetry anomaly in the superconformal Wess-Zumino model}

\author[a,b]{Georgios Katsianis}  
\author[c]{Ioannis Papadimitriou}
\author[a,b]{Kostas Skenderis}
\author[a,b]{Marika Taylor}

\affiliation[a]{STAG Research Centre, Highfield, University of Southampton, SO17 1BJ Southampton, UK}  
\affiliation[b]{Mathematical Sciences, Highfield, University of Southampton, SO17 1BJ Southampton, UK}
\affiliation[c]{School of Physics, Korea Institute for Advanced Study, 85 Hoegi-ro, Seoul 02455, Korea}

\emailAdd{G.Katsianis@soton.ac.uk}
\emailAdd{ioannis@kias.re.kr}
\emailAdd{K.Skenderis@soton.ac.uk}
\emailAdd{M.M.Taylor@soton.ac.uk}

\abstract{We present a comprehensive analysis of supersymmetry anomalies in the free and massless Wess-Zumino (WZ) model in perturbation theory. At the classical level the model possesses ${\mathcal N}=1$ superconformal symmetry, which is partially broken by quantum anomalies. The form of the anomalies and the part of the symmetry they break depend on the multiplet of conserved currents used. It was previously shown that the R-symmetry anomaly of the conformal current multiplet induces an anomaly in Q-supersymmetry, which appears first in 4-point functions. Here we confirm this result by an explicit 1-loop computation using a supersymmetric Pauli-Villars regulator. 

The conformal current multiplet does not exist in the regulated theory because the regulator breaks conformal invariance, R-symmetry and S-supersymmetry explicitly. The minimal massive multiplet is the Ferrara-Zumino (FZ) one and the supersymmetry preserved by the regulator is a specific field dependent combination of Q- and S- supersymmetry of the conformal multiplet. While this supersymmetry is non anomalous, conformal invariance, R-symmetry and the original Q- and S-supersymmetries are explicitly broken by finite contact terms, both in the regulated and renormalized theories.        

A conformal current multiplet does exist for the renormalized theory and may be obtained from the FZ multiplet by a set of finite local counterterms that eliminate the explicit symmetry breaking, thus restoring superconformal invariance up to anomalies. However, this necessarily renders both Q- and S-supersymmetries anomalous, as is manifest starting at 4-point functions of conformal multiplet currents. The paper contains a detailed discussion of a number of issues and subtleties related to Ward identities that may be useful in a wider context.

}


\preprint{KIAS-P20067}

\begin{document}  
\maketitle

\newpage


\section{Introduction and summary of results}
\label{sec:intro}

Anomalies are a cornerstone of modern QFT.  Anomalies of rigid  (sometimes also called global) symmetries are a feature of the theory.  They provide important observables of the theory (for example, the conformal anomaly coefficients in CFTs); consistency conditions for duality relations and more generally may be used as a tool to understand the dynamics of the theory. On the other hand, anomalies of local symmetries lead to inconsistencies, such as lack of unitarity and renormalizability, and they should be canceled. 

The purpose of this paper is to discuss rigid supersymmetry anomalies for the four dimensional massless Wess-Zumino model in perturbation theory. This paper is a companion of  \cite{Katsianis:2019hhg} and we aim to present a comprehensive discussion of the perturbative 1-loop computation.  Since this is a rather technical topic we will present a summary of the methods and results in this section. We begin however with a recap of some of the main properties of anomalies illustrated with the standard chiral anomaly, as this will help to put into context the result on the supersymmetry anomaly. 

Anomalies present a breaking of a symmetry at the quantum level. They appear either because of a 0/0 structure, where a zero due to a symmetry is compensated by a UV infinity, leaving a finite remainder (for example, chiral anomaly or type A conformal anomalies), or because counterterms needed to cancel infinities break some of the symmetries (for example, type B conformal anomalies). In perturbation theory they appear as finite contact terms that violate Ward identities associated to corresponding symmetries\footnote{Ward identities associated with an insertion of (anomalous) currents in correlation functions of elementary fields are in general non-anomalous. In the case of the free massless fermion theory reviewed below this is manifestly the case, as these Ward identities saturate at tree-level. Anomalies arise in correlation functions of symmetry currents with composite operators.}. Anomalies also appear as lack of symmetry conservation in the presence of background fields. These are two sides of the same coin.

Introducing sources that couple to gauge invariant operators, one may encode the correlation functions in the partition function. The rigid symmetry now  translates into invariance of the partition function under corresponding local transformations that act on the sources, and the presence of the anomaly leads to lack of gauge invariance at one loop order. To illustrate the discussion, let us consider the case of a free massless fermion in four dimensions. The theory is invariant under rigid vector and axial $U(1)$ transformations.  Let $\cj_V^\m$ and $\cj_A^\m$ be the vector and axial Noether currents and $V^\m$ and  $A^\m$ the corresponding sources. The partition function $\mathscr{Z}[V,A]$ is classically invariant separately under $\delta_v V^\m = \pa_\m \e_v$ and $\delta_a A^\m = \pa_\m \e_a$, where $\e_v, \e_a$ are the parameters of the vector and axial transformations,  but at one loop order one cannot maintain both invariances. Imposing that the vector invariance holds, one finds that the conservation of the axial current is violated in correlation functions,
\begin{equation}
\mathscr{Z}[V+d \e_v,A] = \mathscr{Z}[V,A], \qquad \mathscr{Z}[V,A+ d \e_a] =\mathscr{Z}[V,A] 
e^{-i \int d^4 x \e_a {\cal A}},
\end{equation}
where ${\cal A} \sim \e^{\m \n \r \s} (F_{\m \n}(V) F_{\r  \s}(V) + \frac{1}{3} F_{\m \n}(A) F_{\r  \s}(A) )$ is the chiral anomaly and $F_{\m \n}(V) = \pa_\m V_{\n} -  \pa_\n V_{\m}, F_{\m \n} (A)= \pa_\m A_{\n} -  \pa_\n A_{\m}$.
This results in the Ward identities
\begin{equation} \label{intro:anom}
\pa_\m \< \cj_A^\m \>_s = {\cal A}, \qquad \pa_\m \< \cj_V^\m \>_s = 0,
\end{equation}
where $\<\cdot\>_s$ denotes connected correlation functions in the presence of sources. Since the anomaly is quadratic in the sources, (\ref{intro:anom}) encodes the fact that triangle diagrams with one or three axial currents are anomalous.

The anomaly can be shifted around by adding finite local terms in the action. For the case of the massless fermion we could consider the partition function,
\begin{equation} \label{intro:ct}
\wt{\mathscr{Z}}[V,A] = \mathscr{Z}[V,A]  \exp \Big(i \a_c \int d^4 x\, \e^{\m \n \r \s} A_\m V_{\n} F_{\r  \s}(V) \Big),
\end{equation}
where $\a_c$ is a constant. By an appropriate choice of $\a_c$ one may arrange for the axial-vector-vector correlator to be conserved on the axial current but then the conservation of the vector current would be anomalous, and the partition function (\ref{intro:ct}) would be invariant  under neither vector nor axial transformations.  The theories with or without the counterterm are physically distinct, as they preserve different symmetries. The standard choice is to keep the vector symmetry non-anomalous (in the original context \cite{Adler:1969gk, Bell:1969ts} the vector symmetry was electromagnetism), but more generally depending on the physics context one may work with either theory. 
In the context of the AdS/CFT correspondence, the finite counterterms correspond to finite boundary terms that should be specified when defining the bulk theory.

Besides shifting the anomaly between symmetries by local counterterms, the anomaly may be `hidden' by introducing additional (background) fields. In the case of the axial anomaly, for example, we may introduce an external scalar field ${\varPhi}$ and modify the partition function as
\begin{equation} \label{intro:comp}
\mathscr{Z}'[V,A, {\varPhi}] = \mathscr{Z}[V,A]  \exp \Big(i \int d^4 x\, {\varPhi} {\cal A} \Big).
\end{equation}
Assigning transformations
\begin{equation} \label{intro:compen}
\delta_v \varPhi =0, \qquad \delta_a \varPhi= \e_a,
\end{equation}
the partition function  $\mathscr{Z}'$ is now gauge invariant under both vector and axial transformations 
\begin{equation}
\mathscr{Z}'[V+d \e_v,A, \varPhi + \delta_v \varPhi] = \mathscr{Z}'[V,A, \varPhi], \qquad 
\mathscr{Z}'[V,A+ d \e_a, \varPhi+\delta_a \varPhi] =\mathscr{Z}'[V,A, \varPhi]. 
\end{equation}
This does not mean that the anomaly has disappeared; the triangle diagrams are not affected by the new terms in (\ref{intro:comp}). 

One could also use the formulation with $\varPhi$ to make the Ward identities look as if there is no anomaly. To see this 
let us rewrite  the coupling of $\varPhi$ in (\ref{intro:comp}) as 
\begin{equation} \label{intro:comp2}
\mathscr{Z}'[V,A, {\varPhi}] = \mathscr{Z}[V,A]  \exp \Big(i \int d^4 x\, \varPhi\, {\cal O}_A \Big),
\end{equation}
where ${\cal O}_A = \pa_\m \cj_A^\m$.  In the classical theory ${\cal O}_A$  is proportional to field equations (it is a null operator) and when we regulate the theory, say with Pauli-Villars regularization,  $\<{\cal O}_A\>_s$ becomes local (and equal to the anomaly, $\< {\cal O}_A \>= {\cal A} $)  as the regulator is removed. 
Working out the Ward identity starting from (\ref{intro:comp2}) one finds,
\begin{equation} \label{intro:anom2}
\pa_\m \< \cj_A^\m \>_s - \< {\cal O}_A \>_s =0.
\end{equation}
In this form the Ward identity appears non-anomalous and one may be tempted to conclude that the theory would be non-anomalous if we include the coupling to the null operator ${\cal O}_A$. Of course, this is just an illusion: (\ref{intro:anom2}) is equal to (\ref{intro:anom}). The only thing that happened was that we moved the anomaly from the r.h.s. to the l.h.s. and gave it a different name.

Fields like ${\varPhi}$ are called `compensators' because they may be used to restore or compensate for broken symmetries, or `gauge-away' fields because one may set them to zero using gauge transformations. Indeed one may use (\ref{intro:compen}) to set $\varPhi$ to zero (and thus also establishing that  $\mathscr{Z}'$ is equivalent to $\mathscr{Z}$). Invariance of the partition function under gauge transformations does not by itself imply absence of anomalies, if gauge away fields are present. One must first set to zero all gauge away fields and then check invariance of the partition function. Similarly, one must set all compensators to zero prior to working out the form of the Ward identities.

Anomalies in supersymmetry have been a recurrent topic of research since the discovery of supersymmetric theories. A set of studies considered the effective action for elementary fields and examined whether it is invariant under supersymmetry including loop effects,  investigated the conservation of the supercurrent inside correlators of elementary fields and/or solved the Wess-Zumino (WZ) consistency conditions relevant for this setup. These works, summarised in the monograph \cite{Piguet:1986ug}, found no supersymmetry anomaly.  As mentioned above, however, one either needs to put the theory on a non-trivial background or consider correlation functions of the supercurrent with other composite operators to probe supersymmetry anomalies. 

Our focus here is on four dimensional ${\cal N}=1$ SCFTs. In addition to conformal symmetry and standard supersymmetry (Q-supersymmetry), these theories also have a $U(1)$ R-symmetry and special conformal supersymmetry (S-supersymmetry). It was realised early on that the trace anomaly, the R-current anomaly and S-anomaly sit in a supermultiplet  \cite{Ferrara:1974pz}  and that one cannot maintain at the same time the conservation of the Q and S supersymmetry  \cite{deWit:1975veh,Abbott:1977in,Abbott:1977xj,Abbott:1977xk,Hieda:2017sqq,Batista:2018zxf}.  This is the standard superconformal anomaly mentioned and is distinct from the anomaly discussed here. Also distinct is the Konishi anomaly \cite{Konishi:1983hf, Konishi:1985tu}, which is a superspace version of the chiral anomaly in supersymmetric gauge theories.  

A supersymmetry anomaly appears in theories with gravitational anomalies \cite{Howe:1985uy, Tanii:1985wy, Itoyama:1985ni}, as one may anticipate based on the fact that the energy momentum tensor and the supercurrent are part of the same supermultiplet. Indeed this supersymmetry anomaly sits in the same multiplet as the gravitational anomaly. A supersymmetry anomaly appears also in super Yang-Mills (SYM) theory in the WZ gauge when there are gauge anomalies \cite{Itoyama:1985qi} (see also \cite{Piguet:1984aa,Guadagnini:1985ea, Zumino:1985vr}). This anomaly has a number of technical similarities with the anomaly we discuss here and has been recently revisited in \cite{Closset:2019ucb, Kuzenko:2019vvi}. In particular,  like the R-symmetry anomaly we discuss here, the gauge anomaly is not invariant under supersymmetry, and so the WZ consistency conditions imply the existence of a supersymmetry anomaly, with the same coefficient as that of the R-symmetry/gauge anomaly.

Many previous works discuss anomalies of dynamical supergravity coupled to matter, and relevant work (mostly in superspace language) may be found in  \cite{Bonora:1984pn,Buchbinder:1986im,Brandt:1993vd, Brandt:1996au, Bonora:2013rta}. Particularly relevant for us  is the discussion in \cite{Gates:1981yc} where it was argued that quantum anomalies in the matter require the old minimal set of auxiliary fields for $\cn=1$ supergravity and the discussion in \cite{deWit:1985bn} about compensators in supergravity.

The anomaly we discuss here was discovered in the context of superconformal theories that can be realised holographically \cite{Papadimitriou:2017kzw}\footnote{Early attempts to compute the supertrace Ward identity can be found in 
\cite{Chaichian:2003kr,Chaichian:2003wm} but these missed contributions
to the anomaly involving the R-symmetry current and the Ricci tensor. }.  In holography, given a bulk action, one can use holographic renormalisation 
\cite{Henningson:1998gx, deHaro:2000vlm} to compute the Ward identities
and anomalies of the dual QFT. AdS/CFT relates ${\cal N} =1$ SCFTs in 
four dimensions to ${\cal N} =2$ gauged supergravity in five dimensions. 
Starting from gauged supergravity in an asymptotically locally AdS$_5$ 
spacetime and turning on sources for all superconformal currents one can 
compute the complete set of superconformal anomalies. 
The holographic anomalies for bosonic currents were computed earlier in \cite{Cassani:2013dba} (reproducing (and correcting) known 
field theory results \cite{Anselmi:1997am}), while \cite{An:2017ihs} obtained the 
superconformal anomalies in the presence of local supersymmetric scalar
couplings.  The holographic computation is available for holographic CFTs with  central charges satisfying $a=c$ as $N \to \infty$  \cite{Henningson:1998gx}. The anomaly for general $a, c$ was obtained in \cite{Papadimitriou:2019gel} by solving the WZ consistency conditions \cite{Wess:1971yu} under the assumption that R-symmetry is only broken by the standard triangle anomaly. 

In this paper we discuss the perturbative computation of the anomaly in the simplest model in which it is realised: the free and massless  Wess-Zumino model \cite{Wess:1973kz}. Since this model is free, the 1-loop computation is exact. The supersymmetry anomaly appears for first time in the 4-point function that involves two supercurrents and either two R-symmetry currents or one R-symmetry current and one energy momentum tensor. In this paper we will focus on the former.

The anomaly appears as a violation of Ward identities. The derivation of the Ward identities is standard by now, but there are a number of subtleties which are perhaps not widely known. Given a theory there are a number of different routes to obtain Ward identities for correlation functions. In the path integral derivation (which is often discussed in textbooks) one starts from insertions of operators in the path integral and makes a change of variables that amount to a symmetry transformation. Making the parameter of the transformation local and using Noether's theorem leads to the Ward identities. This formulation has the advantage that it directly constrains the correlators computed by Feynman diagrams (more on this below). It has the disadvantage that the Ward identities contain theory specific terms. 

An alternative way to obtain the Ward identities is to introduce sources that couple to the operators and then impose gauge invariance of the partition function. Defining $n$-point functions as $n$ functional derivatives of the partition function w.r.t. the sources, the Ward identities take a universal form for all theories that realise the same symmetries. However, the $n$-point functions defined in this way are not the same with the insertion of $n$ operators in the path integral. To emphasise the difference we use different notation for the two types of correlator. Let $J$ denote the source for an operator ${\cal O}$. Then the two definitions of $n$-point functions are
\bsub
\begin{align}
\< {\cal O} (x_1) \cdots {\cal O} (x_n) \> &= \left.\frac{\delta}{\delta J(x_1)} \cdots  \frac{\delta}{\delta J(x_n)} \mathscr{W}[J]\right|_{J=0} \label{func}, \\
<{\cal O} (x_1) \cdots {\cal O} (x_n)> &=- \frac{i}{\mathscr{Z}}\int[\tx d\{\F\}] {\cal O} (x_1) \cdots {\cal O} (x_n) e^{iS[\{\F\}]}, \label{PI}
\end{align}
\esub
where $\{ \F \}$ denotes collectively all elementary fields and  $\mathscr{W}[J]=-i\log \mathscr{Z}[J]$. The functional derivatives defining the wide bracket  correlators in (1.9a) are taken with the operators kept fixed.  In contrast, in the correlators defined in (1.9b) one takes the functional derivatives in the path integral keeping fixed the operator at $J=0$. The chain rule in this case produces additional semi-local correlators involving $\delta {\cal O}(x_i)/\delta J(x_j)$  -- the so-called `seagull terms'. We should emphasize that this dependence of the operator $\co$ on the source $J$ that arises through the classical coupling of the theory to background sources is local and should not be confused with the generically non-local dependence of the 1-point function  $\<\co\>_J$ that is obtained by performing the path integral in the presence of sources. The seagull terms are theory dependent and their contribution drops out when all insertions are at separated points. However, since our purpose is to discuss anomalies, which are local contributions to the Ward identities, we cannot ignore the contribution of seagull terms. 

As mentioned above, the Ward identities derived using path integral manipulations involving  $<\cdot>$ correlators contain theory specific terms. These terms combine with the seagull terms to produce the universal Ward identities written in terms of $\< \cdot \>$ correlators. More precisely, the universal Ward identities split into a number of path integral Ward identities. Apart from the universality of the Ward identities in terms of $\< \cdot \>$ correlators, there is yet another reason to use this definition: the local counterterms required to renormalise $\< \cdot \>$ correlators are universal and depend only on the sources $J$ of the composite operators $\co$. To renormalise the path integral correlators $\lb\cdot\rb$ one needs counterterms that involve additional sources (besides $J$) for the model specific seagull operators $\delta {\cal O}(x_i)/\delta J(x_j)$. Such terms cancel out between the path integral correlators and correlators involving seagull terms, resulting in the universal counterterms that depend only on the sources $J$. A related discussion appeared recently in \cite{Bzowski:2017poo}. For these reasons our discussion in the main text will be phrased in terms of the correlators in (\ref{func}). However, we explain in detail the issues discussed here in a series of appendices.

To derive the Ward identities for symmetry currents we need to couple the currents to corresponding sources in a gauge invariant fashion. In simple cases\footnote{{\it i.e.} when the currents are invariant under the symmetry.} linear couplings suffice, but in general one has to include non-linear terms, and these in turn give rise to the seagull contributions discussed above. One can work out all terms needed using the Noether method, but luckily in our case the answer is already known. The sources comprise an $\cn=1$ conformal supergravity multiplet, so our starting point is the coupling of the $\cn=1$ massless WZ model to $\cn=1$ conformal supergravity.  It is then a straightforward (if tedious) computation to work out the Ward identities that $\<\cq^\m(x)\bar\cq^\s(y)\cj^\k(z)\cj^\a(w)\>$ satisfy, where $\cq^\m$ is the supercurrent and $\cj^\k$ the R-symmetry current. This correlator satisfies three different Ward identities: one due the supercurrent conservation (Q-supersymmetry), one from gamma-trace conservation (S-supersymmetry) and one from the conservation of the R-current. These relate the 4-point function to lower-point correlators, which themselves satisfy their own Ward identities.  The theory would be non-anomalous if the correlators computed at one loop satisfy the superconformal Ward identities.

To do this computation we need to regulate the theory and an important issue is the choice of regulator. Any consistent regulator would suffice but certain regulators are more convenient than others. 
An {\it ad hoc} regularisation of Feynman diagrams may lead to inconsistencies. A regulator is consistent if it can be implemented at the level of the action, so one can formulate a regulated theory from which the Feynman rules follow and the renormalised correlators are obtained after appropriate subtractions are made and the regulator is removed. 

A second important issue is the symmetries that the regulator preserves. 
If a regulator breaks non-anomalous symmetries then this complicates the computation, as one would need to compensate 
the explicit breaking by finite counterterms. On the other hand, it is important to make sure that the computational scheme does not implicitly assume the existence of a regulator that respects the symmetry one investigates, since if a regulator exists that respects a symmetry then this symmetry is non-anomalous. In our context we want to study supersymmetry anomalies, so to avoid such implicit assumptions we will do the 1-loop computation using components rather than superspace graphs. 

In  \cite{Katsianis:2019hhg} we sketched the 1-loop computation using momentum cut-off, following the original computation of the chiral anomaly  \cite{Adler:1969gk, Bell:1969ts} (see also Jackiw's lectures in \cite{Treiman:1986ep}). The 4-point 1-loop diagrams entering the Ward identities are superficially linearly divergent like the triangle diagrams and they have a similar momentum routing 
ambiguity. However, the complete Ward identity involves additional divergences that need to be subtracted with counterterms and such a computation is challenging when carried out with momentum cut-off, since it breaks Lorentz invariance.  

In this paper we will use Pauli-Villars (PV) regularisation \cite{Pauli:1949zm}, which manifestly preserves Lorentz invariance. PV regularisation is often presented as a prescription where propagators are replaced by differences of propagators involving massive PV fields, or entire diagrams are replaced by linear combinations of identical diagrams involving PV fields such that the result has improved UV behaviour. For such prescriptions to be consistent, the new Feynman rules should follow from a regulated action involving the massive PV fields.   In the presence of interactions there is a standard way to introduce Pauli-Villars fields such that all correlation functions of elementary fields are regulated, see for example  \cite{ZinnJustin:2002ru}. In our case the theory is free and we are considering correlation functions of symmetry currents, so this method does not apply automatically.  Following in spirit the discussion of PV regularisation of QED in \cite{Itzykson:1980rh}, however, we found that all diagrams we compute are regulated using as PV fields three massive $\cn=1$ chiral multiplets, one with standard statistics and two with `wrong' statistics, with masses appropriately correlated.

Since the regulated action is supersymmetric, one may wonder whether this already shows that there is no supersymmetry anomaly. To establish the absence of a supersymmetry anomaly we still need to couple the symmetry currents to sources supersymmetrically. Since the PV fields form massive supermultiplets, they break conformal symmetry, R-symmetry and S-supersymmetry and as such they cannot couple supersymmetrically to conformal supergravity. Instead, one can consistently couple the regulated theory to old minimal supergravity. The supersymmetry of old minimal supergravity can be identified with a field dependent linear combination of the Q- and S-supersymmetry of conformal supergravity. However, the algebra that involves this field dependent combination does not close unless additional auxiliary fields are turned on. The auxiliary field required for the closure of the old minimal supergravity algebra is a complex scalar, $M$. Once this additional field is included the algebra closes, but differs from that of conformal supergravity. While the algebra of conformal supergravity is a gauged version of the $\cn=1$ superconformal algebra, old minimal supergravity amounts to gauging the super-Poincar\'e algebra. Having coupled the currents to sources in a gauge invariant fashion one may proceed to obtain the regulated Ward identities, and then use them to find out which symmetries are anomalous. As we will verify, old minimal supersymmetry is non anomalous.

It is instructive to rephrase this discussion in terms of current multiplets. Prior to regularisation the theory can be coupled to conformal supergravity and the corresponding symmetry currents form an $\cn=1$ conformal current multiplet. The Ward identities derived from conformal supergravity refer to this multiplet. The regularisation breaks conformal symmetry and the theory can only couple to old minimal supergravity. The resulting currents constitute the Ferrara-Zumino (FZ) multiplet, which contains all currents of the conformal multiplet (some of which are now not conserved) and, in addition, the complex scalar operator sourced by the auxiliary field $M$. The Ward identities derived using old minimal supergravity refer to the FZ multiplet and (by construction) preserve (old minimal) supersymmetry at the quantum level. 

Once (supersymmetrically) renormalized, the quantum theory can again consistently couple to conformal supergravity, but some of the original symmetries are now broken by quantum anomalies. To determine the superconformal anomalies of this $\cn=1$ SCFT we need to relate the Ward identities for the conformal multiplet of the renormalized theory with those for the FZ multiplet of the regulated (and renormalized) theory. The FZ multiplet Ward identities for the regulated theory contain explicit symmetry breaking terms relative to those of the conformal multiplet. Since the theory is classically superconformal, these terms can at most give rise to contact terms as the regulator is removed. If they are zero the corresponding Ward identity is non-anomalous, but otherwise there is an anomaly. We find that the R-symmetry, S- and Q-supersymmetry Ward identities are all broken at the level of FZ multiplet 3-point functions. Of course, the old minimal combination of Q- and S-supersymmetry is by construction non-anomalous.

As mentioned earlier, one may shift the anomalies from one symmetry to another. We show that one can find finite local counterterms involving the old minimal supergravity fields that bring the anomalies to their superconformal form, including the standard chiral anomaly for R-symmetry. Moreover, they render the scalar operator of the FZ multiplet null, resulting in the conformal multiplet of currents. It should be stressed that the required counterterms necessarily depend on the auxiliary field $M$ of old minimal supergravity, since the WZ consistency conditions preclude the existence of a local counterterm involving only the conformal supergravity fields that cancels the Q-supersymmetry anomaly \cite{Papadimitriou:2019gel}. Once the local counterterms are added, both Q- and S-supersymmetry (and any field independent linear combination of them) are anomalous, with the anomaly for Q-superymmetry appearing first in the $\<\cq^\m(x)\bar\cq^\s(y)\cj^\k(z)\cj^\a(w)\>$ correlator. The results reproduce exactly the structure of anomalies obtained by solving the WZ consistency conditions for conformal supergravity \cite{Papadimitriou:2019gel}. 

{\bf Note added:} As this paper was finalised, we became aware of \cite{Bzowski:2020tue} which has some overlap with this work. We thank the authors of \cite{Bzowski:2020tue} for sharing their draft with us and for discussions.

\bigskip

This paper is organised as follows. In section \ref{sec:WardIDs} we review the Ward identities for the conformal multiplet of currents and the structure of the corresponding superconformal anomalies obtained by solving the WZ consistency conditions for conformal supergravity. Moreover, we explicitly derive the form of these Ward identities for flat space correlation functions of conformal multiplet currents.  
In section \ref{sec:WZmodel} we introduce the WZ model and discuss its coupling to conformal supergravity. In section \ref{sec:PV} we  present the PV regulators, the coupling of the regulated model to old minimal supergravity and the derivation of the regulated (Ferrara-Zumino multiplet) Ward identities. Section \ref{sec:correlators} contains the 1-loop computation and the anomalies, both for the Ferrara-Zumino multiplet and the renormalised  conformal current multiplet. We conclude in section \ref{discussion} with the discussion of our results. The paper contains a series of appendices, where we collect a number of technical results. In appendix \ref{sec:conventions} we present our conventions and tabulate useful spinor identities. Appendix \ref{sec:path-integralWIDs} presents a detailed discussion of the path integral Ward identities. This appendix includes a discussion of subtleties that is not available elsewhere (to our knowledge). Appendix \ref{sec:seagulls} discusses the relation between the two different types of brackets  in \eqref{func} and \eqref{PI}  for the correlators we compute in this paper. In appendix \ref{sec:divergences} we show that the PV regulator properly regulates all correlators entering the Ward identities and in appendix \ref{sec:local-correlators} we collect the results about symmetry breaking terms in the regulated Ward identities.

\section{Conformal multiplet Ward identities and anomalies}
\label{sec:WardIDs}

The superconformal Ward identities can be formulated independently of any specific SCFT in terms of the $a$ and $c$ anomaly coefficients, whose values depend on the specific theory. The current multiplet of $\cn=1$ superconformal theories consists of the stress tensor, $\ct^\m_a$, the R-current, $\cj^\m$, and the supercurrent, $\cq^\m$, which is a Majorana spinor in our conventions. These couple respectively to the vierbein, $e^a_\m$, the graviphoton, $A_\m$, and the gravitino, $\j_\m$, which comprise the field content of $\cn=1$ conformal supergravity \cite{Kaku:1977pa,Kaku:1977rk,Kaku:1978nz,Townsend:1979ki}, which we briefly review in section \ref{sec:confsugra}. The consistent (as opposed to the covariant \cite{Bardeen:1984pm}) current operators are defined accordingly as 
\be\label{currents}
\<\ct^\m_a\>_s\equiv e^{-1}\frac{\d\mathscr{W}}{\d e^a_\m},
\qquad \<\cj^\m\>_s\equiv e^{-1}\frac{\d\mathscr{W}}{\d A_\m},\qquad
\<\cq^\m\>_s\equiv e^{-1}\frac{\d\mathscr{W}}{\d\bar\j_\m},
\ee
where $e\equiv \det(e^a_\m)$, $\mathscr{W}[e,A,\j]$ is the generating function of renormalized connected current correlators, and the notation $\<\cdot\>_s$ denotes connected correlation functions in the presence of sources. In particular, further derivatives of these one-point functions result in higher-point functions. 

The current operators \eqref{currents} are defined independently of whether there exists a Lagrangian description of the theory. If a Lagrangian description exists, $\mathscr{W}[e,A,\j]$ is given by 
\be\label{connected-generator}
\mathscr{W}[e,A,\j]=-i\log \mathscr{Z}[e,A,\j],
\ee
where $\mathscr{Z}[e,A,\j]$ is obtained from the path integral
\be
\mathscr{Z}[e,A,\j]=\int[\tx d\{\F\}] e^{iS[\{\F\};e,A,\j]}\Big|_{\rm ren},
\ee
over the microscopic fields $\{\F\}$, after renormalization. In the model we will focus on in the subsequent sections, $\{\F\}$ will consist of a number of $\cn=1$ chiral multiplets, corresponding to a massless Wess-Zumino (WZ) model and a set of massive Pauli-Villars regulator fields. 

More accurately, the operators \eqref{currents} comprise the so called conformal current multiplet. This multiplet does not exist for massive theories such as the regulated WZ model, even if it is classically conformal. As we will discuss in more detail later on, in order to accommodate massive theories, the current multiplet must be suitably extended to include additional (auxiliary) operators. However, the conformal current multiplet \eqref{currents} exists for all SCFTs {\em after} renormalization and this is the multiplet we discuss in this section.

\subsection{Ward identities for 1-point functions with arbitrary sources}

The superconformal Ward identities and anomalies for arbitrary $\cn=1$ SCFTs in four dimensions were derived in \cite{Papadimitriou:2019gel}, using the local symmetries of $\cn=1$ conformal supergravity and the associated WZ consistency conditions. For theories with equal anomaly coefficients, i.e. $a=c$, these coincide with the superconformal Ward identities derived earlier holographically in \cite{Papadimitriou:2017kzw}. In terms of the currents \eqref{currents}, the superconformal Ward identities take the form
\bbxd
\bal
\label{WardIDs}
&e^a_\m\nabla_\n\<\ct^\n_a\>_s+\nabla_\n(\bar\j_\m  \<\cq^\n\>_s)-\bar\j_\n\overleftarrow D_\m \<\cq^\n\>_s-F_{\m\n}\<\cj^\n\>_s\NO\\
&\hspace{2.cm}+A_\m\big(\nabla_\n\<\cj^\n\>_s+i\bar\j_\n \g^5\<\cq^\n\>_s\big)-\o_\m{}^{ab}\Big(e_{\n [a}\<\ct^\n_{b]}\>_s+\frac14\bar\j_\n\g_{ab}\<\cq^\n\>_s\Big)=0,\NO\\
&e^a_\m\<\ct^\m_a\>_s+\frac12\bar\j_\m \<\cq^\m\>_s=\ca_W,\NO\\
&e_{\m[a} \<\ct^\m_{b]}\>_s+\frac14\bar\j_\m\g_{ab} \<\cq^\m\>_s=0,\NO\\
&\nabla_\m \<\cj^\m\>_s+i\bar\j_\m\g^5 \<\cq^\m\>_s=\ca_R,\NO\\
&D_\m \<\cq^\m\>_s-\frac12\g^a\j_\m \<\ct^\m_a\>_s-\frac{3i}{4}\g^5\f_\m\<\cj^\m\>_s=\ca_Q,\NO\\
&\g_\m \<\cq^\m\>_s-\frac{3i}{4}\g^5\j_\m\<\cj^\m\>_s=\ca_{S}. 
\eal
\ebxd
Here and in the following the spin connection is given by
\be\label{spin-connection}
\o_\m{}_{ab}(e,\j)\equiv\o_\m{}_{ab}(e)+\frac14\big(\bar\j_a\g_\m\j_b+\bar\j_\m\g_a\j_b-\bar\j_\m\g_b\j_a\big),
\ee
with $\o_\m{}_{ab}(e)$ denoting the unique torsion-free part. Moreover, $\f_\m$ is the gravitino fieldstrength
\be\label{phi}
\f_\m\equiv \frac13\g^\n\Big(D_\n\j_\m-D_\m\j_\n-\frac{i}{2}\g^5 \e_{\n\m}{}^{\r\s}D_\r\j_\s\Big)=-\frac16\big(4\d^{[\r}_\m\d^{\s]}_\n+i\g^5 \e_{\m\n}{}^{\r\s}\big)\g^\n D_\r\j_\s.
\ee
$\nabla_\m$ denotes the Levi-Civita connection, while $D_\m$ stands for the spinor covariant derivative, which acts on the gravitino and its fieldstrength as
\bal\label{covD-psi-phi}
D_\m\j_\n=&\;\Big(\pa_\m+\frac14\o_\m{}^{ab}(e,\j)\g_{ab}+i\g^5A_\m\Big)\j_\n-\G^\r_{\m\n}\j_\r\equiv \big(\mathscr{D}_\m+i\g^5A_\m\big)\j_\n,\NO\\
D_\m\f_\n=&\;\Big(\pa_\m+\frac14\o_\m{}^{ab}(e,\j)\g_{ab}-i\g^5A_\m\Big)\f_\n-\G^\r_{\m\n}\f_\r=\big(\mathscr{D}_\m-i\g^5A_\m\big)\f_\n.
\eal
Since the gravitino and the supercurrent have opposite R-charge, the covariant derivative acts on the supercurrent as
\be
D_\m\<\cq^\n\>=\Big(\pa_\m+\frac14\o_\m{}^{ab}(e,\j)\g_{ab}-i\g^5A_\m\Big)\<\cq^\n\>+\G^\n_{\m\r}\cq^\r\equiv\big(\mathscr{D}_\m-i\g^5A_\m\big)\<\cq^\n\>.
\ee

The superconformal anomalies on the r.h.s. of the Ward identities \eqref{WardIDs} are local functions of the background conformal supergravity fields and take the form
\bbxd
\bal\label{anomalies}
\hskip-.3cm\ca_W=&\frac{c}{16\p^2}\Big(W^2-\frac{8}{3}F^2\Big)-\frac{a}{16\p^2} E+\co(\j^2),\NO\\
\hskip-.3cm\ca_R=&\frac{(5a-3c)}{27\p^2}\;\wt F F+\frac{(c-a)}{24\p^2}\cp,\NO\\
\hskip-.3cm\ca_Q=&-\frac{(5a-3c)i}{9\p^2}\wt F^{\m\n}A_\m\g^5\f_\n+\frac{(a-c)}{6\p^2}\nabla_\m\big(A_\r \wt R^{\r\s\m\n} \big)\g_{(\n}\j_{\s)}-\frac{(a-c)}{24\p^2}F_{\m\n} \wt R^{\m\n\r\s} \g_\r\j_\s+\co(\j^3),\NO\\
\hskip-.3cm\ca_{S}=&\frac{(5a-3c)}{6\p^2}\wt F^{\m\n}\Big(D_\m-\frac{2i}{3}A_\m\g^5\Big)\j_{\n}+\frac{ic}{6\p^2} F^{\m\n}\big(\g_{\m}{}^{[\s}\d_{\n}^{\r]}-\d_{\m}^{[\s}\d_{\n}^{\r]}\big)\g^5D_\r\j_\s\NO\\
&\hskip-.3cm+\frac{3(2a-c)}{4\p^2}P_{\m\n}g^{\m[\n}\g^{\r\s]}D_\r\j_\s+\frac{(a-c)}{8\p^2}\Big(R^{\m\n\r\s}\g_{\m\n}-\frac12Rg_{\m\n}g^{\m[\n}\g^{\r\s]}\Big)D_\r\j_\s+\co(\j^3).\hskip.0cm
\eal
\ebxd
The antisymmetric tensor $F_{\m\n}=2\pa_{[\m} A_{\n]}$ denotes the fieldstrength of the graviphoton, while the dual fieldstrength, $\wt F_{\m\n}$, is given by 
\be\label{dualF}
\wt F_{\m\n}\equiv\frac12 \e_{\m\n}{}^{\r\s}F_{\r\s}.
\ee
These are the building blocks of the two independent quadratic curvature invariants 
\be\label{gauge-curvatures}
F^2\equiv F_{\m\n}F^{\m\n},\qquad F\wt F\equiv \frac12 \e^{\m\n\r\s}F_{\m\n}F_{\r\s}.
\ee
The geometric curvature invariants are built out of the Riemann tensor and its dual 
\be
\label{dualR}
\wt R_{\m\n\r\s}\equiv\frac12\e_{\m\n}{}^{\k\l}R_{\k\l\r\s},
\ee
(which is not symmetric under exchange of the first and second pair of indices) as well as  the Schouten tensor, $P_{\m\n}$, which in four dimensions is given by 
\be\label{Schouten}
P_{\m\n}\equiv\frac12\Big(R_{\m\n}-\frac16Rg_{\m\n}\Big).
\ee

The independent quadratic curvature invariants are the square of the Weyl tensor, $W^2$, the Euler density, $E$, and the Pontryagin density, $\cp$,  defined respectively as
\bal\label{metric-curvatures}
W^2\equiv&\; W_{\m\n\r\s}W^{\m\n\r\s}=R_{\m\n\r\s}R^{\m\n\r\s}-2R_{\m\n}R^{\m\n}+\frac13R^2,\NO\\
E=&\;R_{\m\n\r\s}R^{\m\n\r\s}-4R_{\m\n}R^{\m\n}+R^2,\NO\\
\cp\equiv&\;\frac12\e^{\k\l\m\n}R_{\k\l\r\s}R_{\m\n}{}^{\r\s}=\wt R^{\m\n\r\s}R_{\m\n\r\s}.
\eal

Finally, the anomaly coefficients $a$ and $c$ are normalized 
as in  \cite{Anselmi:1997am}, so that for $N_\c$ free chiral multiplets and $N_v$ free vector multiplets they take the form
\be
a=\frac{1}{48}(N_\c+9N_v),\qquad c=\frac{1}{24}(N_\c+3N_v).
\ee
In particular, for the WZ model we have 
\be\label{WZ-anomaly-coefficients}
c=2a=\frac{1}{24}.
\ee  

\subsection{Ward identities for flat space correlation functions}
\label{sec:CS-WIDs-flat}

Differentiating the superconformal Ward identities \eqref{WardIDs} with respect to the background conformal supergravity fields leads to relations among higher-point correlations functions. In particular, applying a sufficient number of derivatives with respect to suitable combinations of background fields on the superconformal anomalies produces contact (i.e. ultralocal) terms that survive once the supergravity fields are set to their Minkowski space value. The fact that the superconformal anomalies \eqref{anomalies} are non trivial solutions of the WZ consistency conditions ensures that there exist no local counterterms that can eliminate the corresponding contact terms in correlation functions. We are interested in the simplest flat space correlation functions of current operators that receive a contribution from  the Q-supersymmetry anomaly, $\ca_Q$. It is straightforward to see that the lowest order flat space current correlation functions where $\ca_Q$ contributes are $\<\cq\bar\cq\cj\cj\>$ and $\<\cq\bar\cq\ct\cj\>$. In this paper, we focus on the 4-point function $\<\cq\bar\cq\cj\cj\>$ and obtain a direct derivation of the Q-supersymmetry anomaly via an 1-loop computation in the WZ model. 

In order to compare the 1-loop calculation with the anomalous superconformal Ward identities \eqref{WardIDs}, we need to determine all constraints these imply for the 4-point function $\<\cq\bar\cq\cj\cj\>$, by applying successive derivatives with respect to the relevant background supergravity fields. We will also derive the flat space constraints for a number of 2- and 3-point functions that enter in the flat space Ward identities of the 4-point function $\<\cq\bar\cq\cj\cj\>$, since these are required in order to determine the Q-supersymmetry anomaly.    

As we discuss in appendix \ref{sec:seagulls}, beyond 1-point functions, correlation functions can be defined either via path integral operator insertions, in which case they will be denoted by $\lb \cdot\rb$, or through functional differentiation, in which case we will use the notation $\<\cdot\>$. Feynman diagrams compute the former, while the latter definition is more general and can be used even for non Lagrangian theories. This is why the discussion in this section is formulated entirely in terms of correlation functions defined by functional differentiation. The two definitions of correlation functions differ by so called `seagull terms', which encode the dependence of the Noether currents on the background fields when a Lagrangian theory is coupled to background supergravity. In particular, the seagull terms correspond to insertions involving derivatives of the Noether currents with respect to the background fields. Those needed for our analysis are determined in appendix \ref{sec:seagulls}.

It will also be useful to note that the functional derivative of correlation functions with respect to the gravitino is unambiguously defined through the identity
\be\label{fermion-variations}
\hskip-.3cm\d_\j\<\co(x)\cdots\>_s=\Big\<\int d^4y\, e\, \bar\cq^\m(y)\d\j_\m(y)\co(x)\cdots\Big\>_s=\Big\<\int d^4y\, e\, \d\bar\j_\m(y) \cq^\m(y)\co(x)\cdots\Big\>_s,\hskip-.3cm
\ee
which holds for any number of local insertions, both bosonic or fermionic. Notice that the second equality follows from the Majorana property of the gravitino and of the supercurrent (see appendix \ref{sec:conventions} for our spinor conventions). Finally, in order to keep the notation as compact as possible, we introduce the covariant Dirac delta function
\be
\d(x,y)\equiv e^{-1}\d^{(4)}(x-y),
\ee
as well as the bidirectional derivative
\be
\d(x,y)\stackrel{\leftrightarrow}{\pa_\m}\d(x,z)\equiv \d(x,y)\pa_\m \d(x,z)-\d(x,z)\pa_\m \d(x,y).
\ee
We will use the covariant delta function even in flat space, with the implicit understanding that $e^{-1}=1$ in that case. All derivatives are taken to be with respect to $x^\m$, unless otherwise indicated.

\subsection*{Q-supersymmetry}

Starting from the Q-supersymmetry Ward identity in \eqref{WardIDs}, applying a functional derivative with respect to $\j_\s(y)$ and subsequently setting the gravitino to zero gives  
\bal\label{Q-susy-WID-2pt-sources}
&D_\m \<\cq^\m(x)\bar\cq^\s(y)\>_s-\frac12 \<\ct^\s_a(x)\>_s\g^a\d(x,y)\NO\\
&+\frac{i}{8}\Big(\<\cj^\n(x)\>_s-\frac{4(5a-3c)}{27\p^2}\wt F^{\m\n}A_\m\Big)\big(4\d^{[\r}_\n\d^{\s]}_\l+i\g^5 \e_{\n\l}{}^{\r\s}\big)\g^5\g^\l (\mathscr D_\r+i\g^5A_\r)\d(x,y)\NO\\
&\;=\frac{(a-c)}{12\p^2}\nabla_\m\big(A_\r \wt R^{\r\s\m\n}+A_\r \wt R^{\r\n\m\s} \big)\g_{\n}\d(x,y)-\frac{(a-c)}{24\p^2}F_{\m\n} \wt R^{\m\n\r\s} \g_\r\d(x,y).
\eal
Notice that, although the gravitino has been set to zero, this identity holds for arbitrary bosonic sources. In particular, by further differentiation with respect to the graviphoton, $A_\k(z)$, we obtain   
\bal\label{Q-susy-WID-3pt-sources}
&D_\m \<\cq^\m(x)\bar\cq^\s(y)\cj^\k(z)\>_s-i\g^5\d(x,z)\<\cq^\k(x)\bar\cq^\s(y)\>_s-\frac12 \<\ct^\s_a(x)\cj^\k(z)\>_s\g^a\d(x,y)\NO\\
&+\frac{i}{8}\Big(\<\cj^\n(x)\cj^\k(z)\>_s-\frac{4(5a-3c)}{27\p^2}\big(\wt F^{\k\n}+\e^{\m\n\t\k}A_\m\nabla_\t \big)\d(x,z)\Big)\big(4\d^{[\r}_\n\d^{\s]}_\l+i\g^5 \e_{\n\l}{}^{\r\s}\big)\times\\
&\g^5\g^\l (\mathscr D_\r+i\g^5A_\r)\d(x,y)+\frac{1}{8}\Big(\<\cj^\n(x)\>_s-\frac{4(5a-3c)}{27\p^2}\wt F^{\m\n}A_\m\Big)\big(4\d^{[\k}_\n\d^{\s]}_\l+i\g^5 \e_{\n\l}{}^{\k\s}\big)\g^\l \d(x,z)\d(x,y)\NO\\
&\;=\frac{(a-c)}{24\p^2}\Big[\nabla_\m\big(\d(x,z)\e^{\k\s}{}_{\a\b}R^{\a\b\m\n}+\d(x,z) \e^{\k\n}{}_{\a\b}R^{\a\b\m\s}\big)+\nabla_\m\d(x,z) \e^{\k\m}{}_{\a\b}R^{\a\b\n\s}\Big]\g_\n\d(x,y),\NO
\eal
which again holds for an arbitrary bosonic supergravity background. 

The 2- and 3-point function constraints \eqref{Q-susy-WID-2pt-sources} and \eqref{Q-susy-WID-3pt-sources} allow us to determine the flat space Ward identities relevant for our analysis. Firstly, setting the bosonic sources in \eqref{Q-susy-WID-2pt-sources} to their Minkowski space values results in the 2-point function Ward identity
\bbxd
\be\label{Q-susy-WID-2pt}
\hskip-.4cm\pa_\m \<\cq^\m(x)\bar\cq^\s(y)\>-\frac12 \<\ct^\s_a(x)\>\g^a\d(x,y)+\frac{i}{8}\<\cj^\n(x)\>\big(4\d^{[\r}_\n\d^{\s]}_\l+i\g^5 \e_{\n\l}{}^{\r\s}\big)\g^5\g^\l \pa_\r\d(x,y)=0.
\ee
\ebxd
Similarly, setting the bosonic background fields in \eqref{Q-susy-WID-3pt-sources} to their Minkowski value gives
\bbxd
\bal
\label{Q-susy-WID-3pt}
&\pa_\m \<\cq^\m(x)\bar\cq^\s(y)\cj^\k(z)\>-i\g^5\d(x,z)\<\cq^\k(x)\bar\cq^\s(y)\>-\frac12\<\ct^\s_a(x)\cj^\k(z)\>\g^a\d(x,y)\NO\\
&+\frac{i}{8}\<\cj^\n(x)\cj^\k(z)\>\big(4\d^{[\r}_\n\d^{\s]}_\l+i\g^5 \e_{\n\l}{}^{\r\s}\big)\g^5\g^\l \pa_\r\d(x,y))\NO\\
&+\frac{1}{8}\<\cj^\n(x)\>\big(4\d^{[\k}_\n\d^{\s]}_\l+i\g^5 \e_{\n\l}{}^{\k\s}\big)\g^\l \d(x,z)\d(x,y)=0.
\eal
\ebxd
As anticipated, there is no trace of the anomalies in the flat space Ward identities \eqref{Q-susy-WID-2pt} and \eqref{Q-susy-WID-3pt}. However, differentiating \eqref{Q-susy-WID-3pt-sources} once more with respect to the graviphoton, $A_\a(w)$, and then setting all sources to their Minkowski space values results in the 4-point function Ward identity 
\bbxd
\bal
\label{Q-susy-WID-4pt}
&\pa_\m \<\cq^\m(x)\bar\cq^\s(y)\cj^\k(z)\cj^\a(w)\>-i\g^5\d(x,w)\<\cq^\a(x)\bar\cq^\s(y)\cj^\k(z)\>\NO\\
&-i\g^5\d(x,z)\<\cq^\k(x)\bar\cq^\s(y)\cj^\a(w)\>-\frac12 \<\ct^\s_a(x)\cj^\k(z)\cj^\a(w)\>\g^a\d(x,y)\NO\\
&+\frac{i}{8}\Big(\<\cj^\n(x)\cj^\k(z)\cj^\a(w)\>-\frac{4(5a-3c)}{27\p^2}\e^{\n\k\a\t}\d(x,z)\stackrel{\leftrightarrow}{\pa_\t}\d(x,w)\Big)\times\NO\\
&\hspace{.2cm}\big(4\d^{[\r}_\n\d^{\s]}_\l+i\g^5 \e_{\n\l}{}^{\r\s}\big)\g^5\g^\l \pa_\r\d(x,y)+\frac{1}{8}\big(4\d^{[\b}_\n\d^{\s]}_\l+i\g^5 \e_{\n\l}{}^{\b\s}\big)\g^\l\times\NO\\
&\hspace{.2cm}\Big(\<\cj^\n(x)\cj^\a(w)\>\d^\k_\b\d(x,z)+\<\cj^\n(x)\cj^\k(z)\>\d^\a_\b\d(x,w)\Big)\d(x,y)=0,
\eal
\ebxd
which contains a contact term proportional to the linear combination $5a-3c$ of the superconformal anomaly coefficients. As we emphasized earlier, this contact term cannot be removed by local counterterms without introducing an anomalous contribution to a different correlation function. Another crucial observation is that the coefficient of this contact term differs from the corresponding term in the 3-point function of consistent R-symmetry currents (see \eqref{R-symmetry-WID-axial} below), which descends from the axial anomaly.\footnote{In fact, the coefficient of the contact term in \eqref{Q-susy-WID-4pt} matches that arising in the 3-point function of {\em covariant} R-symmetry currents \cite{Papadimitriou:2017kzw}.} One of the main goals of the present paper is to reproduce the 4-point function Ward identity \eqref{Q-susy-WID-4pt} from a 1-loop calculation in the WZ model.

\subsection*{S-supersymmetry}

Another anomalous Ward identity for the 4-point function $\<\cq\bar\cq\cj\cj\>$ follows from  S-supersymmetry. Differentiating the S-supersymmetry Ward identity in \eqref{WardIDs} with respect to $\j_\s(y)$ we obtain  
\bal
\label{S-susy-WID-2pt-sources}
&\g_\m \<\cq^\m(x)\bar\cq^\s(y)\>_s-\frac{3i}{4}\Big(\<\cj^\s(x)\>_s-\frac{4(5a-3c)}{27\p^2}\wt F^{\m\s}A_\m\Big)\g^5\d(x,y)\NO\\
&=\Big[\frac{(5a-3c)}{6\p^2}\wt F^{\r\s}+\frac{ic}{6\p^2}F^{\m\n}\big(\g_{\m}{}^{[\s}\d_{\n}^{\r]}-\d_{\m}^{[\s}\d_{\n}^{\r]}\big)\g^5+\frac{3(2a-c)}{4\p^2}P_{\m\n}g^{\m[\n}\g^{\r\s]}\NO\\
&+\frac{(a-c)}{8\p^2}\Big(R^{\m\n\r\s}\g_{\m\n}-\frac12Rg_{\m\n}g^{\m[\n}\g^{\r\s]}\Big)\Big](\mathscr D_\r+i\g^5A_\r)\d(x,y).
\eal
As in \eqref{Q-susy-WID-2pt-sources} above, we have set the gravitino to zero in this expression, but have kept the bosonic background fields arbitrary. Further differentiation with respect to $A_\k(z)$ results in the identity
\bal
\label{S-susy-WID-3pt-sources}
&\g_\m \<\cq^\m(x)\bar\cq^\s(y)\cj^\k(z)\>_s-\frac{3i}{4}\Big(\<\cj^\s(x)\cj^\k(z)\>_s-\frac{4(5a-3c)}{27\p^2}\big(\wt F^{\k\s}+\e^{\m\s\t\k}A_\m\nabla_\t \big)\d(x,z)\Big)\g^5\d(x,y)\NO\\
&=\frac{1}{6\p^2}\Big((5a-3c)\e^{\r\s\t\k}+c\,i\g^5\big(\g^{\k}{}^{[\r}g^{\s]\t}-\g^{\t}{}^{[\r}g^{\s]\k}+2g^{\t[\r}g^{\s]\k}\big)\Big)\nabla_{\t}\d(x,z)(\mathscr D_\r+i\g^5A_\r)\d(x,y)\NO\\
&+\Big[\frac{(5a-3c)}{6\p^2}\wt F^{\k\s}+\frac{ic}{6\p^2}F^{\m\n}\big(\g_{\m}{}^{[\s}\d_{\n}^{\k]}-\d_{\m}^{[\s}\d_{\n}^{\k]}\big)\g^5+\frac{3(2a-c)}{4\p^2}P_{\m\n}g^{\m[\n}\g^{\k\s]}\NO\\
&+\frac{(a-c)}{8\p^2}\Big(R^{\m\n\k\s}\g_{\m\n}-\frac12Rg_{\m\n}g^{\m[\n}\g^{\k\s]}\Big)\Big]i\g^5\d(x,z)\d(x,y).
\eal

The flat space Ward identities we are interested in can be read off these 2- and 3-point function constraints. Setting the bosonic supergravity fields to their Minkowski values in \eqref{S-susy-WID-2pt-sources} we obtain
\bbxd
\vskip.4cm
\be
\label{S-susy-WID-2pt}
\g_\m \<\cq^\m(x)\bar\cq^\s(y)\>-\frac{3i}{4}\g^5\d(x,y)\<\cj^\s(x)\>=0,
\ee
\ebxd
while \eqref{S-susy-WID-3pt-sources} gives
\bbxd
\bal\label{S-susy-WID-3pt}
&\g_\m \<\cq^\m(x)\bar\cq^\s(y)\cj^\k(z)\>-\frac{3i}{4}\<\cj^\s(x)\cj^\k(z)\>\g^5\d(x,y)\NO\\
&=\frac{1}{6\p^2}\Big((5a-3c)\e^{\r\s\t\k}+c\,i\g^5\big(\g^{\k}{}^{[\r}\h^{\s]\t}-\g^{\t}{}^{[\r}\h^{\s]\k}+2\h^{\t[\r}\h^{\s]\k}\big)\Big)\pa_{\t}\d(x,z)\pa_\r\d(x,y).
\eal
\ebxd
Notice that, in contrast to the Q-supersymmetry anomaly, the S-supersymmetry anomaly does affect the flat space 3-point function $\<\cq\bar\cq\cj\>$. Finally, differentiating \eqref{S-susy-WID-3pt-sources} one more time with respect to $A_\a(w)$ leads to the Minkowski space identity
\bbxd
\bal\label{S-susy-WID-4pt}
&\hskip-.3cm\g_\m \<\cq^\m(x)\bar\cq^\s(y)\cj^\k(z)\cj^\a(w)\>\NO\\
&\hskip-.3cm-\frac{3i}{4}\Big(\<\cj^\s(x)\cj^\k(z)\cj^\a(w)\>-\frac{4(5a-3c)}{27\p^2}\e^{\s\k\a\t}\d(x,z)\stackrel{\leftrightarrow}{\pa_\t} \d(x,w)\Big)\g^5\d(x,y)\NO\\
&\hskip-.3cm=\frac{1}{6\p^2}\Big((5a-3c)i\g^5\e^{\a\s\t\k}-c\big(\g^{\k}{}^{[\a}\h^{\s]\t}-\g^{\t}{}^{[\a}\h^{\s]\k}+2\h^{\t[\a}\h^{\s]\k}\big)\Big)\pa_{\t}\d(x,z)\d(x,w)\d(x,y)\hskip-.3cm\NO\\
&\hskip-.3cm+\frac{1}{6\p^2}\Big((5a-3c)i\g^5\e^{\k\s\t\a}-c\big(\g^{\a}{}^{[\k}\h^{\s]\t}-\g^{\t}{}^{[\k}\h^{\s]\a}+2\h^{\t[\k}\h^{\s]\a}\big)\Big)\pa_\t \d(x,w)\d(x,z)\d(x,y).\hskip-.0cm
\eal
\ebxd

\subsection*{R-symmetry}

The last set of flat space Ward identities we need for our analysis follow from R-symmetry. Firstly, functional differentiation of the R-symmetry Ward identity in \eqref{WardIDs} with respect to $A_\n(y)$ gives the 2-point function identity  
\bal\label{R-symmetry-WID-2pt1-sources}
\nabla_\m \<\cj^\m(x)\cj^\n(y)\>_s+i\bar\j_\m\g^5 \<\cq^\m(x)\cj^\n(y)\>_s=\frac{2(5a-3c)}{27\p^2}\;\e^{\k\l\t\n}F_{\k\l}\nabla_\t \d(x,y),
\eal
where both the bosonic and fermionic background fields have been kept arbitrary. Differentiating this expression one more time with respect to $A_\r(z)$ and setting all sources to their flat space value produces the standard Ward identity for the 3-point function of consistent R-currents, namely  
\bbxd
\vskip.4cm
\be
\label{R-symmetry-WID-axial}
\pa_\m \Big(\<\cj^\m(x)\cj^\n(y)\cj^\r(z)\>+\frac{2(5a-3c)}{27\p^2}\;\e^{\m\n\r\t} \d(x,y)\stackrel{\leftrightarrow}\pa_\t \d(x,z)\Big)=0.
\ee
\ebxd

If instead we differentiate the R-symmetry Ward identity in \eqref{WardIDs} with respect to the gravitino, $\bar\j_\n(y)$, we obtain the 2-point function identity 
\bal\label{R-symmetry-WID-2pt2-sources}
\nabla_\m \<\cj^\m(x)\cq^\n(y)\>_s+\d(x,y)i\g^5 \<\cq^\n(x)\>_s+i\<\cq^\n(y)\bar\cq^\m(x)\>_s\g^5\j_\m =0,
\eal
which again holds in the presence of arbitrary sources. Notice that the R-symmetry anomaly does not contribute to this identity because it is independent of the gravitino. Further differentiation with respect to $\j_\r(z)$ leads to the flat space identity  
\bbxd
\vskip-.1cm
\be
\label{R-symmetry-WID-3pt}
\pa_\m \<\cj^\m(x)\cq^\n(y)\bar\cq^\r(z)\>+\d(x,y)i\g^5 \<\cq^\n(x)\bar\cq^\r(z)\>+i\<\cq^\n(y)\bar\cq^\r(x)\>\g^5\d(x,z) =0,
\ee
\ebxd
while differentiating \eqref{R-symmetry-WID-2pt2-sources} with respect to both $\j_\r(z)$ and $A_\s(w)$ gives in flat space  
\bbxd
\bal
\label{R-symmetry-WID-4pt}
&\pa_\m \<\cj^\m(x)\cq^\n(y)\bar\cq^\r(z)\cj^\s(w)\>\NO\\
&+\d(x,y)i\g^5 \<\cq^\n(x)\bar\cq^\r(z)\cj^\s(w)\>+i\<\cq^\n(y)\bar\cq^\r(x)\cj^\s(w)\>\g^5\d(x,z) =0.
\eal
\ebxd
The Ward identities \eqref{Q-susy-WID-4pt}, \eqref{S-susy-WID-4pt} and \eqref{R-symmetry-WID-4pt} constitute three constraints for the flat space 4-point function $\<\cq\bar\cq\cj\cj\>$. The main goal of the rest of this paper is to derive these Ward identities -- including the contact terms due to the superconformal anomalies -- from a 1-loop calculation in the WZ model.

\section{The free and massless Wess-Zumino model}
\label{sec:WZmodel}

In this section we consider the free and massless WZ model. We begin with a description of the flat space theory and its classical symmetries, before discussing its coupling to background conformal supergravity. It section \ref{sec:PV} we will identify a suitable Pauli-Villars regulator for the free and massless WZ model, which we use in section \ref{sec:correlators} and appendix \ref{sec:divergences} for regulating the 1-loop diagrams.  

An off-shell $\cn=1$ chiral multiplet consists of a complex scalar, $\f$, a Grassmann-valued Majorana spinor, $\c$, and an auxiliary complex scalar, $F$. The free and massless WZ model for a chiral multiplet in Minkowski space is described by the Lagrangian\footnote{The auxiliary field $F$ should not be confused with the fieldstrength of the background graviphoton in e.g. eq.~\eqref{gauge-curvatures}. The arbitrary normalization constant $\a$ has been introduced in order to facilitate comparison with different conventions in the literature. The propagators \eqref{WZ-propagators} are canonically normalized for $\a=1/\sqrt{2}$.} 
\be
\label{WZ-lagrangian-flat}
\hat{\mathscr{L}}_{\rm WZ}=-\frac{1}{2\a^2}\big(\pa_\m\f^*\pa^\m\f+\a^2\bar\c\slashed\pa\c-F^*F\big),
\ee
where a hat $\hskip2pt\hat{\cdot}\hskip2pt$ indicates quantities evaluated in a Minkowski background. It will be omitted later on when referring to the corresponding quantities in the presence of background supergravity fields. 

\subsection*{Propagators}

The momentum space propagators following from the Lagrangian \eqref{WZ-lagrangian-flat} are
\bal\label{WZ-contractions}
&\contraction{}{\f}{(p)}{\f}\nomathglue\f(p)\f^*(p')=\,\contraction{}{\f}{^*(p')}{\f}\nomathglue\f^*(p')\f(p)=(2\p)^{4}\d(p+p')P_\f(p),\NO\\
&\contraction{}{\c}{(p)}{\c}\nomathglue\c(p)\bar\c(p')=\,-\contraction{}{\bar\c}{(p')}{\c}\nomathglue{\bar\c}(p')\c(p)=(2\p)^{4}\d(p+p')P_\c(p),\NO\\
&\contraction{}{F}{(p)}{F}\nomathglue F(p)F^*(p')=\,\contraction{}{F}{^*(k')}{F}\nomathglue F^*(p')F(p)=(2\p)^{4}\d(p+p')P_F(p),
\eal
where
\be\label{WZ-propagators}
P_\f(p)=-\frac{2i\a^2}{p^{2}},\qquad P_\c(p)=-\frac{\slashed p}{p^2},\qquad P_F(p)=2i\a^2.
\ee

\subsection{Symmetries and the conformal multiplet of conserved currents}

The free and massless Wess-Zumino model is classically invariant under the superconformal group $SU(2,2|1)$ \cite{Fradkin:1985am, Park:1997bq,Osborn:1998qu,Park:1999pd}. An infinitesimal $SU(2,2|1)$ transformation can be parameterized as 
\be\label{superconformal-transformation}
\hat\d=a^\m P_\m+\ell_{\m\n}M^{\m\n}+b^\m K_\m+\l D+\th_0 R+\bar\ve_0 Q+\bar\h_0 S,
\ee
where $P_\m$, $M_{\m\n}$, $K_\m$, $D$, $R$, $Q$ and $S$ are respectively the generators of spacetime translations, Lorentz, special conformal, scaling, R-symmetry, Q- and S-supersymmetry transformations. The action of these generators on the chiral multiplet fields is given in table \ref{superconformal-transformations}. 
\bbxd
\begin{minipage}{6.2in}
\vskip.8cm
\begin{center}
\begin{tabular}{|c|l|}
\hline
\rule{.0cm}{.6cm} $P_\m$  & $\d_{a}\f=a^\m\pa_\m\f$,\;\; $\d_{a}\c_L=a^\m\pa_\m\c_L$,\;\; $\d_{a} F=a^\m\pa_\m F$  \\
\rule{.0cm}{.6cm} $K_\m$ &  $\d_b\f=b_\m\big((2x^\m x^\n-\h^{\m\n} x^2)\pa_\n\f+2 x^\m\f\big)$ \\
& $\d_b\c_L=b_\m\big((2x^\m x^\n-\h^{\m\n} x^2)\pa_\n\c_L+3 x^\m\c_L+x_{\n}\g^{\m\n}\c_L\big)$\\
& $\d_b F=b_\m\big((2x^\m x^\n-\h^{\m\n} x^2)\pa_\n F+4 x^\m F\big)$\\
\rule{.0cm}{.6cm} $M_{\m\n}$ & $\d_\ell\f=\ell_{\m\n} x^{[\m}\pa^{\n]}\f$,\;\; $\d_\ell\c_L=\ell_{\m\n} \big(x^{[\m}\pa^{\n]}\c_L+\frac14\g^{\m\n}\c_L\big)$,\;\; $\d_\ell F=\ell_{\m\n} x^{[\m}\pa^{\n]} F$\\
\rule{.0cm}{.6cm} $R$ & $\d_{\th_0}\f=iq_R\th_0\f$,\;\; $\d_{\th_0}\c_L=i(q_R+1)\th_0\c_L$,\;\; $\d_{\th_0} F=i(q_R+2)\th_0 F$\\ 
\rule{.0cm}{.6cm} $D$ & $\d_\l\f=\l (x^\m\pa_\m+1)\f$,\;\; $\d_\l\c_L=\l \big(x^\m\pa_\m+\frac32\big)\c_L$,\;\; $\d_\l F=\l (x^\m\pa_\m +2)F$ \vspace{-.2cm}\\&\\
\hline&\vspace{-.2cm}\\
$Q$ & $\d_{\ve_0}\f=\a\bar\ve_{0L}\c_L$,\;\; $\d_{\ve_0}\c_L=\frac{1}{2\a}(\slashed\pa\f\ve_{0R}+F\ve_{0L})$,\;\; $\d_{\ve_0} F=\a\bar\ve_{0R}\slashed\pa\c_L$ \\
\rule{.0cm}{.6cm} $S$ & $\d_{\h_0}\f=-\a x^\m\bar\h_{0R}\g_\m\c_L$,\;\; $\d_{\h_0}\c_L=\frac{1}{2\a}(x^\m\slashed\pa\f\g_\m\h_{0L}+x^\m F\g_\m \h_{0R}+2\f\h_{0L})$ \\
& $\d_{\h_0} F=-\a x^\m\bar\h_{0L}\g_\m \slashed\pa\c_L$ \vspace{-.2cm}\\&\\
\hline
\end{tabular} 
\captionof{table}{$SU(2,2|1)$ action on a chiral multiplet of R-charge $q_R$. Superconformal invariance requires $q_R=-\frac23$.} 
\label{superconformal-transformations}
\end{center}
\vskip0.5cm
\end{minipage}
\ebxd
The $SU(2,2|1)$ generators satisfy the algebra
\bbxd
\bal
\label{superconformal-algebra}
&[D,P_\m]=-P_\m,\qquad [D,K_\m]=K_\m,\qquad 
[P_\m,K_\n]=2(\h_{\m\n}D-2M_{\m\n}),\NO\\
&[M_{\m\n},P_\r]=\h_{\s[\m}\h_{\n]\r}P^\s,\qquad [M_{\m\n},K_\r]=\h_{\s[\m}\h_{\n]\r}K^\s,\NO\\
&[M_{\m\n},M_{\r\s}]=\h_{\l[\m}\h_{\n]\r}M^\l{}_\s-\h_{\l[\m}\h_{\n]\s}M^\l{}_\r,\NO\\
&\{Q^\a,\bar Q_\b\}=\frac12(\g^\m)^\a{}_\b P_\m,\qquad \{S^\a,\bar S_\b\}=-\frac12(\g^\m)^\a{}_\b K_\m,\NO\\
&\{Q^\a,\bar S_\b\}=\frac12\d^\a{}_\b D-\frac{1}{2}(\g^{\m\n})^\a{}_\b M_{\m\n}+\frac{3i}{4}(\g^5)^\a{}_\b R,\NO\\
&[P_\m,S]=-\g_\m Q,\qquad [K_\m,Q]=\g_\m S,\qquad [M_{\m\n},Q]=-\frac14\g^{\m\n}Q,\qquad [M_{\m\n},S]=-\frac14\g^{\m\n}S,\NO\\
&[D,Q]=-\frac12 Q,\qquad [D,S]=\frac12 S,\qquad [R,Q]=i\g^5 Q,\qquad [R,S]=-i\g^5S.
\eal
\ebxd

\subsection*{Noether currents and the conformal multiplet}

Noether's theorem for $SU(2,2|1)$ invariance results in only three independent current operators, corresponding to the conserved currents associated with translations, R-symmetry and Q-supersymmetry transformations. They comprise the conformal current multiplet of the massless WZ model and are given respectively by
\bal
\label{WZ-Noether-currents-Mink}
\hat\ct^\m{}_{\n}=&\;\frac{1}{\a^2}\pa^{(\m}\f^*\pa_{\n)}\f+\frac12\bar\c\g^\m \pa_\n\c-\frac{1}{8}\pa_\r\big(\bar\c\g_\n\g^{\r\m}\c+\bar\c\g^\m\g^{\r}{}_\n\c-\bar\c\g^\r\g^{\m}{}_\n\c\big)\NO\\
&-\frac{1}{6\a^2}\big(\pa^\m\pa_\n-\h^\m_\n\pa^2\big)(\f^*\f)-\frac{1}{2\a^2}\h^\m_\n\big(\pa_\r\f^*\pa^\r\f+\a^2\bar\c\slashed \pa\c-F^*F\big),\NO\\
\rule{0.cm}{.7cm}\hat\cj^\m=&\;\frac{i}{3\a^2}\Big(\f^*\pa^\m\f-\f \pa^\m\f^*+\frac{\a^2}{2}\bar\c\g^\m\g^5\c\Big),\NO\\
\rule{0.cm}{.7cm}\hat\cq^\m=&\;\frac{1}{2\a}(\slashed\pa\f \g^\m\c_R+\slashed\pa\f^*\g^\m\c_L)+\frac{1}{3\a}\g^{\m\n}\pa_\n(\f \c_R+\f^* \c_L).
\eal
Notice that, although the stress tensor can be further simplified using the equations of motion (see Table III in \cite{Katsianis:2019hhg}), only the off-shell form leads to the correct Ward identities.

The currents \eqref{WZ-Noether-currents-Mink} satisfy the on-shell conservation laws 
\be\label{WZ-conserved-currents}
\pa_\m\hat\ct^\m{}_\n=0,\qquad \pa_\m\hat\cj^\m=0,\qquad \pa_\m\hat\cq^\m=0.
\ee
The conserved currents for the remaining $SU(2,2|1)$ symmetries are obtained by contracting the stress tensor and the supercurrent with the corresponding Killing vector or spinor (see \eqref{Killing} below) and result in three algebraic constraints on the currents \eqref{WZ-Noether-currents-Mink}. In particular, Lorentz, scale and S-supersymmetry invariance require that 
\be\label{WZ-algebraic-constraints}
\hat\ct_{[\m\n]}=0,\qquad \hat\ct^\m_\m=0,\qquad \g_\m\hat\cq^\m=0.
\ee
No further condition arises from special conformal transformations, since these are equivalent to a translation, preceded and followed by a discrete inversion. The stress tensor and the supercurrent in \eqref{WZ-Noether-currents-Mink} include suitable improvement terms that do not affect the conservation equations \eqref{WZ-conserved-currents}, but ensure that the algebraic constraints \eqref{WZ-algebraic-constraints} hold on-shell \cite{Ferrara:1974pz}.

In appendix \ref{sec:path-integralWIDs} we provide a path integral derivation of the `naive' (i.e. classical) superconformal Ward identities using the Noether procedure and the conservation laws \eqref{WZ-conserved-currents} and \eqref{WZ-algebraic-constraints} for the massless WZ model. By construction, this procedure involves correlators defined via path integral operator insertions. Identifying the seagull terms from the symmetry transformations of the current operators, the resulting Ward identities can be cast in a more compact form in terms of correlators defined through functional differentiation. This rather tedious procedure reproduces the classical version of the superconformal Ward identities discussed in section \ref{sec:WardIDs}. However, the derivation in appendix B is provided only for completeness and is not needed for our analysis. The seagull terms that connect the Ward identities in section \ref{sec:WardIDs} to the Feynman diagram computation of correlation functions can be obtained more efficiently by coupling the theory to background supergravity, which we discuss next.

\subsection{Coupling to background conformal supergravity}
\label{sec:confsugra}

Coupling a supersymmetric field theory to off-shell background supergravity allows for a simpler and universal description of the global symmetries and their physical consequences, without reference to a specific model. It also facilitates powerful computational techniques, such as supersymmetric localization \cite{Pestun:2016zxk}. Coupling a theory to background supergravity amounts to gauging the global symmetries by turning on appropriate gauge fields -- the supergravity fields -- and suppressing their kinetic terms \cite{Festuccia:2011ws}. Crucially, the local symmetry transformations of off-shell supergravity are independent of any matter multiplets present and the algebra closes without invoking the equations of motion. Clearly, without this property, eliminating the kinetic terms of the supergravity fields would not be consistent. As we saw in section \ref{sec:WardIDs}, the universality of the local symmetry transformations of off-shell supergravity is also what enables the general derivation of the Ward identities and their quantum anomalies, by solving the WZ consistency conditions. 

Classically superconformal theories, such as the massless WZ model, can also be coupled to conformal supergravity, which facilitates an alternative, more efficient formulation of the Noether procedure. For example, the conserved currents that couple to background supergravity are those satisfying the algebraic constraints \eqref{WZ-algebraic-constraints}, and so this formulation of the Noether procedure leads directly to the improved currents. However, massive theories cannot be consistently coupled to conformal supergravity and so it is not possible to quantize a classically superconformal theory on a conformal supergravity background while preserving superconformal symmetry, since any regulator necessarily introduces a mass scale. A suitable background supergravity for massive theories is old minimal supergravity \cite{Ferrara:1978em}, which we will discuss in section \ref{sec:PV}. In the remaining of this section, we review the coupling of the massless WZ model to background conformal supergravity, focusing on the symmetries of the classical theory.    

\subsection*{Symmetries of conformal supergravity}

As we reviewed in section \ref{sec:WardIDs}, the field content of $\cn=1$ conformal supergravity \cite{Kaku:1977pa,Kaku:1977rk,Kaku:1978nz,Townsend:1979ki} consists of 
the vierbein, $e^a_\m$, the graviphoton, $A_\m$, and the gravitino, $\j_\m$. Its local symmetries are diffeomorphisms with infinitesimal parameter $\x^\m(x)$, Weyl transformations $\s(x)$, local Lorentz transformations $\l^{ab}(x)$, axial $U(1)$ gauge transformations $\th(x)$, as well as Q- and S-supersymmetry, parameterized respectively by the local spinors $\ve(x)$ and $\h(x)$. These local symmetries act on the fields of conformal supergravity as   
\bal\label{sugra-trans}
\d e^a_\m=&\;\x^\l\pa_\l e^a_\m+e^a_\l\pa_\m\x^\l-\l^a{}_b e^b_\m+\s e^a_\m-\frac12\lbar\j_\m\g^a\ve,\NO\\
\d\j_\m=&\;\x^\l\pa_\l\j_\m+\j_\l\pa_\m\x^\l-\frac14\l_{ab}\g^{ab}\j_\m+\frac12\s\j_\m+D_\m\ve-\g_\m\h- i\g^5\th\j_\m,\NO\\
\d A_\m=&\;\x^\l\pa_\l A_\m+A_\l\pa_\m\x^\l+\frac{3i}{4}\lbar\f_\m\g^5\ve-\frac{3i}{4}\lbar\j_\m\g^5\h+\pa_\m\th,
\eal
where the covariant derivatives of the spinor parameters, $\ve$ and $\h$, are given by (cf. \eqref{covD-psi-phi})
\bal\label{covD-parameters}
D_\m\ve\equiv&\;\Big(\pa_\m+\frac14\o_\m{}^{ab}(e,\j)\g_{ab}+i\g^5A_\m\Big)\ve\equiv \big(\mathscr{D}_\m+i\g^5A_\m\big)\ve,\NO\\
D_\m\h\equiv&\;\Big(\pa_\m+\frac14\o_\m{}^{ab}(e,\j)\g_{ab}-i\g^5A_\m\Big)\h\equiv \big(\mathscr{D}_\m-i\g^5A_\m\big)\h,
\eal
and the spin connection, $\o_\m{}^{ab}(e,\j)$, was given in \eqref{spin-connection}. These transformations close off-shell and form a so called `soft algebra' or `algebroid', which refers to an algebra with field dependent structure constants. In our conventions, this algebra can be found in \cite{Papadimitriou:2019gel}. 

Before discussing the coupling of the massless WZ model to conformal supergravity, let us briefly recall how the superconformal algebra $SU(2,2|1)$ arises as the Killing symmetry of Minkowski space from the local symmetries of $\cn=1$ conformal supergravity. The Killing symmetries of a conformal supergravity background $(\hat e^a_\m,\hat A_\m, \hat\j_\m)$ correspond to the subset of local symmetry transformations that preserve the background. In the case of Minkowski space, the local transformations \eqref{sugra-trans} determine that the Killing conditions are
\be\label{Killing-conditions}
\hat e^a_\l\pa_\m\hat\x^\l-\hat\l^a{}_b \hat e^b_\m+\hat\s \hat e^a_\m=0,\qquad
\pa_\m\hat\ve-\g_\m\hat\h=0,\qquad
\pa_\m\hat\th=0,
\ee
where $\hat e^a_\l=\h^a_\l$ is the vierbein of Minkowski space. The trace and antisymmetric parts of the first equation determine the parameters $\hat\s$ and $\hat\l_{ab}$ in terms of $\hat\x^\m$ through the relations
\be
\hat\s=-\frac14\pa_\m\hat\x^\m,\qquad \hat\l_{ab}=-\hat e^\m_a \hat e^\n_b\pa_{[\m}\hat\x_{\n]},
\ee
which lead to the conformal Killing vector equation
\be
\pa_\m\hat\x^\n-\pa_{[\m}\hat\x^{\n]}-\frac14\pa_\r\hat\x^\r \d^\n_\m=0.
\ee
The general solution of this system of equations takes the form 
\bal\label{Killing}
&\hat\x^\m=a^\m-\ell^\m{}_\n x^\n+\l x^\m+(2\d^\m_\n b_\r-\h_{\n\r}b^\m)x^\n x^\r,\qquad \hat\s=-\l-2b_\r x^\r,\NO\\ 
&\hat\l_{ab}=-\hat e^\m_a \hat e^\n_b(\ell_{\m\n}+4b_{[\m}x_{\n]}),\qquad \hat\th=\th_0,\qquad \hat\ve=\ve_0+\g_\m x^\m\h_0,\qquad \hat\h=\h_0,
\eal
where $a^\m$, $\ell_{\m\n}$, $\l$, $b^\m$, $\th_0$, $\ve_0$ and $\h_0$ are constants parameterizing the global $SU(2,2|1)$ symmetry transformations in \eqref{superconformal-transformation}.

\subsection*{Wess-Zumino model coupled to conformal supergravity}

Up to quadratic terms in the gravitino, the coupling of the massless WZ model to conformal supergravity takes the form \cite{Kaku:1978ea,Stelle:1978yr,Fradkin:1985am} 
\bbxd
\bal
\label{WZ-action-CS}
e^{-1}\mathscr{L}_{\rm WZ}=&\;-\frac{1}{2\a^2}D_\m\f^*D^\m\f-\frac12\bar\c\slashed D\c+\frac{1}{2\a^2}F^*F-\frac{1}{12\a^2}\f^*\f\; R[\o(e)]\NO\\
&\hspace{-1.2cm}+\frac{1}{2\a}\bar\j_\m(\slashed\pa\f \g^\m\c_R+\slashed\pa\f^*\g^\m\c_L)-\frac{1}{3\a}(\f \bar\c_R+\f^* \bar\c_L)\g^{\m\n}\mathscr{D}_\m\j_\n+\frac{i}{3\a}A_\m\bar\j^\m(\f \c_R-\f^*\c_L)\NO\\
&\hspace{-1.2cm}-\frac{1}{12\a^2}\pa^\m(\f\f^*)\bar\j_\n\g^\n\j_\m-\frac{i}{12\a^2}\f^*\f\; \e^{\m\n\r\s}\bar\j_\m\g^5\g_\n\mathscr{D}_\r\j_\s\\
&\hspace{-1.2cm}+\frac{i}{16\a^2}\e^{\m\n\r\s}\bar\j_\m\g_\n\j_\r\Big(\f^*\pa_\s\f-\f\pa_\s\f^*+\frac{\a^2}{2}\bar\c\g_\s\g^5\c\Big)-\frac{1}{16}(\bar\c\g^5\g^\n\c)(\bar\j_\m\g^5\g_\n\j^\m)+\co(\j^3).\NO
\eal
\ebxd
where the covariant derivatives act on the chiral multiplet fields as
\be
D_\m\f=\Big(\pa_\m+\frac{2i}{3}A_\m\Big)\f,\qquad 
D_\m\c=\Big(\pa_\m+\frac14\o_\m{}^{ab}(e,\j)\g_{ab}-\frac{i}{3}\g^5A_\m\Big)\c.
\ee
Under the local symmetries of $\cn=1$ conformal supergravity, the WZ fields transform as
\bal
\label{WZ-susy-transformations}
\d\f=&\;\x^\m\pa_\m\f-\s\f-\frac{2i}{3}\th\f+\a\bar\ve_L\c_L,\\
\d\c_L=&\;\x^\m\pa_\m\c_L-\frac32\s\c_L+\frac{i}{3}\th\c_L-\frac14\l_{ab}\g^{ab}\c_L+\frac{1}{2\a}\Big(\g^\m\big(D_\m\f-\a\bar\j_{\m L}\c_L\big)\ve_R+F\ve_L+2\f\h_L\Big),\NO\\
\d F=&\;\x^\m\pa_\m F-2\s F+\frac{4i}{3}\th F+\frac12\bar\ve_R\g^\m\Big(2\a D_\m\c_L-\g^\n\big(D_\n\f-\a\bar\j_{\n L}\c_L\big)\j_{\m R}-F\j_{\m L}-2\f\,\f_{\m L}\Big),\NO
\eal
where the gravitino fieldstrength, $\f_\m$ (not to be confused with the lowest component of the chiral multiplet, $\f$), in the transformation of $F$ was defined in \eqref{phi}. Together with the transformations of the supergravity fields in \eqref{sugra-trans}, these leave the Lagrangian \eqref{WZ-action-CS} invariant, up to a total derivative term. Notice that evaluating the transformations \eqref{WZ-susy-transformations} on Minkowski space and replacing the local parameters with their Killing form in \eqref{Killing} leads to the global symmetry transformations in table \ref{superconformal-transformations}. Moreover, the flat space limit of the S-supersymmetry transformation in \eqref{WZ-susy-transformations} coincides with $\wt\d_{\h_0}$ defined in \eqref{shifted-S}.

The variation of the WZ Lagrangian \eqref{WZ-action-CS} with respect to the background supergravity fields determines the corresponding current operators, as discussed in section \ref{sec:WardIDs}. This gives
\bbxd
\bal
\label{WZ-currents-CS}
\ct^\m_a=&\;\frac{1}{\a^2}D^{(\m}\f^*D_{a)}\f+\frac12\bar\c\g^\m D_a\c-\frac{1}{8}\nabla_\r\big(\bar\c\g_a\g^{\r\m}\c+\bar\c\g^\m\g^{\r}{}_a\c-\bar\c\g^\r\g^{\m}{}_a\c\big)\NO\\
&\hspace{-1.cm}+\frac{1}{6\a^2}e^\n_a\big(R^\m_\n-\nabla^\m\nabla_\n+\d^\m_\n\square\big)(\f\f^*)-\frac{1}{2\a^2}e^\m_a\Big(D_\n\f^*D^\n\f+\a^2\bar\c\slashed D\c-FF^*+\frac{1}{6}\f\f^*R\Big)+\co(\j),\hspace{-.2cm}\NO\\
\rule{0.cm}{.9cm}\cj^\m=&\;\frac{i}{3\a^2}\Big(\f^*D^\m\f-\f D^\m\f^*+\frac{\a^2}{2}\bar\c\g^\m\g^5\c+\a\bar\j^\m(\f \c_R-\f^* \c_L)\Big),\NO\\
\rule{0.cm}{.9cm}\cq^\m=&\;\frac{1}{2\a}(\slashed\pa\f \g^\m\c_R+\slashed\pa\f^*\g^\m\c_L)+\frac{1}{3\a}\g^{\m\n}\mathscr{D}_\n(\f \c_R+\f^* \c_L)+\frac{i}{3\a}A^\m(\f \c_R-\f^* \c_L)\NO\\
&-\frac{1}{6\a^2}\g^{[\m}\j_\n\pa^{\n]}(\f\f^*)-\frac{i}{12\a^2}\e^{\m\n\r\s}\big(2\f^*\f\;\g^5\g_\n\mathscr{D}_\r\j_\s+\pa_\r(\f^*\f)\g^5\g_\n\j_\s\big)\NO\\
&+\frac{i}{8\a^2}\e^{\m\n\r\s}\g_\n\j_\r\Big(\f^*\pa_\s\f-\f\pa_\s\f^*+\frac{\a^2}{2}\bar\c\g_\s\g^5\c\Big)-\frac{1}{8}\g^5\g_\n\j^\m(\bar\c\g^5\g^\n\c)+\co(\j^2).
\eal
\ebxd
Notice that the expression for the R-current is exact to all orders in the gravitino, since the qubic and quartic terms in the gravitino in the WZ action do not involve the gauge field, $A_\m$, (see e.g. eqs.~(2.5) and (2.6) in \cite{Kaku:1978ea}). 

Using the equations of motion following from the WZ action \eqref{WZ-action-CS}, it can be shown that the currents \eqref{WZ-currents-CS} satisfy the classical version of the Ward identities \eqref{WardIDs}, namely
\bbxd
\bal
&e^a_\m\nabla_\n\ct^\n_a+\nabla_\n(\bar\j_\m  \cq^\n)-\bar\j_\n\overleftarrow D_\m \cq^\n-F_{\m\n}\cj^\n=0,\NO\\
&\nabla_\m \cj^\m+i\bar\j_\m\g^5 \cq^\m=0,\qquad
D_\m\cq^\m-\frac12\g^a\j_\m \ct^\m_a-\frac{3i}{4}\g^5\f_\m\cj^\m=0,\NO\\
&e^a_\m\ct^\m_a+\frac12\bar\j_\m \cq^\m=0,\qquad
e_{\m[a} \ct^\m_{b]}+\frac14\bar\j_\m\g_{ab} \cq^\m=0,\qquad
\g_\m \cq^\m-\frac{3i}{4}\g^5\j_\m\cj^\m=0. 
\eal
\ebxd
These generalize the flat space conservation equations \eqref{WZ-conserved-currents} and algebraic constraints \eqref{WZ-algebraic-constraints} to a general supergravity background.

The flat space limit of the currents \eqref{WZ-currents-CS} coincides with the improved Noether currents \eqref{WZ-Noether-currents-Mink}. However, they also contain linear and higher order couplings to the background fields. In appendix \ref{sec:path-integralWIDs} we demonstrate that these terms ensure that the symmetry transformations of the current operators involve only currents and are independent of the specific microscopic model. This is not the case for the Noether currents \eqref{WZ-Noether-currents-Mink}, whose transformations are model dependent. The difference between the current operators \eqref{WZ-currents-CS} defined through functional differentiation and the model dependent Noether currents \eqref{WZ-Noether-currents-Mink} gives rise to the seagull terms discussed in appendix \ref{sec:seagulls}. Since Feynman diagram computations involve the Noether currents \eqref{WZ-Noether-currents-Mink}, in order to obtain the Ward identities in the model independent form discussed in section \ref{sec:WardIDs} from a 1-loop computation it is necessary to include suitable seagull terms to replace the Noether currents with the corresponding operators obtained through functional differentiation.    

A 1-loop computation also requires regulating and renormalizing all Feynman diagrams that enter in the evaluation of current multiplet correlation functions. Since any regulator breaks scale invariance, the conformal multiplet of current operators cannot be defined for the regulated theory. The current multiplet that the regulated theory admits and the background supergravity it can be coupled to is the subject of the next section.

\section{Pauli-Villars regularization}
\label{sec:PV}

The 1-loop diagrams that determine the correlation functions of conserved currents in the free and massless WZ model \eqref{WZ-lagrangian-flat} suffer from   
UV divergences that must be regulated and renormalized. In this section we present a supersymmetric Pauli-Villars (PV) regulator that suffices for removing the 1-loop UV divergences from all correlation functions that appear in the Ward identities we examine in section \ref{sec:correlators}. 

Consistency of any PV regulator requires that its contributions to the 1-loop diagrams follow from a local Lagrangian. A supersymmetric PV regulator further demands that this Lagrangian preserves supersymmetry and hence must involve a number of $\cn=1$ multiplets. The PV regulator we use consists of three massive chiral multiplets, one with standard statistics and two with `wrong' statistics. The corresponding PV Lagrangian is a standard massive WZ model, except that terms involving the multiplets with wrong statistics are appropriately modified. 

To help make the discussion of the PV regulator more transparent, it is instructive to discuss in parallel the standard massive WZ Lagrangian  
\be
\label{WZ-lagrangian-flat-massive}
\hat{\mathscr{L}}_{\rm WZ}=-\frac{1}{2\a^2}\big(\pa_\m\f^*\pa^\m\f+\a^2\bar\c\slashed\pa\c-F^*F\big)-\frac12 m \bar\c\c+\frac{m}{2\a^2}(\f F+\f^* F^*),
\ee
as a reference. Despite using the same notation as for the massless WZ model in \eqref{WZ-lagrangian-flat}, we emphasize that \eqref{WZ-lagrangian-flat-massive} is only meant as a generic reference -- it is neither the physical model we are interested in nor the PV Lagrangian we use. We simply use it to elucidate the structure of the PV regulator and of the symmetries it preserves.   

Integrating out the auxiliary field, $F$, using its equation of motion, $F=-m\f^*$, the massive WZ Lagrangian \eqref{WZ-lagrangian-flat-massive} becomes 
\be
\label{WZ-lagrangian-flat-massive-integrated}
-\hat{\mathscr{L}}_{\rm WZ}=\frac{1}{2\a^2}\pa_\m\f^*\pa^\m\f+\frac{m^2}{2\a^2}\f \f^*+\frac12\bar\c\slashed\pa\c+\frac{m}{2} \bar\c\c.
\ee
The PV Lagrangian we consider takes the closely related form
\bbxd
\bal\label{PV-lagrangian-Dirac}
-\hat{\mathscr{L}}_{\rm PV}=&\;\frac{1}{2\a^2}\pa_\m\vf_2^*\pa^\m\vf_2+\frac{m_2^2}{2\a^2}\vf^*_2\vf_2+\frac{1}{2}\bar{\l}_2\slashed\pa \l_2+\frac{m_2}{2}\bar{\l}_2\l_2\\
&+\frac{1}{2\a^2}\pa_\m\vf_1^*\pa^\m\vf_1+\frac{m_1^2}{2\a^2}\vf^*_1\vf_1+\frac{1}{2\a^2}\pa_\m\vth_1\pa^\m\vth_1^*+\frac{m_1^2}{2\a^2}{\vth}_1\vth_1^*+\dbar{\l}_1\slashed\pa \l_1+m_1\dbar{\l}_1\l_1,\NO
\eal
\ebxd
where $(\vf_2,\l_2)$ is a standard massive WZ multiplet, consisting of a commuting complex scalar, $\vf_2$, and an anticommuting Majorana spinor, $\l_2$, while $\vf_1$, $\vth_1$ are anticommuting complex scalars and $\l_1$ is a commuting Dirac spinor. Here, we should emphasize the distinction between the Dirac (e.g. $\dbar\l_1$) and Majorana (e.g. $\bar\l_2$) conjugates of a spinor, both of which are discussed in appendix \ref{sec:conventions}.

In fact, the fields $(\vf_1,\vth_1,\l_1)$ form two chiral multiplets with `wrong' statistics. This can be made manifest by means of the field redefinition 
\be\label{PV-redefinition}
\l_+\equiv\frac12(\l_1+\l_1^C),\quad \l_-\equiv\frac{1}{2i}(\l_1-\l_1^C),\quad \vf_+\equiv\frac{\vf_1+\vth_1}{2},\quad \vf_-\equiv\frac{\vf_1-\vth_1}{2i},
\ee
where the Majorana conjugate, $\l_1^C$, is defined in \eqref{Majorana-conjugate}, $\vf_\pm$ are anticommuting complex scalars and $\l_\pm$ are commuting Majorana spinors. The two chiral multiplets correspond to $(\vf_+,\l_+)$ and $(\vf_-,\l_-)$. 
The advantage of this parameterization is that it greatly simplifies the discussion of the symmetries preserved by the PV regulator. However, once expressed in terms of the fields \eqref{PV-redefinition}, the PV Lagrangian \eqref{PV-lagrangian-Dirac} contains non diagonal terms for the fields of wrong statistics, namely
\bbxd
\bal\label{PV-lagrangian-Majorana}
-\hat{\mathscr{L}}_{\rm PV}=&\;\frac{1}{2\a^2}\pa_\m\vf_2^*\pa^\m\vf_2+\frac{m^2_2}{2\a^2}\vf^*_2\vf_2+\frac{1}{2}\bar{\l}_2\slashed\pa \l_2+\frac{m_2}{2}\bar{\l}_2\l_2\NO\\
&+\frac{i}{\a^2}(\pa_\m\vf_+^*\pa^\m\vf_--\pa_\m\vf_-^*\pa^\m\vf_+)+\frac{im_1^2}{\a^2}(\vf^*_+\vf_--\vf_-^*\vf_+)\NO\\
&+i(\bar\l_+\slashed\pa\l_--\bar\l_-\slashed\pa\l_+)+im_1(\bar\l_+\l_--\bar\l_-\l_+).
\eal
\ebxd
Both forms \eqref{PV-lagrangian-Dirac} and \eqref{PV-lagrangian-Majorana} of the PV Lagrangian will be used in the following to analyze different aspects of the regulator.  

Notice that the PV Lagrangian contains two independently supersymmetric parts, namely the standard WZ action for the massive chiral multiplet with canonical statistics, and the remaining terms for the two massive chiral multiplets of wrong statistics. The non diagonal terms in \eqref{PV-lagrangian-Majorana} imply that the latter do not preserve supersymmetry independently -- it is not possible to write down a supersymmetric Lagrangian for a single massive chiral multiplet of wrong statistics. Supersymmetry does not impose any relation between the mass of the standard WZ multiplet and that of the two chiral multiplets with wrong statistics. However, in appendix \ref{sec:divergences} we show that cancellation of the UV divergences requires that $m_2^2=2m_1^2$. After adding suitable counterterms to renormalize the 1-loop diagrams, this mass parameter will be sent to infinity.

\subsection*{Propagators} 

It is most convenient to express the propagators of the PV fields in diagonal form using the parameterization \eqref{PV-lagrangian-Dirac}. Paying attention to the statistics of the various fields, they take the form 
\bal\label{PV-contractions}
&\contraction{}{\vf_2}{(p)}{\vf}\vf_2(p){}\vf_2^*(p')=\,\contraction{}{\vf_2}{^*(p')}{{}}\vf_2^*(p'){}\vf_2(p)=(2\p)^{4}\d(p+p')P_{\vf_2}(p),\NO\\
&\contraction{}{\l}{_2(p)}{\l}\nomathglue\l_2(p)\bar\l_2(p')=\,-\contraction{}{\bar\l}{_2(p')}{\l}\nomathglue{\bar\l_2}(p')\l_2(p)=(2\p)^{4}\d(p+p')P_{\l_2}(p),\NO\\
&\rule{0cm}{.9cm}\contraction{}{\vf_1}{(p)}{\vf}\vf_1(p){}\vf_1^*(p')=\,-\contraction{}{\vf_1}{^*(p')}{{}}\vf_1^*(p'){}\vf_1(p)=(2\p)^{4}\d(p+p')P_{\vf_1}(p),\NO\\
&\contraction{}{\vth^*_1}{(p)}{\vth}\vth^*_1(p){}\vth_1(p')=\,-\contraction{}{\vth_1}{(p')}{{}}\vth_1(p'){}\vth^*_1(p)=(2\p)^{4}\d(p+p')P_{\vth_1}(p),\NO\\
&\contraction{}{\l}{_1(p)}{\l}\nomathglue\l_1(p)\dbar\l_1(p')=\,\contraction{}{\dbar\l}{_1(p')}{\l}\nomathglue{\dbar\l_1}(p')\l_1(p)=(2\p)^{4}\d(p+p')P_{\l_1}(p),
\eal
where
\bal\label{PV-propagators}
&P_{\vf_2}(p)=-\frac{2i\a^2}{p^2+m^2_2},\qquad P_{\l_2}(p)=-i\frac{(-i\slashed p+m_2)}{p^2+m_2^2},\NO\\
&P_{\vf_1}(p)=-\frac{2i\a^2}{p^2+m_1^2},\qquad P_{\vth_1}(p)=-\frac{2i\a^2}{p^2+m_1^2},\qquad P_{\l_1}(p)=-i\frac{(-i\slashed p+m_1)}{p^2+m_1^2}.
\eal

\subsection{Symmetries and the Ferrara-Zumino current multiplet}

The PV Lagrangian \eqref{PV-lagrangian-Dirac} or \eqref{PV-lagrangian-Majorana} is invariant only under a subset of the superconformal symmetries in table \ref{superconformal-transformations}. Like the standard massive WZ model \eqref{WZ-lagrangian-flat-massive}, it is invariant only under the Poincar\'e symmetries and Q-supersymmetry. After integrating out the auxiliary fields in the chiral multiplets, Q-supersymmetry acts on the PV fields as   
\bal\label{PV-susy-transformations}
&\d_{\ve_0}\vf_2=\a\bar\ve_{0L}\l_2,\qquad \d_{\ve_0}\vf^{*}_2=\a\bar{\ve}_{0R}\l_2,\NO\\
&\d_{\ve_0}\l_2=\frac{1}{2\a}\big(\slashed{\pa}\vf_2\ve_{0R}+\slashed{\pa}\vf^{*}_2\ve_{0L}-m_2\vf_2\ve_{0R}-m_2\vf^{*}_2\ve_{0L}\big),\NO\\&
\d_{\ve_0}\bar{\l}_2=-\frac{1}{2\a}\big(\bar\ve_{0L}\slashed{\pa}\vf^{*}_2+\bar\ve_{0R}\slashed{\pa}\vf_2+m_2\bar\ve_{0R} \vf_2+m_2\vf^{*}_2\bar\ve_{0L}\big),\NO\\
\rule{.0cm}{.7cm}&\d_{\ve_0}\vf_\pm=\a\bar\ve_{0L}\l_\pm,\qquad \d_{\ve_0}\vf_\pm^*=-\a\bar\ve_{0R}\l_\pm,\NO\\
&\d_{\ve_0}\l_\pm=\frac{1}{2\a}\Big(\slashed{\pa}\vf_\pm^*\ve_{0L}-\slashed{\pa}\vf_\pm\ve_{0R}-m_1\vf_\pm^*\ve_{0L}+m_1\vf_\pm\ve_{0R}\Big),
\eal
where any sign differences in the transformations of $(\vf_\pm,\l_\pm)$ relative to the standard transformations of $(\vf_2,\l_2)$ are due to the different statistics.   

Using the field redefinition \eqref{PV-redefinition}, we find that the fields $(\vf_1,\vth_1,\l_1)$ transform as
\bal\label{PV-susy-transformations-Dirac}
&\d_{\ve_0}\vth_1^*=-\a\dbar{\ve}_{0L}\l_1=-\a\bar\ve_{0R}\l_1,\qquad \d_{\ve_0}\vth_1=-\a\dbar{\l}_1\ve_{0L}=\a\bar\ve_{0L}\l_1^C, \NO\\&
\d_{\ve_0}\vf_1=\a\dbar{\ve}_{0R}\l_1=\a\bar{\ve}_{0L}\l_1,\qquad \d_{\ve_0}\vf^{*}_1=\a\dbar{\l}_1\ve_{0R}=-\a\bar\ve_{0R}\l_1^C,\NO\\ 
&\d_{\ve_0}\l_1=\frac{1}{2\a}\big(\slashed{\pa}\vth_1^*\ve_{0L}-\slashed{\pa}\vf_1\ve_{0R}-m_1\vth_1^*\ve_{0L}+m_1\vf_1\ve_{0R}\big),\NO\\&
\d_{\ve_0}\dbar{\l}_1=-\frac{1}{2\a}\big(\bar\ve_{0R}\slashed{\pa}\vth_1-\bar\ve_{0L}\slashed{\pa}\vf_1^*+m_1\bar\ve_{0R} \vth_1-m_1\bar\ve_{0L}\vf^{*}_1\big).
\eal
Once again, in these expressions one must be careful to distinguish between the Dirac and Majorana conjugates of a spinor, both of which are discussed in appendix \ref{sec:conventions}. In particular, for a Majorana spinor, $\c$, the Dirac conjugate, $\dbar\c$, and the Majorana conjugate, $\bar\c$, coincide, i.e. $\dbar\c=\bar\c$. Moreover, the Dirac and Majorana conjugates of Weyl spinors are related as $\dbar\c_{L,R}=\bar\c_{R,L}$, while those of Dirac spinors are unrelated. For example, using the decomposition of the Dirac spinor $\l_1$ into two Majorana spinors $\l_\pm$ as in \eqref{PV-redefinition}, we have   
\be
\l_1=\l_++i\l_-,\qquad \dbar\l_1=\dbar\l_+-i\dbar\l_-=\bar\l_+-i\bar\l_-.
\ee

\subsection*{Ferrara-Zumino multiplet operators}

The Noether procedure for Poincar\'e and Q-supersymmetry invariance of either the massive WZ model \eqref{WZ-lagrangian-flat-massive} or the PV Lagrangian \eqref{PV-lagrangian-Dirac} results in a conserved and symmetric stress tensor, $\hat{\wt\ct}{}^\m{}_{\n}$, and a conserved supercurrent, $\hat{\wt\cq}{}^\m$, i.e.
\be\label{FZ-conserved}
\pa_\m\hat{\wt\ct}{}^\m{}_{\n}=0,\qquad \hat{\wt\ct}_{[\m\n]}=0,\qquad \pa_\m\hat{\wt\cq}{}^\m=0,
\ee
where the tilde indicates that these are operators of a massive theory and, as above, the hat denotes quantities evaluated in Minkowski space. 

Although the conformal multiplet of currents consists only of the classically conserved current operators \eqref{WZ-Noether-currents-Mink}, any current multiplet for a massive theory contains additional operators \cite{Komargodski:2010rb}. The simplest and best known such multiplet is the Ferrara-Zumino (FZ) multiplet \cite{Ferrara:1974pz}, which, as we will show later on, turns out to be particularly relevant for examining the presence of supersymmetry anomalies in superconformal theories. Besides the stress tensor and supercurrent, the FZ multiplet contains a generically non conserved R-current, $\hskip3pt\hat{\hskip-3pt\wt\cj}{}^\m$, as well as a complex scalar operator $\hat{\wt\co}_M$ and its complex conjugate, $\hat{\wt\co}_{M^*}$.   

For the massive WZ model \eqref{WZ-lagrangian-flat-massive}, the FZ multiplet operators take the form
\bal
\label{WZ-FZ-operators-Mink}
\hat{\wt\ct}{}^\m{}_{\n}=&\;\frac{1}{\a^2}\pa^{(\m}\f^*\pa_{\n)}\f+\frac12\bar\c\g^\m \pa_\n\c-\frac{1}{8}\pa_\r\big(\bar\c\g_\n\g^{\r\m}\c+\bar\c\g^\m\g^{\r}{}_\n\c-\bar\c\g^\r\g^{\m}{}_\n\c\big)\NO\\
&-\frac{1}{6\a^2}\big(\pa^\m\pa_\n-\h^\m_\n\pa^2\big)(\f^*\f)-\frac{1}{2\a^2}\h^\m_\n\big(\pa_\r\f^*\pa^\r\f+\a^2\bar\c\slashed \pa\c-F^*F+m\a^2\bar\c\c-m(\f F+\f^* F^*)\big),\NO\\
\rule{0.cm}{.7cm}\hat{\hskip-3pt\wt\cj}{}^\m=&\;\frac{i}{3\a^2}\Big(\f^*\pa^\m\f-\f \pa^\m\f^*+\frac{\a^2}{2}\bar\c\g^\m\g^5\c\Big),\NO\\
\rule{0.cm}{.7cm}\hat{\wt\cq}{}^\m=&\;\frac{1}{2\a}(\slashed\pa\f \g^\m\c_R+\slashed\pa\f^*\g^\m\c_L)+\frac{1}{3\a}\g^{\m\n}\pa_\n(\f \c_R+\f^* \c_L)+\frac{m}{2\a}(\f \g^\m\c_L+\f^* \g^\m\c_R),\NO\\
\rule{0.cm}{.7cm}\hat{\wt\co}_M=&\;\frac{m}{4\a^2}\f^2,\qquad 
\hat{\wt\co}_{M^*}=\;\frac{m}{4\a^2}\f^{*2}.
\eal
Using the equations of motion following from \eqref{WZ-lagrangian-flat-massive}, 
\be
\square\f=m^2\f,\qquad \slashed\pa\c=-m\c,\qquad F=-m\f^*,
\ee
it is straightforward to verify that the stress tensor and the supercurrent satisfy the identities \eqref{FZ-conserved}. Moreover, we find that
\be\label{FZ-breaking}
\hat{\wt\ct}{}^\m_\m=-\frac{m}{2}\bar\c\c-\frac{m^2}{\a^2}\f^*\f,\qquad \pa_\m\hskip3pt\hat{\hskip-3pt\wt\cj}{}^\m=\frac{im}{3}\bar\c\g^5\c,\qquad \g_\m\hat{\wt\cq}{}^\m=\frac{m}{\a}(\f\c_L+\f^*\c_R),
\ee
which reflect the breaking of, respectively, scale invariance, R-symmetry and S-supersymmetry.  

Integrating out the auxiliary fields in the PV multiplets and including suitable improvement terms, the FZ multiplet operators following from the PV Lagrangian \eqref{PV-lagrangian-Dirac} take the form
\bal
\label{PV-FZ-operators-Mink}
\hat{\wt\ct}{}^\m{}_{\n}\big|_{\rm PV}=&\;\frac{1}{\a^2}\pa^{(\m}\vf_2^*\pa_{\n)}\vf_2-\frac{1}{6\a^2}\big(\pa^\m\pa_\n-\h^\m_\n\pa^2\big)(\vf_2^*\vf_2)-\frac{1}{2\a^2}\h^\m_\n\big(\pa_\r\vf_2^*\pa^\r\vf_2+m_2^2\vf_2^*\vf_2\big)\NO\\
&+\frac12\bar\l_2\g^\m \pa_\n\l_2-\frac{1}{2}\h^\m_\n\big(\bar\l_2\slashed \pa\l_2+m_2\bar\l_2\l_2\big)-\frac{1}{8}\pa_\r\big(\bar\l_2\g_\n\g^{\r\m}\l_2+\bar\l_2\g^\m\g^{\r}{}_\n\l_2-\bar\l_2\g^\r\g^{\m}{}_\n\l_2\big)\NO\\
&+\frac{1}{\a^2}\pa^{(\m}\vf_1^*\pa_{\n)}\vf_1-\frac{1}{6\a^2}\big(\pa^\m\pa_\n-\h^\m_\n\pa^2\big)(\vf_1^*\vf_1)-\frac{1}{2\a^2}\h^\m_\n\big(\pa_\r\vf_1^*\pa^\r\vf_1+m_1^2\vf_1^*\vf_1\big)\NO\\
&+\frac{1}{\a^2}\pa^{(\m}\vth_1\pa_{\n)}\vth_1^*-\frac{1}{6\a^2}\big(\pa^\m\pa_\n-\h^\m_\n\pa^2\big)(\vth_1\vth_1^*)-\frac{1}{2\a^2}\h^\m_\n\big(\pa_\r\vth_1\pa^\r\vth_1^*+m_1^2\vth_1\vth_1^*\big)\NO\\
&+\dbar\l_1\g^\m \pa_\n\l_1-\h^\m_\n\big(\dbar\l_1\slashed \pa\l_1+m_1\dbar\l_1\l_1\big)-\frac{1}{4}\pa_\r\big(\dbar\l_1\g_\n\g^{\r\m}\l_1+\dbar\l_1\g^\m\g^{\r}{}_\n\l_1-\dbar\l_1\g^\r\g^{\m}{}_\n\l_1\big),\NO\\
\rule{0.cm}{.7cm}\hat{\hskip-3pt\wt\cj}{}^\m\big|_{\rm PV}=&\;\frac{i}{3\a^2}\Big(\vf_2^*\stackrel{\leftrightarrow}{\pa^\m}\hskip-2pt\vf_2+\frac{\a^2}{2}\bar\l_2\g^\m\g^5\l_2\Big)+\frac{i}{3\a^2}\Big(\vf_1^*\stackrel{\leftrightarrow}{\pa^\m}\hskip-2pt\vf_1-\vth_1^*\stackrel{\leftrightarrow}{\pa^\m}\hskip-2pt\vth_1+\a^2\dbar\l_1\g^\m\g^5\l_1\Big),\NO\\
\rule{0.cm}{.7cm}\hat{\wt\cq}{}^\m\big|_{\rm PV}=&\;\frac{1}{2\a}(\slashed\pa\vf_2 \g^\m\l_{2R}+\slashed\pa\vf_2^*\g^\m\l_{2L})+\frac{1}{3\a}\g^{\m\n}\pa_\n(\vf_2 \l_{2R}+\vf_2^* \l_{2L})+\frac{m_2}{2\a}(\vf_2 \g^\m\l_{2L}+\vf_2^* \g^\m\l_{2R})\NO\\
&+\frac{1}{2\a}(\slashed\pa\vth_1 \g^\m\l_{1R}-\slashed\pa\vf_1^*\g^\m\l_{1L})+\frac{1}{3\a}\g^{\m\n}\pa_\n(\vth_1 \l_{1R}-\vf_1^* \l_{1L})+\frac{m_1}{2\a}(\vth_1 \g^\m\l_{1L}-\vf_1^* \g^\m\l_{1R})\NO\\
&+\frac{1}{2\a}(\slashed\pa\vth_1^*\g^\m\l^C_{1L}-\slashed\pa\vf_1 \g^\m\l^C_{1R})+\frac{1}{3\a}\g^{\m\n}\pa_\n(\vth_1^* \l^C_{1L}-\vf_1 \l^C_{1R})+\frac{m_1}{2\a}(\vth_1^*\g^\m\l^C_{1R}-\vf_1 \g^\m\l^C_{1L}),\NO\\
\rule{0.cm}{.7cm}\hat{\wt\co}_M\big|_{\rm PV}=&\;\frac{m_2}{4\a^2}\vf_2^2+\frac{m_1}{2\a^2}\vf_1\vth_1,\qquad
\hat{\wt\co}_{M^*}\big|_{\rm PV}=\;\frac{m_2}{4\a^2}\vf_2^{*2}+\frac{m_1}{2\a^2}\vth_1^*\vf_1^*.
\eal
We emphasize that the supercurrent for the PV fields remains an anticommuting Majorana fermion, which is essential for coupling the theory to background supergravity. Its Majorana conjugate is
\bal
\hspace{-.2cm}\hat{\wt{\stackrel{\rule{4.5pt}{.5pt}}{\cq}}}{}^\m\big|_{\rm PV}=&\,\frac{1}{2\a}(\bar\l_{2R}\g^\m\slashed\pa\vf_2 +\bar\l_{2L}\g^\m\slashed\pa\vf_2^*)-\frac{1}{3\a}\pa_\n(\vf_2 \bar\l_{2R}+\vf_2^* \bar\l_{2L})\g^{\m\n}-\frac{m_2}{2\a}(\vf_2 \bar\l_{2L}+\vf_2^* \bar\l_{2R})\g^\m\\
&+\frac{1}{2\a}\dbar\l_1^C(P_R\g^\m\slashed\pa\vth_1 -P_L\g^\m\slashed\pa\vf_1^*)-\frac{1}{3\a}\pa_\n(\vth_1 \dbar\l_1^CP_{R}-\vf_1^* \dbar\l_1^CP_{L})\g^{\m\n}-\frac{m_1}{2\a}\dbar\l_1^C(\vth_1 P_{L}-\vf_1^* P_{R})\g^\m\NO\\
&+\frac{1}{2\a}\dbar\l_1(P_L\g^\m\slashed\pa\vth_1^*-P_R\g^\m\slashed\pa\vf_1 )-\frac{1}{3\a}\pa_\n(\vth_1^*\dbar\l_1P_L-\vf_1\dbar\l_1P_R)\g^{\m\n}-\frac{m_1}{2\a}\dbar\l_1(\vth_1^*P_R-\vf_1 P_L)\g^\m.\NO
\eal

The FZ multiplet operators of the full theory, comprising the massless WZ model and the PV fields, are the sum of the conformal currents \eqref{WZ-Noether-currents-Mink} and the PV operators in \eqref{PV-FZ-operators-Mink}, and will be denoted by $\hat{\wt\ct}{}^\m{}_{\n}$, $\;\hat{\hskip-3pt\wt\cj}{}^\m$, $\hat{\wt\cq}{}^\m$ and $\hat{\wt\co}_M$ in the following. Their counterparts without a hat, $\hskip3pt\hat\cdot\hskip3pt$, will again refer to the corresponding operators defined through functional differentiation by coupling the theory to background supergravity, which is the subject of the next subsection. The FZ multiplet operators satisfy the on-shell identities \eqref{FZ-conserved}, while the breaking of scale invariance, R-symmetry and S-supersymmetry is reflected respectively in the relations
\bal\label{PV-non-conservation-laws}
\hat{\wt\ct}{}^\m_\m=&\;-\frac{m_2^2}{\a^2}\vf_2^*\vf_2-\frac{m_2}{2}\bar\l_2\l_2-\frac{m_1^2}{\a^2}(\vf_1^*\vf_1+\vth_1\vth_1^*)-m_1\dbar\l_1\l_1\equiv \hat\cb_W,\NO\\ 
\pa_\m\hskip3pt\hat{\hskip-3pt\wt\cj}{}^\m=&\;\frac{im_2}{3}\bar\l_2\g^5\l_2+\frac{2im_1}{3}\dbar\l_1\g^5\l_1\equiv \hat\cb_R,\NO\\
\g_\m\hat{\wt\cq}{}^\m=&\;\frac{m_2}{\a}(\vf_2\l_{2L}+\vf_2^*\l_{2R})+\frac{m_1}{\a}(\vth_1 \l_{1L}-\vf_1^*\l_{1R})+\frac{m_1}{\a}(\vth_1^*\l^C_{1R}-\vf_1 \l^C_{1L})\equiv\hat\cb_S,
\eal
where we have introduced the notation $\hat\cb_W$, $\hat\cb_R$ and $\hat\cb_S$ for the quantities on the r.h.s. of these identities for later convenience. Notice that these quantities, as well as the scalar operator $\co_M$, receive contributions only from the PV fields. This observation will play a key role in the discussion of anomalies in section \ref{sec:correlators}. As for the Noether currents \eqref{WZ-Noether-currents-Mink} following from superconformal symmetry, in appendix \ref{sec:path-integralWIDs} we provide a path integral derivation of the naive Ward identities associated with the conservation laws \eqref{FZ-conserved} of the massive theory. However, the coupling to background supergravity that we discuss next provides a more efficient derivation of these Ward identities, directly in their model independent form in terms of operators defined by functional differentiation.

\subsection{Coupling to background old minimal supergravity}
\label{sec:omsugra}

As discussed earlier, a massive theory does not admit a conformal current multiplet  due to the broken scale invariance and, hence, it cannot be coupled to conformal supergravity. The off-shell supergravity that sources the FZ multiplet is old minimal supergravity \cite{Ferrara:1978em,Stelle:1978ye,Fradkin:1978jq}. It can be obtained from $\cn=1$ conformal supergravity by adding a compensating superconformal chiral multiplet, $(\wt\f,\wt\c,\wt F)$, and suitable gauge fixing \cite{Kaku:1978ea,Ferrara:1978rk,Kugo:1982cu}. The compensating chiral multiplet  sources a subset of operators in the FZ multiplet \cite{Ferrara:1974pz,Komargodski:2010rb} that form a chiral multiplet. This chiral multiplet comprises a complex scalar operator ($x$ in the notation of \cite{Komargodski:2010rb}) that is sourced by $\wt F$, the gamma trace of the supercurrent sourced by $\wt\c$, and the trace of the stress tensor and the divergence of the R-current, which are sourced by the real and imaginary parts the complex scalar $\wt\f$. 

The local symmetry transformations of the compensating multiplet are exactly those of the chiral multiplet in \eqref{WZ-susy-transformations}, namely   
\bal
\label{Compensator-transformations}
\d\wt\f=&\;\x^\m\pa_\m\wt\f-\s\wt\f-\frac{2i}{3}\th\wt\f+\a\bar\ve_L\wt\c_L,\\
\d\wt\c_L=&\;\x^\m\pa_\m\wt\c_L-\frac32\s\wt\c_L+\frac{i}{3}\th\wt\c_L-\frac14\l_{ab}\g^{ab}\wt\c_L+\frac{1}{2\a}\Big(\g^\m\big(D_\m\wt\f-\a\bar\j_{\m L}\wt\c_L\big)\ve_R+\wt F\ve_L+2\wt\f\h_L\Big),\NO\\
\d \wt F=&\;\x^\m\pa_\m \wt F-2\s \wt F+\frac{4i}{3}\th \wt F+\frac12\bar\ve_R\g^\m\Big(2\a D_\m\wt\c_L-\g^\n\big(D_\n\wt\f-\a\bar\j_{\n L}\c_L\big)\j_{\m R}-\wt F\j_{\m L}-2\wt\f\,\f_{\m L}\Big).\NO
\eal

The compensating multiplet allows us to redefine the supergravity fields so that they are invariant under Weyl, S-supersymmetry and axial gauge transformations. Using the conformal supergravity transformations in \eqref{sugra-trans}, it is straightforward to verify that the redefined fields    
\bbxd
\bal
\label{old-minimal-redefinitions}
e'^a_\m=&\;|\wt\f|e^a_\m,\qquad
\j'_\m=\;\frac{1}{|\wt\f|}\big(\wt\f P_R+\wt\f^*P_L\big)^{3/2}\j_\m+\frac{\a}{|\wt\f|}\g_\m\big(\wt\f P_L+\wt\f^*P_R\big)^{1/2}\wt\c,\NO\\
A'_\m=&\;A_\m-\frac{3i}{4}|\wt\f|^{-2}\Big(\wt\f^*\pa_\m\wt\f-\wt\f\pa_\m\wt\f^*+\frac{\a^2}{2}\bar{\wt\c}\g_\m\g^5\wt\c-\a\bar\j_\m(\wt\f^*\wt\c_L-\wt\f\wt\c_R)\Big).
\eal
\ebxd
do not transform under Weyl, S-supersymmetry and axial gauge transformations \cite{Kaku:1978ea}.

Having defined the invariant supergravity fields $(e'^a_\m,\j'_\m,A'_\m)$, we fix the gauge by setting 
\bbxd
\vskip.2cm
\be\label{old-minimal-gauge-fixing}
\wt\f=1,\qquad \wt\c=0. 
\ee
\ebxd
This gauge choice eliminates the sources of the gamma trace of the supercurrent, the trace of the stress tensor and the divergence of the R-current in the FZ multiplet, all of which are redundant. In this gauge, the field redefinition \eqref{old-minimal-redefinitions} reduces to the identity, so that $(e'^a_\m,\j'_\m,A'_\m)=(e^a_\m,\j_\m,A_\m)$. However, only a subset of the local transformations of conformal supergravity preserve this gauge. From the transformations \eqref{Compensator-transformations} of the compensator multiplet we see that the gauge \eqref{old-minimal-gauge-fixing} is preserved if and only if the local symmetry parameters satisfy the conditions 
\bbxd
\vskip.4cm
\be
\s=0,\qquad \th=0,\qquad \h=\frac{i}{3}\slashed A\g^5\ve-\frac12(\wt F\ve_L+\wt F^*\ve_R).
\ee
\ebxd
The surviving local symmetries are those of old minimal Poincar\'e supergravity with 
\bbxd
\vskip.4cm
\be\label{old-minimal-susy}
\d_\ve^{\rm om}=\d_\ve+\d_{\h(\ve)},\qquad \h(\ve)=\frac{i}{3}\slashed A\g^5\ve-\frac12(\wt F\ve_L+\wt F^*\ve_R),
\ee
\ebxd
where $\d_\ve$, $\d_\h$ are the Q- and S-supersymmetry transformations of $\cn=1$ conformal supergravity. 

The field content of old minimal supergravity, therefore, consists of that of $\cn=1$ 
conformal supergravity, as well as the auxiliary complex scalar, $\wt F$, of the compensator multiplet, which is not fixed by the gauge fixing conditions \eqref{old-minimal-gauge-fixing}. Adopting standard notation \cite{Wess:1992cp}, we denote this field by $M$ in the following.\footnote{In fact, $M$ here is related to $M_{\rm WB}$ in \cite{Wess:1992cp} as $M=-3M_{\rm WB}^*$.} The supersymmetry transformations of old minimal supergravity are  
\bbxd
\bal\label{old-minimal-susy-transformations}
\d_\ve^{\rm om} e^a_\m=&\;-\frac12\lbar\j_\m\g^a\ve,\NO\\
\d_\ve^{\rm om}\j_\m=&\;D_\m\ve-\g_\m\Big(\frac{i}{3}\slashed A\g^5\ve-\frac12(M\ve_L+M^*\ve_R)\Big),\NO\\
\d_\ve^{\rm om} A_\m=&\;\frac{3i}{4}\lbar\f_\m\g^5\ve-\frac{3i}{4}\lbar\j_\m\g^5\Big(\frac{i}{3}\slashed A\g^5\ve-\frac12(M\ve_L+M^*\ve_R)\Big),\NO\\
\d_\ve^{\rm om} M=&-\bar\ve_R\g^\m\Big(\frac{i}{3}\slashed A\j_{\m R}+\frac12M\j_{\m L}+\f_{\m L}\Big).
\eal
\ebxd

\subsection*{Ward identities}

Given the field content and local symmetry transformations of old minimal supergravity, we can define the corresponding current multiplet operators and determine the Ward identities they satisfy. A variation of the generating function of (regulated) connected correlation functions takes the form
\be\label{WvarOM}
\d\wt{\mathscr{W}}[e,A,\j,M]=\int d^4x\,e\big(\<\wt\ct^\m_a\>_s\d e^a_\m+\<\wt\cj^\m\>_s\d A_\m+\d\bar\j_\m\<\wt\cq^\m\>_s+\<\wt\co_M\>_s\d M+\<\wt\co_{M^*}\>_s\d M^*\big),
\ee
where the local operators defined by 
\bal\label{currents-om}
&\<\wt\ct^\m_a\>_s=e^{-1}\frac{\d\wt{\mathscr{W}}}{\d e^a_\m},
\qquad \<\wt\cj^\m\>_s=e^{-1}\frac{\d\wt{\mathscr{W}}}{\d A_\m},\qquad
\<\wt\cq^\m\>_s=e^{-1}\frac{\d\wt{\mathscr{W}}}{\d\bar\j_\m},\NO\\
&\<\wt\co_M\>_s=e^{-1}\frac{\d\wt{\mathscr{W}}}{\d M},\qquad \<\wt\co_{M^*}\>_s=e^{-1}\frac{\d\wt{\mathscr{W}}}{\d M^*},
\eal
comprise the FZ current multiplet \cite{Ferrara:1974pz,Komargodski:2010rb}. Like the currents \eqref{currents} defined from conformal supergravity, this definition of the FZ multiplet is independent of the specific theory and applies even to non Lagrangian theories.

The local symmetries of old minimal supergravity consist of diffeomorphisms, local frame rotations, as well as the local supersymmetry transformations \eqref{old-minimal-susy-transformations}. The algebra of these transformations closes off-shell \cite{Ferrara:1978em,Stelle:1978ye}. Since there exist no gravitational or Lorentz anomalies in four dimensions, diffeomorphisms and local frame rotations are preserved at the quantum level. Whether the old minimal supersymmetry transformations \eqref{old-minimal-susy-transformations} are anomalous can be determined using the associated WZ consistency conditions. We will not perform such an analysis here, but we will show that the four point functions of currents in the free and massless WZ model are compatible with a non anomalous old minimal supersymmetry.\footnote{The claim that there exists no supersymmetry anomaly in old minimal supergravity is implicit in \cite{Bonora:1984pn,Brandt:1993vd,Brandt:1996au,Bonora:2013rta,Butter:2013ura}. However, these works concern the superspace formulation of old minimal supergravity, which contains additional fields that may act as compensators.}  

Inserting the local symmetry transformations of old minimal supergravity in the variation \eqref{WvarOM} and invoking the invariance of $\wt{\mathscr{W}}[e,A,\j,M]$ leads to the three Ward identities 
\bbxd
\bal
\label{WardIDsOM}
&e^a_\m\nabla_\n\<\wt\ct^\n_a\>_s+\nabla_\n(\bar\j_\m  \<\wt\cq^\n\>_s)-\bar\j_\n\overleftarrow D_\m \<\wt\cq^\n\>_s-F_{\m\n}\<\wt\cj^\n\>_s-\pa_\m M\<\wt\co_M\>_s-\pa_\m M^*\<\wt\co_{M^*}\>_s\NO\\
&\hspace{2.cm}+A_\m\big(\nabla_\n\<\wt\cj^\n\>_s+i\bar\j_\n \g^5\<\wt\cq^\n\>_s\big)-\o_\m{}^{ab}\Big(e_{\n [a}\<\wt\ct^\n_{b]}\>_s+\frac14\bar\j_\n\g_{ab}\<\wt\cq^\n\>_s\Big)=0,\NO\\
\rule{.0cm}{.8cm}&e_{\m[a} \<\wt\ct^\m_{b]}\>_s+\frac14\bar\j_\m\g_{ab} \<\wt\cq^\m\>_s=0,\NO\\
\rule{.0cm}{.8cm}&D_\m \<\wt\cq^\m\>_s-\frac12\g^a\j_\m \<\wt\ct^\m_a\>_s-\frac{3i}{4}\g^5\f_\m\<\wt\cj^\m\>_s\NO\\
&+\frac12\Big(M P_L+M^*P_R-\frac{2i}{3}\slashed A\g^5\Big)\Big(\g_\m \<\wt\cq^\m\>_s-\frac{3i}{4}\g^5\j_\m\<\wt\cj^\m\>_s\Big)\\
&+\g^\m\Big(\frac{i}{3}\slashed A\j_{\m R}+\frac12 M\j_{\m L}+\f_{\m L}\Big)\<\wt\co_M\>_s+\g^\m\Big(-\frac{i}{3}\slashed A\j_{\m L}+\frac12 M^*\j_{\m R}+\f_{\m R}\Big)\<\wt\co_{M^*}\>_s=0.\NO
\eal
\ebxd
Since diffeomorphisms and local frame rotations are preserved at the quantum level, the first two Ward identities hold also in the quantum theory. In section \ref{sec:correlators} we will show that the old minimal supersymmetry Ward identity is also maintained at the quantum level, at least for the current correlation functions we examine. 

Differentiating the supersymmetry Ward identity in \eqref{WardIDsOM} with respect to the old minimal sources leads to the flat space Ward identities  
\bbxd
\bal\label{OM-susy-WID-2pt}
&\pa_\m \<\wt\cq^\m(x)\wt{\bar\cq}{}^\s(y)\>-\frac12 \<\wt\ct^\s_a(x)\>\g^a\d(x,y)+\frac{i}{8}\<\wt\cj^\n(x)\>\big(4\d^{[\r}_\n\d^{\s]}_\l+i\g^5 \e_{\n\l}{}^{\r\s}\big)\g^5\g^\l \pa_\r\d(x,y)\NO\\
&+\frac13\big(\<\wt\co_M\>P_R+\<\wt\co_{M^*}\> P_L\big)\g^{\s\r}\pa_\r^x\d(x,y)=0,
\eal
\ebxd
\bbxd
\bal
\label{OM-susy-WID-3pt}
&\pa_\m \<\wt\cq^\m(x)\wt{\bar\cq}{}^\s(y)\wt\cj^\k(z)\>-i\g^5\d(x,z)\<\wt\cq^\k(x)\wt{\bar\cq}{}^\s(y)\>-\frac12\<\wt\ct^\s_a(x)\wt\cj^\k(z)\>\g^a\d(x,y)\NO\\
&+\frac{i}{8}\<\wt\cj^\n(x)\wt\cj^\k(z)\>\big(4\d^{[\r}_\n\d^{\s]}_\l+i\g^5 \e_{\n\l}{}^{\r\s}\big)\g^5\g^\l \pa_\r\d(x,y))\NO\\
&+\frac{1}{8}\<\wt\cj^\n(x)\>\big(4\d^{[\k}_\n\d^{\s]}_\l+i\g^5 \e_{\n\l}{}^{\k\s}\big)\g^\l \d(x,z)\d(x,y)\NO\\
&-\frac{i}{3}\d(x,z)\g^\k\g^5\Big(\g_\m \<\wt\cq^\m(x)\wt{\bar\cq}{}^\s(y)\>-\frac{3i}{4}\g^5\d(x,y)\<\wt\cj^\s(x)\>\Big)\NO\\
&+\frac{i}{3}\h^{\s\k}\d(x,y)\d(x,z)\Big(\<\wt\co_M(x)\>P_R-\<\wt\co_{M^*}(x)\>P_L\Big)\NO\\
&+\frac13\g^{\s\r}\pa_\r\d(x,y)\Big(\<\wt\co_M(x)\wt\cj^\k(z)\>P_R+\<\wt\co_{M^*}(x)\wt\cj^\k(z)\>P_L\Big)=0,
\eal
\ebxd
\bbxd
\bal
\label{OM-susy-WID-4pt}
&\pa_\m \<\wt\cq^\m(x)\wt{\bar\cq}{}^\s(y)\wt\cj^\k(z)\wt\cj^\a(w)\>-i\g^5\d(x,w)\<\wt\cq^\a(x)\wt{\bar\cq}{}^\s(y)\wt\cj^\k(z)\>\NO\\
&-i\g^5\d(x,z)\<\wt\cq^\k(x)\wt{\bar\cq}{}^\s(y)\wt\cj^\a(w)\>-\frac12 \<\wt\ct^\s_a(x)\wt\cj^\k(z)\wt\cj^\a(w)\>\g^a\d(x,y)\NO\\
&+\frac{i}{8}\<\wt\cj^\n(x)\wt\cj^\k(z)\wt\cj^\a(w)\>\big(4\d^{[\r}_\n\d^{\s]}_\l+i\g^5 \e_{\n\l}{}^{\r\s}\big)\g^5\g^\l \pa_\r\d(x,y)\NO\\
&+\frac{1}{8}\big(4\d^{[\b}_\n\d^{\s]}_\l+i\g^5 \e_{\n\l}{}^{\b\s}\big)\g^\l\Big(\<\wt\cj^\n(x)\wt\cj^\a(w)\>\d^\k_\b\d(x,z)+\<\wt\cj^\n(x)\wt\cj^\k(z)\>\d^\a_\b\d(x,w)\Big)\d(x,y)\NO\\
&-\frac{i}{3}\d(x,z)\g^\k\g^5\Big(\g_\m \<\wt\cq^\m(x)\wt{\bar\cq}{}^\s(y)\cj^\a(w)\>-\frac{3i}{4}\<\wt\cj^\s(x)\wt\cj^\a(w)\>\g^5\d(x,y)\Big)\NO\\
&-\frac{i}{3}\d(x,w)\g^\a\g^5\Big(\g_\m \<\wt\cq^\m(x)\wt{\bar\cq}{}^\s(y)\cj^\k(z)\>-\frac{3i}{4}\<\wt\cj^\s(x)\wt\cj^\k(z)\>\g^5\d(x,y)\Big)\NO\\
&+\frac{i}{3}\d(x,y)\d(x,z)\h^{\s\k}\Big(\<\wt\co_M(x)\wt\cj^\a(w)\>P_R-\<\wt\co_{M^*}(x)\wt\cj^\a(w)\>P_L\Big)\NO\\
&+\frac{i}{3}\d(x,y)\d(x,w)\h^{\s\a}\Big(\<\wt\co_M(x)\wt\cj^\k(z)\>P_R-\<\wt\co_{M^*}(x)\wt\cj^\k(z)\>P_L\Big)\NO\\
&+\frac13\g^{\s\r}\pa_\r\d(x,y)\Big(\<\wt\co_M(x)\wt\cj^\k(z)\wt\cj^\a(w)\>P_R+\<\wt\co_{M^*}(x)\wt\cj^\k(z)\wt\cj^\a(w)\>P_L\Big)=0.
\eal
\ebxd
In section \ref{sec:correlators} we will verify that these identities remain true at one loop for both the regulated and renormalized WZ model.

\subsection*{Ward identities for broken symmetries}

Even though Weyl invariance, R-symmetry and S-supersymmetry are not symmetries of old minimal supergravity and are therefore explicitly broken in the FZ current multiplet, there still exist Ward identities associated with these symmetries. This is because the FZ multiplet operators possess well defined transformations under these symmetries. These Ward identities can be derived by temporarily relaxing the gauge fixing conditions \eqref{old-minimal-gauge-fixing}. The operators sourced by the extra components $\wt \f$ and $\wt\c$ of the compensator chiral multiplet do not belong to the FZ multiplet, but result in symmetry breaking terms in the Ward identities.      

Relaxing the gauge fixing conditions \eqref{old-minimal-gauge-fixing}, the generating functional of the FZ multiplet is $\wt{\mathscr{W}}[e',A',\j',M']$, where $e'^a_\m$, $A'_\m$ and $\j'_\m$ are the composite supergravity fields \eqref{old-minimal-redefinitions} and $M'\equiv\wt\f^{*-2}M$, like $e',A',\j'$, is invariant under Weyl, R-symmetry and S-supersymmetry transformations. Hence, an infinitesimal variation of the generating functional under any of these symmetries, collectively denoted by $\d_B$, leads to the identity 
\bal
0=\d_B\wt{\mathscr{W}}[e',A',\j',M']=&\;\int d^4x\,e\big(\<\wt\ct^\m_a\>_s\d_B e^a_\m+\<\wt\cj^\m\>_s\d_B A_\m+\d\bar\j_\m\<\wt\cq^\m\>_s+\<\wt\co_M\>_s\d_B M\NO\\
&+\<\wt\co_{M^*}\>_s\d_B M^*+\<\co_{\wt\f}\>_s\d_B\wt\f+\<\co_{\wt\f^*}\>_s\d_B\wt\f^*+\d_B\wt{\bar\c}\<\co_{\wt\c}\>_s\big),
\eal
where $\<\co_{\wt\f}\>_s$, $\<\co_{\wt\f^*}\>_s$ and $\<\co_{\wt\c}\>_s$ are defined analogously to \eqref{currents-om}. Using the symmetry transformations \eqref{Compensator-transformations} of the compensator multiplet, this identity leads to the following three Ward identities, respectively for Weyl invariance, R-symmetry and S-supersymmetry:    
\bbxd
\bal
\label{BWardIDs-FZ}
&e^a_\m\<\wt\ct^\m_a\>_s+\frac12\bar\j_\m \<\wt\cq^\m\>_s=\<\cb_W\>_s,\NO\\
&\rule{0cm}{.7cm}\nabla_\m \<\wt\cj^\m\>_s+i\bar\j_\m\g^5 \<\wt\cq^\m\>_s=\<\cb_R\>_s,\NO\\
&\rule{0cm}{.7cm}\g_\m \<\wt\cq^\m\>_s-\frac{3i}{4}\g^5\j_\m\<\wt\cj^\m\>_s=\<\cb_{S}\>_s, 
\eal
\ebxd
where, after imposing the gauge fixing conditions \eqref{old-minimal-gauge-fixing}, the symmetry breaking terms on the r.h.s. take the form
\bbxd
\bal\label{Bs-Os}
\cb_W=&\;2\Re\co_{\wt\f}+2(M\wt\co_M+M^*\wt\co_{M^*}),\NO\\ \cb_R=&\;\frac{4}{3}\Im\co_{\wt\f}+\frac{4i}{3}(M\wt\co_M-M^*\wt\co_{M^*}),\NO\\ \cb_{S}=&\;-\frac{1}{\a}\co_{\wt\c}.
\eal
\ebxd
As we will see below, for a classically conformal theory $\cb_W$, $\cb_R$ and $\cb_S$ depend only on the regulator and so they become ultralocal after renormalization.

The S-supersymmetry Ward identity in \eqref{BWardIDs-FZ} implies that flat space correlators satisfy
\bbxd
\rule{0cm}{.4cm}
\be
\label{S-susy-WIDs-2pt-FZ}
\g_\m \<\wt\cq^\m(x)\wt{\bar\cq}{}^\s(y)\>-\frac{3i}{4}\g^5\d(x,y)\<\wt\cj^\s(x)\>=\<\cb_S(x)\wt{\bar\cq}{}^\s(y)\>,
\ee
\ebxd
\bbxd
\rule{0cm}{.4cm}
\be
\label{S-susy-WIDs-3pt-FZ}
\g_\m \<\wt\cq^\m(x)\wt{\bar\cq}{}^\s(y)\wt\cj^\k(z)\>-\frac{3i}{4}\<\wt\cj^\s(x)\wt\cj^\k(z)\>\g^5\d(x,y)=\<\cb_S(x)\wt{\bar\cq}{}^\s(y)\wt\cj^\k(z)\>,
\ee
\ebxd
\bbxd
\rule{0cm}{.0cm}
\be
\label{S-susy-WIDs-4pt-FZ}\g_\m \<\wt\cq^\m(x)\wt{\bar\cq}{}^\s(y)\wt\cj^\k(z)\wt\cj^\a(w)\>-\frac{3i}{4}\<\wt\cj^\s(x)\wt\cj^\k(z)\wt\cj^\a(w)\>\g^5\d(x,y)=\<\cb_S(x)\wt{\bar\cq}{}^\s(y)\wt\cj^\k(z)\wt\cj^\a(w)\>,\hskip-2pt
\ee
\ebxd
while R-symmetry implies that
\bbxd
\rule{0cm}{.25cm}
\be
\label{R-symmetry-WIDs-JJ-FZ}
\pa_\m \<\wt\cj^\m(x)\wt\cj^\n(y)\>=\<\cb_R(x)\wt\cj^\n(y)\>,
\ee
\ebxd
\bbxd
\rule{0cm}{.25cm}
\be
\label{R-symmetry-WIDs-JJJ-FZ}
\pa_\m \<\wt\cj^\m(x)\wt\cj^\n(y)\wt\cj^\r(z)\>=\<\cb_R(x)\wt\cj^\n(y)\wt\cj^\r(z)\>,
\ee
\ebxd
\bbxd
\rule{0cm}{.0cm}
\be
\label{R-symmetry-WIDs-QQJ-FZ}
\hskip-4pt\pa_\m \<\wt\cj^\m(x)\wt\cq^\n(y)\wt{\bar\cq}{}^\r(z)\>+\d(x,y)i\g^5 \<\wt\cq^\n(x)\wt{\bar\cq}{}^\r(z)\>+i\<\wt\cq^\n(y)\wt{\bar\cq}{}^\r(x)\>\g^5\d(x,z)=\<\cb_R(x)\wt\cq^\n(y)\wt{\bar\cq}{}^\r(z)\>,
\ee
\ebxd
\bbxd
\rule{0cm}{.0cm}
\bal
\label{R-symmetry-WIDs-QQJJ-FZ}
&\pa_\m \<\wt\cj^\m(x)\wt\cq^\n(y)\wt{\bar\cq}{}^\r(z)\wt\cj^\s(w)\>+\d(x,y)i\g^5 \<\wt\cq^\n(x)\wt{\bar\cq}{}^\r(z)\wt\cj^\s(w)\>\NO\\
&+i\<\wt\cq^\n(y)\wt{\bar\cq}{}^\r(x)\wt\cj^\s(w)\>\g^5\d(x,z) =\<\cb_R(x)\wt\cq^\n(y)\wt{\bar\cq}{}^\r(z)\wt\cj^\s(w)\>.
\eal
\ebxd

Using the identities for the broken symmetries, the FZ multiplet supersymmetry Ward identities may be written in a simpler form that is more suitable for the analysis in section \ref{sec:correlators}. Namely, inserting \eqref{S-susy-WIDs-2pt-FZ} and \eqref{S-susy-WIDs-3pt-FZ} respectively in \eqref{OM-susy-WID-3pt} and \eqref{OM-susy-WID-4pt} we get 
\bbxd
\bal
\label{OM-susy-WID-3pt-simple}
&\pa_\m \<\wt\cq^\m(x)\wt{\bar\cq}{}^\s(y)\wt\cj^\k(z)\>-i\g^5\d(x,z)\<\wt\cq^\k(x)\wt{\bar\cq}{}^\s(y)\>-\frac12\<\wt\ct^\s_a(x)\wt\cj^\k(z)\>\g^a\d(x,y)\NO\\
&+\frac{i}{8}\<\wt\cj^\n(x)\wt\cj^\k(z)\>\big(4\d^{[\r}_\n\d^{\s]}_\l+i\g^5 \e_{\n\l}{}^{\r\s}\big)\g^5\g^\l \pa_\r\d(x,y))\NO\\
&+\frac{1}{8}\<\wt\cj^\n(x)\>\big(4\d^{[\k}_\n\d^{\s]}_\l+i\g^5 \e_{\n\l}{}^{\k\s}\big)\g^\l \d(x,z)\d(x,y)\NO\\
&=\frac{i}{3}\d(x,z)\g^\k\g^5\<\cb_S(x)\wt{\bar\cq}{}^\s(y)\>-\frac{i}{3}\h^{\s\k}\d(x,y)\d(x,z)\Big(\<\wt\co_M(x)\>P_R-\<\wt\co_{M^*}(x)\>P_L\Big)\NO\\
&-\frac13\g^{\s\r}\pa_\r\d(x,y)\Big(\<\wt\co_M(x)\wt\cj^\k(z)\>P_R+\<\wt\co_{M^*}(x)\wt\cj^\k(z)\>P_L\Big)=0,
\eal
\ebxd
\bbxd
\bal
\label{OM-susy-WID-4pt-simple}
&\pa_\m \<\wt\cq^\m(x)\wt{\bar\cq}{}^\s(y)\wt\cj^\k(z)\wt\cj^\a(w)\>-i\g^5\d(x,w)\<\wt\cq^\a(x)\wt{\bar\cq}{}^\s(y)\wt\cj^\k(z)\>\NO\\
&-i\g^5\d(x,z)\<\wt\cq^\k(x)\wt{\bar\cq}{}^\s(y)\wt\cj^\a(w)\>-\frac12 \<\wt\ct^\s_a(x)\wt\cj^\k(z)\wt\cj^\a(w)\>\g^a\d(x,y)\NO\\
&+\frac{i}{8}\<\wt\cj^\n(x)\wt\cj^\k(z)\wt\cj^\a(w)\>\big(4\d^{[\r}_\n\d^{\s]}_\l+i\g^5 \e_{\n\l}{}^{\r\s}\big)\g^5\g^\l \pa_\r\d(x,y)\NO\\
&+\frac{1}{8}\big(4\d^{[\b}_\n\d^{\s]}_\l+i\g^5 \e_{\n\l}{}^{\b\s}\big)\g^\l\Big(\<\wt\cj^\n(x)\wt\cj^\a(w)\>\d^\k_\b\d(x,z)+\<\wt\cj^\n(x)\wt\cj^\k(z)\>\d^\a_\b\d(x,w)\Big)\d(x,y)\NO\\
&=\frac{i}{3}\d(x,z)\g^\k\g^5\<\cb_S(x)\wt{\bar\cq}{}^\s(y)\cj^\a(w)\>+\frac{i}{3}\d(x,w)\g^\a\g^5 \<\cb_S(x)\wt{\bar\cq}{}^\s(y)\cj^\k(z)\>\NO\\
&-\frac{i}{3}\d(x,y)\d(x,z)\h^{\s\k}\Big(\<\wt\co_M(x)\wt\cj^\a(w)\>P_R-\<\wt\co_{M^*}(x)\wt\cj^\a(w)\>P_L\Big)\NO\\
&-\frac{i}{3}\d(x,y)\d(x,w)\h^{\s\a}\Big(\<\wt\co_M(x)\wt\cj^\k(z)\>P_R-\<\wt\co_{M^*}(x)\wt\cj^\k(z)\>P_L\Big)\NO\\
&-\frac13\g^{\s\r}\pa_\r\d(x,y)\Big(\<\wt\co_M(x)\wt\cj^\k(z)\wt\cj^\a(w)\>P_R+\<\wt\co_{M^*}(x)\wt\cj^\k(z)\wt\cj^\a(w)\>P_L\Big)=0.
\eal
\ebxd

\subsection*{Massive chiral multiplet coupled to old minimal supergravity} 

The coupling of a massive chiral multiplet to old minimal supergravity is discussed e.g. in \cite{Cremmer:1982en,West:1990tg,Wess:1992cp} (see also \cite{Festuccia:2011ws}). However, the analysis in the literature typically involves field redefinitions that are not suitable for our purposes here and so it is easier to derive the relevant results directly in our conventions.\footnote{For example, comparing the old minimal supersymmetry transformations in \eqref{old-minimal-chiral-transformations} with eq.~(19.21) in \cite{Wess:1992cp} or eq.~(3.19) in \cite{Cremmer:1982en}, one sees that the top component of the chiral multiplet there has been shifted by a multiple of $\f M$. Such a shift is related to an improvement term of the FZ multiplet (see \cite{Festuccia:2011ws} for a discussion of this shift) that allows old minimal supersymmetric actions to be expressed in terms of a generic holomorphic superpotential, but obscures the relation between the old minimal and conformal supergravity supersymmetry transformations. } 

As for the supergravity fields, the supersymmetry transformation of a chiral multiplet in old minimal supergravity corresponds to the combination \eqref{old-minimal-susy} of Q- and and S-supersymmetry in conformal supergravity, given in \eqref{WZ-susy-transformations}. Namely, 
\bal
\label{old-minimal-chiral-transformations}
\d_\ve^{\rm om}\f=&\;\x^\m\pa_\m\f+\a\bar\ve_L\c_L,\NO\\
\d_\ve^{\rm om}\c_L=&\;\x^\m\pa_\m\c_L-\frac14\l_{ab}\g^{ab}\c_L+\frac{1}{2\a}\Big(\g^\m\big(D_\m\f-\a\bar\j_{\m L}\c_L\big)\ve_R+F\ve_L+2\f\h_L\Big),\NO\\
\d_\ve^{\rm om} F=&\;\x^\m\pa_\m F+\frac12\bar\ve_R\g^\m\Big(2\a D_\m\c_L-\g^\n\big(D_\n\f-\a\bar\j_{\n L}\c_L\big)\j_{\m R}-F\j_{\m L}-2\f\,\f_{\m L}\Big).
\eal
It follows that the massless WZ model \eqref{WZ-action-CS} preserves old minimal supersymmetry. The interaction terms for a superconformal chiral multiplet coupled to conformal supergravity \cite{Kaku:1978nz}, namely
\bbxd
\bal
\label{WZ-action-int}
e^{-1}\mathscr{L}^{\rm int}_{\rm WZ}=&\; g (\f\bar\c\c_L+\f^*\bar\c\c_R)-\frac{g}{2\a^2}(\f^2 F+\f^{*2} F^*)\\
&-\frac{g}{2\a}(\f^2\bar\j_\m \g^\m\c_L+\f^{*2}\bar\j_\m \g^\m\c_R)
-\frac{g}{12\a^2}(\f^3\bar\j_\m\g^{\m\n}\j_{\n R}+\f^{*3}\bar\j_\m\g^{\m\n}\j_{\n L})+\co(\j^3),\NO
\eal
\ebxd
where, the overall factor, $g$, is the arbitrary coupling constant, also preserve old minimal supersymmetry independently. Although we do not consider the interacting WZ model in our analysis, we will see shortly that these terms can help us determine the supersymmetric local counterterms required to renormalize the non interacting WZ model. 

The only term that breaks the conformal supergravity symmetries to those of old minimal supergravity is the supersymmetric mass term\footnote{Once again, we use the generic massive WZ model \eqref{WZ-lagrangian-flat-massive} in order to simplify the discussion. In our analysis this is eventually replaced with the massive WZ model for the PV fields.}
\bbxd
\bal
\label{WZ-action-mass}
e^{-1}\mathscr{L}^{\rm mass}_{\rm WZ}=&\;-\frac12 m \bar\c\c+\frac{m}{2\a^2}(\f F+\f^* F^*)+\frac{m}{4\a^2}(\f^2M + \f^{*2}M^*)+\frac{m}{2\a}(\f\bar\j_\m \g^\m\c_L+\f^*\bar\j_\m \g^\m\c_R)\NO\\
&+\frac{m}{8\a^2}(\f^2\bar\j_\m\g^{\m\n}\j_{\n R}+\f^{*2}\bar\j_\m\g^{\m\n}\j_{\n L})+\co(\j^3),
\eal
\ebxd
where $m$ is an arbitrary mass parameter. This mass term is invariant under the old minimal supersymmetry transformation, but not under the individual Q- and S-supersymmetry transformations of conformal supergravity. It also breaks R-symmetry and local Weyl invariance explicitly. The general coupling of a chiral multiplet to old minimal supergravity therefore takes the form
\be\label{WZ-action-OM}
\mathscr{L}^{\rm om}_{\rm WZ}=\mathscr{L}_{\rm WZ}+\mathscr{L}^{\rm int}_{\rm WZ}+\mathscr{L}^{\rm mass}_{\rm WZ},
\ee
where the three terms are given respectively in \eqref{WZ-action-CS}, \eqref{WZ-action-int} and \eqref{WZ-action-mass}.

The Lagrangian \eqref{WZ-action-OM} determines that the FZ multiplet operators of a massive non interacting WZ model take the form
\bbxd
\bal
\label{WZ-currents-OM}
\wt\ct^\m_a=&\;\ct^\m_a+\frac12e^\m_a\Big(-m\bar\c\c+\frac{m}{\a^2}(\f F+\f^* F^*)+\frac{m}{2\a^2}(\f^2M + \f^{*2}M^*)+\co(\j)\Big),\NO\\
\rule{0.cm}{.cm}\wt\cq^\m=&\;\cq^\m+\frac{m}{2\a}(\f \g^\m\c_L+\f^* \g^\m\c_R)+\frac{m}{4\a^2}\g^{\m\n}(\f^2\j_{\n R}+\f^{*2}\j_{\n L})+\co(\j^2),\NO\\
\rule{0.cm}{.cm}\wt\cj^\m=&\;\cj^\m,\qquad \wt\co_M=\;\frac{m}{4\a^2}\f^2,\qquad 
\wt\co_{M^*}=\;\frac{m}{4\a^2}\f^{*2}.
\eal
\ebxd
where $\ct^\m_a$, $\cj^\m$ and $\cq^\m$ are the currents \eqref{WZ-currents-CS} of the conformal multiplet. In the flat space limit these reduce to the operators \eqref{WZ-FZ-operators-Mink} obtained earlier, but as is shown again in appendix \ref{sec:path-integralWIDs}, the terms linear and higher in the background fields ensure that the symmetry transformations of the currents \eqref{WZ-currents-OM} are model independent, contrary to the transformations of the operators \eqref{WZ-FZ-operators-Mink}. The dependence of the operators \eqref{WZ-currents-OM} on the background fields also determines the seagull terms discussed in appendix \ref{sec:seagulls}, which relate correlation functions of these operators with Feynman diagrams. 

The FZ multiplet operators \eqref{WZ-currents-OM} satisfy the classical version of the Ward identities \eqref{WardIDsOM}, which generalize the flat space conservation laws \eqref{FZ-conserved} to an arbitrary supergravity background. They also satisfy the classical version of the identities \eqref{BWardIDs-FZ} for the broken symmetries with  
\bbxd
\bal\label{Bs}
\cb_W=&\;-\frac{m}{2}\bar\c\c+\frac{m}{2\a^2}(\f F+\f^* F^*)+3(M\wt\co_M +M^*\wt\co_{M^*})+\co(\j),\NO\\
\cb_R=&\;\frac{im}{3}\bar\c\g^5\c-\frac{im}{3\a^2}(\f F-\f^* F^*)+\frac{4i}{3}(M\wt\co_M-M^*\wt\co_{M^*})\NO\\
&-\frac{im}{3\a}(\f\bar\j_\m \g^\m\c_L-\f^*\bar\j_\m \g^\m\c_R)-\frac{i}{3}(\wt\co_M\bar\j_\m\g^{\m\n}\j_{\n R}-\wt\co_{M^*}\bar\j_\m\g^{\m\n}\j_{\n L})+\co(\j^3),\NO\\
\cb_S=&\;\frac{m}{\a}(\f \c_L+\f^* \c_R)+\frac{3}{2}\g^{\n}(\wt\co_M\j_{\n R}+\wt\co_{M^*}\j_{\n L})+\co(\j^2).
\eal
\ebxd
These again generalize the flat space identities \eqref{PV-non-conservation-laws} to a generic supergravity background. Notice that the expressions \eqref{Bs} for the symmetry breaking terms are fully off-shell. They can be obtained either from the corresponding symmetry transformations of the massive WZ model action, or equivalently using the relations \eqref{Bs-Os} in terms of functional derivatives with respect to the lowest components of the compensating multiplet. Since infinitesimal variations of the compensator fields $\wt\f$ and $\wt\c$ around the gauge-fixing values \eqref{old-minimal-gauge-fixing} are equivalent to infinitesimal Weyl, R-symmetry and S-supersymmetry transformations, only the symmetry breaking mass terms \eqref{WZ-action-mass} contribute.

\subsection{Local counterterms}
\label{sec:counterterms}

The local counterterms required to renormalize correlation functions of the FZ multiplet operators are local functionals of the background supergravity fields and preserve the symmetries of the regulated theory. In particular, the local counterterms $\wt{\mathscr{W}}_{\rm ct}[e,A,\j,M]$ that remove the UV divergences are a local functional invariant under the symmetries of old minimal supergravity and allow us to define  the generating function of renormalized correlation functions as 
\be\label{Wren}
\wt{\mathscr{W}}_{\rm ren}[e,A,\j,M]\equiv\wt{\mathscr{W}}[e,A,\j,M]+\wt{\mathscr{W}}_{\rm ct}[e,A,\j,M],
\ee
where $\wt{\mathscr{W}}[e,A,\j,M]$ is the generating function of regulated correlators.

The counterterms can be decomposed into terms of increasing mass dimension as
\be\label{counterterms}
\wt{\mathscr{W}}_{\rm ct}[e,A,\j,M]=\wt{\mathscr{W}}^{(1)}_{\rm ct}+\wt{\mathscr{W}}^{(2)}_{\rm ct}+\wt{\mathscr{W}}^{(3)}_{\rm ct}+\wt{\mathscr{W}}^{(4)}_{\rm ct},
\ee
each of which is separately invariant under the symmetries of old minimal supergravity. Dimensional analysis determines that for $n<4$, $\wt{\mathscr{W}}_{\rm ct}^{(n)}$ must be of order $4-n$ in the regulator mass, while $\wt{\mathscr{W}}^{(4)}_{\rm ct}$ is logarithmic. Finite counterterms are order zero in the PV masses and will be discussed separately below. In appendix \ref{sec:divergences} we show that, at least to 1-loop, there are no UV divergences proportional to odd powers of the PV mass and so only even mass dimension counterterms are required. 

At each mass level, the local counterterms can be determined by writing down the most general local expression involving the fields of old minimal supergravity using the mass dimensions $[\pa_\m]=[A_\m]=[M]=1$, $[\j_\m]=1/2$, and $[e^a_\m]=0$ of the building blocks, and imposing old minimal supersymmetry. However, in practice, most of the counterterms can be obtained directly from known results using conformal calculus. For example, although not required for the WZ model, the mass dimension 1 counterterm follows from the interaction Lagrangian \eqref{WZ-action-int} for the compensating chiral multiplet of old minimal supergravity. Gauge fixing the compensating multiplet as in \eqref{old-minimal-gauge-fixing} leads to a mass dimension 1 invariant action 
\bbxd
\vskip.4cm
\be
\label{OMlocal-1}
\ci_1=\;\int d^4x\,e\Big(M+M^*+\frac16\bar\j_\m\g^{\m\n}\j_{\n} +\co(\j^4)\Big).
\ee
\ebxd
This is the unique old minimal local invariant at mass dimension 1 and so
\be
\label{counterterms-1}
\wt{\mathscr{W}}^{(1)}_{\rm ct}=a_1\,\ci_1,
\ee
where the constant $a_1$ is cubic in the regulator mass, but vanishes for the WZ model.  

Similarly, the mass dimension 2 counterterm can be obtained from the free and massless WZ action \eqref{WZ-action-CS} for the compensating chiral multiplet using the gauge fixing condition \eqref{old-minimal-gauge-fixing}. This leads to the well known Poincar\'e supergravity action \cite{Ferrara:1978em,Stelle:1978ye,Fradkin:1978jq}    
\bbxd
\vskip.4cm
\be
\label{OMlocal-2}
\ci_2=\; \int d^4x\,e\Big(R[\o(e)]+\frac{8}{3}A^\m A_\m-6M^*M+i\e^{\m\n\r\s}\bar\j_\m\g^5\g_\n\mathscr{D}_\r\j_\s+\co(\j^4)\Big).
\ee
\ebxd
This is again the unique local invariant at mass dimension 2 and so
\be
\label{counterterms-2}
\wt{\mathscr{W}}^{(2)}_{\rm ct}=a_2\,\ci_2,
\ee
where the constant $a_2$ is quadratic in the PV masses and will be determined by canceling the corresponding UV divergences in the current multiplet correlators. 

The mass dimension 3 counterterm requires a bit more effort to determine using conformal calculus, but this is not necessary here since there are no UV divergences linear in the PV mass in the free and massless WZ model. Moving to mass dimension 4, 
there exist three local densities that are invariant under old minimal supergravity, given in full generality in \cite{Ferrara:1988qx}. Two of these correspond to the supersymmetrized Weyl squared and Euler densities and are in fact invariant under the full symmetries of conformal supergravity, while the third is only invariant under the symmetries of old minimal supergravity. The supersymmetrized Euler density is locally a total derivative and so we focus on the other two local invariants. 

The supersymmetric Weyl squared invariant is the conformal supergravity action 
\cite{Kaku:1977pa,Kaku:1977rk,Kaku:1978nz,Townsend:1979ki,Fradkin:1985am} 
\bbxd
\bal
\label{OMlocal-4}
\ci_4=&\;\int d^4x\,e\Big(W^2-\frac{8}{3}F_{\m\n}F^{\m\n}+8i\e^{\m\n\r\s}\bar\f_\m\g^5\g_\n D_\r\f_\s+4\bar\j^\m\g^\r\j^\s\nabla_\m P_{\r\s}\NO\\
&-2R^{\m\n}\Big(\bar\j^\r\g_{\r\n}\f_\m-\bar\j_\m\g_{\r\n}\f^\r-\frac{2}{3}g_{\m\n}\bar\j^\r\g_{\r\s}\f^\s+2\bar\j^\r\g_\n(D_{[\m}\j_{\r]}-\g_{[\m}\f_{\r]})\Big)\NO\\
&+\frac{4i}{3}F^{\r\s}\bar\j_\m(2\h^{[\m}_\r \h^{\n]}_\s-i\g^5\e^{\m\n}{}_{\r\s})\g^5\f_\n+\co(\j^4)\Big).
\eal
\ebxd
Notice that this action does not involve the auxiliary field $M$ of old minimal supergravity and it is separately invariant under the Q- and S-supersymmetry transformations of conformal supergravity. The second invariant at mass dimension 4 relevant for our analysis takes the form \cite{Ferrara:1978rk,Ferrara:1988qx}
\bbxd
\bal
\label{OMlocal-4-prime}
\ci_4'=&\;\int d^4x\,e\Big(R^2-36\pa_\m M\pa^\m M^*+16(\nabla^\m A_\m)^2+\frac{16}{3}R A_\r A^\r+36\bar\f_\m\g^\m \slashed{\mathscr D}\g^\n\f_\n\NO\\
&-12i\bar\f_\m\g^\m\slashed A\g^5\g^\n\f_\n+24\e^{\m\n\r\s}A_\m\bar\j_\n\mathscr D_\r\f_\s\NO\\
&+48i\bar\j_\r \g^5\slashed A\g^{(\r} \mathscr D^{\s)}\f_\s-48iA^\r\bar\j_\m\g^5\g^{\m\s}\mathscr D_{[\r}\f_{\s]}+24iA_\m\bar\j^\m\g^5\g^{\r\s}\mathscr D_\r\f_\s-24i\bar\j^\s\g^5\slashed A\slashed{\mathscr D}\f_\s\NO\\
&+8\big(g_{\n\r}A_\m\nabla^\m A_{\s}-g_{\n\r}A_{\s}\nabla^\m A_\m-A_\r\nabla_\n A_{\s}\big)\bar\j^\r\g^\n\j^{\s}\NO\\
&+\frac{16}{3}A_\m A^\m\bar\j^{\r}\g^{\s}\mathscr D_{[\r}\j_{\s]}+\frac{16}{3}A^\m\bar\j_\m\slashed A\g^{\r\s}\mathscr D_{\r}\j_{\s}-\frac{8}{3}A_\m A^\m\bar\j_\n\g^{\n\r\s}\mathscr D_{\r}\j_{\s}\NO\\
&-8A_\m A_\n\bar\j^\n\g^{\m\r\s}\mathscr D_\r\j_\s+16A^\r A_\m\bar\j_\n\g^{\m\n\s}\mathscr D_{[\r}\j_{\s]}+4A_\n A^\n\bar\j_\m\g^{\m\r\s}\mathscr D_\r\j_\s+\cdots\Big),
\eal
\ebxd
where the ellipses in this expression stand for terms that are not relevant for our analysis (not only $\co(\j^4)$ terms). The full expression can be found in eq.~(4.4) of \cite{Ferrara:1988qx}.\footnote{The notation of \cite{Ferrara:1988qx} translates to that of this paper under the following substitutions: 
\be
ie^{-1}\ve^{\m\n\r\s}\to \e^{\m\n\r\s},\quad A_\m\to 2 A_\m,\quad S\to \frac32 (M+M^*),\quad P\to \frac{3i}{2}(M-M^*),\quad \h\to\frac{i}{3}\slashed A\g^5-\frac12(MP_L+M^* P_R).  
\ee
Moreover, the notation for the covariant derivatives on spinors is reversed: 
$D_\m\to \mathscr D_\m,\quad \mathscr D_\m \to D_\m$, and the Riemann curvature is defined with opposite sign: $R^\m{}_{\n\r\s}\to -R^\m{}_{\n\r\s}$. } The general form of the local counterterms at mass dimension 4 therefore is
\be
\label{counterterms-4}
\wt{\mathscr{W}}^{(4)}_{\rm ct}=a_4\,\ci_4+a_4'\,\ci_4',
\ee
where the constants $a_4$ and $a_4'$ are logarithmic in the PV mass.

\subsection*{Finite local counterterms}

The structure of the local counterterms that cancel the UV divergences of FZ multiplet correlators is uniquely determined by the symmetries of old minimal supergravity. However, there remains the freedom of adding {\em finite} local counterterms, on top of those that cancel the UV divergences. Such finite local counterterms may or may not preserve the symmetries of the theory. For example, one could add an arbitrary linear combination of the two old minimal invariants \eqref{OMlocal-4} and \eqref{OMlocal-4-prime} with coefficients that are independent of the PV masses. Such a finite counterterm would preserve all symmetries of the regulated theory, with the coefficients parameterizing different supersymmetric renormalization schemes. As we discuss in section \ref{sec:correlators}, a specific finite counterterm of this form (given in \eqref{fin-counterterm-inv}) is required to bring all 2-point functions computed with our PV regulator to a form compatible with the corresponding conformal multiplet 2-point functions.

Finite local counterterms that do not preserve all symmetries may be used to move quantum anomalies from one symmetry to another. In particular, we claim that there exists a finite local counterterm, $\wt{\mathscr{W}}_{\rm fin}[e,A,\j,M]$, that can be added to the renormalized generating function, $\wt{\mathscr{W}}_{\rm ren}[e,A,\j,M]$, of the FZ multiplet, such that 
\bbxd
\vskip.25cm
\be\label{Wren-CS}
\mathscr{W}[e,A,\j]=\wt{\mathscr{W}}_{\rm ren}[e,A,\j,M]+\wt{\mathscr{W}}_{\rm fin}[e,A,\j,M],
\ee
\ebxd
where $\mathscr{W}[e,A,\j]$ depends only on the conformal supergravity fields, and the current operators 
\be\label{currents-CS}
\<\ct^\m_a\>=\<\wt\ct^\m_a\>_{\rm ren}+e^{-1}\frac{\d\wt{\mathscr{W}}_{\rm fin}}{\d e^a_\m},
\quad \<\cj^\m\>=\<\wt\cj^\m\>_{\rm ren}+e^{-1}\frac{\d\wt{\mathscr{W}}_{\rm fin}}{\d A_\m},\quad
\<\cq^\m\>=\<\wt\cq^\m\>_{\rm ren}+e^{-1}\frac{\d\wt{\mathscr{W}}_{\rm fin}}{\d\bar\j_\m},
\ee
satisfy the superconformal Ward identities \eqref{WardIDs}, with the superconformal anomalies given in \eqref{anomalies}. Moreover, since $\mathscr{W}[e,A,\j]$ depends only on the conformal supergravity fields, 
\be
\<\wt\co_M\>_{\rm ren}+e^{-1}\frac{\d\wt{\mathscr{W}}_{\rm fin}}{\d M}=0,\qquad \<\wt\co_{M^*}\>_{\rm ren}+e^{-1}\frac{\d\wt{\mathscr{W}}_{\rm fin}}{\d M^*}=0.
\ee
This implies that, for a classically superconformal theory, $\wt\co_M$ and $\wt\co_{M^*}$ are {\em ultralocal} operators. The finite local counterterm $\wt{\mathscr{W}}_{\rm fin}$, therefore, relates the conformal and FZ multiplets of any renormalized $\cn=1$ SCFT in four dimensions. The role of this counterterm on the way superconformal symmetry is broken at the quantum level in each of these multiplets will be discussed in detail in section \ref{sec:correlators}.

Since $\wt{\mathscr{W}}_{\rm ren}$ is invariant under old minimal supersymmetry, the relation \eqref{Wren-CS} implies that the counterterm $\wt{\mathscr{W}}_{\rm fin}$ transforms exactly as the conformal multiplet generating function $\mathscr{W}$, namely
\bbxd
\vskip.4cm
\be\label{Wfin-def}
\d_\ve^{\rm om}\wt{\mathscr W}_{\rm fin}=-\int d^4x\, e\,\bar\ve\Big[\ca_Q+\frac12\Big(M P_L+M^*P_R-\frac{2i}{3}\slashed A\g^5\Big)\ca_S\Big],
\ee
\ebxd
where the Q- and S-supersymmetry anomalies are given in \eqref{anomalies}. This may be taken as the defining property of $\wt{\mathscr{W}}_{\rm fin}$ and can be expressed in the form of a functional partial differential equation as
\bbxd
\bal
&D_\m \Big(\frac{\d\wt{\mathscr{W}}_{\rm fin}}{\d\bar\j_\m}\Big)-\frac12\g^a\j_\m \frac{\d\wt{\mathscr{W}}_{\rm fin}}{\d e^a_\m}-\frac{3i}{4}\g^5\f_\m\frac{\d\wt{\mathscr{W}}_{\rm fin}}{\d A_\m}\NO\\
&+\frac12\Big(M P_L+M^*P_R-\frac{2i}{3}\slashed A\g^5\Big)\Big(\g_\m \frac{\d\wt{\mathscr{W}}_{\rm fin}}{\d\bar\j_\m}-\frac{3i}{4}\g^5\j_\m\frac{\d\wt{\mathscr{W}}_{\rm fin}}{\d A_\m}\Big)\NO\\
&+\g^\m\Big(\frac{i}{3}\slashed A\j_{\m R}+\frac12 M\j_{\m L}+\f_{\m L}\Big)\frac{\d\wt{\mathscr{W}}_{\rm fin}}{\d M}+\g^\m\Big(-\frac{i}{3}\slashed A\j_{\m L}+\frac12 M^*\j_{\m R}+\f_{\m R}\Big)\frac{\d\wt{\mathscr{W}}_{\rm fin}}{\d M^*}=\NO\\
&e\ca_Q+\frac12\Big(M P_L+M^*P_R-\frac{2i}{3}\slashed A\g^5\Big)e\ca_S.
\eal
\ebxd
This can be used in order to systematically determine the general form of the finite counterterm $\wt{\mathscr{W}}_{\rm fin}$, but we will not pursue this problem here. Instead, in section \ref{sec:correlators} we determine certain parts of $\wt{\mathscr{W}}_{\rm fin}$ that are relevant for the specific current correlators we consider.

\section{Ferrara-Zumino and conformal multiplet Ward identities at one loop}
\label{sec:correlators}

We are now in a position to prove our main result. We begin by demonstrating that the FZ multiplet supersymmetry remains non anomalous at one loop, at least up to the level of 4-point functions. However, R-symmetry, conformal invariance, as well as the original Q- and S-supersymmetry of the conformal multiplet are broken explicitly. Evaluating the local symmetry breaking terms in the corresponding Ward identities, we show that there exists a local counterterm that removes the explicit breaking from all correlators, except for terms that reproduce exactly the superconformal anomalies of the conformal multiplet discussed in section \ref{sec:WardIDs}.       

The model independent form of the Ward identities suitable for discussing quantum anomalies involves current multiplet correlators defined through functional differentiation with respect to the corresponding sources, i.e. the background supergravity fields. Such correlators are indicated throughout the paper by the wide brackets $\<\cdot\>$. As we have already mentioned and discuss in more detail in appendix \ref{sec:seagulls}, these differ from correlation functions defined by operator insertions in the path integral by seagull terms, which encode the dependence of the current multiplet operators on the background supergravity fields. Individual correlators defined via path integral operator insertions, denoted by $\lb\cdot\rb$, generically contain additional UV divergences and cannot be renormalized by local counterterms that depend only on the background supergravity fields. These model dependent UV divergences cancel in the combinations of path path integral correlators corresponding to correlation functions defined by functional differentiation. The remaining UV divergences are model independent and can be regulated and renormalized using local counterterms that involve only background supergravity fields.  

Although we will organize the analysis of the Ward identities in terms of universal correlation functions defined through functional differentiation, actual Feynman diagrams compute correlators corresponding to path integral operator insertions. The dictionary between the two definitions of correlators is provided explicitly in appendix \ref{sec:seagulls} and is used extensively in the subsequent analysis. Another subtlety in the evaluation of the 1-loop correlation functions concerns the presence of an overall phase factor due to the definition of the generating function of connected correlators in \eqref{connected-generator}, and the normalization of the current multiplet operators that is chosen to coincide with derivatives of the classical action. Namely, for connected path integral $n$-point functions
\be\label{phase-factor}
\lb\co_1\co_2\cdots\co_n\rb\, \sim -i\int[\tx d\{\F\}] \Big(i\frac{\d S}{\d J_1}\Big)\cdots\Big(i\frac{\d S}{\d J_n}\Big)e^{iS[\{\F\};\{J\}]}\sim i^{n-1}\times \text{Feynman diagrams},
\ee
where the Feynman diagrams are evaluated using standard Wick contractions.  

Finally, the following generic observations help organize and simplify the computation of the current multiplet correlation functions. Firstly, it is useful to notice that the fermion propagators in   
\eqref{WZ-propagators} and \eqref{PV-propagators} may be expressed in terms of their bosonic counterparts as 
\be\label{fermion-propagators}
P_{\c}(p)=\frac{-i\slashed p}{2\a^2}P_{\f}(p),\qquad P_{\l_1}(p)=\frac{-i\slashed p+m_1}{2\a^2}P_{\vf_1}(p),\qquad P_{\l_2}(p)=\frac{-i\slashed p+m_2}{2\a^2}P_{\vf_2}(p).
\ee
Using these relations, supersymmetry together with the PV mass condition \eqref{PV-mass-condition} give rise to certain universal structures that render the 1-loop integrals UV finite. In particular, in the computation of all $n$-point functions one encounters the two homogeneous propagator polynomials 
\bal\label{Gs}
G^{(n)}_1(q_1,q_2,\ldots,q_n)\equiv&\;P_{\f}(q_1)\cdots P_{\f}(q_n)+P_{\vf_2}(q_1)\cdots P_{\vf_2}(q_n)-2P_{\vf_1}(q_1)\cdots P_{\vf_1}(q_n),\NO\\
G^{(n)}_2(q_1,q_2,\ldots,q_n)\equiv&\;P_{\vf_2}(q_1)\cdots P_{\vf_2}(q_n)-P_{\vf_1}(q_1)\cdots P_{\vf_1}(q_n).
\eal
Imposing the mass condition \eqref{PV-mass-condition}, the large momentum behavior of these polynomials  is 
\be
G^{(n)}_1(q,\ldots,q)\sim q^{-2n-4},\qquad G^{(n)}_2(q,\ldots,q)\sim q^{-2n-2}.
\ee 
In appendix \ref{sec:divergences} we show that this suffices to ensure the UV finiteness of all relevant  correlators.

\subsection{Ward identities for the Ferrara-Zumino multiplet}
\label{sec:FZ-WardIDs}

Since the PV regulator introduced in section \ref{sec:PV} manifestly preserves old minimal supersymmetry, it follows that the FZ multiplet Ward identities \eqref{OM-susy-WID-2pt}-\eqref{OM-susy-WID-4pt}, as well as \eqref{S-susy-WIDs-2pt-FZ}-\eqref{S-susy-WIDs-4pt-FZ} and \eqref{R-symmetry-WIDs-JJ-FZ}-\eqref{R-symmetry-WIDs-QQJJ-FZ} for the broken symmetries, hold at the quantum level provided all relevant Feynman diagrams are properly regulated. In   appendix \ref{sec:divergences} we verify explicitly that the UV divergences of all Feynman diagrams required to compute correlation functions that involve only FZ multiplet operators are canceled at one loop. For correlators with insertions of the symmetry breaking operators \eqref{Bs} that involve PV fields only, UV finiteness is demonstrated in appendix \ref{sec:local-correlators}. Hence, all FZ multiplet Ward identities for the regulated theory remain valid at one loop. This remains true for the renormalized theory, since all divergences of FZ multiplet correlators as the PV masses are sent to infinity are canceled by the supersymmetric counterterms \eqref{counterterms-2}-\eqref{counterterms-4}, with the coefficients $a_2$ and $a_4$ specified respectively in \eqref{a2} and \eqref{a4}.   

Comparing the FZ multiplet Ward identities \eqref{S-susy-WIDs-2pt-FZ}-\eqref{S-susy-WIDs-4pt-FZ}, \eqref{R-symmetry-WIDs-JJ-FZ}-\eqref{R-symmetry-WIDs-QQJJ-FZ} and \eqref{OM-susy-WID-3pt-simple}-\eqref{OM-susy-WID-4pt-simple} with the corresponding ones for the conformal current multiplet in section \ref{sec:WardIDs}, one sees that they coincide exactly, except for the contact terms related to superconformal anomalies in the conformal multiplet Ward identities, and the terms involving the symmetry breaking operators $\cb_W$, $\cb_R$ and $\cb_S$ in the FZ multiplet ones. In order to provide an independent perturbative calculation of the anomalies, therefore, it suffices to evaluate all correlation functions with insertions of the symmetry breaking operators $\cb_W$, $\cb_R$ and $\cb_S$. We will see, however, that evaluating these correlation functions does not immediately reproduce the contact terms of the conformal multiplet Ward identities. For example, the 2-point functions $\<\cb_R\wt\cj\>$ and $\<\cb_S\wt{\bar\cq}{}\,\>$ turn out to be non zero, while there are no contact terms in the corresponding conformal multiplet Ward identities arising due the superconformal anomalies. In the next subsection we show that all 2-point functions of FZ multiplet operators can be mapped to conformal multiplet 2-point functions using {\em supersymmetric} local counterterms. Moreover, there exists a finite local counterterm that depends on the old minimal fields (but not supersymmetric) that maps all contact terms in 3- and higher-point functions with $\cb_W$, $\cb_R$ and $\cb_S$ insertions to precisely those of the conformal multiplet in section \ref{sec:WardIDs} due to the superconformal anomalies. The explicit 1-loop computation of all contact terms arising from FZ multiplet correlation functions with $\cb_W$, $\cb_R$ and $\cb_S$ insertions is carried out in appendix \ref{sec:local-correlators}.

\subsection{Finite counterterm and the Ward identities for the conformal multiplet}
\label{sec:Conformal-WardIDs}

Having obtained all breaking terms we now proceed to add finite counterterms to restore the conformal multiplet Ward identities to the extend possible. For the correlators we analyze the relevant counterterms $\wt{\mathscr{W}}^*_{\rm fin}\subset \wt{\mathscr{W}}_{\rm fin}$ are:
\begin{align}
\label{fin-counterterm}
&\wt{\mathscr{W}}^*_{\rm fin}=-\frac{1}{2^43^3\pi^2}\bigg[\frac{3i}{8} \e^{\m\n\r\s}\bar{\j}_\m \g_\s\g^5 \mathscr{D}_\r\mathscr{D}^2\j_\n+ A^\r \nabla^2  A_\r -\frac{1}{2}R A_\r A^\r+\frac{9}{4}\pa_\m M\pa^\m M^*\NO\\
&-iA^\r\big( \pa^{[\n}\bar{\j}^{\m]}\g_\m\g^5 \pa_\r\j_\n+\pa_{[\r}\bar{\j}^{\m]}\g_\m\g^5 \pa_\n\j^\n+\pa^{[\m}\bar{\j}^{\n]}\g_\r\g^5 \pa_\n\j_\m	+\pa^{[\m}\bar{\j}^{\n]}\g_\m\g^5 \pa_\n\j_\r+2\pa_\n\pa_{[\r}\bar{\j}_{\m]}\g^\m\g^5 \j_\n\big)\NO\\
&+\frac{1}{8}A_\r\Big(7\e^{\r\t\m\n}\pa_\m\bar{\j}^\s\g_\t\pa_\n\j_\s+2\e^{\r\s\t\m}\pa_\m\bar{\j}_\s\g_\t\pa_\n\j^\n+14\e^{\r\s\t\m}\pa_\n\bar{\j}_\s\g_\t\pa_\m\j^\n+7\e^{\r\t\m\n}\pa_\s\bar{\j}_\m\g_\t\pa^\s\j_\n\NO\\
&+4\e^{\r\s\t\m}\pa_\n\pa_\m\bar{\j}_\s\g_\t\j^\n-2\e^{\r\s\m\n}\pa_\m\bar{\j}_\s\g_\t\pa_\n\j^\t-2\e^{\r\s\m\n}\pa_\n\bar{\j}_\s\g_\t\pa^\t\j_\m\Big)\NO\\
&+\frac{1}{3}A_\m A^\n\pa_\t\bar{\j}^\m\g^\t\j_\n -\frac{5}{12}A_\n A^\n\pa_\t\bar{\j}^\m\g^\t\j_\m+\frac{1}{2}\pa_\m A^\t A^\n \bar{\j}^\m \g_\t\j_\n\NO\\&+\frac{1}{3}A_\m A^\n\bar{\j}^\m\g^\t\pa_\n\j_\t-\frac{1}{4}A_\n A^\n\pa_\m\bar{\j}^\m\g^\t\j_\t-\frac{5}{12}A_\n A^\n\bar{\j}^\m\g^\t\pa_\m\j_\t\NO\\
&
-\frac{1}{2}\pa_\t A_\m A_\n \bar{\j}^\m\g^\t\j^\n-A^\t A^\m\pa_\m\bar{\j}^\n\g_\t\j_\n+A^\t A^\n\pa_\m\bar{\j}^\m\g_\t\j_\n
+\frac{3}{2}\pa_\m A^\n A^\t \bar{\j}^\m \g_\t\j_\n\NO\\
&
+\frac{5i}{24}A_\r A_\s \e^{\s\m\n\t}\bar{\j}_\m\g^\r\g^5\pa_\t\j_\n+\frac{i}{4}\pa_\n A_\s A_\t \e^{\s\m\r\t}\bar{\j}_\m\g_\r\g^5\j^\n\NO\\
&
+\frac{23i}{24} A^\s A_\t \e^{\t\m\n\r}\pa_\s\bar{\j}_\m\g_\r\g^5\j_\n	+\frac{i}{8}\pa_\n A_\s A_\t \e^{\t\s\r\n}\bar{\j}_\m\g_\r\g^5\j^\m\NO\\
&
-\frac{5i}{24} A^\s A_\t \e^{\t\m\r\n}\pa_\n\bar{\j}_\m\g_\r\g^5\j_\s-\frac{23i}{24}A_\t A_\s \e^{\t\m\r\n}\bar{\j}_\m\g_\r\g^5\pa_\n\j^\s\bigg].
\end{align}	
The terms quadratic in the sources renormalize (finitely) the 2-point functions of the R-current and  the supercurrent, which are now conserved (and gamma-trace conserved for the supercurrent).  The finite renormalization of the 2-point functions now enters in the 3-point function Ward identities and the contribution of the cubic counterterm is such that the 3-point function $\<\cq\bar\cq\cj\>$
preserves Q-supersymmetry and R-symmetry and the anomaly of S-supersymmetry exactly matches that presented in section \ref{sec:WardIDs}. Similarly, the counterterms ensure that the  $\<\ct\cj\cj\>$ correlator preserves R-symmetry and has the correct trace anomaly. Finally, moving to the Ward identity for $\<\cq\bar\cq\cj\cj\>$, collecting all finite terms from the finite renormalisation of the lower-point functions and the quartic contribution of the finite counterterm one finds that R-symmetry Ward identity is non-anomalous but both Q- and S-supersymmetry are anomalous with the anomalies exactly matching those of section \ref{sec:WardIDs}. It should be emphasized that the local counterterms \eqref{fin-counterterm} deduced from the correlation functions we computed are ambiguous up to an arbitrary multiple of the superconformal invariant $\ci_4$ in \eqref{OMlocal-4}. This ambiguity corresponds to the usual scheme dependence of the conformal multiplet and does not affect the form of the anomalies. In writing down \eqref{fin-counterterm} we have made an implicit choice for the coefficient of the superconformal invariant $\ci_4$.  

In fact, all terms in $\wt{\mathscr{W}}^*_{\rm fin}$ that contribute to 2-point functions are actually supersymmetric. Using the form of the supersymmetric invariants $\ci_4$ and $\ci_4'$ respectively in \eqref{OMlocal-4} and \eqref{OMlocal-4-prime}, we get
\bal
\label{fin-counterterm-inv}
&-\frac{1}{2^83^2\p^2}\Big(\ci_4-\frac13\ci_4'\Big)=\NO\\
&-\frac{1}{2^83^2\p^2}\int d^4x\,e\Big(R_{\m\n\r\s}R^{\m\n\r\s}-2R_{\m\n}R^{\m\n}+\frac{16}{3}A_\m\Big(\nabla^2 A^\m-\frac12 RA^\m-2P^{\m\n} A_\n\Big)+12\pa_\m M\pa^\m M^*\NO\\
&+8i\e^{\m\n\r\s}\bar\f_\m\g^5\g_\n D_\r\f_\s-12\bar\f_\m\g^\m \slashed D\g^\n\f_\n+16i\bar\f_\m\g^\m \slashed A\g^5\g^\n\f_\n-8\e^{\m\n\r\s}A_\m\bar\j_\n\mathscr D_\r\f_\s\NO\\
&-16i\bar\j_\r \g^5\slashed A\g^{(\r} \mathscr D^{\s)}\f_\s+16iA^\r\bar\j_\m\g^5\g^{\m\s}\mathscr D_{[\r}\f_{\s]}-8iA_\m\bar\j^\m\g^5\g^{\r\s}\mathscr D_\r\f_\s+8i\bar\j^\s\g^5\slashed A\slashed{\mathscr D}\f_\s\NO\\
&+\frac{4i}{3}F^{\r\s}\bar\j_\m(2\h^{[\m}_\r \h^{\n]}_\s-i\g^5\e^{\m\n}{}_{\r\s})\g^5\f_\n-\frac{8}{3}\big(g_{\n\r}A_\m\nabla^\m A_{\s}-g_{\n\r}A_{\s}\nabla^\m A_\m-A_\r\nabla_\n A_{\s}\big)\bar\j^\r\g^\n\j^{\s}\NO\\
&-\frac{16}{9}A_\m A^\m\bar\j^{\r}\g^{\s}\mathscr D_{[\r}\j_{\s]}-\frac{16}{9}A^\m\bar\j_\m\slashed A\g^{\r\s}\mathscr D_{\r}\j_{\s}+\frac{8}{9}A_\m A^\m\bar\j_\n\g^{\n\r\s}\mathscr D_{\r}\j_{\s}\NO\\
&+\frac{8}{3}A_\m A_\n\bar\j^\n\g^{\m\r\s}\mathscr D_\r\j_\s-\frac{16}{3}A^\r A_\m\bar\j_\n\g^{\m\n\s}\mathscr D_{[\r}\j_{\s]}-\frac{4}{3}A_\n A^\n\bar\j_\m\g^{\m\r\s}\mathscr D_\r\j_\s+\cdots\Big),
\eal
where the ellipses stand for terms that do not contribute to the correlators we are considering. While the coefficient of $\ci_4'$ is unambiguous, as we mentioned above, the coefficient of $\ci_4$ corresponds to the superconformal scheme dependence of the conformal multiplet and is a priory arbitrary. The coefficient in \eqref{fin-counterterm-inv} is chosen to match the choice made implicitly in \eqref{fin-counterterm}.

It is clear that the contribution of this supersymmetric counterterm to the 2-point functions $\<\cj\cj\>$ and $\<\co_M\co_{M^*}\>$ agrees with the corresponding ones in \eqref{fin-counterterm}. Verifying that the contributions to the 2-point function $\<\bar\cq\cq\>$ agree requires a bit more work. Using the identities 
\be
\g^\m\f_\m=-\frac13\g^{\r\s}D_\r\j_\s,
\ee
and 
\be
i\e^{\m\n\r\s}\bar\f_\m\g^5\g_\n D_\r\f_\s=2\bar\f_\m\g^\m D^\n\f_\n-\bar\f_\m \slashed D\f^\m-\bar\f_\m\g^\m \slashed D\g^\n\f_\n,
\ee
one finds that, up to a total derivative, 
\bal
&8i\e^{\m\n\r\s}\bar\f_\m\g^5\g_\n D_\r\f_\s+12\bar\f_\m\g^\m \slashed D\g^\n\f_\n=16\bar\f_\m\g^\m D^\n\f_\n-8\bar\f_\m \slashed D\f^\m-20\bar\f_\m\g^\m \slashed D\g^\n\f_\n\NO\\
=&-\frac13\bar\j_\r\big(3\g^{\r\s}\g^{\k\l\t}+\g^{\t\r\s}\g^{\k\l}-6\g^\r\g^\t\g^{[\k} g^{\l]\s}+6\g^\s\g^\t\g^{[\k} g^{\l]\r}\big)D_\s D_\t D_\k\j_\l.
\eal
Isolating the part that is totally symmetric in the indices $\s$, $\k$ and $\t$ it is straightforward to show that the contributions of \eqref{fin-counterterm} and \eqref{fin-counterterm-inv} to the 2-point function $\<\bar\cq\cq\>$ coincide up to a local term that is divergence and gamma-trace free. The supersymmetric counterterm \eqref{fin-counterterm-inv} should also bring the 2-point function of the stress tensor to its conformal multiplet form, but we cannot verify this directly here because our PV regulator does not regulate this 2-point function. Assuming this is indeed the case, we deduce that   
\be
\wt{\mathscr{W}}^*_{\rm fin}=-\frac{1}{2^83^2\p^2}\Big(\ci_4-\frac13\ci_4'\Big)+\wt{\mathscr{W}}^{**}_{\rm fin},
\ee
where the symmetry breaking term $\wt{\mathscr{W}}^{**}_{\rm fin}$ contributes to 3- and higher-point correlation functions. As in the conformal multiplet, therefore, all symmetry breaking terms in the FZ multiplet appear first in 3-point functions. 

This finishes the construction of part of $\wt{\mathscr{W}}_{\rm fin}$. In principle, performing a similar analysis  involving (a suitable set of) other correlators one should be able to construct the full counterterm $\wt{\mathscr{W}}_{\rm fin}$ (though this would be tedious in practice).
We should emphasise that the existence of a finite counterterm $\wt{\mathscr{W}}_{\rm fin}$ satisfying \eqref{Wren-CS} does not imply that Q-supersymmetry in conformal supergravity is non anomalous. Indeed, the symmetry algebra of conformal supergravity implies that there exists no local counterterm -- {\em in conformal supergravity} -- that removes the Q-supersymmetry anomaly, without breaking diffeomorphism and/or Lorentz invariance \cite{Papadimitriou:2019gel}. The existence of $\wt{\mathscr{W}}_{\rm fin}$ instead implies that there exists a field-dependent combination of Q- and S-supersymmetry that is non anomalous. No such combination can be a symmetry of pure conformal supergravity, however, since the algebra does not close off-shell without additional auxiliary fields. The unique combination of Q- and S-supersymmetry that is non anomalous is precisely the old minimal supersymmetry \eqref{old-minimal-susy} transformation. Closure of the algebra requires the additional auxiliary field $M$ of old minimal supergravity \cite{Ferrara:1978em,Stelle:1978ye,Fradkin:1978jq}.         

While the existence of a finite counterterm $\wt{\mathscr{W}}_{\rm fin}$ does not contradict the results of \cite{Papadimitriou:2017kzw,Papadimitriou:2019gel,Katsianis:2019hhg}, it may be in agreement with the superspace analysis in \cite{Kuzenko:2019vvi}. In particular, $\wt{\mathscr{W}}_{\rm ren}[e,A,\j,M]$ may be identified with what was termed the `minimal Q-supersymmetric scheme', while $\mathscr{W}[e,A,\j]$ was referred to as the `Wess-Zumino scheme' in \cite{Kuzenko:2019vvi}.\footnote{We note that in this respect there is a similarity with an analogous discussion of the supersymmetry anomaly related to gauge/flavor anomalies, where the anomaly is manifest in the WZ gauge \cite{Closset:2019ucb, Kuzenko:2019vvi}. There are however also important differences. For example, the WZ gauge for the gauge multiplet is obtained by eliminating the so called ‘gauge away’ fields, while the relation between conformal and old minimal supergravity is more subtle, since they do not differ only by gauge away fields. A detailed discussion of these issues is left for future work.} However, a more detailed comparison between the two analyses, including the general form of the finite counterterm $\wt{\mathscr{W}}_{\rm fin}$, is beyond the scope of the present work.

\section{Discussion}
\label{discussion}

We have presented a comprehensive analysis of the 1-loop computation of supersymmetry anomalies in the free and massless WZ model. As for any $\cn=1$ SCFT in four dimensions, the renormalized theory admits both a conformal and a Ferrara-Zumino multiplet of currents, with the latter inherited from the regulated theory.\footnote{Other multiplets such as the R- and S-multiplets are also admissible but were not considered in this paper. The supersymmetry anomaly of the R-multiplet was studied recently in \cite{An:2019zok,Papadimitriou:2019yug}. See also \cite{Gates:1981yc}. Like the FZ multiplet, the S-multiplet does not suffer from a supersymmetry anomaly.} The two multiplets are related by a set of local counterterms that shift the anomalies between different symmetries. The conformal multiplet possesses the standard superconformal anomalies of $\cn=1$ SCFTs. In particular, R-symmetry is anomalous because of the standard triangle diagram, but both Q- and S-supersymmetries are necessarily also anomalous. On the contrary, the Poincar\'e supersymmetry of the FZ multiplet (which corresponds to a specific field-dependent linear combination of the Q- and S-supersymmetries of the conformal multiplet) is non anomalous, but R-symmetry and S-supersymmetry are explicitly broken. The FZ multiplet is more natural if one wishes to view the massless model as the zero mass limit of a massive WZ model, while the conformal multiplet is more natural if one wishes to view the massless WZ model as an example of an $\cn=1$ SCFT. Indeed, in the context of the AdS/CFT correspondence only the conformal multiplet is available and it is in this context that the anomaly was first discovered \cite{Papadimitriou:2017kzw}. 

Our analysis focused on the $\<\cq\bar\cq\cj\cj\>$ correlator, which is where the supersymmetry anomaly first arises in flat space. We demonstrated that, if the currents belong to the FZ multiplet, this 4-point function (as well as all lower-point functions entering in the supersymmetry Ward identity) is regulated by a suitable PV regulator that preserves old minimal supersymmetry. Hence the FZ multiplet supersymmetry Ward identity for this correlator is non anomalous. The PV regulator we used does not regulate all correlation functions of the FZ multiplet (for example it does not regulate all correlators that enter in the supersymmetry Ward identity of $\<\cq\bar\cq\cj\ct\>$, which is expected to be anomalous for the conformal multiplet), but there is hardly any doubt that such a regulator exists. Identifying it remains an interesting open problem, since it would allow one to demonstrate explicitly that old minimal supersymmetry is preserved at the quantum level for all FZ multiplet correlators.   
The second main aspect of our analysis was demonstrating the existence of local counterterms that interpolate between the FZ and superconformal multiplets. In the present analysis we do this at the level of correlation functions, but it would be prohibitively complicated and inefficient to try to generalize this to all current multiplet correlators. The general form of the local counterterm that interpolates between conformal and old minimal supergravity  may be determined instead directly from the structure of the solution of the WZ consistency conditions, which again we hope to address in future work.      

The presence of a supersymmetry anomaly in the conformal and R-multiplet \cite{An:2019zok,Papadimitriou:2019yug} is an important caveat one should keep in mind in the context of supersymmetric localization, especially when the results are compared with holographic computations, as noted in \cite{Papadimitriou:2017kzw,An:2017ihs,An:2018roi,Papadimitriou:2019gel} in relation to the analysis of \cite{Genolini:2016ecx}. In particular, in the presence of anomalies, physical observables depend on the choice of current multiplet and one should make sure that only results specific to a given multiplet are compared. Given that different multiplets are often used for field theory and holographic computations, failing to do so may result in a superficial mismatch.  We anticipate that a local counterterm analogous to that relating the conformal and FZ multiplets interpolates between the R-multiplet, which couples to new minimal supergravity, and the S-multiplet, corresponding to 16+16 supergravity, enabling one to remove the supersymmetry anomaly of the R-multiplet. Determining this counterterm would be particularly interesting for supersymmetric localization applications.

Since current multiplets describing SCFTs are related by finite local counterterms, such counterterms can be used to match the computation of physical observables using different multiplets. Indeed, it was through the identification of a non-covariant local counterterm (specific to a class of rigid supersymmetric backgrounds) that the authors of \cite{Genolini:2016ecx} managed to reconcile their holographic computation with the expected field theory result. Understanding the general structure of supersymmetry anomalies in different multiplets allows one to explicitly determine the local counterterms that interpolate between them. Of course, such counterterms are not unique since it is always possible to add further `trivial' local counterterms that preserve all the symmetries. In particular, the local counterterms interpolating between the conformal and FZ multiplets that we determined here through the 1-loop computation may agree with the superspace results in \cite{Kuzenko:2019vvi} up to such trivial terms. 

Finally, in this paper supergravity was viewed as non-dynamical and it would be interesting to extend the analysis to include dynamical supergravity. Our results imply that only the FZ multiplet can be consistently coupled to (old minimal) dynamical supergravity, since the Poincar\'e supersymmetry of old minimal supergravity is non anomalous. This is in line with earlier work \cite{Gates:1981yc} where it was argued that quantum anomalies in the matter sector require the use of old minimal supergravity. However, the conformal multiplet that suffers from a supersymmetry anomaly may still be coupled to dynamical supergravity in the context of effective field theory \cite{Cecotti:1987nw,Preskill:1990fr,LopesCardoso:1991ifk,Ferrara:1993yj,Ferrara:1996wv}. In that context, the anomalies are canceled either by fields with a mass above the cut-off through a generalized Green-Schwarz mechanism, or more generally by supersymmetric anomaly inflow \cite{inflow}.

\section*{Acknowledgments}

We thank Sergei Kuzenko, Peter van Nieuwenhuizen, Martin Ro\v{c}ek, Adam Schwimmer, Nati Seiberg, Misha Shifman, Stefan Theisen, Arkady Vainshtein and Peter West for insightful discussions. GK would like to thank Manthos Karydas for helpful discussions and Stanislav Schmidt for sharing expertise in Mathematica. GK was supported by the Onassis Foundation - Scholarship ID: F ZL 040-2/2019-2020. The work of IP is supported by a KIAS Individual Grant (PG064402) at the Korea Institute for Advanced Study. KS and MMT are supported in part by the Science and Technology Facilities Council (Consolidated Grants
ST/P000711/1 and ST/T000775/1). We acknowledge use of the symbolic manipulation package FeynCalc \cite{MERTIG1991345,Shtabovenko:2016sxi,Shtabovenko:2020gxv}.

\appendix

\renewcommand{\theequation}{\Alph{section}.\arabic{equation}}

\setcounter{section}{0}

\section*{Appendices}
\setcounter{section}{0}

\section{Spinor conventions and identities}
\label{sec:conventions}

We largely follow the spinor conventions of \cite{Freedman:2012zz}. We use the Minkowski metric $\h=\diag(-1,1,1,1)$ and the Levi-Civita symbol $\ve_{\m\n\r\s}=\pm 1$ satisfies $\ve_{0123}=1$. This is related to the Levi-Civita {\em tensor} as $\e_{\m\n\r\s}=\sqrt{-g}\;\ve_{\m\n\r\s}=e\;\ve_{\m\n\r\s}$, where $e\equiv \det(e^a_\m)$ is the determinant of the vierbein. We also use the convention that complex conjugation reverses the order of Grassmann fields (spinors or scalars), e.g. $(AB)^*\equiv B^*A^*$.

\paragraph{Gamma matrices} The gamma matrices satisfy the Hermiticity properties 
\be
\g^{\m\dag}=\g^0\g^\m\g^0,\qquad \g^{5\dag}=\g^5,
\ee
where the chirality matrix in four dimensions is given by
\be
\g^5=i\g_0\g_1\g_2\g_3.
\ee
The antisymmetrized products of gamma matrices are defined as
\be
\g^{\m_1\m_2\ldots\m_n}\equiv\g^{[\m_1}\g^{\m_2}\cdots\g^{\m_n]},
\ee
where antisymmetrization with weight one in understood, e.g. $\g^{\m\n}=\frac12[\g^\m,\g^\n]$.

The following is a list of identities in $d$ dimensions that the antisymmetrized products of gamma matrices satisfy, several of which we use repeatedly in this paper (see also section 3 of \cite{Freedman:2012zz}):  
\bal\label{d-gamma-ids} 
\g^{\m\n\r}&=\frac12\{\g^\m,\g^{\n\r}\},\NO\\
\g^{\m\n\r\s}&=\frac12[\g^\m,\g^{\n\r\s}],\NO\\
\g^{\m\n}\g_{\r\s}&=\g^{\m\n}{}_{\r\s}+4\g^{[\m}{}_{[\s}\d^{\n]}{}_{\r]}+2\d^{[\m}{}_{[\s}\d^{\n]}{}_{\r]},\NO\\
\g_{\m}\g^{\n_1\ldots\n_p}&=\g_\m{}^{\n_1\ldots\n_p}+p\d ^{[\n_1}_{\m}\g^{\n_2\ldots\n_p]},\NO\\
\g^{\n_1\ldots\n_p}\g_{\m}&=\g^{\n_1\ldots\n_p}{}_{\m}+p\g^{[\n_1\ldots\n_{p-1}}\d ^{\n_p]}_{\m},\NO\\
\g^{\m\n\r}\g_{\s\t}&=\g^{\m\n\r}{}_{\s\t}+6\g^{[\m\n}{}_{[\t}\d ^{\r]}{}_{\s]}+6\g^{[\m}\d^\n{}_{[\t}\d^{\r]}{}_{\s]},\NO\\
\g^{\m\n\r\s}\g_{\t\l}&=\g^{\m\n\r\s}{}_{\t\l}+8\g^{[\m\n\r}{}_{[\l}\d^{\s]}{}_{\t]}+12\g^{[\m\n}\d^\r{}_{[\l}\d^{\s]}{}_{\t]},\NO\\
\g^{\m\n\r}\g_{\s\t\l}&=\g^{\m\n\r}{}_{\s\t\l}+9\g^{[\m\n}{}_{[\t\l}\d^{\r]}{}_{\s]}+18\g^{[\m}{}_{[\l}\d^\n{}_{\t}\d^{\r]}{}_{\s]}+6\d^{[\m}{}_{[\l}\d^\n{}_{\t}\d^{\r]}{}_{\s]},\NO\\
\g^{\m_1\ldots\m_r\n_1\ldots\n_s}\g_{\n_s\ldots\n_1}&=\frac{(d-r)!}{(d-r-s)!}\g^{\m_1\ldots\m_r},\NO\\
\g^{\m\r}\g_{\r\n}&=(d-2)\g^{\m}{}_\n+(d-1)\d^\m_\n,\NO\\
\g^{\m\n\r}\g_{\r\s}&=(d-3)\g^{\m\n}{}_\s+2(d-2)\g^{[\m}\d^{\n]}{}_\s,\NO\\
\g_{\m\n}\g^{\n\r\s}&=(d-3)\g_{\m}{}^{\r\s}+2(d-2)\d^{[\r}_\m\g^{\s]},\NO\\
\g^{\m\n\l}\g_{\l\r\s}&=(d-4)\g^{\m\n}{}_{\r\s}+4(d-3)\g^{[\m}{}_{[\s}\d^{\n]}{}_{\r]}+2(d-2)\d^{[\m}{}_{[\s}\d^{\n]}{}_{\r]},\NO\\
\g_{\m\r}\g^{\r\s\t}\g_{\t\n}&=(d-4)^2\g_\m{}^\s{}_\n+(d-4)(d-3)\big(\g_\m\d_\n^\s-\g^\s g_{\m\n}\big)\NO\\
&\hskip0.5cm+(d-3)(d-2)\d_\m^\s\g_\n-(d-3)\g^\s\g_{\m\n},\NO\\
\g_\r \g^{\m_1\m_2\ldots \m_p}\g^\r&=(-1)^p(d-2p)\g^{\m_1\m_2\ldots \m_p}.
\eal 

For $d=4$ specifically, we have the gamma matrix identities  
\bal\label{d=4-gamma-ids}
&\g^\r\g^\m\g^\s+\g^\s\g^\m\g^\r=2(g^{\m\r}\g^\s+g^{\m\s}\g^\r-g^{\r\s}\g^\m),\NO\\
&\g^\r\g^\m\g^\s-\g^\s\g^\m\g^\r=2\g^{\r\m\s},\NO\\
&\g^\r\g^\m\g^\s=g^{\m\r}\g^\s+g^{\m\s}\g^\r-g^{\r\s}\g^\m+\g^{\r\m\s}\NO\\
&\g^{\m\r}\g^\s=\g^{\m\r\s}+\g^\m g^{\r\s}-\g^\r g^{\m\s}\NO\\
&\g^{[\m}\g_{\r\s}\g^{\n]}=-i\e^{\m\n}{}_{\r\s}\g^5+2g^{[\m}_\r g^{\n]}_\s,\NO\\
&\g^{\m\n\r\s}=-i\e^{\m\n\r\s}\g^5,\NO\\
&\g^{\m\r\s}=i\e^{\m\r\s\n}\g_\n\g^5,\NO\\
&\g^{\m\n}=\frac{i}{2}\e^{\m\n\r\s}\g_{\r\s}\g^5,
\eal
as well as the trace relations
\bal
\label{d=4-gamma-traces}
&\tr(\text{any odd number of gamma matrices})=0,\NO\\
&\tr(\g^\m\g^\n\g^5)=0,\NO\\
&\tr(\g^\m\g^\n)=4\h^{\m\n},\NO\\
&\tr(\g^\m\g^\n\g^\r\g^\s)=4(\h^{\m\n}\h^{\r\s}-\h^{\m\r}\h^{\n\s}+\h^{\m\s}\h^{\n\r}),\NO\\ &\tr(\g^\m\g^\n\g^\r\g^\s\g^5)=-4i\e^{\m\n\r\s}.
\eal

\paragraph{Dirac conjugate} The Dirac conjugate of a Dirac spinor, $\c$, is defined as 
\be\label{D-conjugate}
\dbar\c\equiv i\c^\dag\g^0,
\ee
and is denoted with a thick overbar in order to distinguish it from the Majorana conjugate. 

\paragraph{Majorana conjugate and spinors} The Majorana conjugate of a Dirac spinor, $\c$, is defined as 
\be\label{M-conjugate}
\bar\c\equiv \c^TC,
\ee
where $C$ is the charge conjugation matrix (see section 3.1.8 of \cite{Freedman:2012zz}). A spinor $\c$ is said to be Majorana if it equals its charge conjugate, or equivalently, if its Dirac and Majorana conjugates coincide, i.e.  
\bbxd\vskip.2cm
\be\label{Majorana-conjugate}
\c^C\equiv B^{-1}\c^*=\c\quad\Leftrightarrow\quad \dbar\c=\bar \c, 
\ee
\ebxd
where the unitary matrix $B$ is related to the charge conjugation matrix, $C$, as in eq.~(3.47) of \cite{Freedman:2012zz}.   

Dirac spinor bilinears involving Majorana conjugation in four dimensions satisfy the identity  
\be
\bar\l\g^{\m_1}\g^{\m_2}\cdots\g^{\m_p}\c=(-1)^p\bar\c\g^{\m_p}\cdots\g^{\m_2}\g^{\m_1}\l,\qquad \text{[eq.~(3.53) in \cite{Freedman:2012zz}]}.
\ee
which also implies that
\be
\bar\l\g^{\m_1}\g^{\m_2}\cdots\g^{\m_p}\g^5\c=(-1)^p\bar\c\g^5\g^{\m_p}\cdots\g^{\m_2}\g^{\m_1}\l=\bar\c\g^{\m_p}\cdots\g^{\m_2}\g^{\m_1}\g^5\l.
\ee
Majorana fermion bilinears possess in addition the reality property 
\be
(\bar\c\g_{\m_1\ldots\m_r}\l)^*=\bar\c\g_{\m_1\ldots\m_r}\l,\qquad \text{[eq.~(3.82) in \cite{Freedman:2012zz}]}.
\ee

\paragraph{Chirality projectors and Weyl spinors} The Weyl projections of a generic Dirac spinor, $\c$, are defined as 
\be\label{Weyl}
\c_L\equiv P_L\c \equiv\frac12(1+\g^5)\c,\qquad \c_R\equiv P_R\c\equiv\frac12(1-\g^5)\c.
\ee
Notice that, since there are no Majorana-Weyl spinors in four dimensions, the Weyl projection of a Majorana spinor is Weyl but not Majorana. Another potential source of confusion we should emphasize is the following relation between the Dirac and Majorana conjugates of Weyl spinors:
\bbxd
\vskip.2cm
\be
\dbar \c_L \equiv i\c_L^\dag\g^0=i\c^\dag P_L^\dag\g^0=\dbar\c P_R=\bar\c_R.
\ee
\ebxd

\paragraph{Fierz identities} Finally, we make extensive use of the following Fierz relations in four dimensions 
\bal
\c_L\bar\l_L=&\;-\frac12P_L(\bar\l\c_L)+\frac18P_L\g^{\m\n}(\bar\l\g_{\m\n}\c_L),\NO\\
\c_L\bar\l_R=&\;-\frac12P_L\g^\m(\bar\l\g_\m\c_L).
\eal

\section{Path integral derivation of the naive Ward identities}
\label{sec:path-integralWIDs}

In this appendix we provide a path integral derivation of the `naive' (i.e. classical) Ward identities for both the conformal and regulated WZ model, using the standard Noether procedure. This derivation does not rely on input from the coupling of the theory to background supergravity, but it is considerably more laborious than the approach adopted in the main body of the paper. In particular, while the Ward identities obtained from background supergravity involve only operators defined through functional differentiation and are therefore model independent, the path integral derivation involves model dependent operators that are directly related to Feynman diagrams. In order to derive the Ward identities in their model independent form using the path integral approach, therefore, one must carefully keep track of all seagull terms that relate the two definitions of the currents. The general structure of the seagull terms for the current multiplet of the WZ model is discussed separately in appendix \ref{sec:seagulls}.

\subsection{Ward identities for the conformal Wess-Zumino model}

Applying Noether's theorem to the path integral of the WZ model \eqref{WZ-lagrangian-flat} turns the three conservation laws \eqref{WZ-conserved-currents} and the algebraic identities \eqref{WZ-algebraic-constraints} into constraints for flat space correlation functions. In particular, using the variation of the WZ action under infinitesimal translations, R-symmetry and Q-supersymmetry transformations with a local parameter, namely\footnote{These relations hold for the unimproved currents. However, the improvement terms do not affect the l.h.s. of the Ward identities  \eqref{PathIntegralWIDs} and so they can be included at the end of the argument. See e.g. section 2 of \cite{DiFrancesco:1997nk}.}
\be
\d_a \hat S_{\rm WZ}=-\int d^4x\,\pa_\m a^\n\hat\ct^\m{}_\n,\quad \d_{\th_0} \hat S_{\rm WZ}=-\int d^4x\,\pa_\m \th_0\hat\cj^\m,\quad \d_{\ve_0} \hat S_{\rm WZ}=-\int d^4x\,\pa_\m \bar\ve_0\hat\cq^\m, 
\ee
and ignoring potential anomalies from the path integral measure, one obtains the identities
\bal\label{PathIntegralWIDs}
&-\int d^4x\,a^\n(x)\pa^x_\m\lb\hat\ct^\m{}_\n(x)\co_1(x_1)\cdots\co_n(x_n)\rb\,=\sum_{i=1}^n\lb\co_1(x_1)\cdots\d_a\co_i(x_i)\cdots\co_n(x_n)\rb,\NO\\
&-\int d^4x\,\th_0(x)\pa^x_\m\lb\hat\cj^\m(x)\co_1(x_1)\cdots\co_n(x_n)\rb\,=\sum_{i=1}^n\lb\co_1(x_1)\cdots\d_{\th_0}\co_i(x_i)\cdots\co_n(x_n)\rb,\NO\\
&-\int d^4x\,\bar\ve_0(x)\pa^x_\m\lb\hat\cq^\m(x)\co_1(x_1)\cdots\co_n(x_n)\rb\,=\sum_{i=1}^n\lb\co_1(x_1)\cdots\d_{\ve_0}\co_i(x_i)\cdots\co_n(x_n)\rb,
\eal
where the brackets $<\cdot>$ stand for connected correlators defined through operator insertions (see appendix \ref{sec:seagulls}) and all spacetime indices in the operators $\co_i$ have been suppressed.   

The transformations of $\co_i$ on the r.h.s. of these identities admit an expansion of the form
\bal\label{OpTransExp}
\d_a\co_i(x)=&\;a^\m(\ct|\co_i)^{[1]}_{\m}+\pa^\n a^\m(\ct|\co_i)^{[2]}_{\m|\n}+\pa^\n\pa^\r a^\m(\ct|\co_i)^{[3]}_{\m|\n\r}+\cdots,\NO\\
\d_{\th_0}\co_i(x)=&\;\th_0(\cj|\co_i)^{[1]}+\pa^\n\th_0(\cj|\co_i)^{[2]}_{\n}+\pa^\n\pa^\r \th^0(\cj|\co_i)^{[3]}_{\n\r}+\cdots,\NO\\
\d_{\ve_0}\co_i(x)=&\;\bar\ve_0(\cq|\co_i)^{[1]}+\pa^\n \bar\ve_0(\cq|\co_i)^{[2]}_{\n}+\pa^\n\pa^\r \bar\ve_0{}(\cq|\co_i)^{[3]}_{\n\r}+\cdots,
\eal
where the coefficients depend on the structure of the specific operator. In particular, the highest order derivatives in $\co_i$ determine the order at which these expansions terminate. For the supercurrent and R-symmetry current in \eqref{WZ-Noether-currents-Mink}, for example, these expansions truncate at the second order, while for the stress tensor they terminate at the third order. Inserting the expansions of the operator transformations back in \eqref{PathIntegralWIDs} and using the fact that the local symmetry parameters are arbitrary leads to local (unintegrated) versions of these Ward identities, which depend explicitly on the coefficients in the expansions \eqref{OpTransExp}. We will write the explicit form of those Ward identities relevant for our analysis, after first computing the symmetry transformations of the currents \eqref{WZ-Noether-currents-Mink}.  

Analogous identities can be derived from the conserved currents associated with Lorentz, scale, special conformal and S-supersymmetry invariance by promoting the corresponding parameters to local functions, namely
\bal
&\d_\ell \hat S_{\rm WZ}=-\int d^4x\,\pa_\m \ell^{\r\s}x_{\r}\hat\ct^\m{}_{\s},\qquad \d_{\l} \hat S_{\rm WZ}=-\int d^4x\,\pa_\m \l\, x^\n\hat\ct^\m{}_\n,\NO\\
&\d_{b} \hat S_{\rm WZ}=-\int d^4x\,\pa_\m b_\r(2x^\r x^\n-\h^{\r\n} x^2)\hat\ct^\m{}_\n,\qquad \d_{\ve_0} \hat S_{\rm WZ}=\int d^4x\,\pa_\m \bar\h_0\,x^\n\g_\n\hat\cq^\m. 
\eal
Inserting these in the path integral and ignoring potential anomalies results in the identities
\bal\label{PathIntegralWIDs-algebraic}
&\hskip-.1cm -\int\hskip-.05cm d^4x\,\ell^{\r\s}(x)\pa^x_\m\lb x_{[\r}\hat\ct^\m_{\s]}(x)\co_1(x_1)\cdots\co_n(x_n)\rb\,=\sum_{i=1}^n\lb\co_1(x_1)\cdots\d_\ell\co_i(x_i)\cdots\co_n(x_n)\rb,\NO\\
&\hskip-.1cm -\int\hskip-.05cm d^4x\,\l(x)\pa^x_\m\lb x^\n\hat\ct^\m_\n(x)\co_1(x_1)\cdots\co_n(x_n)\rb\,=\sum_{i=1}^n\lb\co_1(x_1)\cdots\d_{\l}\co_i(x_i)\cdots\co_n(x_n)\rb,\NO\\
&\hskip-.1cm -\int\hskip-.05cm d^4x\,b_\r(x)\pa^x_\m\lb (2x^\r x^\n-\h^{\r\n} x^2)\hat\ct^\m_\n(x)\co_1(x_1)\cdots\co_n(x_n)\rb\,=\sum_{i=1}^n\lb\co_1(x_1)\cdots\d_{b}\co_i(x_i)\cdots\co_n(x_n)\rb,\NO\\
&\hskip-.1cm \int\hskip-.05cm d^4x\,\bar\h_0(x)\pa^x_\m\lb x^\n\g_\n\hat\cq^\m(x)\co_1(x_1)\cdots\co_n(x_n)\rb\,=\sum_{i=1}^n\lb\co_1(x_1)\cdots\d_{\h_0}\co_i(x_i)\cdots\co_n(x_n)\rb\hskip-3pt.
\eal

\subsection*{Symmetry transformations of the Noether currents}

Since we are interested in correlation functions of current multiplet operators, we will focus on the case where all operators $\co_i$ are currents. In order to derive the local (unintegrated) Ward identities following from \eqref{PathIntegralWIDs} and \eqref{PathIntegralWIDs-algebraic}, we need to evaluate the classical transformations \eqref{OpTransExp} using the expressions for the currents in \eqref{WZ-conserved-currents} and the symmetry transformations in table \ref{superconformal-transformations}.   

The R-symmetry transformations of the currents take the form 
\bal\label{current-R-transformations-flat}
\d_{\th_0}\hat\ct^\m{}_\n=&\;\orangebox{(\h^{\m\r}\h^\s_\n+\h^{\m\s}\h^\r_\n-\h^\m_\n\h^{\r\s})\hat\cj_\r\pa_\s\th_0-\frac{i}{6}\bar\c\g_\n\g^5\c\pa^\m\th_0},\NO\\
\d_{\th_0}\hat\cj^\m=&\;\orangebox{\frac{4}{9\a^2}\f^*\f\,\pa^\m\th_0},\NO\\
\d_{\th_0} \hat\cq^\m=&\;i\th_0\g^5\hat\cq^\m+\orangebox{\frac{i}{3\a}\pa^\m\th_0(\f^*\c_L-\f\c_R)},
\eal
where the terms in pink/oval frames are model dependent and correspond to seagull terms, discussed in more detail in appendix \ref{sec:seagulls}. As we will see shortly, they cancel against the transformations of the background supergravity fields in the corresponding current operators defined through functional differentiation. 

Similarly, the Q-supersymmetry transformations of the R-current and of the supercurrent are
\bal\label{current-Q-transformations-flat}
\rule{0.cm}{.7cm}\d_{\ve_0}\hat\cj^\m=&\;-i\bar\ve_0\g^5\hat\cq^\m+\orangebox{\textcolor{black}{\frac{i}{3\a}(\f^*\bar\c_L-\f\bar\c_R)\pa^\m\ve_0}}+\cyanbox{\textcolor{black}{\frac{i}{3\a}\bar\ve_0\g^\m(\f^*\slashed\pa\c_L-\f\slashed\pa\c_R)+\frac{i}{6\a}\bar\ve_0\g^\m(F^*\c_L-F\c_R)}},\NO\\
\rule{0.cm}{.7cm}\d_{\ve_0} \hat\cq^\m=&\;\frac12\hat\ct^\m{}_{\n}\g^\n\ve_0+\frac{i}{8}\pa_\r\big[\hat\cj_{\s}(i\e^{\m\n\r\s}\g^5+2\h^{\m\n}\h^{\r\s}-2\h^{\r\n}\h^{\m\s})\g_\n\g^5\ve_0\big]-\orangebox{\textcolor{black}{\frac{3}{8}\e^{\m\n\r\s}\hat\cj_{\s}\g_\n\pa_\r\ve_0}}\NO\\
&+\orangebox{\textcolor{black}{\frac18(\bar\c\g_\s\g^5\c)\g^\s\g^5\pa^\m\ve_0+\frac{1}{12\a^2}\pa_\r(\f\f^*)(i\e^{\m\n\r\s}\g^5+\h^{\m\n}\h^{\r\s}-\h^{\m\r}\h^{\n\s})\g_\n\pa_\s\ve_0}}\NO\\
&+\cyanbox{\textcolor{black}{\frac{1}{8}\big(\bar\c(\g^{\m\n}-\h^{\m\n})\g^5\slashed\pa\c\big)\g_\n\g^5\ve_0-\frac{1}{8}\big(\bar\c(\g^{\m\n}-\h^{\m\n})\slashed\pa\c\big)\g_\n\ve_0}}\NO\\
&+\cyanbox{\textcolor{black}{\frac{1}{4\a^2}(F^*\slashed\pa\f \g^\m\ve_{0R}+F\slashed\pa\f^*\g^\m\ve_{0L})+\frac{1}{6\a^2}\g^{\m\n}\pa_\n(\f F^*\ve_{0R}+\f^*F \ve_{0L})}},
\eal
where seagull terms appear again in pink/oval frames, while those in blue/rectangular frames vanish on-shell.  

Finally, the S-supersymmetry transformations of the R-current and of the supercurrent are
\be
\d_{\h_0}\hat\cj^\m=(\d_{\ve_0=x^\m\g_\m\h_0}+\wt\d_{\h_0})\hat\cj^\m,\qquad \d_{\h_0} \hat\cq^\m=(\d_{\ve_0= x^\k\g_\k\h_0}+\wt\d_{\h_0}) \hat\cq^\m,
\ee
where $\d_{\ve_0= x^\k\g_\k\h_0}$ denotes a Q-supersymmetry transformation with parameter $\ve_0= x^\m\g_\m\h_0$ and 
\bal\label{current-S-transformations-flat}
\wt\d_{\h_0}\hat\cj^\m=&\;\orangebox{\frac{i}{3\a}\bar\h_0\g^\m(\f^*\c_L-\f\c_R)},\NO\\
\wt\d_{\h_0} \hat\cq^\m=&\;\frac{3i}{4}\g^5\h_0\hat\cj^\m-\orangebox{\frac{3i}{4}\g^5\h_0\hat\cj^\m+\frac{1}{3\a^2}\g^{\m\n}\pa_\n(\f^*\f\h_0)+\frac{1}{2\a^2}(\f\slashed\pa\f^*\g^\m\h_{0L}+\f^*\slashed\pa\f\g^\m\h_{0R})}.
\eal

\subsection*{Symmetry transformations of the model independent currents}

In order to clarify the significance of the model dependent terms in the above transformations of the Noether currents, it is instructive to determine the corresponding transformations of the conformal multiplet currents \eqref{WZ-currents-CS}, which are obtained by coupling the theory to background supergravity.

The flat space limit of the Q-supersymmetry transformation of the supercurrent in \eqref{WZ-currents-CS} is
\bal
\label{S-var}
\d_\ve \cq^\m=&\;\d_\ve \hat\cq^\m+\frac38\e^{\m\n\r\s}\g_\n\pa_\r\ve\hat\cj_{\s}\NO\\
&-\frac{1}{12\a^2}\pa_{\r}(\f\f^*)(i\e^{\m\n\r\s}\g^5+\h^{\m\n}\h^{\r\s}-\h^{\m\r}\h^{\n\s})\g_{\n}\pa_\s\ve-\frac{1}{8}(\bar\c\g^\s\g^5\c)\g_\s\g^5\pa^\m\ve,
\eal
where $\hat \cj^\m$, $\hat\cq^\m$ are the flat space currents in \eqref{WZ-Noether-currents-Mink} and $\d_\ve \hat\cq^\m$ is the supersymmetry transformation of the flat space supercurrent in \eqref{current-Q-transformations-flat}. Notice that the flat space limit of $\d_\ve \cq^\m$ differs from $\d_\ve \hat\cq^\m$ by terms arising from the supersymmetry transformation of the gravitino, which cancel the model dependent terms in pink/oval frames in $\d_\ve \hat\cq^\m$. Hence, on-shell, the flat space limit of $\d_\ve \cq^\m$ is
\bbxd
\vskip.4cm
\be
\label{Q-Q-transformation}
\d_\ve \cq^\m=\frac12\hat\ct^\m_{a}\g^a\ve+\frac{i}{4}\pa_\r\Big[\hat\cj_{\s}\Big(\frac{i}{2}\e^{\m\n\r\s}\g^5+\h^{\m\n}\h^{\r\s}-\h^{\r\n}\h^{\m\s}\Big)\g_\n\g^5\ve\Big].
\ee
\ebxd
Contrary to the transformation of the Noether current $\hat\cq^\m$, the transformation of the supercurrent $\cq^\m$ that is defined through functional differentiation involves only current multiplet operators and its form is independent of the specific theory. The seagull terms eliminate the model dependence of the Noether currents, resulting in the same universal transformations.  

Similarly, the Q-supersymmetry transformation of the R-current in \eqref{WZ-currents-CS} takes the form
\be
\d_\ve\cj^\m=\d_\ve\hat\cj^\m+\frac{i}{3\a}(\f \bar\c_R-\f^* \bar\c_L)\pa^\m\ve,
\ee
where $\d_\ve \hat\cj^\m$ is the supersymmetry transformation of the flat space current in \eqref{current-Q-transformations-flat}, and the second term on the r.h.s. arises from the supersymmetry transformation of the gravitino. This cancels the term in the pink/oval frame in $\d_\ve \hat\cj^\m$, so that, on-shell, the flat space limit of $\d_\ve\cj^\m$ is  
\bbxd
\vskip.25cm
\be
\label{J-Q-transformation}
\d_\ve \cj^\m=-i\bar\ve\g^5\hat\cq^\m.
\ee
\ebxd

The flat space limit of the R-symmetry and S-supersymmetry transformations of the currents \eqref{WZ-currents-CS} can be determined similarly. We find respectively 
\bbxd
\bal\label{R-transformations}
\d_{\th}\ct^\m{}_\n=&\;\d_{\th}\hat\ct^\m{}_\n-(\h^{\m\r}\h^\s_\n+\h^{\m\s}\h^\r_\n-\h^\m_\n\h^{\r\s})\hat\cj_\r\pa_\s\th+\frac{i}{6}\bar\c\g_\n\g^5\c\pa^\m\th=0,\NO\\
\d_{\th}\cj^\m=&\;\d_{\th}\hat\cj^\m-\frac{4}{9\a^2}\f^*\f\,\pa^\m\th=0,\NO\\
\d_{\th} \cq^\m=&\;\d_{\th} \hat\cq^\m-\frac{i}{3\a}\pa^\m\th(\f^*\c_L-\f\c_R)=i\th\g^5\hat\cq^\m,
\eal
\ebxd
\bbxd
\bal\label{S-transformations}
\d_\h\cj^\m=&\;\wt\d_\h\hat\cj^\m-\frac{i}{3\a}\bar\h\g^\m(\f^*\c_L-\f\c_R)=0,\\
\d_\h\cq^\m=&\;\wt\d_\h\hat\cq^\m+\frac{3i}{4}\g^5\h\hat\cj^\m-\frac{1}{3\a^2}\g^{\m\n}\pa_\n(\f^*\f\h)-\frac{1}{2\a^2}(\f\slashed\pa\f^*\g^\m\h_L+\f^*\slashed\pa\f\g^\m\h_R)=\frac{3i}{4}\g^5\h\hat\cj^\m,\NO
\eal
\ebxd
where $\wt\d_\h$ was introduced in \eqref{current-S-transformations-flat} and denotes the flat space limit of the S-supersymmetry transformation of conformal supergravity.

In all cases, the model dependent seagull terms in the Noether currents cancel against terms in the current multiplet operators defined by functional differentiation that are linear in the background supergravity fields, i.e. quadratic at the level of the Lagrangian. The resulting transformations involve only current multiplet operators and are independent of the specific theory. In fact, they can be obtained directly from the symmetry transformations of the background supergravity fields and the general definition of the currents in \eqref{currents}, without any reference to a specific theory. This approach also determines the quantum corrections to the transformations arising from the superconformal anomalies. For theories coupled to $\cn=1$ conformal supergravity, the current transformations were derived using this approach in \cite{Papadimitriou:2017kzw,Papadimitriou:2019gel}.

\subsection*{Naive superconformal Ward identities}

Having clarified the role of the seagull terms in the transformation of the Noether currents, we return to the computation of the path integral derivation of the naive superconformal Ward identities for the massless WZ model. Inserting the transformations of the Noether currents in the relations \eqref{PathIntegralWIDs}, \eqref{PathIntegralWIDs-algebraic} and eliminating the arbitrary local transformation parameters determines the corresponding classical Ward identities. In order to simplify the resulting expressions it is convenient to  introduce the `seagull operators' 
\be\label{seagull-operators-WZ}
\hat s_{(1|0)}=\f^*\f,\qquad \hat{s}{}_{(2|1)}^{\m}=\bar\c\g^\m\g^5\c,\qquad 
\hat s_{\(3|\frac12\)}=i(\f^*\c_L-\f\c_R),\qquad \hat s^\m_{(4|1)}=\f\pa^\m\f^*, 
\ee
and the `null operators'
\bal\label{null-operators-WZ}
\hat n^{\m\n}_{(1|2)}=&\;\bar\c(\g^{\m\n}-\h^{\m\n})\g^5\slashed\pa\c,\qquad 
\hat n^{\m\n}_{(2|2)}=\;\bar\c(\g^{\m\n}-\h^{\m\n})\slashed\pa\c,\qquad
\hat n_{\(3|\frac12\)}=\;i(\f^*\slashed\pa\c_L-\f\slashed\pa\c_R),\NO\\
\hat n_{(4|0)}=&\;F^*\f,\qquad 
\hat n^\m_{(5|1)}=\;F^*\pa^\m\f,\qquad 
\hat n_{\(6|\frac12\)}=\;i(F^*\c_L-F\c_R),
\eal
which are proportional to the classical equations of motion. We use small letters to denote these operators, as opposed to capital script letters, in order to emphasize that they are model dependent. Moreover, the first entry in the subscript $(\cdot|\cdot)$ simply labels the operator, while the second entry indicates its spin. As is shown in appendix \ref{sec:seagulls}, the derivatives of the current multiplet operators with respect to background supergravity fields can be expressed uniquely in terns of the seagull operators \eqref{seagull-operators-WZ}. However, there is no unique way of writing the operators \eqref{seagull-operators-WZ} in terms of derivatives of currents.   

From the conservation of the R-current in \eqref{PathIntegralWIDs}, we obtain the three naive identities 
\bal\label{R-symmetry-WID-axial-PI}
&\pa^x_\m\hskip-.1cm\lb\hat\cj^\m(x)\hat\cj^\n(y)\hat\cj^\r(z)\rb=\frac{4}{9\a^2}\big(\pa_x^\n\d(x,y)\lb\orangebox{\hat s_{(1|0)}(y)}\hat\cj^\r(z)\rb+\pa_x^\r\d(x,z)\lb\hat\cj^\n(y)\orangebox{\hat s_{(1|0)}(z)}\rb\hskip-.1cm\big),\NO\\
&\rule{.0cm}{.9cm}\pa^x_\m\lb\hat\cj^\m(x)\hat\cq^\n(y)\hat{\bar\cq}^\r(z)\rb+\d(x,y)i\g^5\lb\hat\cq^\n(y)\hat{\bar\cq}^\r(z)\rb+i\d(x,z)\lb\hat\cq^\n(y)\hat{\bar\cq}^\r(z)\rb\g^5\NO\\
&-\frac{1}{3\a}\pa_x^\n\d(x,y)\lb\orangebox{\hat n_{\(3|\frac12\)}(y)}\hat{\bar\cq}^\r(z)\rb-\frac{1}{3\a}\pa_x^\r\d(x,z)\lb\hat\cq^\n(y)\orangebox{\hat{\bar n}_{\(3|\frac12\)}(z)}\rb=0,\NO\\
&\rule{.0cm}{.9cm}\pa^x_\m\lb\hat\cj^\m(x)\hat\cq^\n(y)\hat{\bar\cq}^\r(z)\hat\cj^\s(w)\rb+\d(x,y)i\g^5\lb\hat\cq^\n(y)\hat{\bar\cq}^\r(z)\hat\cj^\s(w)\rb\NO\\
&+i\d(x,z)\lb\hat\cq^\n(y)\hat{\bar\cq}^\r(z)\hat\cj^\s(w)\rb\g^5-\frac{1}{3\a}\pa_x^\n\d(x,y)\lb\orangebox{\hat n_{\(3|\frac12\)}(y)}\hat{\bar\cq}^\r(z)\hat\cj^\s(w)\rb\NO\\
&-\frac{1}{3\a}\pa_x^\r\d(x,z)\lb\hat\cq^\n(y)\orangebox{\hat{\bar n}_{\(3|\frac12\)}(z)}\hat\cj^\s(w)\rb-\frac{4}{9\a^2}\pa_x^\s\d(x,w)\lb\bar\cq^\n(y)\hat{\bar\cq}^\r(z)\orangebox{\hat s_{(1|0)}(w)}\rb=0,
\eal
which should be compared respectively with the fully quantum identities \eqref{R-symmetry-WID-axial}, \eqref{R-symmetry-WID-3pt} and \eqref{R-symmetry-WID-4pt}. Except for the superconformal anomalies that we have not accounted for in the path integral derivation of the naive Ward identities, the only difference between the two sets of identities are the terms in pink/oval frames, which correspond to seagull terms that are absorbed in the definition of the correlation functions $\<\cdot \>$ used in section \ref{sec:WardIDs} (see appendix \ref{sec:seagulls}).   

Similarly, the conservation of the supercurrent in \eqref{PathIntegralWIDs} leads to the naive identities
\bal\label{Q-susy-WID-2pt-PI}
&\pa^x_\m\lb\hat\cq^\m(x)\hat{\bar\cq}^\s(y)\rb-\frac12\lb\hat\ct^\s{}_{\l}(x)\rb\g^\l\d(x,y)+\frac{i}{8}\lb\hat\cj^\n(x)\rb\big(i\g^5\e^{\r\s}{}_{\n\l}+4\h^{[\r}_\n\h^{\s]}_\l\big)\g^5\g^\l\pa^x_\r\d(x,y)\NO\\
&+\frac{3}{8}\lb\orangebox{\hat\cj_{\n}(y)}\rb\e^{\r\s\n\l}\g_\l\pa^x_\r\d(x,y)+\frac{1}{12\a^2}\pa_y^\n\lb\orangebox{\hat s_{(1|0)}(y)}\rb\big(-i\g^5\e^{\r\s}{}_{\n\l}+2\h^{[\r}_\n\h^{\s]}_\l\big)\g^\l\pa^x_\r\d(x,y)\NO\\
&+\frac18\lb\orangebox{\hat{s}{}_{(2|1)}^{\l}(y)}\rb\g^5\g_\l\pa_x^\s\d(x,y)+\frac{1}{8}\lb\cyanbox{\hat n^{\s\l}_{(1|2)}(x)}\rb\g_\l\g^5\d(x,y)+\frac{1}{8}\lb\cyanbox{\hat n^{\s\l}_{(2|2)}(x)}\rb\g_\l\d(x,y)\NO\\
&+\frac{1}{4\a^2}\big(\lb\cyanbox{\hat n^\r_{(5|1)}(x)}\rb P_R +\lb \cyanbox{\hat n^{\r*}_{(5|1)}(x)}\rb P_L\big)\g^\s\g_\r\d(x,y)\NO\\
&+\frac{1}{6\a^2}\big(\lb\cyanbox{\hat n_{(4|0)}(x)}\rb P_{R}+\lb\cyanbox{\hat n_{(4|0)}^*(x)}\rb P_{L}\big)\g^{\s\n}\pa^x_\n\d(x,y)=0,
\eal
\bal\label{Q-susy-WID-3pt-PI} 
&\pa^x_\m\lb\hat\cq^\m(x)\hat{\bar\cq}^\s(y)\hat\cj^\k(z)\rb-i\g^5\d(x,z)\lb\hat\cq^\k(x)\hat{\bar\cq}^\s(y)\rb-\frac12\lb\hat\ct^\s{}_{\l}(x)\hat\cj^\k(z)\rb\g^\l\d(x,y)\NO\\
&+\frac{i}{8}\lb\hat\cj^\n(x)\hat\cj^\k(z)\rb\big(i\g^5\e^{\r\s}{}_{\n\l}+4\h^{[\r}_\n\h^{\s]}_\l\big)\g^5\g^\l\pa^x_\r\d(x,y)+\frac{3}{8}\lb\orangebox{\hat\cj_{\n}(y)}\hat\cj^\k(z)\rb\e^{\r\s\n\l}\g_\l\pa^x_\r\d(x,y)\NO\\
&+\frac{1}{12\a^2}\pa_y^\n\lb\orangebox{\hat s_{(1|0)}(y)}\hat\cj^\k(z)\rb\big(-i\g^5\e^{\r\s}{}_{\n\l}+2\h^{[\r}_\n\h^{\s]}_\l\big)\g^\l\pa^x_\r\d(x,y)\NO\\
&+\frac18\lb\orangebox{\hat{s}{}_{(2|1)}^{\l}(y)}\hat\cj^\k(z)\rb\g^5\g_\l\pa_x^\s\d(x,y)-\frac{1}{3\a}\pa^\k_x\d(x,z)\lb\orangebox{\hat s_{\(3|\frac12\)}(z)}\hat{\bar\cq}^\s(y)\rb\NO\\
&+\frac{1}{8}\lb\cyanbox{\hat n^{\s\l}_{(1|2)}(x)}\hat\cj^\k(z)\rb\g_\l\g^5\d(x,y)+\frac{1}{8}\lb\cyanbox{\hat n^{\s\l}_{(2|2)}(x)}\hat\cj^\k(z)\rb\g_\l\d(x,y)\NO\\
&+\frac{1}{3\a}\d(x,z)\g^\k\lb\cyanbox{\hat n_{\(3|\frac12\)}(x)}\hat{\bar\cq}^\s(y)\rb+\frac{1}{6\a}\d(x,z)\g^\k\lb\cyanbox{\hat n_{\(6|\frac12\)}(x)}\hat{\bar\cq}^\s(y)\rb\NO\\
&+\frac{1}{4\a^2}\big(\lb\cyanbox{\hat n^\r_{(5|1)}(x)}\hat\cj^\k(z)\rb P_R +\lb \cyanbox{\hat n^{\r*}_{(5|1)}(x)}\hat\cj^\k(z)\rb P_L\big)\g^\s\g_\r\d(x,y)\NO\\
&+\frac{1}{6\a^2}(\lb\cyanbox{\hat n_{(4|0)}(x)}\hat\cj^\k(z)\rb P_{R}+\lb\cyanbox{\hat n^*_{(4|0)}(x)}\hat\cj^\k(z)\rb P_{L})\g^{\s\n}\pa^x_\n\d(x,y)=0,
\eal
\bal\label{Q-susy-WID-4pt-PI}
&\pa^x_\m\lb\hat\cq^\m(x)\hat{\bar\cq}^\s(y)\hat\cj^\k(z)\hat\cj^\a(w)\rb-i\g^5\d(x,z)\lb\hat\cq^\k(x)\hat{\bar\cq}^\s(y)\hat\cj^\a(w)\rb-i\g^5\d(x,w)\lb\hat\cq^\a(x)\hat{\bar\cq}^\s(y)\hat\cj^\k(z)\rb\NO\\
&-\frac12\lb\hat\ct^\s{}_{\l}(x)\hat\cj^\k(z)\hat\cj^\a(w)\rb\g^\l\d(x,y)+\frac{i}{8}\lb\hat\cj^\n(x)\hat\cj^\k(z)\hat\cj^\a(w)\rb\big(i\g^5\e^{\r\s}{}_{\n\l}+4\h^{[\r}_\n\h^{\s]}_\l\big)\g^5\g^\l\pa^x_\r\d(x,y)\NO\\
&+\frac{3}{8}\lb\orangebox{\hat\cj_{\n}(y)}\hat\cj^\k(z)\hat\cj^\a(w)\rb\e^{\r\s\n\l}\g_\l\pa^x_\r\d(x,y)+\frac18\lb\orangebox{\hat{s}{}_{(2|1)}^{\l}(y)}\hat\cj^\k(z)\hat\cj^\a(w)\rb\g^5\g_\l\pa_x^\s\d(x,y)\NO\\
&+\frac{1}{12\a^2}\pa_y^\n\lb\orangebox{\hat s_{(1|0)}(y)}\hat\cj^\k(z)\hat\cj^\a(w)\rb\big(-i\g^5\e^{\r\s}{}_{\n\l}+2\h^{[\r}_\n\h^{\s]}_\l\big)\g^\l\pa^x_\r\d(x,y)\NO\\
&-\frac{1}{3\a}\Big(\pa^\k_x\d(x,z)\lb\orangebox{\hat s_{\(3|\frac12\)}(z)}\hat{\bar\cq}^\s(y)\hat\cj^\a(w)\rb+\pa_x^\a\d(x,w)\lb\orangebox{\hat s_{\(3|\frac12\)}(w)}\hat{\bar\cq}^\s(y)\hat\cj^\k(z)\rb\Big)\NO\\
&+\frac{1}{8}\Big(\lb\cyanbox{\hat n^{\s\l}_{(1|2)}(x)}\hat\cj^\k(z)\hat\cj^\a(w)\rb\g_\l\g^5+\lb\cyanbox{\hat n^{\s\l}_{(2|2)}(x)}\hat\cj^\k(z)\hat\cj^\a(w)\rb\g_\l\Big)\d(x,y)\NO\\
&+\frac{1}{3\a}\Big(\d(x,z)\g^\k\lb\cyanbox{\hat n_{\(3|\frac12\)}(x)}\hat{\bar\cq}^\s(y)\hat\cj^\a(w)\rb+\d(x,w)\g^\a\lb\cyanbox{\hat n_{\(3|\frac12\)}(x)}\hat{\bar\cq}^\s(y)\hat\cj^\k(z)\rb\Big)\NO\\
&+\frac{1}{6\a}\Big(\d(x,z)\g^\k\lb\cyanbox{\hat n_{\(6|\frac12\)}(x)}\hat{\bar\cq}^\s(y)\hat\cj^\a(w)\rb+\d(x,w)\g^\a\lb\cyanbox{\hat n_{\(6|\frac12\)}(x)}\hat{\bar\cq}^\s(y)\hat\cj^\k(z)\rb\Big)\NO\\
&+\frac{1}{4\a^2}\Big(\lb\cyanbox{\hat n^\r_{(5|1)}(x)}\hat\cj^\k(z)\hat\cj^\a(w)\rb P_R +\lb \cyanbox{\hat n^{\r*}_{(5|1)}(x)}\hat\cj^\k(z)\hat\cj^\a(w)\rb P_L\Big)\g^\s\g_\r\d(x,y)\NO\\
&+\frac{1}{6\a^2}(\lb\cyanbox{\hat n_{(4|0)}(x)}\hat\cj^\k(z)\hat\cj^\a(w)\rb P_{R}+\lb\cyanbox{\hat n^*_{(4|0)}(x)}\hat\cj^\k(z)\hat\cj^\a(w)\rb P_{L})\g^{\s\n}\pa^x_\n\d(x,y)=0,
\eal
which should be compared respectively with \eqref{Q-susy-WID-2pt}, \eqref{Q-susy-WID-3pt} and \eqref{Q-susy-WID-4pt}. Besides the superconformal anomalies and the terms in pink/oval frame that correspond to seagull terms, the two sets of identities now differ also by the terms in blue/rectangular frame. Classically, these terms vanish on-shell, but they contribute inside correlation functions. In particular, using the propagators \eqref{WZ-propagators} and the Noether currents \eqref{WZ-Noether-currents-Mink}, it can be shown that (up to further seagull terms)
\bal\label{current-contraction-1pt}
&\hskip-.cm\big(\bar\c(\g^{\s\l}-\h^{\s\l})\g^5\contraction{}{\slashed\pa\c}{\big)(x)}{\hat\cj}\slashed\pa\c\big)(x)\hat\cj^\k(z)\g_\l\g^5\d(x,y)+\big(\bar\c(\g^{\s\l}-\h^{\s\l})\contraction{}{\slashed\pa\c}{\big)(x)}{\hat\cj}\slashed\pa\c\big)(x)\hat\cj^\k(z)\g_\l\d(x,y)\\
&\hskip-.cm+\frac{8i}{3\a}\d(x,z)\g^\k\big(\f^*\contraction{}{\slashed\pa\c_L}{)(x)}{\hat}\slashed\pa\c_L(x)\hat{\bar\cq}^\s(y)-\f\contraction{}{\slashed\pa\c_R}{)(x)}{\hat}\slashed\pa\c_R(x)\hat{\bar\cq}^\s(y)\big)=\hat\cj^\n(x)\big(i\g^5\e^{\k\s}{}_{\n\l}+4\h^{[\k}_\n\h^{\s]}_\l\big)\g^\l\d(x,y)\d(x,z),\NO
\eal
and similarly,
\bal\label{current-contraction-2pt}
&\big(\bar\c(\g^{\s\l}-\h^{\s\l})\g^5\contraction{}{\slashed\pa\c}{\big)(x)}{\hat\cj}\slashed\pa\c\big)(x)\hat\cj^\k(z)\hat\cj^\a(w)\g_\l\g^5\d(x,y)+\big(\bar\c(\g^{\s\l}-\h^{\s\l})\g^5\contraction{}{\slashed\pa\c}{\big)(x)\hat\cj^\k(z)}{\hat\cj}\slashed\pa\c\big)(x)\hat\cj^\k(z)\hat\cj^\a(w)\g_\l\g^5\d(x,y)\NO\\
&+\big(\bar\c(\g^{\s\l}-\h^{\s\l})\contraction{}{\slashed\pa\c}{\big)(x)}{\hat\cj}\slashed\pa\c\big)(x)\hat\cj^\k(z)\hat\cj^\a(w)\g_\l\d(x,y)+\big(\bar\c(\g^{\s\l}-\h^{\s\l})\contraction{}{\slashed\pa\c}{\big)(x)\hat\cj^\k(z)}{\hat\cj}\slashed\pa\c\big)(x)\hat\cj^\k(z)\hat\cj^\a(w)\g_\l\d(x,y)\NO\\
&+\frac{8i}{3\a}\d(x,z)\g^\k\Big(\f^*\contraction{}{\slashed\pa\c_L}{)(x)}{\hat}\slashed\pa\c_L(x)\hat{\bar\cq}^\s(y)\hat\cj^\a(w)-\f\contraction{}{\slashed\pa\c_R}{)(x)}{\hat}\slashed\pa\c_R(x)\hat{\bar\cq}^\s(y)\hat\cj^\a(w)\Big)\NO\\
&+\frac{8i}{3\a}\d(x,z)\g^\k\Big(\f^*\contraction{}{\slashed\pa\c_L}{)(x)\hat{\bar\cq}^\s(y)}{\hat}\slashed\pa\c_L(x)\hat{\bar\cq}^\s(y)\hat\cj^\a(w)-\f\contraction{}{\slashed\pa\c_R}{)(x)\hat{\bar\cq}^\s(y)}{\hat}\slashed\pa\c_R(x)\hat{\bar\cq}^\s(y)\hat\cj^\a(w)\Big)\NO\\
&+\frac{8i}{3\a}\d(x,w)\g^\a\Big(\f^*\contraction{}{\slashed\pa\c_L}{)(x)}{\hat}\slashed\pa\c_L(x)\hat{\bar\cq}^\s(y)\hat\cj^\k(z)-\f\contraction{}{\slashed\pa\c_R}{)(x)}{\hat}\slashed\pa\c_R(x)\hat{\bar\cq}^\s(y)\hat\cj^\k(z)\Big)\NO\\
&+\frac{8i}{3\a}\d(x,w)\g^\a\Big(\f^*\contraction{}{\slashed\pa\c_L}{)(x)\hat{\bar\cq}^\s(y)}{\hat}\slashed\pa\c_L(x)\hat{\bar\cq}^\s(y)\hat\cj^\k(z)-\f\contraction{}{\slashed\pa\c_R}{)(x)\hat{\bar\cq}^\s(y)}{\hat}\slashed\pa\c_R(x)\hat{\bar\cq}^\s(y)\hat\cj^\k(z)\Big)\NO\\
&=\big(4\d^{[\b}_\n\d^{\s]}_\l+i\g^5 \e_{\n\l}{}^{\b\s}\big)\g^\l\Big(\hat\cj^\n(x)\hat\cj^\a(w)\d^\k_\b\d(x,z)+\hat\cj^\n(x)\hat\cj^\k(z)\d^\a_\b\d(x,w)\Big)\d(x,y).
\eal
Moreover, all correlation functions involving the auxiliary field $F$ can be shown to vanish identically.
It follows that the naive Ward identities \eqref{Q-susy-WID-2pt-PI}, \eqref{Q-susy-WID-3pt-PI} and \eqref{Q-susy-WID-4pt-PI} agree with their counterparts in section \ref{sec:WardIDs}, up to seagull terms  and quantum anomalies.

Finally, the S-supersymmetry identity in \eqref{PathIntegralWIDs-algebraic} can be decomposed as 
\bal
&\int d^4x\,\bar\h_0(x)x^\n\g_\n\pa^x_\m\lb \hat\cq^\m(x)\co_1(x_1)\cdots\co_n(x_n)\rb+\int d^4x\,\bar\h_0(x)\g_\m\lb\hat\cq^\m(x)\co_1(x_1)\cdots\co_n(x_n)\rb\NO\\
&=\sum_{i=1}^n\lb\co_1(x_1)\cdots\d_{\ve_0= x^\m\g_\m\h_0}\co_i(x_i)\cdots\co_n(x_n)\rb+\sum_{i=1}^n\lb\co_1(x_1)\cdots\wt\d_{\h_0}\co_i(x_i)\cdots\co_n(x_n)\rb,
\eal
where, as above $\d_{\ve_0= x^\m\g_\m\h_0}$ is a Q-supersymmetry transformation with parameter $\ve_0= x^\m\g_\m\h_0$ and 
\be\label{shifted-S}
\wt\d_{\h_0}\equiv \d_{\h_0}-\d_{\ve_0= x^\m\g_\m\h_0}.
\ee 
As we will see in section \ref{sec:confsugra}, the flat space limit of the S-supersymmetry transformation in conformal supergravity coincides with $\wt\d_{\h_0}$ -- not with $\d_{\h_0}$, i.e. $\hat\d^{\rm sugra}_{\h_0}=\wt\d_{\h_0}$.  Utilizing the Q-supersymmetry identity in \eqref{PathIntegralWIDs}, we therefore arrive at the gamma trace constraint 
\be\label{PathIntegralWIDs-gamma-constraint}
\int d^4x\,\bar\h_0(x)\g_\m\lb\hat\cq^\m(x)\co_1(x_1)\cdots\co_n(x_n)\rb
=\sum_{i=1}^n\lb\co_1(x_1)\cdots\wt\d_{\h_0}\co_i(x_i)\cdots\co_n(x_n)\rb.
\ee

Inserting the transformations \eqref{current-S-transformations-flat} in \eqref{PathIntegralWIDs-gamma-constraint} leads to the naive gamma trace Ward identities
\bal\label{S-susy-WID-2pt-PI}
&\g_\m\lb\hat\cq^\m(x)\bar\cq^\s(y)\rb=\frac{3i}{4}\d(x,y)\g^5\lb\hat\cj^\s(x)\rb-\frac{3i}{4}\d(x,y)\g^5\lb\orangebox{\hat\cj^\s(x)}\rb+\frac{1}{3\a^2}\pa^x_\r\d(x,y)\g^{\s\r}\lb\orangebox{\hat s_{(1|0)}(y)}\rb\NO\\
&+\frac{1}{2\a^2}\d(x,y)\g^\s\g_\r(\lb\orangebox{\hat s^\r_{(4|1)}(x)}\rb P_{L}+\lb\orangebox{\hat s^{\r*}_{(4|1)}(x)}\rb P_{R}),\NO\\
&\rule{.0cm}{.9cm}\g_\m\lb\hat\cq^\m(x)\bar\cq^\s(y)\hat\cj^\k(z)\rb=\frac{3i}{4}\d(x,y)\g^5\lb\hat\cj^\s(x)\hat\cj^\k(z)\rb-\frac{3i}{4}\d(x,y)\g^5\lb\orangebox{\hat\cj^\s(x)}\hat\cj^\k(z)\rb\NO\\
&+\frac{1}{2\a^2}\d(x,y)\g^\s\g_\r(\lb\orangebox{\hat s^\r_{(4|1)}(x)}\hat\cj^\k(z)\rb P_{L}+\lb\orangebox{\hat s^{\r*}_{(4|1)}(x)}\hat\cj^\k(z)\rb P_{R})\NO\\
&+\frac{1}{3\a^2}\pa^x_\r\d(x,y)\g^{\s\r}\lb\orangebox{\hat s_{(1|0)}(y)}\hat\cj^\k(z)+\frac{1}{3\a}\d(x,z)\g^\k\lb\orangebox{\hat s_{\(3|\frac12\)}(z)}\bar\cq^\s(y)\rb,\NO\\
&\rule{.0cm}{.9cm}\g_\m\lb\hat\cq^\m(x)\bar\cq^\s(y)\hat\cj^\k(z)\hat\cj^\a(w)\rb=\frac{3i}{4}\d(x,y)\g^5\lb\hat\cj^\s(x)\hat\cj^\k(z)\hat\cj^\a(w)\rb\NO\\
&-\frac{3i}{4}\d(x,y)\g^5\lb\orangebox{\hat\cj^\s(x)}\hat\cj^\k(z)\hat\cj^\a(w)\rb+\frac{1}{3\a^2}\pa^x_\r\d(x,y)\g^{\s\r}\lb\orangebox{\hat s_{(1|0)}(y)}\hat\cj^\k(z)\hat\cj^\a(w)\rb\NO\\
&+\frac{1}{2\a^2}\d(x,y)\g^\s\g_\r(\lb\orangebox{\hat s^\r_{(4|1)}(x)}\hat\cj^\k(z)\hat\cj^\a(w)\rb P_{L}+\lb\orangebox{\hat s^{\r*}_{(4|1)}(x)}\hat\cj^\k(z)\hat\cj^\a(w)\rb P_{R})\NO\\
&+\frac{1}{3\a}\d(x,z)\g^\k\lb\orangebox{\hat s_{\(3|\frac12\)}(x)}\bar\cq^\s(y)\hat\cj^\a(w)\rb+\frac{1}{3\a}\d(x,w)\g^\a\lb\orangebox{\hat s_{\(3|\frac12\)}(x)}\bar\cq^\s(y)\hat\cj^\k(z)\rb,
\eal
which should be compared respectively with \eqref{S-susy-WID-2pt}, \eqref{S-susy-WID-3pt} and \eqref{S-susy-WID-4pt}. Again, we see that these agree with their counterparts in section \ref{sec:WardIDs}, up to seagull terms and quantum anomalies.

\subsection{Ward identities for the regulated Wess-Zumino model}

The classical action of the regulated massless WZ model is  
\be\label{PV-reg-theory}
\hat S_{\rm reg}\equiv \hat S_{\rm WZ}+\hat S_{\rm PV},
\ee   
where $\hat S_{\rm WZ}$ is the massless WZ model action \eqref{WZ-lagrangian-flat} and $\hat S_{\rm PV}$ is the PV action \eqref{PV-lagrangian-Dirac}. The total action is invariant only under the symmetries preserved by the regulator, namely spacetime translations, Lorentz transformations and Q-supersymmetry, so that   
\be
\d_a \hat S_{\rm reg}=-\int d^4x\,\pa_\m a^\n\hat{\wt\ct}{}^\m{}_\n,\quad \d_\ell \hat S_{\rm reg}=-\int d^4x\,\pa_\m \ell^{\r\s}x_{\r}\hat{\wt\ct}{}^\m{}_{\s},\quad \d_{\ve_0} \hat S_{\rm reg}=-\int d^4x\,\pa_\m \bar\ve_0\hat{\wt\cq}{}^\m.
\ee

As for the bare theory, inserting these variations in the path integral for the regulated theory leads to the integrated identities
\bal\label{PathIntegralWIDs-massive}
&-\int d^4x\,a^\n(x)\pa^x_\m\lb\hat{\wt\ct}{}^\m{}_\n(x)\wt\co_1(x_1)\cdots\wt\co_n(x_n)\rb\,=\sum_{i=1}^n\lb\wt\co_1(x_1)\cdots\d_a\wt\co_i(x_i)\cdots\wt\co_n(x_n)\rb,\NO\\
&-\int d^4x\,\ell^{\r\s}(x)\pa^x_\m\lb x_{[\r}\hat{\wt\ct}{}^\m_{\s]}(x)\wt\co_1(x_1)\cdots\wt\co_n(x_n)\rb\,=\sum_{i=1}^n\lb\wt\co_1(x_1)\cdots\d_\ell\wt\co_i(x_i)\cdots\wt\co_n(x_n)\rb,\NO\\
&-\int d^4x\,\bar\ve_0(x)\pa^x_\m\lb\hat{\wt\cq}{}^\m(x)\wt\co_1(x_1)\cdots\wt\co_n(x_n)\rb\,=\sum_{i=1}^n\lb\wt\co_1(x_1)\cdots\d_{\ve_0}\wt\co_i(x_i)\cdots\wt\co_n(x_n)\rb,
\eal
where again we have suppressed any spacetime indices in the operators $\wt\co_i$.

\subsection*{Symmetry transformations of the Noether currents}

As for the conformal case, in order to turn the integrated identities \eqref{PathIntegralWIDs-massive} into local constraints for flat space correlation functions we need to evaluate the classical symmetry transformations of the operator insertions. We are only interested in the supersymmetry Ward identity for correlation functions involving the supercurrent and the R-current and, hence, we need only determine the naive supersymmetry transformations of these two operators. 

Restoring the auxiliary fields in the PV multiplets, we determine that the off shell supersymmetry transformations of the PV R-current and of the supercurrent take respectively the form 
\bal\label{PV-R-current-Q-transformation-flat}
\d_{\ve_0}\hskip2pt\hat{\hskip-2pt\wt\cj}{}^\m\big|_{\rm PV}=&\;-\frac{i}{3}\bar\ve_0\g^5(2\h^\m_\n-\g^\m{}_\n)\hat{\wt\cq}{}^\n\big|_{\rm PV}\NO\\
&+\orangebox{\frac{i}{3\a}\bar\ve_0\stackrel{\leftarrow}{\pa^\m\hskip-2pt}(\vf_2^*\l_{2L}-\vf_2\l_{2R})}-\orangebox{\frac{i}{3\a}\bar\ve_0\stackrel{\leftarrow}{\pa^\m\hskip-2pt}\big(\vf_1^*\l_{1L}-\vf_1\l_{1R}^C+\vth_1\l_{1R}-\vth_1^*\l_{1L}^C\big)}\NO\\
&+\cyanbox{\frac{i}{3\a}\bar\ve_0\g^\m\big(\vf_2^*(\slashed\pa\l_{2L}+m_2\l_{2R})-\vf_2(\slashed\pa\l_{2R}+m_2\l_{2L})\big)}\NO\\
&-\cyanbox{\frac{i}{3\a}\bar\ve_0\g^\m\big(\vf_1^*(\slashed\pa\l_{1L}+m_1\l_{1R})+\vth_1(\slashed\pa\l_{1R}+m_1\l_{1L})\big)}\NO\\
&+\cyanbox{\frac{i}{3\a}\bar\ve_0\g^\m\big(\vf_1(\slashed\pa\l_{1R}^C+m_1\l_{1L}^C)+\vth_1^*(\slashed\pa\l_{1L}^C+m_1\l_{1R}^C)\big)}\NO\\
&+\cyanbox{\frac{i}{6\a}\bar\ve_0\g^\m\big((F_{\vf_2}^*+m_2\vf_2)\l_{2L}-(F_{\vf_2}+m_2\vf_2^*)\l_{2R}\big)}\NO\\
&+\cyanbox{\frac{i}{6\a}\bar\ve_0\g^\m\big((F_{\vf_1}+m_1\vf_1^*)\l_{1R}+(F_{\vth_1}^*+m_1\vth_1)\l_{1L}\big)}\NO\\
&-\cyanbox{\frac{i}{6\a}\bar\ve_0\g^\m\big((F_{\vf_1}^*+m_1\vf_1)\l_{1L}^C+(F_{\vth_1}+m_1\vth_1^*)\l_{1R}^C\big)},
\eal
\bal\label{PV-supercurrent-Q-transformation-flat}
\d_{\ve_0} \hat{\wt\cq}{}^\m\big|_{\rm PV}=&\;\frac12\hat{\wt\ct}{}^\m{}_{\n}\big|_{\rm PV}\g^\n\ve_0+\frac{i}{8}\pa_\r\big(\hskip2pt\hat{\hskip-2pt\wt\cj}{}_{\s}\big|_{\rm PV}(i\e^{\m\n\r\s}\g^5+2\h^{\m\n}\h^{\r\s}-2\h^{\r\n}\h^{\m\s})\g_\n\g^5\ve_0\big)\NO\\
&+\frac{1}{3}\g^{\m\n}\pa_\n\big(\hat{\wt\co}_M\big|_{\rm PV}\ve_{0R}+\hat{\wt\co}_{M^*}\big|_{\rm PV}\ve_{0L}\big)-\orangebox{\frac{3}{8}\e^{\m\n\r\s}\hskip2pt\hat{\hskip-2pt\wt\cj}{}_{\s}\big|_{\rm PV}\g_\n\pa_\r\ve_0}\NO\\
&-\orangebox{\g^{\m\n}\big(\hat{\wt\co}_M\big|_{\rm PV}\pa_\n\ve_{0R}+\hat{\wt\co}_{M^*}\big|_{\rm PV}\pa_\n\ve_{0L}\big)}+\orangebox{\frac18(\bar\l_2\g_\s\g^5\l_2+2\dbar\l_1\g_\s\g^5\l_1)\g^\s\g^5\pa^\m\ve_0}\NO\\
&+\orangebox{\frac{1}{12\a^2}\pa_\r(\vf^*_2\vf_2+\vf_1^*\vf_1+\vth_1\vth^*_1)(i\e^{\m\n\r\s}\g^5+\h^{\m\n}\h^{\r\s}-\h^{\m\r}\h^{\n\s})\g_\n\pa_\s\ve_0}\NO\\
&+\cyanbox{\frac{1}{8}\big(\bar\l_2(\g^{\m\n}-\h^{\m\n})\g^5(\slashed\pa+m_2)\l_2\big)\g_\n\g^5\ve_0-\frac{1}{8}\big(\bar\l_2(\g^{\m\n}-\h^{\m\n})(\slashed\pa+m_2)\l_2\big)\g_\n\ve_0}\NO\\
&+\cyanbox{\frac{1}{4}\big(\dbar\l_1(\g^{\m\n}-\h^{\m\n})\g^5(\slashed\pa+m_1)\l_1\big)\g_\n\g^5\ve_0-\frac{1}{4}\big(\dbar\l_1(\g^{\m\n}-\h^{\m\n})(\slashed\pa+m_1)\l_1\big)\g_\n\ve_0}\NO\\
&+\cyanbox{\frac{1}{4\a^2}\big((F_{\vf_2}^*+m_2\vf_2)\slashed\pa\vf_2 \g^\m\ve_{0R}+(F_{\vf_2}+m_2\vf_2^*)\slashed\pa\vf_2^*\g^\m\ve_{0L}\big)}\NO\\
&+\cyanbox{\frac{1}{6\a^2}\g^{\m\n}\pa_\n\big(\vf_2 (F_{\vf_2}^*+m_2\vf_2)\ve_{0R}+\vf_2^*(F_{\vf_2}+m_2\vf_2^*) \ve_{0L}\big)}\NO\\
&+\cyanbox{\frac{1}{4\a^2}\big((F_{\vf_1}^*+m_1\vf_1)\slashed\pa\vth_1 \g^\m\ve_{0R}+(F_{\vth_1}+m_1\vth_1^*)\slashed\pa\vf_1^*\g^\m\ve_{0L}}\NO\\
&\hskip2.cm-\cyanbox{(F_{\vth_1}^*+m_1\vth_1)\slashed\pa\vf_1\g^\m\ve_{0R}-(F_{\vf_1}+m_1\vf_1^*)\slashed\pa\vth_1^*\g^\m\ve_{0L}\big)}\NO\\
&+\cyanbox{\frac{1}{6\a^2}\g^{\m\n}\pa_\n\big((F_{\vf_1}^*+m_1\vf_1)\vth_1\ve_{0R}+(F_{\vth_1}+m_1\vth_1^*)\vf_1^*\ve_{0L}}\NO\\
&\hskip2.cm-\cyanbox{(F_{\vth_1}^*+m_1\vth_1)\vf_1\ve_{0R}-(F_{\vf_1}+m_1\vf_1^*)\vth_1^*\ve_{0L}\big)},
\eal
where again seagull terms are indicated by an pink/oval frame, while terms in a blue/rectangular frame vanish on-shell.

\subsection*{Symmetry transformations of the model independent FZ currents}

The model dependent seagull terms in the transformations \eqref{PV-R-current-Q-transformation-flat}-\eqref{PV-supercurrent-Q-transformation-flat} again cancel against the transformation of terms linear in supergravity fields in the FZ currents. In particular, generalizing the FZ multiplet operators \eqref{WZ-currents-OM} to the PV-regulated theory described by the action \eqref{PV-reg-theory} and using the on-shell form of the Noether current transformations \eqref{PV-R-current-Q-transformation-flat}-\eqref{PV-supercurrent-Q-transformation-flat} gives 
\bbxd
\bal\label{OM-current-Q-transformations-flat}
\hskip-.15cm\d_\ve^{\rm om}\wt\cj^\m
=&-\frac{i}{3}\bar\ve\g^5(2\h^\m_\n-\g^\m{}_\n)\hat{\wt\cq}{}^\n,\hskip-.15cm\\
\hskip-.15cm\rule{0.cm}{.7cm}\d^{\rm om}_\ve \wt\cq^\m=&\frac12\hat{\wt\ct}{}^\m_{\n}\g^\n\ve+\frac{i}{4}\pa_\r\Big[\hskip3pt\hat{\hskip-3pt\wt\cj}_{\s}\Big(\frac{i}{2}\e^{\m\n\r\s}\g^5+\h^{\m\n}\h^{\r\s}-\h^{\r\n}\h^{\m\s}\Big)\g_\n\g^5\ve\Big]+\frac{1}{3}\g^{\m\n}\pa_\n\big(\hat{\wt\co}_M\ve_{R}+\hat{\wt\co}_{M^*}\ve_{L}\big).\hskip-.15cm\NO
\eal
\ebxd
These transformations depend only on current multiplet operators and follow directly from the general definition of the FZ current multiplet operators in \eqref{currents-om}, together with the old minimal supergravity transformations  \eqref{old-minimal-susy-transformations}.

\subsection*{Naive supersymmetry Ward identities}

Returning to the path integral derivation of the naive Ward identities for the regulated WZ model, it is convenient to introduce a shorthand notation for the seagull and null operators that enter the Noether current transformations \eqref{PV-R-current-Q-transformation-flat} and \eqref{PV-supercurrent-Q-transformation-flat}. Namely, the analogues of the seagull operators \eqref{seagull-operators-WZ} in the regulated theory are
\bbxd
\bal\label{seagull-operators}
\hat{\wt s}_{(1|0)}=&\;\f^*\f+\vf^*_2\vf_2+\vf_1^*\vf_1+\vth_1\vth^*_1,\NO\\
\hat{\wt s}{}_{(2|1)}^{\m}=&\;\bar\c\g^\m\g^5\c+\bar\l_2\g^\m\g^5\l_2+2\dbar\l_1\g^\m\g^5\l_1,\NO\\
\hat{\wt s}_{\(3|\frac12\)}=&\;i\big(\f^*\c_L-\f\c_R+\vf_2^*\l_{2L}-\vf_2\l_{2R}-\vf_1^*\l_{1L}+\vf_1\l_{1R}^C-\vth_1\l_{1R}+\vth_1^*\l_{1L}^C\big),
\eal
\ebxd
while the null operators \eqref{null-operators-WZ} become
\bbxd
\bal\label{null-operators}
\hat{\wt n}^{\m\n}_{(1|2)}=&\;\bar\c(\g^{\m\n}-\h^{\m\n})\g^5\slashed\pa\c+\bar\l_2(\g^{\m\n}-\h^{\m\n})\g^5(\slashed\pa+m_2)\l_2+2\dbar\l_1(\g^{\m\n}-\h^{\m\n})\g^5(\slashed\pa+m_1)\l_1,\NO\\
\hat{\wt n}^{\m\n}_{(2|2)}=&\;\bar\c(\g^{\m\n}-\h^{\m\n})\slashed\pa\c+\bar\l_2(\g^{\m\n}-\h^{\m\n})(\slashed\pa+m_2)\l_2+2\dbar\l_1(\g^{\s\n}-\h^{\m\n})(\slashed\pa+m_1)\l_1,\NO\\
\hat{\wt n}_{\(3|\frac12\)}=&\;i\big(\f^*\slashed\pa\c_L-\f\slashed\pa\c_R+\vf_2^*(\slashed\pa\l_{2L}+m_2\l_{2R})-\vf_2(\slashed\pa\l_{2R}+m_2\l_{2L})\NO\\
&-\vf_1^*(\slashed\pa\l_{1L}+m_1\l_{1R})+\vf_1(\slashed\pa\l_{1R}^C+m_1\l_{1L}^C)-\vth_1(\slashed\pa\l_{1R}+m_1\l_{1L})+\vth_1^*(\slashed\pa\l_{1L}^C+m_1\l_{1R}^C)\big),\NO\\
\hat{\wt n}^\m_{(5|0)}=&\;F^*\pa^\m\f +(F_{\vf_2}^*+m_2\vf_2)\pa^\m\vf_2+(F_{\vf_1}^*+m_1\vf_1)\pa^\m\vth_1+(F_{\vth_1}+m_1\vth_1^*)\pa^\m\vf_1^*,\NO\\
\hat{\wt n}_{(4|0)}=&\;F^*\f+(F_{\vf_2}^*+m_2\vf_2)\vf_2+(F_{\vf_1}^*+m_1\vf_1)\vth_1+(F_{\vth_1}+m_1\vth_1^*)\vf_1^*,\NO\\
\hat{\wt n}_{\(6|\frac12\)}=&\;i\big(F^*\c_L-F\c_R+(F_{\vf_2}^*+m_2\vf_2)\l_{2L}-(F_{\vf_2}+m_2\vf_2^*)\l_{2R}\\
&+(F_{\vf_1}+m_1\vf_1^*)\l_{1R}+(F_{\vth_1}^*+m_1\vth_1)\l_{1L}-(F_{\vf_1}^*+m_1\vf_1)\l_{1L}^C-(F_{\vth_1}+m_1\vth_1^*)\l_{1R}^C\big).\NO
\eal
\ebxd

Inserting the supercurrent transformation \eqref{PV-supercurrent-Q-transformation-flat} in the integrated supersymmetry Ward identity \eqref{PathIntegralWIDs-massive} results in the following local Ward identity for the 2-point function of the supercurrent: 
\bal\label{Reg-Q-susy-WID-2pt-PI}
&\pa^x_\m\lb\hat{\wt\cq}{}^\m(x)\hat{\wt{\bar{\cq}}}{\hskip7pt}^\s(y)\rb-\frac12\lb\hat{\wt\ct}{}^\s{}_{\t}(x)\rb\g^\t\d(x,y)
+\frac{i}{8}\lb\hskip2pt\hat{\hskip-2pt\wt\cj}{}^{\n}(x)\rb(i\e^{\s\r}{}_{\t\n}\g^5+4\h^{[\s}_\t\h^{\r]}_\n)\g^5\g^\t\pa^x_\r\d(x,y)\NO\\
&+\frac{1}{3}\big(\lb\hat{\wt\co}_M(x)\rb P_{R}+\lb\hat{\wt\co}_{M^*}(x)\rb P_{L}\big)\g^{\s\t}\pa_\t^x\d(x,y)+\frac{3}{8}\lb\orangebox{\hskip2pt\hat{\hskip-2pt\wt\cj}{}_{\n}(y)}\rb\e^{\r\s\n\t}\g_\t\pa^x_\r\d(x,y)\NO\\
&-\big(\lb\orangebox{\hat{\wt\co}_M(y)}\rb P_R+\lb\orangebox{\hat{\wt\co}_{M^*}(y)}\rb P_L\big)\g^{\s\t}\pa_\t^x\d(x,y)
+\frac18\lb\orangebox{\hat{\wt s}{}_{(2|1)}^{\t}(y)}\rb\g^5\g_\t\pa^\s_x\d(x,y)\NO\\
&+\frac{1}{12\a^2}\pa^\n_y\lb\orangebox{\hat{\wt s}_{(1|0)}(y)}\rb\big(2\h^{[\s}_\t\h_\n^{\r]}-i\e^{\s\r}{}_{\t\n}\g^5\big)\g^\t\pa^x_\r\d(x,y)
+\frac{1}{8}\lb\cyanbox{\hat{\wt n}^{\s\n}_{(1|2)}(x)}\rb\g_\n\g^5\d(x,y)\NO\\
&+\frac{1}{8}\lb\cyanbox{\hat{\wt n}^{\s\n}_{(2|2)}(x)}\rb\g_\n\d(x,y)
+\frac{1}{4\a^2}\big(\lb\cyanbox{\hat{\wt n}^\r_{(5|1)}(x)}\rb P_R+\lb\cyanbox{\hat{\wt n}^{\r*}_{(5|1)}(x)}\rb P_L \big)\g^\s\g_\r\d(x,y)\NO\\
&+\frac{1}{6\a^2}\big(\lb\cyanbox{\hat{\wt n}_{(4|0)}(x)}\rb P_R+\lb\cyanbox{\hat{\wt n}_{(4|0)}^*(x)}\rb P_L\big) \g^{\s\n}\pa^x_\n\d(x,y)=0.
\eal
In terms of composite operators, this identity coincides with \eqref{Q-susy-WID-2pt-PI} for the bare theory, except for terms involving the complex scalar operator $\co_M$ of the FZ multiplet. Moreover, using the fact that the 1-point functions of all null operators vanish and absorbing the seagull terms in the definition of the current correlators, it is straightforward to verify that it reproduces the Ward identity \eqref{OM-susy-WID-2pt} obtained from old minimal supergravity.

For the 3- and 4-point functions we obtain similarly
\bal\label{Reg-Q-susy-WID-3pt-PI}
&\pa^x_\m\lb\hat{\wt\cq}{}^\m(x)\hat{\wt{\bar{\cq}}}{\hskip7pt}^\s(y)\hskip2pt\hat{\hskip-2pt\wt\cj}{}^{\k}(z)\rb-\frac{i}{3}\d(x,z)\g^5(2\h^\k_\n-\g^\k{}_\n)\lb\hat{\wt\cq}{}^\n(x)\hat{\wt{\bar{\cq}}}{\hskip7pt}^\s(y)\rb-\frac12\lb\hat{\wt\ct}{}^\s{}_{\t}(x)\hskip2pt\hat{\hskip-2pt\wt\cj}{}^{\k}(z)\rb\g^\t\d(x,y)\NO\\
&+\frac{i}{8}\lb\hskip2pt\hat{\hskip-2pt\wt\cj}{}^{\n}(x)\hskip2pt\hat{\hskip-2pt\wt\cj}{}^{\k}(z)\rb(i\e^{\s\r}{}_{\t\n}\g^5+4\h^{[\s}_\t\h^{\r]}_\n)\g^5\g^\t\pa^x_\r\d(x,y)+\frac{3}{8}\lb\orangebox{\hskip2pt\hat{\hskip-2pt\wt\cj}{}_{\n}(y)}\hskip2pt\hat{\hskip-2pt\wt\cj}{}^{\k}(z)\rb\e^{\r\s\n\t}\g_\t\pa^x_\r\d(x,y)\NO\\
&+\frac{1}{3}\big(\lb\hat{\wt\co}_M(x)\hskip2pt\hat{\hskip-2pt\wt\cj}{}^{\k}(z)\rb P_R+\lb\hat{\wt\co}_{M^*}(x) \hskip2pt\hat{\hskip-2pt\wt\cj}{}^{\k}(z)\rb P_{L}\big)\g^{\s\t}\pa_\t^x\d(x,y)\NO\\
&-\big(\lb\orangebox{\hat{\wt\co}_M(y) }\hskip2pt\hat{\hskip-2pt\wt\cj}{}^{\k}(z)\rb P_R+\lb\orangebox{\hat{\wt\co}_{M^*}(y)}\hskip2pt\hat{\hskip-2pt\wt\cj}{}^{\k}(z)\rb P_L\big)\g^{\s\t}\pa_\t^x\d(x,y)-\frac{1}{3\a}\pa_x^\k\d(x,z)\lb\orangebox{\hat{\wt s}_{\(3|\frac12\)}(z)}\hat{\wt{\bar{\cq}}}{\hskip7pt}^\s(y)\rb\NO\\
&+\frac18\lb\orangebox{\hat{\wt s}{}_{(2|1)}^{\t}(y)}\hskip2pt\hat{\hskip-2pt\wt\cj}{}^{\k}(z)\rb\g^5\g_\t\pa^\s_x\d(x,y)
+\frac{1}{12\a^2}\pa^\n_y\lb\orangebox{\hat{\wt s}_{(1|0)}(y)}\hskip2pt\hat{\hskip-2pt\wt\cj}{}^{\k}(z)\rb\big(2\h^{[\s}_\t\h_\n^{\r]}-i\e^{\s\r}{}_{\t\n}\g^5\big)\g^\t\pa^x_\r\d(x,y)\NO\\
&+\frac{1}{3\a}\d(x,z)\g^\k\lb\cyanbox{\hat{\wt n}_{\(3|\frac12\)}(x)}\hat{\wt{\bar{\cq}}}{\hskip7pt}^\s(y)\rb+\frac{1}{6\a}\d(x,z)\g^\k\lb\cyanbox{\hat{\wt n}_{\(6|\frac12\)}(x)}\hat{\wt{\bar{\cq}}}{\hskip7pt}^\s(y)\rb\NO\\
&+\frac{1}{8}\lb\cyanbox{\hat{\wt n}^{\s\n}_{(1|2)}(x)}\hskip2pt\hat{\hskip-2pt\wt\cj}{}^{\k}(z)\rb\g_\n\g^5\d(x,y)
+\frac{1}{8}\lb\cyanbox{\hat{\wt n}^{\s\n}_{(2|2)}(x)}\hskip2pt\hat{\hskip-2pt\wt\cj}{}^{\k}(z)\rb\g_\n\d(x,y)\NO\\
&+\frac{1}{4\a^2}\big(\lb\cyanbox{\hat{\wt n}^\r_{(5|1)}(x)}\hskip2pt\hat{\hskip-2pt\wt\cj}{}^{\k}(z)\rb P_R+\lb\cyanbox{\hat{\wt n}^{\r*}_{(5|1)}(x)}\hskip2pt\hat{\hskip-2pt\wt\cj}{}^{\k}(z)\rb P_L\big) \g^\s\g_\r\d(x,y)\NO\\
&+\frac{1}{6\a^2}\big(\lb\cyanbox{\hat{\wt n}_{(4|0)}(x)}\hskip2pt\hat{\hskip-2pt\wt\cj}{}^{\k}(z)\rb P_R+\lb\cyanbox{\hat{\wt n}_{(4|0)}^*(x)}\hskip2pt\hat{\hskip-2pt\wt\cj}{}^{\k}(z)\rb P_L\big) \g^{\s\n}\pa^x_\n\d(x,y)=0,
\eal
\bal\label{Reg-Q-susy-WID-4pt-PI}
&\pa^x_\m\lb\hat{\wt\cq}{}^\m(x)\hat{\wt{\bar{\cq}}}{\hskip7pt}^\s(y)\hskip2pt\hat{\hskip-2pt\wt\cj}{}^{\k}(z)\hskip2pt\hat{\hskip-2pt\wt\cj}{}^{\a}(w)\rb-\frac{i}{3}\d(x,z)\g^5(2\h^\k_\n-\g^\k{}_\n)\lb\hat{\wt\cq}{}^\n(x)\hat{\wt{\bar{\cq}}}{\hskip7pt}^\s(y)\hskip2pt\hat{\hskip-2pt\wt\cj}{}^{\a}(w)\rb\NO\\
&-\frac{i}{3}\d(x,w)\g^5(2\h^\a_\n-\g^\a{}_\n)\lb\hat{\wt\cq}{}^\n(x)\hat{\wt{\bar{\cq}}}{\hskip7pt}^\s(y)\hskip2pt\hat{\hskip-2pt\wt\cj}{}^{\k}(z)\rb-\frac12\lb\hat{\wt\ct}{}^\s{}_{\t}(x)\hskip2pt\hat{\hskip-2pt\wt\cj}{}^{\k}(z)\hskip2pt\hat{\hskip-2pt\wt\cj}{}^{\a}(w)\rb\g^\t\d(x,y)\NO\\
&+\frac{i}{8}\lb\hskip2pt\hat{\hskip-2pt\wt\cj}{}^{\n}(x)\hskip2pt\hat{\hskip-2pt\wt\cj}{}^{\k}(z)\hskip2pt\hat{\hskip-2pt\wt\cj}{}^{\a}(w)\rb(i\e^{\s\r}{}_{\t\n}\g^5+4\h^{[\s}_\t\h^{\r]}_\n)\g^5\g^\t\pa^x_\r\d(x,y)\NO\\
&+\frac{1}{3}\big(\lb\hat{\wt\co}_M(x) \hskip2pt\hat{\hskip-2pt\wt\cj}{}^{\k}(z)\hskip2pt\hat{\hskip-2pt\wt\cj}{}^{\a}(w)\rb P_R +\lb\hat{\wt\co}_{M^*}(x)\hskip2pt\hat{\hskip-2pt\wt\cj}{}^{\k}(z)\hskip2pt\hat{\hskip-2pt\wt\cj}{}^{\a}(w)\rb P_{L}\big)\g^{\s\t}\pa_\t^x\d(x,y)\NO\\
&-\big(\lb\orangebox{\hat{\wt\co}_M (y)}\hskip2pt\hat{\hskip-2pt\wt\cj}{}^{\k}(z)\hskip2pt\hat{\hskip-2pt\wt\cj}{}^{\a}(w)\rb P_R+\lb\orangebox{\hat{\wt\co}_{M^*}(y)}\hskip2pt\hat{\hskip-2pt\wt\cj}{}^{\k}(z)\hskip2pt\hat{\hskip-2pt\wt\cj}{}^{\a}(w)\rb P_L\big)\g^{\s\t}\pa_\t^x\d(x,y)\NO\\
&+\frac{3}{8}\lb\orangebox{\hskip2pt\hat{\hskip-2pt\wt\cj}{}_{\n}(y)}\hskip2pt\hat{\hskip-2pt\wt\cj}{}^{\k}(z)\hskip2pt\hat{\hskip-2pt\wt\cj}{}^{\a}(w)\rb\e^{\r\s\n\t}\g_\t\pa^x_\r\d(x,y)
+\frac18\lb\orangebox{\hat{\wt s}{}_{(2|1)}^{\t}(y)}\hskip2pt\hat{\hskip-2pt\wt\cj}{}^{\k}(z)\hskip2pt\hat{\hskip-2pt\wt\cj}{}^{\a}(w)\rb\g^5\g_\t\pa^\s_x\d(x,y)\NO\\
&+\frac{1}{12\a^2}\pa^\n_y\lb\orangebox{\hat{\wt s}_{(1|0)}(y)}\hskip2pt\hat{\hskip-2pt\wt\cj}{}^{\k}(z)\hskip2pt\hat{\hskip-2pt\wt\cj}{}^{\a}(w)\rb\big(2\h^{[\s}_\t\h_\n^{\r]}-i\e^{\s\r}{}_{\t\n}\g^5\big)\g^\t\pa^x_\r\d(x,y)\NO\\
&-\frac{1}{3\a}\pa_x^\k\d(x,z)\lb\orangebox{\hat{\wt s}_{\(3|\frac12\)}(z)}\hat{\wt{\bar{\cq}}}{\hskip7pt}^\s(y)\hskip2pt\hat{\hskip-2pt\wt\cj}{}^{\a}(w)\rb
-\frac{1}{3\a}\pa_x^\a\d(x,w)\lb\orangebox{\hat{\wt s}_{\(3|\frac12\)}(w)}\hat{\wt{\bar{\cq}}}{\hskip7pt}^\s(y)\hskip2pt\hat{\hskip-2pt\wt\cj}{}^{\k}(z)\rb\NO\\
&+\frac{1}{8}\lb\cyanbox{\hat{\wt n}^{\s\n}_{(1|2)}(x)}\hskip2pt\hat{\hskip-2pt\wt\cj}{}^{\k}(z)\hskip2pt\hat{\hskip-2pt\wt\cj}{}^{\a}(w)\rb\g_\n\g^5\d(x,y)
+\frac{1}{8}\lb\cyanbox{\hat{\wt n}^{\s\n}_{(2|2)}(x)}\hskip2pt\hat{\hskip-2pt\wt\cj}{}^{\k}(z)\hskip2pt\hat{\hskip-2pt\wt\cj}{}^{\a}(w)\rb\g_\n\d(x,y)\NO\\
&+\frac{1}{3\a}\d(x,z)\g^\k\lb\cyanbox{\hat{\wt n}_{\(3|\frac12\)}(x)}\hat{\wt{\bar{\cq}}}{\hskip7pt}^\s(y)\hskip2pt\hat{\hskip-2pt\wt\cj}{}^{\a}(w)\rb
+\frac{1}{3\a}\d(x,w)\g^\a\lb\cyanbox{\hat{\wt n}_{\(3|\frac12\)}(x)}\hat{\wt{\bar{\cq}}}{\hskip7pt}^\s(y)\hskip2pt\hat{\hskip-2pt\wt\cj}{}^{\k}(z)\rb\NO\\
&+\frac{1}{4\a^2}\big(\lb\cyanbox{\hat{\wt n}^\r_{(5|1)}(x)}\hskip2pt\hat{\hskip-2pt\wt\cj}{}^{\k}(z)\hskip2pt\hat{\hskip-2pt\wt\cj}{}^{\a}(w)\rb P_R+\lb\cyanbox{\hat{\wt n}^{\r*}_{(5|1)}(x)}\hskip2pt\hat{\hskip-2pt\wt\cj}{}^{\k}(z)\hskip2pt\hat{\hskip-2pt\wt\cj}{}^{\a}(w)\rb P_L\big) \g^\s\g_\r\d(x,y)\NO\\
&+\frac{1}{6\a^2}\big(\lb\cyanbox{\hat{\wt n}_{(4|0)}(x)}\hskip2pt\hat{\hskip-2pt\wt\cj}{}^{\k}(z)\hskip2pt\hat{\hskip-2pt\wt\cj}{}^{\a}(w)\rb P_R+\lb\cyanbox{\hat{\wt n}_{(4|0)}^*(x)}\hskip2pt\hat{\hskip-2pt\wt\cj}{}^{\k}(z)\hskip2pt\hat{\hskip-2pt\wt\cj}{}^{\a}(w)\rb P_L\big) \g^{\s\n}\pa^x_\n\d(x,y)\NO\\
&+\frac{1}{6\a}\d(x,z)\g^\k\lb\cyanbox{\hat{\wt n}_{\(6|\frac12\)}(x)}\hat{\wt{\bar{\cq}}}{\hskip7pt}^\s(y)\hskip2pt\hat{\hskip-2pt\wt\cj}{}^{\a}(w)\rb
+\frac{1}{6\a}\d(x,w)\g^\a\lb\cyanbox{\hat{\wt n}_{\(6|\frac12\)}(x)}\hat{\wt{\bar{\cq}}}{\hskip7pt}^\s(y)\hskip2pt\hat{\hskip-2pt\wt\cj}{}^{\k}(z)\rb=0.
\eal
These Ward identities coincide with those of the bare theory, respectively \eqref{Q-susy-WID-3pt-PI} and \eqref{Q-susy-WID-4pt-PI}, except for terms involving the complex scalar operator $\co_M$ and the gamma trace of the supercurrent, both of which vanish in the bare theory. Moreover, evaluating the 2- and 3-point functions involving null operators as in \eqref{current-contraction-1pt} and \eqref{current-contraction-2pt} and absorbing the seagull terms in the definition of the correlators, one finds that these identities reproduce the Ward identities \eqref{OM-susy-WID-3pt} and \eqref{OM-susy-WID-4pt} that follow from the coupling to old minimal supergravity.

\section{Functional differentiation versus operator insertions}
\label{sec:seagulls}

In this appendix we discuss the relation between correlation functions defined through functional differentiation and those obtained by path integral operator insertions. The two differ by the so called `seagull terms', which can be expressed as derivatives of operators with respect to sources. The distinction is important because the structure of ultraviolet divergences is different in the two cases. In particular, only the ultraviolet divergences of current multiplet correlators defined through functional differentiation can be canceled by counterterms that depend on the background supergravity fields. Feynman diagram computations, however, result in correlation functions involving operator insertions. The two different definitions also affect the form of the Ward identities. 

Throughout this manuscript correlation functions defined through functional differentiation are denoted by wide brackets, $\< \cdot \>$, while those involving operator insertions by $\lb\cdot\rb$. For the current multiplet correlators of the PV regulated WZ model we discuss in the main text, the relation between these two definitions simplifies because several operator derivatives vanish. In particular, using the form of the currents in \eqref{WZ-currents-CS} and \eqref{WZ-currents-OM} when the theory is coupled to background supergravity and the seagull operators defined in \eqref{seagull-operators}, we determine that the only non zero operator derivatives are 
\bbxd
\bal\label{operator-derivatives}
\frac{\d\wt\ct^{\m\n}(x)}{\d A_\r(y)}=&\;-(\h^{\m\r}\h^{\n\s}+\h^{\m\s}\h^{\n\r}-\h^{\m\n}\h^{\r\s})\d(x,y)\wt\cj_\s+\frac{i}{6}\h^{\m\r}\d(x,y)\hat{\wt s}{}_{(2|1)}^{\n},\NO\\
\frac{\d\wt\ct^{\m\n}(x)}{\d A_\r(y)\d A_\s(z)}=&\;\frac{4}{9\a^2}(\h^{\m\r}\h^{\n\s}+\h^{\m\s}\h^{\n\r}-\h^{\m\n}\h^{\r\s})\d(x,y)\d(x,z)\hat{\wt s}_{(1|0)},\NO\\
\frac{\d\wt\cj^\m(x)}{\d A_\n(y)}=&\;-\frac{4}{9\a^2}\h^{\m\n}\d(x,y)\hat{\wt s}_{(1|0)},\NO\\
\frac{\d\wt\cj^\m(x)}{\d \bar\j_\n(y)}=&\;\frac{\d\wt\cq^\m(x)}{\d A_\n(y)}=-\frac{1}{3\a}\h^{\m\n}\d(x,y)\hat{\wt s}_{\(3|\frac12\)},\\
\frac{\d\wt\cq^\m(x)}{\d \j_\n(y)}=&\;\frac38\e^{\m\n\r\s}\hskip2pt\hat{\hskip-2pt\wt\cj}{}_\r\g_\s\d(x,y)+\g^{\m\n}(\wt\co_M P_R+\wt\co_{M^*}P_L)\d(x,y)+\frac18\hat{\wt s}{}_{(2|1)}^{\r}\g^5\g_\r\h^{\m\n}\d(x,y)\NO\\
&+\frac{1}{12\a^2}\pa^\r\hat{\wt s}_{(1|0)}\big(2\h_\r^{[\m}\h^{\n]}_\s+i\e^{\m\n}{}_{\r\s}\g^5\big)\g^\s\d(x,y)+\frac{i}{6\a^2}\e^{\m\n\r\s}\hat{\wt s}_{(1|0)}\g^5\g_\s\pa^x_\r\d(x,y).\NO
\eal
\ebxd

It follows that the two definitions of current multiplet correlation functions of the regulated WZ model that are relevant for our analysis are related as
\bal\label{correlator-relations}
&\<\wt\cj^\m(x)\wt\cj^\n(y)\>=\;\lb\wt\cj^\m(x)\wt\cj^\n(y)\rb+\lb\frac{\d\wt\cj^\m(x)}{\d A_\n(y)}\rb,\NO\\
&\rule{0.cm}{.7cm}\<\wt\cj^\m(x)\wt\cj^\n(y)\wt\cj^\k(z)\>=\;\lb\wt\cj^\m(x)\wt\cj^\n(y)\wt\cj^\k(z)\rb\NO\\
&\;\hskip1.cm+\lb\frac{\d\wt\cj^\m(x)}{\d A_\k(z)}\wt\cj^\n(y)\rb+\lb\wt\cj^\m(x)\frac{\d\wt\cj^\n(y)}{\d A_\k(z)}\rb+\lb\frac{\d\wt\cj^\m(x)}{\d A_\n(y)}\wt\cj^\k(z)\rb,\NO\\
&\rule{0.cm}{.9cm}\<\wt\ct^\m_\n(x)\wt\cj^\r(y)\>=\;\lb\wt\ct^\m_\n(x)\wt\cj^\r(y)\rb+\lb\frac{\d\wt\ct^\m_\n(x)}{\d A_\r(y)}\rb,\NO\\
&\rule{0.cm}{.7cm}\<\wt\ct^\m_\n(x)\wt\cj^\r(y)\wt\cj^\s(z)\>=\;\lb\wt\ct^\m_\n(x)\wt\cj^\r(y)\wt\cj^\s(z)\rb\NO\\
&\;\hskip1.cm+\lb\frac{\d\wt\ct^\m_\n(x)}{\d A_\r(y)}\wt\cj^\s(z)\rb+\lb\frac{\d\wt\ct^\m_\n(x)}{\d A_\s(z)}\wt\cj^\r(y)\rb+\lb\wt\ct^\m_\n(x)\frac{\d\wt\cj^\r(y)}{\d A_\s(z)}\rb+\lb\frac{\d^2\wt\ct^\m_\n(x)}{\d A_\r(y)\d A_\s(z)}\rb,\NO\\
&\rule{0.cm}{.9cm}\<\wt\cq^\m(x)\wt{\bar\cq}{}^\n(y)\>=\;\lb\wt\cq^\m(x)\wt{\bar\cq}{}^\n(y)\rb+\lb\frac{\d\wt\cq^\m(x)}{\d \j_\n(y)}\rb,\NO\\
&\rule{0.cm}{.7cm}\<\wt\cq^\m(x)\wt{\bar\cq}{}^\n(y)\wt\cj^\r(z)\>=\;\lb\wt\cq^\m(x)\wt{\bar\cq}{}^\n(y)\wt\cj^\r(z)\rb\NO\\
&\;\hskip1.cm+\lb\frac{\d\wt\cq^\m(x)}{\d \j_\n(y)}\wt\cj^\r(z)\rb+\lb\frac{\d\wt\cq^\m(x)}{\d A_\r(z)}\wt{\bar\cq}{}^\n(y)\rb+\lb\wt\cq^\m(x)\frac{\d\wt{\bar\cq}{}^\n(y)}{\d A_\r(z)}\rb,\NO\\
&\rule{0.cm}{.7cm}\<\wt\cq^\m(x)\wt{\bar\cq}{}^\n(y)\wt\cj^\r(z)\wt\cj^\s(w)\>=\;\lb\wt\cq^\m(x)\wt{\bar\cq}{}^\n(y)\wt\cj^\r(z)\wt\cj^\s(w)\rb\NO\\
&\;\hskip1.cm+\lb\frac{\d\wt\cq^\m(x)}{\d \j_\n(y)}\wt\cj^\r(z)\wt\cj^\s(w)\rb+\lb\frac{\d\wt\cq^\m(x)}{\d A_\r(z)}\wt{\bar\cq}{}^\n(y)\wt\cj^\s(w)\rb+\lb\frac{\d\wt\cq^\m(x)}{\d A_\s(w)}\wt{\bar\cq}{}^\n(y)\wt\cj^\r(z)\rb\NO\\
&\;\hskip1.cm+\lb\frac{\d\wt\cq^\m(x)}{\d \j_\n(y)}\frac{\d\wt\cj^\r(z)}{\d A_\s(w)}\rb+\lb\frac{\d\wt\cq^\m(x)}{\d A_\r(z)}\frac{\d\wt{\bar\cq}{}^\n(y)}{A_\s(w)}\rb+\lb\frac{\d\wt\cq^\m(x)}{\d A_\s(w)}\frac{\d\wt{\bar\cq}{}^\n(y)}{\d A_\r(z)}\rb\NO\\
&\;\hskip1.cm+\lb\wt\cq^\m(x)\frac{\d\wt{\bar\cq}{}^\n(y)}{\d A_\r(z)}\wt\cj^\s(w)\rb+\lb\wt\cq^\m(x)\frac{\d\wt{\bar\cq}{}^\n(y)}{\d A_\s(w)}\wt\cj^\r(z)\rb+\lb\wt\cq^\m(x)\wt{\bar\cq}{}^\n(y)\frac{\wt\cj^\r(z)}{\d A_\s(w)}\rb.
\eal
Analogous expressions can be derived for any other current multiplet correlator. Explicit evaluation of the correlation functions \eqref{correlator-relations} in appendix \ref{sec:divergences} shows that individual correlators on the r.h.s. of these relations contain additional UV divergences that cancel among different path integral correlators. The UV divergences that survive in the linear combination corresponding to correlators defined through functional differentiation can be canceled by the counterterms \eqref{counterterms-1}-\eqref{counterterms-4} that depend only on the background supergravity fields.      

Clearly, the difference between correlation functions defined via functional differentiation and operator insertions affects the form of the Ward identities. In sections \ref{sec:CS-WIDs-flat} and \ref{sec:omsugra} we present the Ward identities in terms of correlation functions defined through functional differentiation. This form of the Ward identities is universal and follows directly from the symmetries of the background supergravity the current multiplet couples to. When expressed in terms of correlators defined through operator insertions, however, these identities contain additional seagull terms, as can be seen explicitly in the path integral derivation of the Ward identities in appendix \ref{sec:path-integralWIDs}. In fact, the universal form of the Ward identities in terms of correlators obtained by functional differentiation can be expressed as a linear combination of path integral Ward identities involving operator insertions.   

To illustrate this point, let us consider the Ward identities \eqref{OM-susy-WID-2pt}-\eqref{OM-susy-WID-4pt} for the FZ multiplet that follow from old minimal supergravity. Using the relations \eqref{correlator-relations} between the two alternative definitions of correlation functions, \eqref{OM-susy-WID-2pt} becomes
\bal\label{OM-susy-WID-2pt-OI}
&\pa_\m \lb\wt\cq^\m(x)\wt{\bar\cq}{}^\s(y)\rb-\frac12 \lb\wt\ct^\s_a(x)\rb\g^a\d(x,y)+\frac{i}{8}\lb\wt\cj^\n(x)\rb\big(4\d^{[\r}_\n\d^{\s]}_\l+i\g^5 \e_{\n\l}{}^{\r\s}\big)\g^5\g^\l \pa_\r\d(x,y)\NO\\
&+\frac13\big(\lb\wt\co_M\rb P_R+\lb\wt\co_{M^*}\rb P_L\big)\g^{\s\r}\pa_\r^x\d(x,y)+\pa_\m\lb\orangebox{\frac{\d\wt\cq^\m(x)}{\d \j_\s(y)}}\rb=0,
\eal
which contains an additional seagull term. This Ward identity is identical to  
\eqref{Reg-Q-susy-WID-2pt-PI}, obtained from the path integral through the Noether procedure. 

Similarly, in terms of path integral correlators \eqref{OM-susy-WID-3pt} takes the form 
\bal
\label{OM-susy-WID-3pt-OI}
&\pa_\m \lb \wt\cq^\m(x)\wt{\bar\cq}{}^\s(y)\wt\cj^\k(z)\rb -i\g^5\d(x,z)\lb \wt\cq^\k(x)\wt{\bar\cq}{}^\s(y)\rb -\frac12\lb \wt\ct^\s_a(x)\wt\cj^\k(z)\rb \g^a\d(x,y)\NO\\
&+\frac{i}{8}\lb \wt\cj^\n(x)\wt\cj^\k(z)\rb \big(4\d^{[\r}_\n\d^{\s]}_\l+i\g^5 \e_{\n\l}{}^{\r\s}\big)\g^5\g^\l \pa_\r\d(x,y))\NO\\
&+\frac{1}{8}\lb \wt\cj^\n(x)\rb \big(4\d^{[\k}_\n\d^{\s]}_\l+i\g^5 \e_{\n\l}{}^{\k\s}\big)\g^\l \d(x,z)\d(x,y)\NO\\
&-\frac{i}{3}\d(x,z)\g^\k\g^5\Big(\g_\m \lb \wt\cq^\m(x)\wt{\bar\cq}{}^\s(y)\rb -\frac{3i}{4}\g^5\d(x,y)\lb \wt\cj^\s(x)\rb \Big)\NO\\
&+\frac{i}{3}\h^{\s\k}\d(x,y)\d(x,z)\Big(\lb \wt\co_M(x)\rb P_R-\lb \wt\co_{M^*}(x)\rb P_L\Big)\NO\\
&+\frac13\g^{\s\r}\pa_\r\d(x,y)\Big(\lb \wt\co_M(x)\wt\cj^\k(z)\rb P_R+\lb \wt\co_{M^*}(x)\wt\cj^\k(z)\rb P_L\Big)\NO\\
&+\pa_\m\lb\orangebox{\frac{\d\wt\cq^\m(x)}{\d \j_\s(y)}}\wt\cj^\k(z)\rb+\pa_\m\lb\orangebox{\frac{\d\wt\cq^\m(x)}{\d A_\k(z)}}\wt{\bar\cq}{}^\s(y)\rb\NO\\
&+\pa_\m\lb\wt\cq^\m(x)\frac{\d\wt{\bar\cq}{}^\s(y)}{\d A_\k(z)}\rb-i\g^5\d(x,z)\lb\frac{\d\wt\cq^\k(x)}{\d \j_\s(y)}\rb -\frac12\lb\frac{\d\wt\ct^\s_\r(x)}{\d A_\k(z)}\rb \g^\r\d(x,y)\NO\\
&+\frac{i}{8}\lb\frac{\d\wt\cj^\n(x)}{\d A_\k(z)}\rb \big(4\d^{[\r}_\n\d^{\s]}_\l+i\g^5 \e_{\n\l}{}^{\r\s}\big)\g^5\g^\l \pa_\r\d(x,y))-\frac{i}{3}\d(x,z)\g^\k\g^5\g_\m \lb\frac{\d\wt\cq^\m(x)}{\d \j_\s(y)}\rb =0.
\eal
This coincides with \eqref{Reg-Q-susy-WID-3pt-PI} obtained using the path integral Noether procedure, except for the last two lines that comprise an independent path integral Ward identity. The 4-point function Ward identity \eqref{Reg-Q-susy-WID-4pt-PI} admits an analogous decomposition in terms of path integral Ward identities.

\section{Cancellation of UV divergences in FZ current multiplet correlators}
\label{sec:divergences}

In this appendix we demonstrate that the supersymmetric PV regulator \eqref{PV-lagrangian-Dirac} removes the UV divergences from all 1-loop correlation functions necessary for the analysis of the 4-point function supersymmetry Ward identity. This must be checked for all Feynman diagrams that enter in the computation of the relevant path integral correlators, including those that involve seagull terms. However, as we discuss in appendix \ref{sec:seagulls}, the only universal UV divergences that can be renormalized by the background supergravity counterterms \eqref{counterterms-1}-\eqref{counterterms-4} are those of current multiplet correlators defined through functional differentiation. These are also the correlators that appear in the model independent form of the Ward identities, and hence determine the quantum anomalies. 

The analysis in this appendix, as well as in section \ref{sec:correlators}, is therefore organized in terms of current multiplet correlators defined by functional differentiation. For all relevant 1- and 2-point functions we obtain their full renormalized form, which allows us to determine the coefficients of the supersymmetric counterterms \eqref{counterterms-1}-\eqref{counterterms-4}. For the 3- and 4-point functions we compute their regulated form and show that the PV regulator removes all UV divergences. All calculations in this appendix involve flat space operators only and so we drop the hat $\hskip2pt\hat\cdot\hskip2pt$ throughout to simplify the notation.         

\subsection{1-point functions}

The 1-point functions of the current multiplet operators all vanish identically, as long as the regulator preserves supersymmetry. However, a number of 1-point functions of seagull terms are not identically zero and require renormalization. Such 1-point functions are related to certain current multiplet 2-point functions and so we defer their discussion until we consider 2-point functions.   

\begin{figure}[h]
\captionsetup{justification=raggedright,
singlelinecheck=false
}
\begin{eqnarray*}
&\begin{tikzpicture}
\begin{feynman}
\vertex (v1) {$\wt\cj^\m$};
\vertex [right=.5cm of v1] (v2);
\vertex [right=1.5cm of v2] (v3);
\vertex [right=.3cm of v3] (v4);
\diagram* {
(v1) -- [white]
(v2) -- [charged scalar, thick, half right, looseness=1.7, insertion={[style=black, size=2pt]0}, black, edge label=$\f$] (v3) 
-- [white] (v4),
(v3) -- [charged scalar, thick, half right, looseness=1.7, black, edge label=$\rule{0pt}{10pt}$] (v2),
};
\end{feynman}
\end{tikzpicture}
\raisebox{1.2cm}{+} 
\begin{tikzpicture}
\begin{feynman}
\vertex (v1) {$\wt\cj^\m$};
\vertex [right=.5cm of v1] (v2);
\vertex [right=1.5cm of v2] (v3);
\vertex [right=.3cm of v3] (v4);
\diagram* {
(v1) -- [white]
(v2) -- [majorana, thick, half right, looseness=1.7, insertion={[style=black, size=2pt]0}, black, edge label=$\c$] (v3) 
-- [white] (v4),
(v3) -- [majorana, thick, half right, looseness=1.7, black, edge label=$\rule{0pt}{10pt}$] (v2),
};
\end{feynman}
\end{tikzpicture}
\raisebox{1.2cm}{+}
\begin{tikzpicture}
\begin{feynman}
\vertex (v1) {$\wt\cj^\m$};
\vertex [right=.5cm of v1] (v2);
\vertex [right=1.5cm of v2] (v3);
\vertex [right=.3cm of v3] (v4);
\diagram* {
(v1) -- [white]
(v2) -- [charged scalar, thick, half right, looseness=1.7, insertion={[style=black, size=2pt]0}, red, edge label=$\vf_2$] (v3) 
-- [white] (v4),
(v3) -- [charged scalar, thick, half right, looseness=1.7, red, edge label=$\rule{0pt}{10pt}$] (v2),
};
\end{feynman}
\end{tikzpicture}
\raisebox{1.2cm}{+}
\begin{tikzpicture}
\begin{feynman}
\vertex (v1) {$\wt\cj^\m$};
\vertex [right=.5cm of v1] (v2);
\vertex [right=1.5cm of v2] (v3);
\vertex [right=.3cm of v3] (v4);
\diagram* {
(v1) -- [white]
(v2) -- [majorana, thick, half right, looseness=1.7, insertion={[style=black, size=2pt]0}, purple, edge label=$\l_2$] (v3) 
-- [white] (v4),
(v3) -- [majorana, thick, half right, looseness=1.7, purple, edge label=$\rule{0pt}{10pt}$] (v2),
};
\end{feynman}
\end{tikzpicture}\\
&\raisebox{1.2cm}{+} 
\begin{tikzpicture}
\begin{feynman}
\vertex (v1) {$\wt\cj^\m$};
\vertex [right=.5cm of v1] (v2);
\vertex [right=1.5cm of v2] (v3);
\vertex [right=.3cm of v3] (v4);
\diagram* {
(v1) -- [white]
(v2) -- [charged scalar, thick, half right, looseness=1.7, insertion={[style=black, size=2pt]0}, orange, edge label=$\vf_1$] (v3) 
-- [white] (v4),
(v3) -- [charged scalar, thick, half right, orange, edge label=$\rule{0pt}{10pt}$] (v2),
};
\end{feynman}
\end{tikzpicture}
\raisebox{1.2cm}{+}
\begin{tikzpicture}
\begin{feynman}
\vertex (v1) {$\wt\cj^\m$};
\vertex [right=.5cm of v1] (v2);
\vertex [right=1.5cm of v2] (v3);
\vertex [right=.3cm of v3] (v4);
\diagram* {
(v1) -- [white]
(v2) -- [charged scalar, thick, half right, looseness=1.7, insertion={[style=black, size=2pt]0}, blue, edge label=$\vth_1$] (v3) 
-- [white] (v4),
(v3) -- [charged scalar, thick, half right, looseness=1.7, blue, edge label=$\rule{0pt}{10pt}$] (v2),
};
\end{feynman}
\end{tikzpicture}
\raisebox{1.2cm}{+} 
\begin{tikzpicture}
\begin{feynman}
\vertex (v1) {$\wt\cj^\m$};
\vertex [right=.5cm of v1] (v2);
\vertex [right=1.5cm of v2] (v3);
\vertex [right=.3cm of v3] (v4);
\diagram* {
(v1) -- [white]
(v2) -- [fermion, thick, half right, looseness=1.7, insertion={[style=black, size=2pt]0}, olive, edge label=$\l_1$] (v3) 
-- [white] (v4),
(v3) -- [fermion, thick, half right, looseness=1.7, olive, edge label=$\rule{0pt}{10pt}$] (v2),
};
\end{feynman}
\end{tikzpicture}
\end{eqnarray*}
\vskip-.5cm
\caption{One-loop diagrams that contribute to the 1-point function of the R-current. The 1-point function of the stress tensor is determined by exactly the same diagrams.}
\label{J-1pt-diags}
\end{figure}
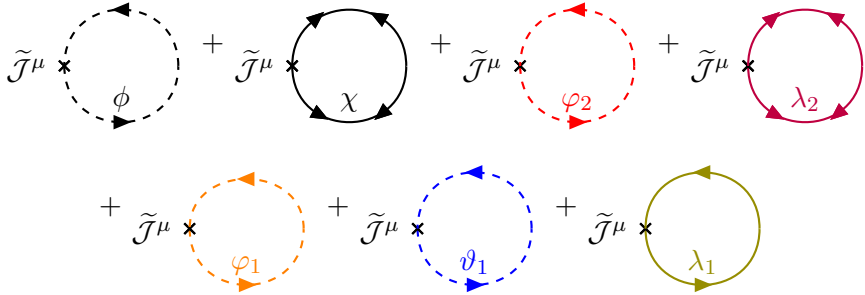

The 1-point functions of the supercurrent and of the scalar operators $\co_M$, $\co_{M^*}$ vanish trivially due to the absence of possible self contractions. The R-current 1-point function receives contributions from the 1-loop diagrams in fig.~\ref{J-1pt-diags}, all of which vanish individually due to parity. In particular, using the propagators \eqref{WZ-contractions}, \eqref{WZ-propagators} and  \eqref{PV-contractions}, \eqref{PV-propagators} together with the expressions for the currents in \eqref{WZ-Noether-currents-Mink} and \eqref{PV-FZ-operators-Mink}, we obtain 
\bal
\<\wt\cj^\m(p)\>=&\;\frac{i}{3\a^2}\int\frac{d^4q}{(2\p)^4}\Big(2iq^\m \big(P_{\f}(q)+P_{\vf_2}(q)-P_{\vf_1}(q)-P_{\vth_1}(q)\big)\NO\\
&\;-\frac{\a^2}{2}\tr\big[\g^\m\g^5\big(P_{\c}(q)+P_{\l_2}(q)-2P_{\l_1}(q)\big)\big]\Big)=0,
\eal
where in the last step we have used the trace identities \eqref{d=4-gamma-traces}.

The stress tensor 1-point function receives contributions from the same diagrams as the R-current in fig.~\ref{J-1pt-diags} and it is straightforward to show that it too vanishes: 
\bal
\<\wt\ct^{\m\n}(p)\>=&\;\int\frac{d^4q}{(2\p)^4}\Big(\frac{1}{\a^2}q^\m q^\n\big(P_{\f}(q)+P_{\vf_2}(q)-P_{\vf_1}(q)-P_{\vth_1}(q)\big)+i\h^{\m\n}(1+1-1-1)\NO\\
&-\frac12q^\n\tr\big[i\g^\m \big(P_{\c}(q)+P_{\l_2}(q)-2P_{\l_1}(q)\big)\big]-\frac {i}{2}\h^{\m\n}\tr(1+1-2)\Big)=0,
\eal
where the last equality follows from the relation between the scalar and spinor propagators \eqref{fermion-propagators} and the trace identities \eqref{d=4-gamma-traces}. Notice that the 1-point function of the stress tensor vanishes due to the presence of the regulator multiplets. In particular, surprisingly, it does not vanish for each individual multiplet.\footnote{This result apparently contradicts the conclusion of \cite{Barvinsky:2018lyi} in the case when the auxiliary fields are integrated out. A possible source for the discrepancy is the values for the auxiliary fields in eq.~(22) of \cite{Barvinsky:2018lyi}, which seem incorrect.} However, multiplet-wise cancellation does occur if one includes the auxiliary fields in the chiral multiplets. 

Using the auxiliary field propagators \cite{Wess:1973kz,Iliopoulos:1974zv} (see also eq.~(9.11) in \cite{Wess:1992cp})
\be
P_F(q)=-q^2P_\f(q),\qquad P_{\f F}=0,\qquad P_{F_{\vf_2}}(q)=-q^2P_{\vf_2}(q),\qquad P_{\vf_2F_{\vf_2}}=-m_2P_{\vf_2}(q),
\ee
and so on, the 1-point function of the stress tensor is given by
\bal
\<\wt\ct^{\m\n}(p)\>=&\;\int\frac{d^4q}{(2\p)^4}\Big(\frac{1}{\a^2}q^\m q^\n\big(P_{\f}(q)+P_{\vf_2}(q)-P_{\vf_1}(q)-P_{\vth_1}(q)\big)\NO\\
&-\frac{1}{2\a^2}\h^{\m\n}q^2\big(P_{\f}(q)+P_{\vf_2}(q)-P_{\vf_1}(q)-P_{\vth_1}(q)\big)\NO\\
&+\frac{1}{2\a^2}\h^{\m\n}\big(P_{F}(q)+P_{F_{\vf_2}}(q)-P_{F_{\vf_1}}(q)-P_{F_{\vth_1}}(q)\big)\NO\\
&+\frac{1}{2\a^2}\h^{\m\n}\big(2m_2P_{\vf_2F_{\vf_2}}(q)-2m_1P_{\vf_1F_{\vf_1}}(q)-2m_1P_{\vth_1F_{\vth_1}}(q)\big)\NO\\
&-\frac12q^\n\tr\big[i\g^\m \big(P_{\c}(q)+P_{\l_2}(q)-2P_{\l_1}(q)\big)\big]-\frac {i}{2}\h^{\m\n}\tr(1+1-2)\Big)\NO\\
=&\;\int\frac{d^4q}{(2\p)^4}\Big(\frac{1}{\a^2}q^\m q^\n\big(P_{\f}(q)+P_{\vf_2}(q)-P_{\vf_1}(q)-P_{\vth_1}(q)\big)+2i\h^{\m\n}(1+1-1-1)\NO\\
&-\frac12q^\n\tr\big[i\g^\m \big(P_{\c}(q)+P_{\l_2}(q)-2P_{\l_1}(q)\big)\big]-\frac {i}{2}\h^{\m\n}\tr(1+1-2)\Big)=0.
\eal
This shows that the 1-point function of the stress tensor vanishes separately for each multiplet once the auxiliary fields are included.

\subsection{2-point functions}

\begin{flushleft}
\noindent\rule{\textwidth}{0.8pt}
\raisebox{-0.1cm}{$\<\co_M\co_{M^*}\>$}
\noindent\rule{\textwidth}{0.8pt}
\end{flushleft}

The simplest 2-point function of current multiplet operators is that between $\wt\co_M$ and $\wt\co_{M^*}$. For massless theories like the massless WZ model we consider here, this 2-point function depends only on the regulator fields and hence its renormalized form is ultralocal. Renormalizing this 2-point function is the fastest way to compute the coefficient $a_2$ of the supersymmetric counterterm \eqref{counterterms-2}.

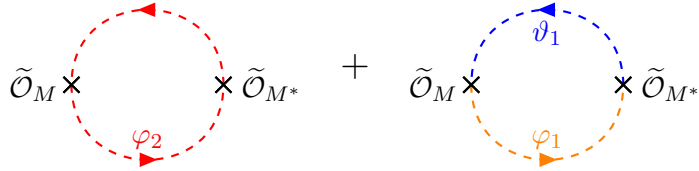
\begin{figure}[h]
\captionsetup{justification=raggedright,
singlelinecheck=false
}
\begin{eqnarray*}
\begin{tikzpicture}
\begin{feynman}
\vertex (v1) {$\wt\co_M$};
\vertex [right=.5cm of v1] (v2);
\vertex [right=2cm of v2] (v3);
\vertex [right=.1cm of v3] (v4) {$\wt\co_{M^*}$};
\diagram* {
(v1) -- [white]
(v2) -- [charged scalar, thick, half right, looseness=1.7, insertion={[style=black, size=3pt]0}, red, edge label=$\vf_2$] (v3) 
-- [white] (v4),
(v3) -- [charged scalar, thick, half right, looseness=1.7, insertion={[style=black, size=3pt]0}, red, edge label=$\rule{0pt}{10pt}$] (v2),
};
\end{feynman}
\end{tikzpicture}
\quad \raisebox{1.45cm}{\Large+}\quad 
\begin{tikzpicture}
\begin{feynman}
\vertex (v1) {$\wt\co_M$};
\vertex [right=.5cm of v1] (v2);
\vertex [right=2cm of v2] (v3);
\vertex [right=.1cm of v3] (v4) {$\wt\co_{M^*}$};
\diagram* {
(v1) -- [white]
(v2) -- [charged scalar, thick, half right, looseness=1.7, insertion={[style=black, size=3pt]0}, orange, edge label=$\vf_1$] (v3) 
-- [white] (v4),
(v3) -- [charged scalar, thick, half right, looseness=1.7, insertion={[style=black, size=3pt]0}, blue, edge label=$\vth_1$] (v2),
};
\end{feynman}
\end{tikzpicture}
\end{eqnarray*}
\vskip-.5cm
\caption{One-loop diagrams that contribute to the 2-point function $\<\wt\co_M(p)\wt\co_{M^*}(-p)\>$.}
\label{OO-2pt-diags}
\end{figure}

Only the two diagrams shown in fig.~\ref{OO-2pt-diags} contribute to this 2-point function. Moreover, there is no difference between the functional differentiation and operator insertion form of this 2-point function since no seagull terms contribute. Using the expressions for $\wt\co_M$, $\wt\co_{M^*}$ in \eqref{PV-FZ-operators-Mink}, we obtain 
\bal
&\<\wt\co_M(p)\wt\co_{M^*}(-p)\>=\lb\wt\co_M(p)\wt\co_{M^*}(-p)\rb\NO\\
&=2i\times\Big(\frac{m_2}{4\a^2}\Big)^2\int \frac{d^{4}q}{(2\p)^4}P_{\vf_2}(p+q)P_{\vf_2}(q)-i\Big(\frac{m_1}{2\a^2}\Big)^2\int \frac{d^{4}q}{(2\p)^4}P_{\vf_1}(p+q)P_{\vth_1}(q)\NO\\
&=-\frac{im_2^2}{2}\int \frac{d^{4}q}{(2\p)^4}\frac{1}{\big((p+q)^2+m_2^2\big)(q^2+m_2^2)}+im_1^2\int \frac{d^{4}q}{(2\p)^4}\frac{1}{\big((p+q)^2+m_1^2\big)(q^2+m_1^2)}.
\eal
The factor of 2 in the first diagram reflects the number of possible Wick contractions, while the overall factor of $i$ is due to \eqref{phase-factor}.
These integrals are individually logarithmically divergent for large loop momentum $q^\m$, but this divergence cancels provided the PV masses satisfy the condition
\be\label{PV-mass-condition}
m_2^2=2m_1^2.
\ee

With this choice of PV masses, the above integrals can be evaluated using Feynman parameters:
\be
\frac{1}{\big((p+q)^2+m_1^2\big)(q^2+m_1^2)}=\int_0^1du\frac{1}{(\ell^2+\D_1)^2},
\ee
where $\ell^\m=q^\m+u p^\m$ and $\D_1=u(1-u)p^2+m_1^2$, and similarly for the first integral. Hence, 
\bal\label{OOreg}
\<\wt\co_M(p)\wt\co_{M^*}(-p)\>=&\;im_1^2\int_0^1du\int \frac{d^{4}\ell}{(2\p)^4}\bigg(\frac{1}{(\ell^2+\D_1)^2}-\frac{1}{(\ell^2+\D_2)^2}\bigg)\NO\\
=&\;-m_1^2\frac{2\p^2}{(2\p)^4}\int_0^1du\int_0^\infty d\ell_E\ell_E^3\bigg(\frac{1}{(\ell_E^2+\D_1)^2}-\frac{1}{(\ell_E^2+\D_2)^2}\bigg)\NO\\
=&\;-\frac{m_1^2}{16\p^2}\int_0^1du\log\Big(\frac{\D_2}{\D_1}\Big)=-\frac{1}{16\p^2}\Big(m_1^2\log 2-\frac{p^2}{12}+\co(m_1^{-2})\Big),
\eal
where we have Wick rotated $\ell^\m$ according to $\ell^0\to i\ell^0_E$ and used that ${\rm Vol}(S^3)=2\p^2$. The divergence of this 2-point function as the PV mass $m_1$ is sent to infinity is canceled by the supersymmetric counterterm \eqref{counterterms-2} provided the constant $a_2$ takes the value 
\be\label{a2}
a_2=-\frac{\log2}{96\p^2}m_1^2.
\ee
This results in the renormalized 2-point function
\bbxd
\vskip.4cm
\be\label{OOren}
\<\wt\co_M(p)\wt\co_{M^*}(-p)\>_{\rm ren}=\frac{p^2}{192\p^2}.
\ee
\ebxd
As we anticipated, this is indeed ultralocal, i.e. polynomial in the momentum $p^\m$, which reflects the fact that the theory is classically conformal.


\begin{flushleft}
\noindent\rule{\textwidth}{0.8pt}
\raisebox{-0.1cm}{$\<\cj\cj\>$}
\noindent\rule{\textwidth}{0.8pt}
\end{flushleft}

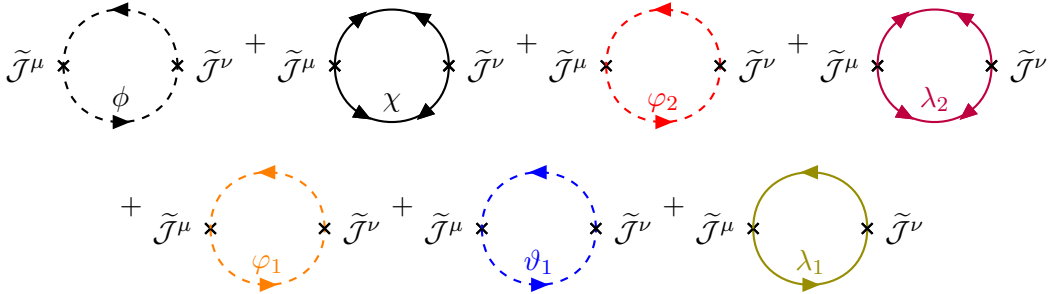
\begin{figure}[h]
\captionsetup{justification=raggedright,
singlelinecheck=false
}
\begin{eqnarray*}
&\begin{tikzpicture}
\begin{feynman}
\vertex (v1) {$\wt\cj^\m$};
\vertex [right=.5cm of v1] (v2);
\vertex [right=1.5cm of v2] (v3);
\vertex [right=.1cm of v3] (v4) {$\wt\cj^\n$};
\diagram* {
(v1) -- [white]
(v2) -- [charged scalar, thick, half right, looseness=1.7, insertion={[style=black, size=2pt]0}, black, edge label=$\f$] (v3) 
-- [white] (v4),
(v3) -- [charged scalar, thick, half right, looseness=1.7, insertion={[style=black, size=2pt]0}, black, edge label=$\rule{0pt}{10pt}$] (v2),
};
\end{feynman}
\end{tikzpicture}
\raisebox{1.2cm}{+} 
\begin{tikzpicture}
\begin{feynman}
\vertex (v1) {$\wt\cj^\m$};
\vertex [right=.5cm of v1] (v2);
\vertex [right=1.5cm of v2] (v3);
\vertex [right=.1cm of v3] (v4) {$\wt\cj^\n$};
\diagram* {
(v1) -- [white]
(v2) -- [majorana, thick, half right, looseness=1.7, insertion={[style=black, size=2pt]0}, black, edge label=$\c$] (v3) 
-- [white] (v4),
(v3) -- [majorana, thick, half right, looseness=1.7, insertion={[style=black, size=2pt]0}, black, edge label=$\rule{0pt}{10pt}$] (v2),
};
\end{feynman}
\end{tikzpicture}
\raisebox{1.2cm}{+}
\begin{tikzpicture}
\begin{feynman}
\vertex (v1) {$\wt\cj^\m$};
\vertex [right=.5cm of v1] (v2);
\vertex [right=1.5cm of v2] (v3);
\vertex [right=.1cm of v3] (v4) {$\wt\cj^\n$};
\diagram* {
(v1) -- [white]
(v2) -- [charged scalar, thick, half right, looseness=1.7, insertion={[style=black, size=2pt]0}, red, edge label=$\vf_2$] (v3) 
-- [white] (v4),
(v3) -- [charged scalar, thick, half right, looseness=1.7, insertion={[style=black, size=2pt]0}, red, edge label=$\rule{0pt}{10pt}$] (v2),
};
\end{feynman}
\end{tikzpicture}
\raisebox{1.2cm}{+}
\begin{tikzpicture}
\begin{feynman}
\vertex (v1) {$\wt\cj^\m$};
\vertex [right=.5cm of v1] (v2);
\vertex [right=1.5cm of v2] (v3);
\vertex [right=.1cm of v3] (v4) {$\wt\cj^\n$};
\diagram* {
(v1) -- [white]
(v2) -- [majorana, thick, half right, looseness=1.7, insertion={[style=black, size=2pt]0}, purple, edge label=$\l_2$] (v3) 
-- [white] (v4),
(v3) -- [majorana, thick, half right, looseness=1.7, insertion={[style=black, size=2pt]0}, purple, edge label=$\rule{0pt}{10pt}$] (v2),
};
\end{feynman}
\end{tikzpicture}\\
&\raisebox{1.2cm}{+} 
\begin{tikzpicture}
\begin{feynman}
\vertex (v1) {$\wt\cj^\m$};
\vertex [right=.5cm of v1] (v2);
\vertex [right=1.5cm of v2] (v3);
\vertex [right=.1cm of v3] (v4) {$\wt\cj^\n$};
\diagram* {
(v1) -- [white]
(v2) -- [charged scalar, thick, half right, looseness=1.7, insertion={[style=black, size=2pt]0}, orange, edge label=$\vf_1$] (v3) 
-- [white] (v4),
(v3) -- [charged scalar, thick, half right, looseness=1.7, insertion={[style=black, size=2pt]0}, orange, edge label=$\rule{0pt}{10pt}$] (v2),
};
\end{feynman}
\end{tikzpicture}
\raisebox{1.2cm}{+}
\begin{tikzpicture}
\begin{feynman}
\vertex (v1) {$\wt\cj^\m$};
\vertex [right=.5cm of v1] (v2);
\vertex [right=1.5cm of v2] (v3);
\vertex [right=.1cm of v3] (v4) {$\wt\cj^\n$};
\diagram* {
(v1) -- [white]
(v2) -- [charged scalar, thick, half right, looseness=1.7, insertion={[style=black, size=2pt]0}, blue, edge label=$\vth_1$] (v3) 
-- [white] (v4),
(v3) -- [charged scalar, thick, half right, looseness=1.7, insertion={[style=black, size=2pt]0}, blue, edge label=$\rule{0pt}{10pt}$] (v2),
};
\end{feynman}
\end{tikzpicture}
\raisebox{1.2cm}{+} 
\begin{tikzpicture}
\begin{feynman}
\vertex (v1) {$\wt\cj^\m$};
\vertex [right=.5cm of v1] (v2);
\vertex [right=1.5cm of v2] (v3);
\vertex [right=.1cm of v3] (v4) {$\wt\cj^\n$};
\diagram* {
(v1) -- [white]
(v2) -- [fermion, thick, half right, looseness=1.7, insertion={[style=black, size=2pt]0}, olive, edge label=$\l_1$] (v3) 
-- [white] (v4),
(v3) -- [fermion, thick, half right, looseness=1.7, insertion={[style=black, size=2pt]0}, olive, edge label=$\rule{0pt}{10pt}$] (v2),
};
\end{feynman}
\end{tikzpicture}
\end{eqnarray*}
\vskip-.5cm
\caption{One-loop diagrams that contribute to the 2-point function of the R-current.}
\label{JJ-2pt-diags}
\end{figure}
Next, we consider the 2-point function of R-currents, which receives contributions from the diagrams in fig.~\ref{JJ-2pt-diags}. As we discuss in appendix \ref{sec:seagulls}, the functional derivative and operator insertion versions of this 2-point function differ by the 1-point function of a seagull term, namely  
\be
\<\wt\cj^\m(p)\wt\cj^\n(-p)\>=\lb\wt\cj^\m(p)\wt\cj^\n(-p)\rb-\frac{4}{9\a^2}\h^{\m\n}\lb\wt s_{(1|0)}(p)\rb,
\ee
where $\wt s_{(1|0)}$ is given in \eqref{seagull-operators}. In terms of loop integrals, its 1-point function takes the form
\be\label{JJ-seagull}
\lb\wt s_{(1|0)}(p)\rb=\int \frac{d^{4}q}{(2\p)^4}\big(P_{\f}(q)+P_{\vf_2}(q)-P_{\vf_1}(q)-P_{\vth_1}(q)\big)=\frac{\a^2}{4\p^2}m_1^2\log2.
\ee
Moreover, the 2-point function defined via operator insertions is given by 
\bal
&\lb\wt\cj^\m(p)\wt\cj^\n(-p)\rb=-i\Big(\frac{i}{3\a^2}\Big)^2\int \frac{d^{4}q}{(2\p)^4}(q+q')^\m(q+q')^\n G^{(2)}_1(q,q')\\
&-2i\Big(\frac{i}{6}\Big)^2\int \frac{d^{4}q}{(2\p)^4}\tr\Big(\g^\m\g^5 P_{\c}(q)\g^\n\g^5P_\c(q')+\g^\m\g^5 P_{\l_2}(q)\g^\n\g^5P_{\l_2}(q')-2\g^\m\g^5 P_{\l_1}(q)\g^\n\g^5P_{\l_1}(q')\Big),\NO
\eal
where we have introduced the shorthand notation $q'^\m\equiv q^\m+p^\m$ and $G^{(2)}_1(q,q')$ was defined in \eqref{Gs}.

Using the Feynman parameterization of the momentum integral, this expression reduces to  
\bal
\lb\wt\cj^\m(p)\wt\cj^\n(-p)\rb=&\,-\frac{i}{9}\int_0^1 du\int\frac{d^4\ell}{(2\p)^4}\Big(5\ell^2\h^{\m\n}+4(1-3u+3u^2)p^\m p^\n-2u(1-u)p^2\h^{\m\n}\Big)\times\NO\\
&\hskip.cm\bigg(\frac{1}{(\ell^2+\D)^2}+\frac{1}{(\ell^2+\D_2)^2}-\frac{2}{(\ell^2+\D_1)^2}\bigg)-\frac{4}{9}\h^{\m\n}\<\wt\co_M(p)\wt\co_{M^*}(-p)\>\NO\\
=&\,\frac{2\p^2}{9(2\p)^4}\int_0^1 du\Big[5\h^{\m\n}\big(\D\log\D+\D_2\log\D_2-2\D_1\log\D_1\big)\\
&\,-\Big(2(1-3u+3u^2)p^\m p^\n-u(1-u)p^2\h^{\m\n}\Big)\log\Big(\frac{\D\D_2}{\D_1^2}\Big)+2m_1^2\h^{\m\n}\log\Big(\frac{\D_2}{\D_1}\Big)\Big],\NO
\eal
where $\D=u(1-u)p^2$, $\D_1=\D+m_1^2$, $\D_2=\D+2m_1^2$ and we have replaced $\ell^\m\ell^\n\to \frac14\ell^2\h^{\m\n}$ in the momentum integral. Combining this with \eqref{JJ-seagull} and expanding for large PV mass gives
\be\label{JJ}
\<\wt\cj^\m(p)\wt\cj^\n(-p)\>
=\frac{2\p^2}{9(2\p)^4}\bigg[4m_1^2\log2\,\h^{\m\n}+(p^\m p^\n-\h^{\m\n}p^2)\Big(\log m_1^2-\log(2p^2)+\frac83\Big)-\frac{1}{3}p^\m p^\n\bigg],
\ee
where terms that vanish in the limit $m_1\to\infty$ have been dropped.

This regulated 2-point function has two divergent terms as $m_1\to\infty$. It is straightforward to verify that the quadratic divergence is removed by the supersymmetric counterterm \eqref{counterterms-2} with the same coefficient $a_2$ as the one obtained from the scalar 2-point function in \eqref{a2}. Moreover, the logarithmic divergence is canceled by the supersymmetric counterterm \eqref{counterterms-4} provided the coefficient $a_4$ takes the value
\be\label{a4}
a_4=-\frac{1}{768\p^2}\Big(\log (m_1^2/\m^2)-\log2+\frac73\Big),
\ee
where the arbitrary scale $\m$ is independent of the PV mass and parameterizes the choice of supersymmetric renormalization scheme. The resulting renormalized 2-point function is  
\bbxd
\vskip.4cm
\be\label{JJren}
\<\wt\cj^\m(p)\wt\cj^\n(-p)\>_{\rm ren}=-\frac{2\p^2}{9(2\p)^4}\Big((p^\m p^\n-\h^{\m\n}p^2)\log (p^2/\m^2)+\frac{1}{3}p^2\h^{\m\n}\Big).
\ee
\ebxd


\begin{flushleft}
\noindent\rule{\textwidth}{0.8pt}
\raisebox{-0.1cm}{$\<\ct\cj\>$}
\noindent\rule{\textwidth}{0.8pt}
\end{flushleft}

Another 2-point function that we need to determine is that between an R-current and a stress tensor. The diagrams that contribute to this 2-point function are identical to the corresponding ones for the 2-point function of two R-currents shown in fig.~\ref{JJ-2pt-diags}, except that one of the two R-current insertions is replaced with a stress tensor. From appendix \ref{sec:seagulls} follows that the $\ct\cj$ 2-point function potentially receives a contribution from the 1-point function of a seagull term. However, using the explicit form of this seagull term in \eqref{operator-derivatives}, one sees that its 1-point function is a linear combination of $\lb\wt\cj^\m(p)\rb$ and $\lb{\wt s}{}_{(2|1)}^{\n}(p)\rb$, both of which vanish. Hence, 
\bal
&\<\wt\ct^{\m\n}(p)\wt\cj^\r(-p)\>=\lb\wt\ct^{\m\n}(p)\wt\cj^\r(-p)\rb\\
&=-\frac{i}{3\a^4}\int \frac{d^{4}q}{(2\p)^4}(q+q')^\r\Big[\Big(q^{(\m}q'^{\n)}-\frac12\h^{\m\n}q\cdot q'+\frac16(p^\m p^\n-\h^{\m\n}p^2)\Big)G^{(2)}_1(q,q')-m_1^2\h^{\m\n}G^{(2)}_2(q,q')\Big]\NO\\
&+\frac{i}{96\a^4}\int \frac{d^{4}q}{(2\p)^4}\Big[G^{(2)}_1(q,q')\tr\Big(\big(4\h^{\n[\m}\g^{\s]}(q'+q)_\s+i\e^{\m\n\k\s}\g_\k\g^5p_\s\big)\slashed q\g^\r\g^5\slashed q'\Big)+8im_1^2\e^{\m\n\r\s}p_\s G^{(2)}_2(q,q')\Big],\NO
\eal
where $q'^\m\equiv q^\m-p^\m$. Since $G^{(2)}_1(q,q')\sim q^{-8}$ as $q^2\to\infty$ and $G^{(2)}_2(q,q')\sim q^{-6}$, the loop integrals are properly regulated. 

Since the loop integrals are regulated, we may transform the loop momentum as $q\to p-q$. This changes the sign of all terms that are odd and symmetric in $q$ and $q'$, which therefore vanish identically. Using the trace identities \eqref{d=4-gamma-traces}, the remaining terms give
\bal
&\<\wt\ct^{\m\n}(p)\wt\cj^\r(-p)\>=\frac{m_1^2}{3}\e^{\m\n\r\s}p_\s \int_0^1du\int \frac{d^{4}\ell}{(2\p)^4}\bigg(\frac{1}{(\ell^2+\D_2)^2}-\frac{1}{(\ell^2+\D_1)^2}\bigg)\NO\\
&+\frac{1}{12}\e^{\m\n\r\s}p_\s\int_0^1du\int \frac{d^{4}\ell}{(2\p)^4}\bigg(\frac{1}{(\ell^2+\D)^2}+\frac{1}{(\ell^2+\D_2)^2}-\frac{2}{(\ell^2+\D_1)^2}\bigg)\big(\ell^2-2u(u-1)p^2\big),
\eal
where now $\ell=q-up$ and $\D$, $\D_1$ and $\D_2$ are as above. Evaluating the loop integrals and taking the limit $m_1\to\infty$ we find that this expression is identically zero. Hence, 
\bbxd
\vskip.25cm
\be\label{TJren}
\<\wt\ct^{\m\n}(p)\wt\cj^\r(-p)\>_{\rm ren}=0.
\ee
\ebxd


\begin{flushleft}
\noindent\rule{\textwidth}{0.8pt}
\raisebox{-0.1cm}{$\<\cq\bar\cq\>$}
\noindent\rule{\textwidth}{0.8pt}
\end{flushleft}

\begin{figure}[h]
\captionsetup{justification=raggedright,
singlelinecheck=false
}
\begin{eqnarray*}
&\begin{tikzpicture}
\begin{feynman}
\vertex (v1) {$\wt\cq^\m$};
\vertex [right=.5cm of v1] (v2);
\vertex [right=1.5cm of v2] (v3);
\vertex [right=.1cm of v3] (v4) {$\wt{\bar\cq}{}^\n$};
\diagram* {
(v1) -- [white]
(v2) -- [charged scalar, thick, half right, looseness=1.7, insertion={[style=black, size=2pt]0}, black, edge label=$\f$] (v3) 
-- [white] (v4),
(v3) -- [majorana, thick, half right, looseness=1.7, insertion={[style=black, size=2pt]0}, black, edge label=$\c$] (v2),
};
\end{feynman}
\end{tikzpicture}
\raisebox{1.2cm}{+} 
\begin{tikzpicture}
\begin{feynman}
\vertex (v1) {$\wt\cq^\m$};
\vertex [right=.5cm of v1] (v2);
\vertex [right=1.5cm of v2] (v3);
\vertex [right=.1cm of v3] (v4) {$\wt{\bar\cq}{}^\n$};
\diagram* {
(v1) -- [white]
(v2) -- [charged scalar, thick, half right, looseness=1.7, insertion={[style=black, size=2pt]0}, red, edge label=$\vf_2$] (v3) 
-- [white] (v4),
(v3) -- [majorana, thick, half right, looseness=1.7, insertion={[style=black, size=2pt]0}, purple, edge label=$\l_2$] (v2),
};
\end{feynman}
\end{tikzpicture}
\raisebox{1.2cm}{+}
\begin{tikzpicture}
\begin{feynman}
\vertex (v1) {$\wt\cq^\m$};
\vertex [right=.5cm of v1] (v2);
\vertex [right=1.5cm of v2] (v3);
\vertex [right=.1cm of v3] (v4) {$\wt{\bar\cq}{}^\n$};
\diagram* {
(v1) -- [white]
(v2) -- [charged scalar, thick, half right, looseness=1.7, insertion={[style=black, size=2pt]0}, orange, edge label=$\vf_1$] (v3) 
-- [white] (v4),
(v3) -- [fermion, thick, half right, looseness=1.7, insertion={[style=black, size=2pt]0}, olive, edge label=$\l_1$] (v2),
};
\end{feynman}
\end{tikzpicture}
\raisebox{1.2cm}{+}
\begin{tikzpicture}
\begin{feynman}
\vertex (v1) {$\wt\cq^\m$};
\vertex [right=.5cm of v1] (v2);
\vertex [right=1.5cm of v2] (v3);
\vertex [right=.1cm of v3] (v4) {$\wt{\bar\cq}{}^\n$};
\diagram* {
(v1) -- [white]
(v2) -- [fermion, thick, half right, looseness=1.7, insertion={[style=black, size=2pt]0}, olive, edge label=$\l_1$] (v3) 
-- [white] (v4),
(v3) -- [charged scalar, thick, half right, looseness=1.7, insertion={[style=black, size=2pt]0}, blue, edge label=$\vth_1$] (v2),
};
\end{feynman}
\end{tikzpicture}
\end{eqnarray*}
\vskip-.5cm
\caption{One-loop diagrams that contribute to the supercurrent 2-point function.}
\label{QQ-2pt-diags}
\end{figure}
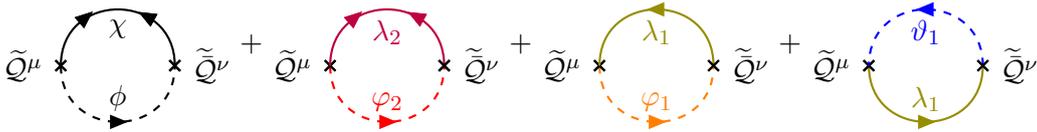

Finally, we evaluate the 2-point function of two supercurrents, corresponding to the 1-loop diagrams shown in fig.~\ref{QQ-2pt-diags}. We see from appendix~\ref{sec:seagulls} that in this case too there is a contribution from the 1-point function of a seagull term. The only non zero contribution to this 1-point function is proportional to  \eqref{JJ-seagull}, from which we obtain
\be\label{QQ-seagull}
\<\wt\cq^\m(p)\wt{\bar\cq}{}^\n(-p)\>=\lb\wt\cq^\m(p)\wt{\bar\cq}{}^\n(-p)\rb+\frac{m_1^2}{24\p^2}\log2\,\e^{\m\n\r\s}\g_\s\g^5 p_\r.
\ee
Moreover, a straightforward calculation determines that 
\bal
&\lb\wt\cq^\m(p)\wt{\bar\cq}{}^\n(-p)\rb=\frac{m_1^2}{2\a^4}\int \frac{d^{4}q}{(2\p)^4}\Big(\frac{1}{2}\big(\g^\m\g^\n\slashed q+\slashed q\g^\m\g^\n+\g^\m\slashed q'\g^\n\big)+\frac{1}{3}p_\r\big(\g^{\m\r}\g^\n+\g^\m\g^{\r\n}\big)\Big)G^{(2)}_2(q,q')\NO\\
&-\frac{1}{2\a^4}\int \frac{d^{4}q}{(2\p)^4}\Big(\frac14\slashed q\g^\m\slashed q'\g^\n\slashed q+\frac{1}{6}p_\r\big(\g^{\m\r}\slashed q'\g^\n\slashed q-\slashed q\g^\m\slashed q'\g^{\n\r}\big)-\frac{1}{9}p_\r p_\s\g^{\m\r}\slashed q'\g^{\n\s}\Big)G^{(2)}_1(q,q'),
\eal
where now $q'\equiv q-p$. Since $G^{(2)}_1(q,q')\sim q^{-8}$ as $q^2\to\infty$ and $G^{(2)}_2(q,q')\sim q^{-6}$, we see again that the loop integrals are properly regulated. 

Evaluating the loop integrals we obtain
\bal\label{QQ}
&\lb\wt\cq^\m(p)\wt{\bar\cq}{}^\n(-p)\rb=\int_0^1du\int \frac{d^{4}\ell}{(2\p)^4}\bigg(\frac{1}{(\ell^2+\D)^2}+\frac{1}{(\ell^2+\D_2)^2}-\frac{2}{(\ell^2+\D_1)^2}\bigg)\times\NO\\
&\;\Big[-\frac14\ell^2\Big(\frac12\big(\g^\m\g^\n\slashed p+\slashed p\g^\m\g^\n-\g^\m\slashed p\g^\n\big)+\frac{2}{3}p_\r\big(\g^{\m\r}\g^\n+\g^\m\g^{\r\n}\big)-i\e^{\m\n\r\s}\g_\s\g^5p_\r\Big)\NO\\
&\;-\frac14u(1-u)\slashed p\g^\m\slashed p\g^\n\slashed p-\frac13 u(1-u)p_\r(\g^{\m\r}\slashed p\g^\n\slashed p-\slashed p\g^\m\slashed p\g^{\n\r})+\frac{1}{9}p_\r p_\s\g^{\m\r}\slashed p\g^{\n\s}\Big]\NO\\
&\;-i\<\wt\co_M(p)\wt\co_{M^*}(-p)\>\Big(\frac12\big(\g^\m\g^\n\slashed p+\slashed p\g^\m\g^\n-\g^\m\slashed p\g^\n\big)+\frac{2}{3}p_\r\big(\g^{\m\r}\g^\n+\g^\m\g^{\r\n}\big)\Big)\NO\\
&\rule{.0cm}{.9cm}=\frac{2i\p^2}{(2\p)^4}\int_0^1du\bigg\{\frac16\big(2p^{(\m}\g^{\n)}+\h^{\m\n}\slashed p+i\e^{\m\n\r\s}\g_\s\g^5p_\r\big)\Big[\frac{m_1^2}{2}\log\Big(\frac{\D_2}{\D_1}\Big)-\frac14\log\Big(\frac{\D^\D\D_2^{\D_2}}{\D_1^{2\D_1}}\Big)\Big]\NO\\
&\;-\frac{p^2}{6}\Big[\Big(u(1-u)-\frac13\Big)\frac{p^\m p^\n}{p^2}\slashed p-\frac12u(1-u)p^{(\m}\g^{\n)}+\Big(\frac13-\frac{5}{4}u(1-u)\Big)\big(\h^{\m\n}\slashed p+i\e^{\m\n\r\s}\g_\s\g^5p_\r\big)\Big]\times\NO\\
&\;\log\Big(\frac{\D\D_2}{\D_1^2}\Big)+\frac{i}{4}\e^{\m\n\r\s}\g_\s\g^5p_\r\log\Big(\frac{\D^\D\D_2^{\D_2}}{\D_1^{2\D_1}}\Big)\bigg\}\NO\\
&\rule{.0cm}{.9cm}=-\frac{2i\p^2}{(2\p)^4}\frac{1}{72}\Big[\Big(\log m_1^2-\log(2p^2)+\frac{7}{3}\Big)\big(2\slashed p(p^\m p^\n-\h^{\m\n}p^2)+i\e^{\m\n\r\s}\g_\s\g^5p_\r p^2\big)+i\e^{\m\n\r\s}\g_\s\g^5p_\r p^2\Big]\NO\\
&\;+\frac{2i\p^2}{(2\p)^4}\frac12m_1^2\log2\,i\e^{\m\n\r\s}\g_\s\g^5p_\r,
\eal
where $\ell=q-up$ and $\D$, $\D_1$ and $\D_2$ are as above. Combining this expression with the contribution from the seagull term in \eqref{QQ-seagull}, one can verify that the quadratic and logarithmic divergences in $m_1$ cancel respectively against the contributions from the supersymmetric counterterms \eqref{counterterms-2} and \eqref{counterterms-4}, with the values of $a_2$ and $a_4$ given in \eqref{a2} and \eqref{a4}. It follows that the renormalized supercurrent 2-point function takes the form
\bbxd
\be\label{QQren}
\hskip-.cm\<\wt\cq^\m(p)\wt{\bar\cq}{}^\n(-p)\>_{\rm ren}=\frac{2i\p^2}{72(2\p)^4}\Big(\big(2\slashed p(p^\m p^\n-\h^{\m\n}p^2)+i\e^{\m\n\r\s}\g_\s\g^5p_\r p^2\big)\log(p^2/\m^2)-i\e^{\m\n\r\s}\g_\s\g^5p_\r p^2\Big).
\ee
\ebxd
Notice that the tensor multiplying the non local term vanishes when contracted with either $p_\m$ or $\g_\m$, but the last term is local and gives zero only when contracted with $p_\m$.

\subsection{3-point functions}


\begin{flushleft}
\noindent\rule{\textwidth}{0.8pt}
\raisebox{-0.1cm}{$\<\cj\cj\cj\>$}
\noindent\rule{\textwidth}{0.8pt}
\end{flushleft}

Moving to 3-point functions, we consider first the correlation function of three R-currents, which determines the R-symmetry anomaly. From the analysis in appendix~\ref{sec:seagulls} follows that this 3-point function receives contributions from the 2-point function between the scalar seagull operator $\wt s_{(1|0)}$ (see \eqref{seagull-operators}) and an R-current. However, this 2-point function takes the form
\be
\lb\wt s_{(1|0)}(p)\wt\cj^\m(-p)\rb=\frac{i^3}{3\a^2}\int \frac{d^{4}q}{(2\p)^4}(q+q')^\m G^{(2)}_1(q,q'),
\ee
where $q'=q+p$, and can be easily shown to vanish identically. Hence, 
\be
\<\wt\cj^\m(p_1)\wt\cj^\n(p_2)\wt\cj^\r(p_3)\>=\lb\wt\cj^\m(p_1)\wt\cj^\n(p_2)\wt\cj^\r(p_3)\rb.
\ee

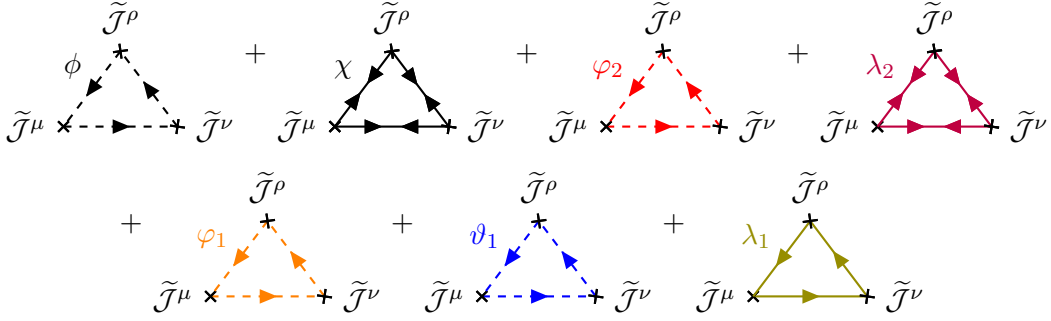
\begin{figure}[h]
\captionsetup{justification=raggedright,
singlelinecheck=false
}
\begin{eqnarray*}
&\begin{tikzpicture}
\begin{feynman}
\vertex (v1) {$\wt\cj^\m$};
\vertex [right=.5cm of v1] (v2);
\vertex [right=1.5cm of v2] (v3);
\vertex [right=.1cm of v3] (v4) {$\wt\cj^\n$};
\vertex at ($(v2)!.5!(v3)!1.cm!90:(v3)$) (v5);
\vertex [above=.1cm of v5] (v6) {$\wt\cj^\r$};
\diagram* {
(v1) -- [white]
(v2) -- [charged scalar, thick, insertion={[style=black, size=2pt]0}, black, edge label=$\rule{0pt}{10pt}$] (v3) -- [white] (v4),
(v3) -- [charged scalar, thick, insertion={[style=black, size=2pt]0}, black, edge label=$\rule{0pt}{10pt}$] (v5),
(v5) -- [charged scalar, thick, insertion={[style=black, size=2pt]0}, black, edge label'=$\f$] (v2),
};
\end{feynman}
\end{tikzpicture}
\raisebox{1.2cm}{+} 
\begin{tikzpicture}
\begin{feynman}
\vertex (v1) {$\wt\cj^\m$};
\vertex [right=.5cm of v1] (v2);
\vertex [right=1.5cm of v2] (v3);
\vertex [right=.1cm of v3] (v4) {$\wt\cj^\n$};
\vertex at ($(v2)!.5!(v3)!1.cm!90:(v3)$) (v5);
\vertex [above=.1cm of v5] (v6) {$\wt\cj^\r$};
\diagram* {
(v1) -- [white]
(v2) -- [majorana, thick, insertion={[style=black, size=2pt]0}, black, edge label=$\rule{0pt}{10pt}$] (v3) -- [white] (v4),
(v3) -- [majorana, thick, insertion={[style=black, size=2pt]0}, black, edge label=$\rule{0pt}{10pt}$] (v5),
(v5) -- [majorana, thick, insertion={[style=black, size=2pt]0}, black, edge label'=$\c$] (v2),
};
\end{feynman}
\end{tikzpicture}
\raisebox{1.2cm}{+}
\begin{tikzpicture}
\begin{feynman}
\vertex (v1) {$\wt\cj^\m$};
\vertex [right=.5cm of v1] (v2);
\vertex [right=1.5cm of v2] (v3);
\vertex [right=.1cm of v3] (v4) {$\wt\cj^\n$};
\vertex at ($(v2)!.5!(v3)!1.cm!90:(v3)$) (v5);
\vertex [above=.1cm of v5] (v6) {$\wt\cj^\r$};
\diagram* {
(v1) -- [white]
(v2) -- [charged scalar, thick, insertion={[style=black, size=2pt]0}, red, edge label=$\rule{0pt}{10pt}$] (v3) -- [white] (v4),
(v3) -- [charged scalar, thick, insertion={[style=black, size=2pt]0}, red, edge label=$\rule{0pt}{10pt}$] (v5),
(v5) -- [charged scalar, thick, insertion={[style=black, size=2pt]0}, red, edge label'=$\vf_2$] (v2),
};
\end{feynman}
\end{tikzpicture}
\raisebox{1.2cm}{+}
\begin{tikzpicture}
\begin{feynman}
\vertex (v1) {$\wt\cj^\m$};
\vertex [right=.5cm of v1] (v2);
\vertex [right=1.5cm of v2] (v3);
\vertex [right=.1cm of v3] (v4) {$\wt\cj^\n$};
\vertex at ($(v2)!.5!(v3)!1.cm!90:(v3)$) (v5);
\vertex [above=.1cm of v5] (v6) {$\wt\cj^\r$};
\diagram* {
(v1) -- [white]
(v2) -- [majorana, thick, insertion={[style=black, size=2pt]0}, purple, edge label=$\rule{0pt}{10pt}$] (v3) -- [white] (v4),
(v3) -- [majorana, thick, insertion={[style=black, size=2pt]0}, purple, edge label=$\rule{0pt}{10pt}$] (v5),
(v5) -- [majorana, thick, insertion={[style=black, size=2pt]0}, purple, edge label'=$\l_2$] (v2),
};
\end{feynman}
\end{tikzpicture}\\
&\raisebox{1.2cm}{+} 
\begin{tikzpicture}
\begin{feynman}
\vertex (v1) {$\wt\cj^\m$};
\vertex [right=.5cm of v1] (v2);
\vertex [right=1.5cm of v2] (v3);
\vertex [right=.1cm of v3] (v4) {$\wt\cj^\n$};
\vertex at ($(v2)!.5!(v3)!1.cm!90:(v3)$) (v5);
\vertex [above=.1cm of v5] (v6) {$\wt\cj^\r$};
\diagram* {
(v1) -- [white]
(v2) -- [charged scalar, thick, insertion={[style=black, size=2pt]0}, orange, edge label=$\rule{0pt}{10pt}$] (v3) -- [white] (v4),
(v3) -- [charged scalar, thick, insertion={[style=black, size=2pt]0}, orange, edge label=$\rule{0pt}{10pt}$] (v5),
(v5) -- [charged scalar, thick, insertion={[style=black, size=2pt]0}, orange, edge label'=$\vf_1$] (v2),
};
\end{feynman}
\end{tikzpicture}
\raisebox{1.2cm}{+}
\begin{tikzpicture}
\begin{feynman}
\vertex (v1) {$\wt\cj^\m$};
\vertex [right=.5cm of v1] (v2);
\vertex [right=1.5cm of v2] (v3);
\vertex [right=.1cm of v3] (v4) {$\wt\cj^\n$};
\vertex at ($(v2)!.5!(v3)!1.cm!90:(v3)$) (v5);
\vertex [above=.1cm of v5] (v6) {$\wt\cj^\r$};
\diagram* {
(v1) -- [white]
(v2) -- [charged scalar, thick, insertion={[style=black, size=2pt]0}, blue, edge label=$\rule{0pt}{10pt}$] (v3) -- [white] (v4),
(v3) -- [charged scalar, thick, insertion={[style=black, size=2pt]0}, blue, edge label=$\rule{0pt}{10pt}$] (v5),
(v5) -- [charged scalar, thick, insertion={[style=black, size=2pt]0}, blue, edge label'=$\vth_1$] (v2),
};
\end{feynman}
\end{tikzpicture}
\raisebox{1.2cm}{+} 
\begin{tikzpicture}
\begin{feynman}
\vertex (v1) {$\wt\cj^\m$};
\vertex [right=.5cm of v1] (v2);
\vertex [right=1.5cm of v2] (v3);
\vertex [right=.1cm of v3] (v4) {$\wt\cj^\n$};
\vertex at ($(v2)!.5!(v3)!1.cm!90:(v3)$) (v5);
\vertex [above=.1cm of v5] (v6) {$\wt\cj^\r$};
\diagram* {
(v1) -- [white]
(v2) -- [fermion, thick, insertion={[style=black, size=2pt]0}, olive, edge label=$\rule{0pt}{10pt}$] (v3) -- [white] (v4),
(v3) -- [fermion, thick, insertion={[style=black, size=2pt]0}, olive, edge label=$\rule{0pt}{10pt}$] (v5),
(v5) -- [fermion, thick, insertion={[style=black, size=2pt]0}, olive, edge label'=$\l_1$] (v2),
};
\end{feynman}
\end{tikzpicture}
\end{eqnarray*}
\vskip-.5cm
\caption{One-loop diagrams that contribute to the 3-point function of R-currents.}
\label{JJJ-3pt-diags}
\end{figure}

The triangle diagrams that contribute to this 3-point function are shown in fig.~\ref{JJJ-3pt-diags}. They give
\bal
&\lb\wt\cj^\m(p_1)\wt\cj^\n(p_2)\wt\cj^\r(p_3)\rb=i^2\Big(\frac{i}{3\a^2}\Big)^3\int \frac{d^{4}q}{(2\p)^4}i^3(q+q')^\m(q'+q'')^\n(q''+q)^\r G_1^{(3)}(q,q',q'')\NO\\
&-4i^2\Big(\frac{i}{6}\Big)^3\int \frac{d^{4}q}{(2\p)^4}\tr\Big(\g^\m\g^5 P_{\c}(q')\g^\n\g^5P_\c(q'')\g^\r\g^5P_\c(q)+\g^\m\g^5 P_{\l_2}(q')\g^\n\g^5P_{\l_2}(q'')\g^\r\g^5P_{\l_2}(q)\NO\\
&-2\g^\m\g^5 P_{\l_1}(q')\g^\n\g^5P_{\l_1}(q'')\g^\r\g^5P_{\l_1}(q)\Big)+\n\leftrightarrow \r,\,p_2\leftrightarrow p_3,
\eal
where $q'=q+p_1$, $q''=q+p_1+p_2$, and $p_1+p_2+p_3=0$. It follows that this 3-point function is also properly regulated by the PV fields. 

The bosonic contribution vanishes identically by symmetry. After some manipulations, the fermionic contribution can be expressed in terms of an integral over Feynman parameters as
\bal
&\lb\wt\cj^\m(p_1)\wt\cj^\n(p_2)\wt\cj^\r(p_3)\rb=-4i^2\Big(\frac{i}{6}\Big)^3\frac{2i\p^2}{(2\p)^4}\bigg[-\frac{i}{3}\e^{\m\n\r\s}p_{13\s}-2i\int_0^1dudvdw\d(u+v+w-1)\times\NO\\
&\frac{1}{\D}\Big(\big(v^2p_2^2+u(1-u)p_3^2-\D\big)\e^{\m\n\r\s}\big(p_{12}+3(t-up_3)\big)_\s\NO\\
&+2\e^{\n\r\s\t}\big(v(1-v)p_1+vwp_3\big)^\m p_{1\s}p_{3\t}+2\e^{\m\r\s\t}\big(v(1-v)p_1+vwp_3\big)^\n p_{1\s}p_{3\t}\NO\\
&+2\e^{\m\n\s\t}p_{1\s}p_{3\t}\big(-v(u-w)p_1+(1+u-w)wp_3\big)^\r+2(\D+vp_1\cdot p_2)\e^{\m\n\r\s}(p_1+t)_\s\Big)\bigg],
\eal 
where $p_{23}\equiv p_2-p_3$, $\D=uvp_1^2+vwp_2^2+uwp_3^2+m_1^2$, $\D_1=\D+m_1^2$, $\D_2=\D+2m_1^2$ and $t\equiv-v p_1+w p_3$.


\begin{flushleft}
\noindent\rule{\textwidth}{0.8pt}
\raisebox{-0.1cm}{$\<\ct\cj\cj\>$}
\noindent\rule{\textwidth}{0.8pt}
\end{flushleft}

Another 3-point function that enters the supersymmetry Ward identity is that between two R-currents and a stress tensor. This correlation function receives contributions from a number of 2- and 1-point functions involving seagull operators, as well as from the R-current 2-point function. In particular, the results in appendix~\ref{sec:seagulls} determine that 
\bal
&\<\wt\ct^{\m\n}(p_1)\wt\cj^\r(p_2)\wt\cj^\s(p_3)\>=\lb\wt\ct^{\m\n}(p_1)\wt\cj^\r(p_2)\wt\cj^\s(p_3)\rb\NO\\
&-(\h^{\m\r}\h^{\n\k}+\h^{\m\k}\h^{\n\r}-\h^{\m\n}\h^{\r\k})\lb\wt\cj_\k(-p_3)\wt\cj^\s(p_3)\rb+\frac{i}{6}\h^{\m\r}\lb{\wt s}{}_{(2|1)}^{\n}(-p_3)\wt\cj^\s(p_3)\rb\NO\\
&-(\h^{\m\s}\h^{\n\k}+\h^{\m\k}\h^{\n\s}-\h^{\m\n}\h^{\s\k})\lb\wt\cj_\k(-p_2)\wt\cj^\r(p_2)\rb+\frac{i}{6}\h^{\m\s}\lb{\wt s}{}_{(2|1)}^{\n}(-p_2)\wt\cj^\r(p_2)\rb\NO\\
&-\frac{4}{9\a^2}\h^{\r\s}\lb\wt\ct^{\m\n}(p_1){\wt s}_{(1|0)}(-p_1)\rb+\frac{4}{9\a^2}(\h^{\m\r}\h^{\n\s}+\h^{\m\s}\h^{\n\r}-\h^{\m\n}\h^{\r\s})\lb{\wt s}_{(1|0)}(0)\rb,
\eal
where all seagull operators are defined in \eqref{seagull-operators}. Since the R-current 2-point function was already computed in \eqref{JJ-seagull} and the 1-point function $\lb{\wt s}_{(1|0)}(0)\rb$ in \eqref{JJ}, we need only determine the two 2-point functions $\lb{\wt s}{}_{(2|1)}^{\n}\wt\cj^\s\rb$ and $\lb\wt\ct^{\m\n}{\wt s}_{(1|0)}\rb$, as well as the Feynman diagram contribution to the 3-point function.

The seagull operator ${\wt s}{}_{(2|1)}^{\n}$ is proportional to the fermionic part of the R-current, and so the first 2-point function can be read off the computation of the R-current 2-point function, namely
\bal
&\lb{\wt s}{}_{(2|1)}^{\n}(p)\wt\cj^\s(-p)\rb=\\
&-2i\Big(\frac{i}{6}\Big)\int \frac{d^{4}q}{(2\p)^4}\tr\Big(\g^\m\g^5 P_{\c}(q)\g^\n\g^5P_\c(q')+\g^\m\g^5 P_{\l_2}(q)\g^\n\g^5P_{\l_2}(q')-2\g^\m\g^5 P_{\l_1}(q)\g^\n\g^5P_{\l_1}(q')\Big),\NO
\eal
where $q'=q+p$. Hence, this 2-point function is properly regulated by our choice of PV fields. The second 2-point function can also be easily evaluated to obtain   
\be
\lb\wt\ct^{\m\n}(p){\wt s}_{(1|0)}(-p)\rb=\frac{i}{\a^2}\int \frac{d^{4}q}{(2\p)^4}\Big[\Big(q^{(\m}q'^{\n)}-\frac12\h^{\m\n}q\cdot q'+\frac16(p^\m p^\n-\h^{\m\n}p^2)\Big)G^{(2)}_1(q,q')
-m_1^2\h^{\m\n}G^{(2)}_2(q,q')\Big],
\ee
which is therefore also fully regulated. 

The Feynman diagrams that contribute to the 3-point function of a stress tensor and two R-currents are the same as those in Fig.~\ref{JJJ-3pt-diags}, with an R-current replaced by a stress tensor. We get
\bal
&\lb\wt\ct^{\m\n}(p_1)\wt\cj^\r(p_2)\wt\cj^\s(p_3)\rb=i^2\Big(\frac{i}{3\a^2}\Big)^2\frac{1}{\a^2}\int \frac{d^{4}q}{(2\p)^4}i^2(q'+q'')^\r(q''+q)^\s \times\NO\\
&\Big[\Big(q^{(\m}q'^{\n)}-\frac12\h^{\m\n}q\cdot q'+\frac16(p^\m p^\n-\h^{\m\n}p^2)\Big)G^{(3)}_1(q,q',q'')
-m_1^2\h^{\m\n}G^{(3)}_2(q,q',q'')\Big]\NO\\
&-4i^2\Big(\frac{i}{6}\Big)^2\int \frac{d^{4}q}{(2\p)^4}\bigg(\frac{i}{8}\tr\Big[\big(4\h^{\m[\k}\h^{\n]\l}\g_\k(q+q')_\l+i\e^{\m\n\k\l}\g_\k\g^5p_{1\k}\big)\times\\
&\Big(P_\c(q')\g^\r\g^5P_\c(q'')\g^\s\g^5P_\c(q)+P_{\l_2}(q')\g^\r\g^5P_{\l_2}(q'')\g^\s\g^5P_{\l_2}(q)-2P_{\l_1}(q')\g^\r\g^5P_{\l_1}(q'')\g^\s\g^5P_{\l_1}(q)\Big)\Big]\NO\\
&-\frac12\h^{\m\n}\tr\Big(m_2P_{\l_2}(q')\g^\r\g^5P_{\l_2}(q'')\g^\s\g^5P_{\l_2}(q)-2m_1P_{\l_1}(q')\g^\r\g^5P_{\l_1}(q'')\g^\s\g^5P_{\l_1}(q)\Big)\bigg)+\begin{matrix}\r\leftrightarrow \s\\p_2\leftrightarrow p_3\end{matrix},\NO
\eal
where again $q'=q+p_1$, $q''=q+p_1+p_2$, and $p_1+p_2+p_3=0$. The bosonic part of this 3-point function is manifestly regulated by simple power counting. Straightforward manipulations of the fermionic part show that it too is properly regulated.


\begin{flushleft}
\noindent\rule{\textwidth}{0.8pt}
\raisebox{-0.1cm}{$\<\cq\bar\cq\cj\>$}
\noindent\rule{\textwidth}{0.8pt}
\end{flushleft}

The last 3-point function of current multiplet operators that enters in the supersymmetry Ward identity is that between two supercurrents and an R-current. Using appendix~\ref{sec:seagulls} we determine   
\bal
&\<\wt\cq^\m(p_1)\wt{\bar\cq}{}^\n(p_2)\wt\cj^\r(p_3)\>=\lb\wt\cq^\m(p_1)\wt{\bar\cq}{}^\n(p_2)\wt\cj^\r(p_3)\rb\NO\\
&+\frac38\e^{\m\n\k\l}\lb\wt\cj_\k(-p_3)\wt\cj^\r(p_3)\rb\g_\l+\frac18\lb{\wt s}{}_{(2|1)}^{\k}(-p_3)\wt\cj^\r(p_3)\rb\g^5\g_\k\h^{\m\n}\NO\\
&+\frac{i}{12\a^2}\lb{\wt s}_{(1|0)}(-p_3)\wt\cj^\r(p_3)\rb\big(2\h_\k^{[\m}\h^{\n]}_\l p_3^\k+i\e^{\m\n}{}_{\k\l}\g^5 p_{21}^\k\big)\g^\l\NO\\
&-\frac{1}{3\a}\h^{\m\r}\lb{\wt s}_{\(3|\frac12\)}(-p_2)\wt{\bar\cq}{}^\n(p_2)\rb-\frac{1}{3\a}\h^{\n\r}\lb\wt\cq^\m(p_1)\hat{\wt{\bar s}}\hspace{6pt}{}_{\(3|\frac12\)}(-p_1)\rb,
\eal
where we have used the fact that correlation functions with a single insertion of $\co_M$ or $\co_{M^*}$ vanish trivially in the free WZ model due to the absence of possible contractions. The only terms that we have not already determined are the seagull 2-point functions $\lb{\wt s}_{(1|0)}\wt\cj^\r\rb$, $\lb{\wt s}_{\(3|\frac12\)}\wt{\bar\cq}{}^\n\rb$, and the Feynman diagram contribution to the 3-point function. 

The two 2-point functions can be easily evaluated and take the form
\be
\lb{\wt s}_{(1|0)}(p)\wt\cj^\r(-p)\rb=-\frac{i}{3\a^2}\int \frac{d^{4}q}{(2\p)^4}(q+q')^\r G^{(2)}_1(q,q'),
\ee
\be
\lb{\wt s}_{\(3|\frac12\)}(p)\wt{\bar\cq}{}^\m(-p)\rb=-\frac{1}{4\a^3}\g^5\int \frac{d^{4}q}{(2\p)^4}\Big[G^{(2)}_1(q,q')\slashed q'\Big(\g^\m\slashed q+\frac{2}{3}\g^{\m\n}p_\n\Big)-2m_1^2G^{(2)}_2(q,q')\g^\m\Big],
\ee
where again $q'=q+p$. These too, therefore, are regulated by the PV fields. In fact, using the transformation $q\to -p-q$ of the loop momentum, we see that the first of these vanishes identically.

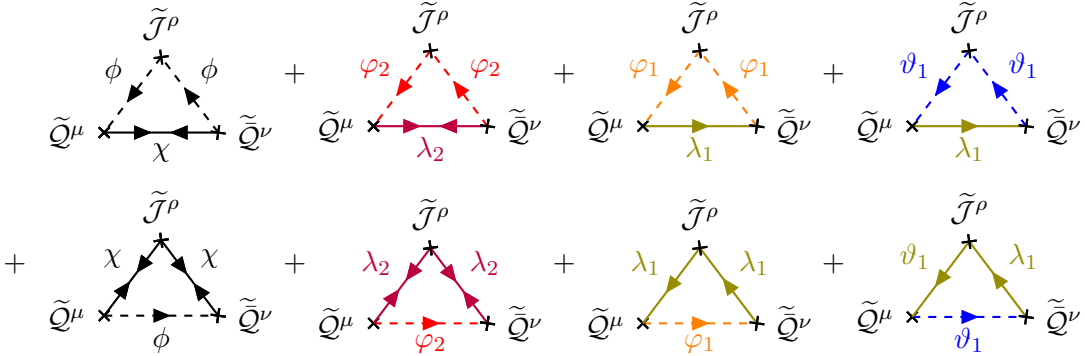
\begin{figure}[h]
\captionsetup{justification=raggedright,
singlelinecheck=false
}
\begin{eqnarray*}
&\begin{tikzpicture}
\begin{feynman}
\vertex (v1) {$\wt\cq^\m$};
\vertex [right=.5cm of v1] (v2);
\vertex [right=1.5cm of v2] (v3);
\vertex [right=.1cm of v3] (v4) {$\wt{\bar\cq}{}^\n$};
\vertex at ($(v2)!.5!(v3)!1.cm!90:(v3)$) (v5);
\vertex [above=.1cm of v5] (v6) {$\wt\cj^\r$};
\diagram* {
(v1) -- [white]
(v2) -- [majorana, thick, insertion={[style=black, size=2pt]0}, black, edge label'=$\c$] (v3) -- [white] (v4),
(v3) -- [charged scalar, thick, insertion={[style=black, size=2pt]0}, black, edge label'=$\f$] (v5),
(v5) -- [charged scalar, thick, insertion={[style=black, size=2pt]0}, black, edge label'=$\f$] (v2),
};
\end{feynman}
\end{tikzpicture}
\raisebox{1.2cm}{+} 
\begin{tikzpicture}
\begin{feynman}
\vertex (v1) {$\wt\cq^\m$};
\vertex [right=.5cm of v1] (v2);
\vertex [right=1.5cm of v2] (v3);
\vertex [right=.1cm of v3] (v4) {$\wt{\bar\cq}{}^\n$};
\vertex at ($(v2)!.5!(v3)!1.cm!90:(v3)$) (v5);
\vertex [above=.1cm of v5] (v6) {$\wt\cj^\r$};
\diagram* {
(v1) -- [white]
(v2) -- [majorana, thick, insertion={[style=black, size=2pt]0}, purple, edge label'=$\l_2$] (v3) -- [white] (v4),
(v3) -- [charged scalar, thick, insertion={[style=black, size=2pt]0}, red, edge label'=$\vf_2$] (v5),
(v5) -- [charged scalar, thick, insertion={[style=black, size=2pt]0}, red, edge label'=$\vf_2$] (v2),
};
\end{feynman}
\end{tikzpicture}
\raisebox{1.2cm}{+}
\begin{tikzpicture}
\begin{feynman}
\vertex (v1) {$\wt\cq^\m$};
\vertex [right=.5cm of v1] (v2);
\vertex [right=1.5cm of v2] (v3);
\vertex [right=.1cm of v3] (v4) {$\wt{\bar\cq}{}^\n$};
\vertex at ($(v2)!.5!(v3)!1.cm!90:(v3)$) (v5);
\vertex [above=.1cm of v5] (v6) {$\wt\cj^\r$};
\diagram* {
(v1) -- [white]
(v2) -- [fermion, thick, insertion={[style=black, size=2pt]0}, olive, edge label'=$\l_1$] (v3) -- [white] (v4),
(v3) -- [charged scalar, thick, insertion={[style=black, size=2pt]0}, orange, edge label'=$\vf_1$] (v5),
(v5) -- [charged scalar, thick, insertion={[style=black, size=2pt]0}, orange, edge label'=$\vf_1$] (v2),
};
\end{feynman}
\end{tikzpicture}
\raisebox{1.2cm}{+}
\begin{tikzpicture}
\begin{feynman}
\vertex (v1) {$\wt\cq^\m$};
\vertex [right=.5cm of v1] (v2);
\vertex [right=1.5cm of v2] (v3);
\vertex [right=.1cm of v3] (v4) {$\wt{\bar\cq}{}^\n$};
\vertex at ($(v2)!.5!(v3)!1.cm!90:(v3)$) (v5);
\vertex [above=.1cm of v5] (v6) {$\wt\cj^\r$};
\diagram* {
(v1) -- [white]
(v2) -- [fermion, thick, insertion={[style=black, size=2pt]0}, olive, edge label'=$\l_1$] (v3) -- [white] (v4),
(v3) -- [charged scalar, thick, insertion={[style=black, size=2pt]0}, blue, edge label'=$\vth_1$] (v5),
(v5) -- [charged scalar, thick, insertion={[style=black, size=2pt]0}, blue, edge label'=$\vth_1$] (v2),
};
\end{feynman}
\end{tikzpicture}\\
\raisebox{1.2cm}{+}
&\begin{tikzpicture}
\begin{feynman}
\vertex (v1) {$\wt\cq^\m$};
\vertex [right=.5cm of v1] (v2);
\vertex [right=1.5cm of v2] (v3);
\vertex [right=.1cm of v3] (v4) {$\wt{\bar\cq}{}^\n$};
\vertex at ($(v2)!.5!(v3)!1.cm!90:(v3)$) (v5);
\vertex [above=.1cm of v5] (v6) {$\wt\cj^\r$};
\diagram* {
(v1) -- [white]
(v2) -- [charged scalar, thick, insertion={[style=black, size=2pt]0}, black, edge label'=$\f$] (v3) -- [white] (v4),
(v3) -- [majorana, thick, insertion={[style=black, size=2pt]0}, black, edge label'=$\c$] (v5),
(v5) -- [majorana, thick, insertion={[style=black, size=2pt]0}, black, edge label'=$\c$] (v2),
};
\end{feynman}
\end{tikzpicture}
\raisebox{1.2cm}{+} 
\begin{tikzpicture}
\begin{feynman}
\vertex (v1) {$\wt\cq^\m$};
\vertex [right=.5cm of v1] (v2);
\vertex [right=1.5cm of v2] (v3);
\vertex [right=.1cm of v3] (v4) {$\wt{\bar\cq}{}^\n$};
\vertex at ($(v2)!.5!(v3)!1.cm!90:(v3)$) (v5);
\vertex [above=.1cm of v5] (v6) {$\wt\cj^\r$};
\diagram* {
(v1) -- [white]
(v2) -- [charged scalar, thick, insertion={[style=black, size=2pt]0}, red, edge label'=$\vf_2$] (v3) -- [white] (v4),
(v3) -- [majorana, thick, insertion={[style=black, size=2pt]0}, purple, edge label'=$\l_2$] (v5),
(v5) -- [majorana, thick, insertion={[style=black, size=2pt]0}, purple, edge label'=$\l_2$] (v2),
};
\end{feynman}
\end{tikzpicture}
\raisebox{1.2cm}{+}
\begin{tikzpicture}
\begin{feynman}
\vertex (v1) {$\wt\cq^\m$};
\vertex [right=.5cm of v1] (v2);
\vertex [right=1.5cm of v2] (v3);
\vertex [right=.1cm of v3] (v4) {$\wt{\bar\cq}{}^\n$};
\vertex at ($(v2)!.5!(v3)!1.cm!90:(v3)$) (v5);
\vertex [above=.1cm of v5] (v6) {$\wt\cj^\r$};
\diagram* {
(v1) -- [white]
(v2) -- [charged scalar, thick, insertion={[style=black, size=2pt]0}, orange, edge label'=$\vf_1$] (v3) -- [white] (v4),
(v3) -- [fermion, thick, insertion={[style=black, size=2pt]0}, olive, edge label'=$\l_1$] (v5),
(v5) -- [fermion, thick, insertion={[style=black, size=2pt]0}, olive, edge label'=$\l_1$] (v2),
};
\end{feynman}
\end{tikzpicture}
\raisebox{1.2cm}{+}
\begin{tikzpicture}
\begin{feynman}
\vertex (v1) {$\wt\cq^\m$};
\vertex [right=.5cm of v1] (v2);
\vertex [right=1.5cm of v2] (v3);
\vertex [right=.1cm of v3] (v4) {$\wt{\bar\cq}{}^\n$};
\vertex at ($(v2)!.5!(v3)!1.cm!90:(v3)$) (v5);
\vertex [above=.1cm of v5] (v6) {$\wt\cj^\r$};
\diagram* {
(v1) -- [white]
(v2) -- [charged scalar, thick, insertion={[style=black, size=2pt]0}, blue, edge label'=$\vth_1$] (v3) -- [white] (v4),
(v3) -- [fermion, thick, insertion={[style=black, size=2pt]0}, olive, edge label'=$\l_1$] (v5),
(v5) -- [fermion, thick, insertion={[style=black, size=2pt]0}, olive, edge label'=$\vth_1$] (v2),
};
\end{feynman}
\end{tikzpicture}
\end{eqnarray*}
\vskip-.5cm
\caption{One-loop diagrams that contribute to the 3-point function of two supercurrents and an R-current.}
\label{QQJ-3pt-diags}
\end{figure}

Turning to the Feynman diagram contribution to the 3-point function, shown in fig.~\ref{QQJ-3pt-diags}, we get
\bal
&\lb\wt\cq^\m(p_1)\wt{\bar\cq}{}^\n(p_2)\wt\cj^\r(p_3)\rb=\NO\\
&\frac{i}{48\a^6}\int \frac{d^{4}q}{(2\p)^4}G_1^{(3)}(q,q',q'')\g^5\big(2(q+q'')^\r\O_1^\m(-q)\slashed q'\O_2^\n(q'')-\O^\m_1(q')\slashed q\g^\r\slashed q''\O_2^\n(-q')\big)\NO\\
&+\frac{im_1^2}{24\a^6}\int \frac{d^{4}q}{(2\p)^4}G_2^{(3)}(q,q',q'')\g^5\Big(2(q+q'')^\r\big(\g^\m\O_2^\n(q'')-\O_1^\m(-q)\g^\n+\g^\m\slashed q'\g^\n\big)\NO\\
&+\O_1^\m(q')(\g^\r\slashed q''-\slashed q\g^\r)\g^\n-\g^\m(\slashed q\g^\r-\g^\r\slashed q'')\O_2^\n(-q')-\O_1^\m(q')\g^\r\O_2^\n(-q')+\g^\m\slashed q\g^\r\slashed q''\g^\n\Big)\NO\\
&+\frac{im_1^4}{12\a^6}\g^5\g^\m\g^\r\g^\n\int \frac{d^{4}q}{(2\p)^4}\Big(P_{\vf_2}(q)P_{\vf_2}(q')P_{\vf_2}(q'')-\frac12P_{\vf_1}(q)P_{\vf_1}(q')P_{\vf_1}(q'')\Big),
\eal
where we have introduced the shorthand notation
\be\label{Omega}
\O_1^\m(q)\equiv \slashed q\g^\m+\frac{3}{2}\g^{\m\r}p_{1\r},\qquad \O_2^\m(q)=\g^\m\slashed q-\frac{2}{3}\g^{\m\r}p_{2\r}.
\ee

\subsection{4-point functions}


\begin{flushleft}
\noindent\rule{\textwidth}{0.8pt}
\raisebox{-0.1cm}{$\<\cq\bar\cq\cj\cj\>$}
\noindent\rule{\textwidth}{0.8pt}
\end{flushleft}

There is only one 4-point function we need to consider, namely the one between two supercurrents and two R-currents. From the results in appendix~\ref{sec:seagulls} we determine that it takes the form
\bal
&\<\wt\cq^\m(p_1)\wt{\bar\cq}{}^\n(p_2)\wt\cj^\r(p_3)\wt\cj^\s(p_4)\>=\lb\wt\cq^\m(p_1)\wt{\bar\cq}{}^\n(p_2)\wt\cj^\r(p_3)\wt\cj^\s(p_4)\rb\NO\\
&+\frac38\e^{\m\n\k\l}\g_\l\lb\wt\cj_\k(-p_3-p_4)\wt\cj^\r(p_3)\wt\cj^\s(p_4)\rb+\frac18\h^{\m\n}\g^5\g_\k\lb{\wt s}{}_{(2|1)}^{\k}(-p_3-p_4)\wt\cj^\r(p_3)\wt\cj^\s(p_4)\rb\NO\\
&+\frac{i}{12\a^2}\big(2\h_\k^{[\m}\h^{\n]}_\l(p_3+p_4)^\k+i\e^{\m\n}{}_{\k\l}\g^5p_{21}^\k\big) \g^\l\lb{\wt s}_{(1|0)}(-p_3-p_4)\wt\cj^\r(p_3)\wt\cj^\s(p_4)\rb\NO\\
&-\frac{1}{3\a}\h^{\m\r}\lb\hat{\wt s}_{\(3|\frac12\)}(-p_2-p_4)\wt{\bar\cq}{}^\n(p_2)\wt\cj^\s(p_4)\rb-\frac{1}{3\a}\h^{\m\s}\lb{\wt s}_{\(3|\frac12\)}(-p_2-p_3)\wt{\bar\cq}{}^\n(p_2)\wt\cj^\r(p_3)\rb\NO\\
&-\frac{1}{3\a}\h^{\n\r}\lb\wt\cq^\m(p_1){\wt{\bar s}}\hspace{0pt}{}_{\(3|\frac12\)}(-p_1-p_4)\wt\cj^\s(p_4)\rb-\frac{1}{3\a}\h^{\n\s}\lb\wt\cq^\m(p_1){\wt{\bar s}}\hspace{0pt}{}_{\(3|\frac12\)}(-p_1-p_3)\wt\cj^\r(p_3)\rb\NO\\
&-\frac{4}{9\a^2}\h^{\r\s}\lb\wt\cq^\m(p_1)\wt{\bar\cq}{}^\n(p_2){\wt s}_{(1|0)}(-p_1-p_2)\rb-\frac{1}{6\a^2}\h^{\r\s}\e^{\m\n\k\l}\g_\l\lb\wt\cj_\k(-p_3-p_4){\wt s}_{(1|0)}(p_3+p_4)\rb\NO\\
&-\frac{i}{27\a^4}\h^{\r\s}\big(2\h_\k^{[\m}\h^{\n]}_\l(p_3+p_4)^\k+i\e^{\m\n}{}_{\k\l}\g^5p^\k_{21}\big)\g^\l\lb{\wt s}_{(1|0)}(-p_3-p_4){\wt s}_{(1|0)}(p_3+p_4)\rb\\
&+\frac{1}{9\a^2}\h^{\m\r}\h^{\n\s}\lb{\wt s}_{\(3|\frac12\)}(-p_2-p_4){\wt{\bar s}}\hspace{0pt}{}_{\(3|\frac12\)}(p_2+p_4)\rb+\frac{1}{9\a^2}\h^{\m\s}\h^{\n\r}\lb{\wt s}_{\(3|\frac12\)}(-p_2-p_3){\wt{\bar s}}\hspace{0pt}{}_{\(3|\frac12\)}(p_2+p_3)\rb,\NO
\eal
where again we have used the fact that correlation functions with a single insertion of $\co_M$ or $\co_{M^*}$ vanish trivially, as does the 2-point function $\lb{\wt s}{}_{(2|1)}^{\k}{\wt s}_{(1|0)}\rb=0$, due to lack of possible contractions. Besides the Feynman diagram contribution to the 4-point function, there are two 2-point functions and four 3-point functions involving seagull operators that we have not already determined. 

The two 2-point functions are easily evaluated with the result 
\be
\lb{\wt s}_{(1|0)}(p){\wt s}_{(1|0)}(-p)\rb=i\int \frac{d^{4}q}{(2\p)^4}G^{(2)}_1(q,q'),
\ee
\be
\lb{\wt s}_{\(3|\frac12\)}(p){\wt{\bar s}}\hspace{0pt}{}_{\(3|\frac12\)}(-p)\rb=\frac{1}{2\a^2}\int \frac{d^{4}q}{(2\p)^4}G^{(2)}_1(q,q')\slashed q',
\ee
where again $q'=q+p$. Both of these are therefore properly regulated. Moreover, the transformation $q\to -p-q$ shows that the second 2-point function is identically zero. 

Since the seagull operator ${\wt s}{}_{(2|1)}^{\m}$ is proportional to the fermionic part of the R-current and the R-current 3-point function receives contributions only from the fermionic part, it follows that
\be
\lb{\wt s}{}_{(2|1)}^{\m}(p_1)\wt\cj^\n(p_2)\wt\cj^\r(p_3)\rb=-6i\lb\wt\cj^\m(p_1)\wt\cj^\n(p_2)\wt\cj^\r(p_3)\rb,
\ee
which is properly regulated and has already been evaluated above. Next, we find that
\be
\lb{\wt s}_{(1|0)}(p_1)\wt\cj^\r(p_2)\wt\cj^\s(p_3)\rb=-\frac{1}{9\a^4}\int \frac{d^{4}q}{(2\p)^4}(q'+q'')^\r(q''+q)^\s G_1^{(3)}(q,q',q'')+\begin{matrix}\r\leftrightarrow \s\\p_2\leftrightarrow p_3\end{matrix},
\ee
where $q'=q+p_1$, $q''=q+p_1+p_2$, and $p_1+p_2+p_3=0$. Hence, this 3-point function is properly regulated by the PV fields as well. The remaining two 3-point functions take the form
\bal
&\lb{\wt s}_{\(3|\frac12\)}(p_1)\wt{\bar\cq}{}^\m(p_2)\wt\cj^\n(p_3)\rb=\frac{i}{24\a^5}\int \frac{d^{4}q}{(2\p)^4}G_1^{(3)}(q,q',q'')\big(2(q+q'')^\n\slashed q'\O_2^\m(q'')-\slashed q\g^\n\slashed q''\O_2^\m(-q')\big)\NO\\
&-\frac{im_1^2}{12\a^5}\int \frac{d^{4}q}{(2\p)^4}G_2^{(3)}(q,q',q'')\big(2(q+q'')^\n\g^\m+\g^\n\O_2^\m(-q')+(\slashed q\g^\n-\g^\n\slashed q'')\g^\m\big),
\eal
and
\bal
&\lb\wt\cq^\m(p_1)\wt{\bar\cq}{}^\n(p_2){\wt s}_{(1|0)}(p_3)\rb=-\frac{i}{8\a^4}\int \frac{d^{4}q}{(2\p)^4}G_1^{(3)}(q,q',q'')\O_1^\m(-q)\slashed q'\O_2^\n(q'')\NO\\
&-\frac{im_1^2}{4\a^4}\int \frac{d^{4}q}{(2\p)^4}G_2^{(3)}(q,q',q'')\big(\g^\m\O_2^\n(q'')-\O_1^\m(-q)\g^\n+\g^\m\slashed q'\g^\n\big),
\eal
and are therefore properly regulates as well.

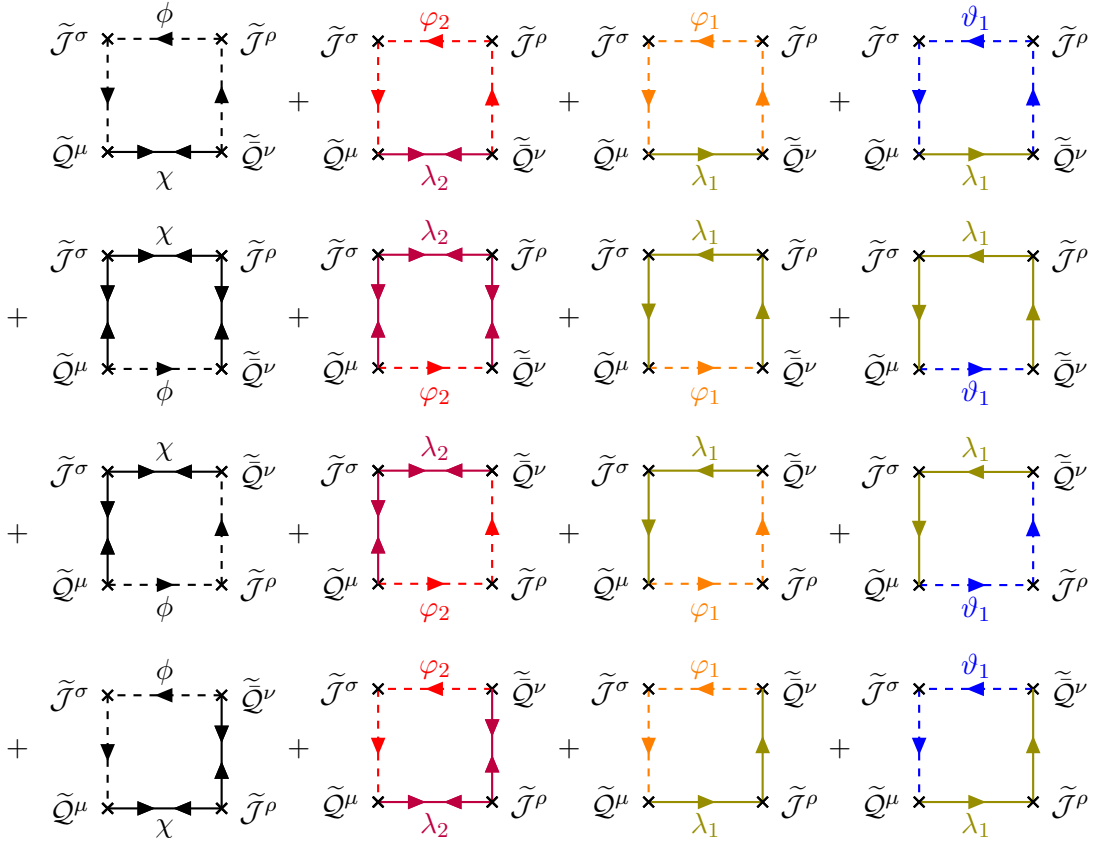
\begin{figure}[h]
\captionsetup{justification=raggedright,
singlelinecheck=false
}
\begin{eqnarray*}
&\begin{tikzpicture}
\begin{feynman}
\vertex (v1) {$\wt\cq^\m$};
\vertex [right=.5cm of v1] (v2);
\vertex [right=1.5cm of v2] (v3);
\vertex [right=.1cm of v3] (v4) {$\wt{\bar\cq}{}^\n$};
\vertex [above=1.5cm of v3] (v5);
\vertex [right=.1cm of v5] (v6) {$\wt\cj^\r$};
\vertex [above=1.5cm of v2] (v7);
\vertex [left=.1cm of v7] (v8) {$\wt\cj^\s$};
\diagram* {
(v1) -- [white]
(v2) -- [majorana, thick, insertion={[style=black, size=2pt]0}, black, edge label'=$\c\rule{0pt}{8pt}$] (v3) -- [white] (v4),
(v3) -- [charged scalar, thick, insertion={[style=black, size=2pt]0}, black, edge label=$\rule{0pt}{10pt}$] (v5),
(v5) -- [charged scalar, thick, insertion={[style=black, size=2pt]0}, black, edge label'=$\f$] (v7),
(v7) -- [charged scalar, thick, insertion={[style=black, size=2pt]0}, black, edge label'=$\rule{0pt}{10pt}$] (v2),
};
\end{feynman}
\end{tikzpicture}
\raisebox{1.2cm}{+} 
\begin{tikzpicture}
\begin{feynman}
\vertex (v1) {$\wt\cq^\m$};
\vertex [right=.5cm of v1] (v2);
\vertex [right=1.5cm of v2] (v3);
\vertex [right=.1cm of v3] (v4) {$\wt{\bar\cq}{}^\n$};
\vertex [above=1.5cm of v3] (v5);
\vertex [right=.1cm of v5] (v6) {$\wt\cj^\r$};
\vertex [above=1.5cm of v2] (v7);
\vertex [left=.1cm of v7] (v8) {$\wt\cj^\s$};
\diagram* {
(v1) -- [white]
(v2) -- [majorana, thick, insertion={[style=black, size=2pt]0}, purple, edge label'=$\l_2$] (v3) -- [white] (v4),
(v3) -- [charged scalar, thick, insertion={[style=black, size=2pt]0}, red, edge label=$\rule{0pt}{10pt}$] (v5),
(v5) -- [charged scalar, thick, insertion={[style=black, size=2pt]0}, red, edge label'=$\vf_2$] (v7),
(v7) -- [charged scalar, thick, insertion={[style=black, size=2pt]0}, red, edge label'=$\rule{0pt}{10pt}$] (v2),
};
\end{feynman}
\end{tikzpicture}
\raisebox{1.2cm}{+}
\begin{tikzpicture}
\begin{feynman}
\vertex (v1) {$\wt\cq^\m$};
\vertex [right=.5cm of v1] (v2);
\vertex [right=1.5cm of v2] (v3);
\vertex [right=.1cm of v3] (v4) {$\wt{\bar\cq}{}^\n$};
\vertex [above=1.5cm of v3] (v5);
\vertex [right=.1cm of v5] (v6) {$\wt\cj^\r$};
\vertex [above=1.5cm of v2] (v7);
\vertex [left=.1cm of v7] (v8) {$\wt\cj^\s$};
\diagram* {
(v1) -- [white]
(v2) -- [fermion, thick, insertion={[style=black, size=2pt]0}, olive, edge label'=$\l_1$] (v3) -- [white] (v4),
(v3) -- [charged scalar, thick, insertion={[style=black, size=2pt]0}, orange, edge label'=$\rule{0pt}{10pt}$] (v5),
(v5) -- [charged scalar, thick, insertion={[style=black, size=2pt]0}, orange, edge label'=$\vf_1$] (v7),
(v7) -- [charged scalar, thick, insertion={[style=black, size=2pt]0}, orange, edge label'=$\rule{0pt}{10pt}$] (v2),
};
\end{feynman}
\end{tikzpicture}
\raisebox{1.2cm}{+}
\begin{tikzpicture}
\begin{feynman}
\vertex (v1) {$\wt\cq^\m$};
\vertex [right=.5cm of v1] (v2);
\vertex [right=1.5cm of v2] (v3);
\vertex [right=.1cm of v3] (v4) {$\wt{\bar\cq}{}^\n$};
\vertex [above=1.5cm of v3] (v5);
\vertex [right=.1cm of v5] (v6) {$\wt\cj^\r$};
\vertex [above=1.5cm of v2] (v7);
\vertex [left=.1cm of v7] (v8) {$\wt\cj^\s$};
\diagram* {
(v1) -- [white]
(v2) -- [fermion, thick, insertion={[style=black, size=2pt]0}, olive, edge label'=$\l_1$] (v3) -- [white] (v4),
(v3) -- [charged scalar, thick, insertion={[style=black, size=2pt]0}, blue, edge label'=$\rule{0pt}{10pt}$] (v5),
(v5) -- [charged scalar, thick, insertion={[style=black, size=2pt]0}, blue, edge label'=$\vth_1$] (v7),
(v7) -- [charged scalar, thick, insertion={[style=black, size=2pt]0}, blue, edge label'=$\rule{0pt}{10pt}$] (v2),
};
\end{feynman}
\end{tikzpicture}\\
\raisebox{1.2cm}{+}
&\begin{tikzpicture}
\begin{feynman}
\vertex (v1) {$\wt\cq^\m$};
\vertex [right=.5cm of v1] (v2);
\vertex [right=1.5cm of v2] (v3);
\vertex [right=.1cm of v3] (v4) {$\wt{\bar\cq}{}^\n$};
\vertex [above=1.5cm of v3] (v5);
\vertex [right=.1cm of v5] (v6) {$\wt\cj^\r$};
\vertex [above=1.5cm of v2] (v7);
\vertex [left=.1cm of v7] (v8) {$\wt\cj^\s$};
\diagram* {
(v1) -- [white]
(v2) -- [charged scalar, thick, insertion={[style=black, size=2pt]0}, black, edge label'=$\f$] (v3) -- [white] (v4),
(v3) -- [majorana, thick, insertion={[style=black, size=2pt]0}, black, edge label'=$\rule{0pt}{10pt}$] (v5),
(v5) -- [majorana, thick, insertion={[style=black, size=2pt]0}, black, edge label'=$\c$] (v7),
(v7) -- [majorana, thick, insertion={[style=black, size=2pt]0}, black, edge label'=$\rule{0pt}{10pt}$] (v2),
};
\end{feynman}
\end{tikzpicture}
\raisebox{1.2cm}{+} 
\begin{tikzpicture}
\begin{feynman}
\vertex (v1) {$\wt\cq^\m$};
\vertex [right=.5cm of v1] (v2);
\vertex [right=1.5cm of v2] (v3);
\vertex [right=.1cm of v3] (v4) {$\wt{\bar\cq}{}^\n$};
\vertex [above=1.5cm of v3] (v5);
\vertex [right=.1cm of v5] (v6) {$\wt\cj^\r$};
\vertex [above=1.5cm of v2] (v7);
\vertex [left=.1cm of v7] (v8) {$\wt\cj^\s$};
\diagram* {
(v1) -- [white]
(v2) -- [charged scalar, thick, insertion={[style=black, size=2pt]0}, red, edge label'=$\vf_2\rule{0pt}{8pt}$] (v3) -- [white] (v4),
(v3) -- [majorana, thick, insertion={[style=black, size=2pt]0}, purple, edge label=$\rule{0pt}{10pt}$] (v5),
(v5) -- [majorana, thick, insertion={[style=black, size=2pt]0}, purple, edge label'=$\l_2$] (v7),
(v7) -- [majorana, thick, insertion={[style=black, size=2pt]0}, purple, edge label'=$\rule{0pt}{10pt}$] (v2),
};
\end{feynman}
\end{tikzpicture}
\raisebox{1.2cm}{+}
\begin{tikzpicture}
\begin{feynman}
\vertex (v1) {$\wt\cq^\m$};
\vertex [right=.5cm of v1] (v2);
\vertex [right=1.5cm of v2] (v3);
\vertex [right=.1cm of v3] (v4) {$\wt{\bar\cq}{}^\n$};
\vertex [above=1.5cm of v3] (v5);
\vertex [right=.1cm of v5] (v6) {$\wt\cj^\r$};
\vertex [above=1.5cm of v2] (v7);
\vertex [left=.1cm of v7] (v8) {$\wt\cj^\s$};
\diagram* {
(v1) -- [white]
(v2) -- [charged scalar, thick, insertion={[style=black, size=2pt]0}, orange, edge label'=$\vf_1\rule{0pt}{8pt}$] (v3) -- [white] (v4),
(v3) -- [fermion, thick, insertion={[style=black, size=2pt]0}, olive, edge label=$\rule{0pt}{10pt}$] (v5),
(v5) -- [fermion, thick, insertion={[style=black, size=2pt]0}, olive, edge label'=$\l_1$] (v7),
(v7) -- [fermion, thick, insertion={[style=black, size=2pt]0}, olive, edge label'=$\rule{0pt}{10pt}$] (v2),
};
\end{feynman}
\end{tikzpicture}
\raisebox{1.2cm}{+}
\begin{tikzpicture}
\begin{feynman}
\vertex (v1) {$\wt\cq^\m$};
\vertex [right=.5cm of v1] (v2);
\vertex [right=1.5cm of v2] (v3);
\vertex [right=.1cm of v3] (v4) {$\wt{\bar\cq}{}^\n$};
\vertex [above=1.5cm of v3] (v5);
\vertex [right=.1cm of v5] (v6) {$\wt\cj^\r$};
\vertex [above=1.5cm of v2] (v7);
\vertex [left=.1cm of v7] (v8) {$\wt\cj^\s$};
\diagram* {
(v1) -- [white]
(v2) -- [charged scalar, thick, insertion={[style=black, size=2pt]0}, blue, edge label'=$\vth_1\rule{0pt}{8pt}$] (v3) -- [white] (v4),
(v3) -- [fermion, thick, insertion={[style=black, size=2pt]0}, olive, edge label=$\rule{0pt}{10pt}$] (v5),
(v5) -- [fermion, thick, insertion={[style=black, size=2pt]0}, olive, edge label'=$\l_1$] (v7),
(v7) -- [fermion, thick, insertion={[style=black, size=2pt]0}, olive, edge label'=$\rule{0pt}{10pt}$] (v2),
};
\end{feynman}
\end{tikzpicture}\\
\raisebox{1.2cm}{+}
&\begin{tikzpicture}
\begin{feynman}
\vertex (v1) {$\wt\cq^\m$};
\vertex [right=.5cm of v1] (v2);
\vertex [right=1.5cm of v2] (v3);
\vertex [right=.1cm of v3] (v4) {$\wt\cj^\r$};
\vertex [above=1.5cm of v3] (v5);
\vertex [right=.1cm of v5] (v6) {$\wt{\bar\cq}{}^\n$};
\vertex [above=1.5cm of v2] (v7);
\vertex [left=.1cm of v7] (v8) {$\wt\cj^\s$};
\diagram* {
(v1) -- [white]
(v2) -- [charged scalar, thick, insertion={[style=black, size=2pt]0}, black, edge label'=$\f$] (v3) -- [white] (v4),
(v3) -- [charged scalar, thick, insertion={[style=black, size=2pt]0}, black, edge label'=$\rule{0pt}{10pt}$] (v5),
(v5) -- [majorana, thick, insertion={[style=black, size=2pt]0}, black, edge label'=$\c$] (v7),
(v7) -- [majorana, thick, insertion={[style=black, size=2pt]0}, black, edge label'=$\rule{0pt}{10pt}$] (v2),
};
\end{feynman}
\end{tikzpicture}
\raisebox{1.2cm}{+} 
\begin{tikzpicture}
\begin{feynman}
\vertex (v1) {$\wt\cq^\m$};
\vertex [right=.5cm of v1] (v2);
\vertex [right=1.5cm of v2] (v3);
\vertex [right=.1cm of v3] (v4) {$\wt\cj^\r$};
\vertex [above=1.5cm of v3] (v5);
\vertex [right=.1cm of v5] (v6) {$\wt{\bar\cq}{}^\n$};
\vertex [above=1.5cm of v2] (v7);
\vertex [left=.1cm of v7] (v8) {$\wt\cj^\s$};
\diagram* {
(v1) -- [white]
(v2) -- [charged scalar, thick, insertion={[style=black, size=2pt]0}, red, edge label'=$\vf_2\rule{0pt}{8pt}$] (v3) -- [white] (v4),
(v3) -- [charged scalar, thick, insertion={[style=black, size=2pt]0}, red, edge label=$\rule{0pt}{10pt}$] (v5),
(v5) -- [majorana, thick, insertion={[style=black, size=2pt]0}, purple, edge label'=$\l_2$] (v7),
(v7) -- [majorana, thick, insertion={[style=black, size=2pt]0}, purple, edge label'=$\rule{0pt}{10pt}$] (v2),
};
\end{feynman}
\end{tikzpicture}
\raisebox{1.2cm}{+}
\begin{tikzpicture}
\begin{feynman}
\vertex (v1) {$\wt\cq^\m$};
\vertex [right=.5cm of v1] (v2);
\vertex [right=1.5cm of v2] (v3);
\vertex [right=.1cm of v3] (v4) {$\wt\cj^\r$};
\vertex [above=1.5cm of v3] (v5);
\vertex [right=.1cm of v5] (v6) {$\wt{\bar\cq}{}^\n$};
\vertex [above=1.5cm of v2] (v7);
\vertex [left=.1cm of v7] (v8) {$\wt\cj^\s$};
\diagram* {
(v1) -- [white]
(v2) -- [charged scalar, thick, insertion={[style=black, size=2pt]0}, orange, edge label'=$\vf_1\rule{0pt}{8pt}$] (v3) -- [white] (v4),
(v3) -- [charged scalar, thick, insertion={[style=black, size=2pt]0}, orange, edge label=$\rule{0pt}{10pt}$] (v5),
(v5) -- [fermion, thick, insertion={[style=black, size=2pt]0}, olive, edge label'=$\l_1$] (v7),
(v7) -- [fermion, thick, insertion={[style=black, size=2pt]0}, olive, edge label'=$\rule{0pt}{10pt}$] (v2),
};
\end{feynman}
\end{tikzpicture}
\raisebox{1.2cm}{+}
\begin{tikzpicture}
\begin{feynman}
\vertex (v1) {$\wt\cq^\m$};
\vertex [right=.5cm of v1] (v2);
\vertex [right=1.5cm of v2] (v3);
\vertex [right=.1cm of v3] (v4) {$\wt\cj^\r$};
\vertex [above=1.5cm of v3] (v5);
\vertex [right=.1cm of v5] (v6) {$\wt{\bar\cq}{}^\n$};
\vertex [above=1.5cm of v2] (v7);
\vertex [left=.1cm of v7] (v8) {$\wt\cj^\s$};
\diagram* {
(v1) -- [white]
(v2) -- [charged scalar, thick, insertion={[style=black, size=2pt]0}, blue, edge label'=$\vth_1\rule{0pt}{8pt}$] (v3) -- [white] (v4),
(v3) -- [charged scalar, thick, insertion={[style=black, size=2pt]0}, blue, edge label=$\rule{0pt}{10pt}$] (v5),
(v5) -- [fermion, thick, insertion={[style=black, size=2pt]0}, olive, edge label'=$\l_1$] (v7),
(v7) -- [fermion, thick, insertion={[style=black, size=2pt]0}, olive, edge label'=$\rule{0pt}{10pt}$] (v2),
};
\end{feynman}
\end{tikzpicture}\\
\raisebox{1.2cm}{+}
&\begin{tikzpicture}
\begin{feynman}
\vertex (v1) {$\wt\cq^\m$};
\vertex [right=.5cm of v1] (v2);
\vertex [right=1.5cm of v2] (v3);
\vertex [right=.1cm of v3] (v4) {$\wt\cj^\r$};
\vertex [above=1.5cm of v3] (v5);
\vertex [right=.1cm of v5] (v6) {$\wt{\bar\cq}{}^\n$};
\vertex [above=1.5cm of v2] (v7);
\vertex [left=.1cm of v7] (v8) {$\wt\cj^\s$};
\diagram* {
(v1) -- [white]
(v5) -- [charged scalar, thick, insertion={[style=black, size=2pt]0}, black, edge label'=$\f$] (v7) -- [white] (v8),
(v7) -- [charged scalar, thick, insertion={[style=black, size=2pt]0}, black, edge label'=$\rule{0pt}{10pt}$] (v2),
(v2) -- [majorana, thick, insertion={[style=black, size=2pt]0}, black, edge label'=$\c$] (v3),
(v3) -- [majorana, thick, insertion={[style=black, size=2pt]0}, black, edge label'=$\rule{0pt}{10pt}$] (v5),
};
\end{feynman}
\end{tikzpicture}
\raisebox{1.2cm}{+} 
\begin{tikzpicture}
\begin{feynman}
\vertex (v1) {$\wt\cq^\m$};
\vertex [right=.5cm of v1] (v2);
\vertex [right=1.5cm of v2] (v3);
\vertex [right=.1cm of v3] (v4) {$\wt\cj^\r$};
\vertex [above=1.5cm of v3] (v5);
\vertex [right=.1cm of v5] (v6) {$\wt{\bar\cq}{}^\n$};
\vertex [above=1.5cm of v2] (v7);
\vertex [left=.1cm of v7] (v8) {$\wt\cj^\s$};
\diagram* {
(v1) -- [white]
(v5) -- [charged scalar, thick, insertion={[style=black, size=2pt]0}, red, edge label'=$\vf_2\rule{0pt}{8pt}$] (v7) -- [white] (v8),
(v7) -- [charged scalar, thick, insertion={[style=black, size=2pt]0}, red, edge label=$\rule{0pt}{10pt}$] (v2),
(v2) -- [majorana, thick, insertion={[style=black, size=2pt]0}, purple, edge label'=$\l_2$] (v3),
(v3) -- [majorana, thick, insertion={[style=black, size=2pt]0}, purple, edge label'=$\rule{0pt}{10pt}$] (v5),
};
\end{feynman}
\end{tikzpicture}
\raisebox{1.2cm}{+}
\begin{tikzpicture}
\begin{feynman}
\vertex (v1) {$\wt\cq^\m$};
\vertex [right=.5cm of v1] (v2);
\vertex [right=1.5cm of v2] (v3);
\vertex [right=.1cm of v3] (v4) {$\wt\cj^\r$};
\vertex [above=1.5cm of v3] (v5);
\vertex [right=.1cm of v5] (v6) {$\wt{\bar\cq}{}^\n$};
\vertex [above=1.5cm of v2] (v7);
\vertex [left=.1cm of v7] (v8) {$\wt\cj^\s$};
\diagram* {
(v1) -- [white]
(v5) -- [charged scalar, thick, insertion={[style=black, size=2pt]0}, orange, edge label'=$\vf_1\rule{0pt}{8pt}$] (v7) -- [white] (v8),
(v7) -- [charged scalar, thick, insertion={[style=black, size=2pt]0}, orange, edge label=$\rule{0pt}{10pt}$] (v2),
(v2) -- [fermion, thick, insertion={[style=black, size=2pt]0}, olive, edge label'=$\l_1$] (v3),
(v3) -- [fermion, thick, insertion={[style=black, size=2pt]0}, olive, edge label'=$\rule{0pt}{10pt}$] (v5),
};
\end{feynman}
\end{tikzpicture}
\raisebox{1.2cm}{+}
\begin{tikzpicture}
\begin{feynman}
\vertex (v1) {$\wt\cq^\m$};
\vertex [right=.5cm of v1] (v2);
\vertex [right=1.5cm of v2] (v3);
\vertex [right=.1cm of v3] (v4) {$\wt\cj^\r$};
\vertex [above=1.5cm of v3] (v5);
\vertex [right=.1cm of v5] (v6) {$\wt{\bar\cq}{}^\n$};
\vertex [above=1.5cm of v2] (v7);
\vertex [left=.1cm of v7] (v8) {$\wt\cj^\s$};
\diagram* {
(v1) -- [white]
(v5) -- [charged scalar, thick, insertion={[style=black, size=2pt]0}, blue, edge label'=$\vth_1\rule{0pt}{8pt}$] (v7) -- [white] (v8),
(v7) -- [charged scalar, thick, insertion={[style=black, size=2pt]0}, blue, edge label=$\rule{0pt}{10pt}$] (v2),
(v2) -- [fermion, thick, insertion={[style=black, size=2pt]0}, olive, edge label'=$\l_1$] (v3),
(v3) -- [fermion, thick, insertion={[style=black, size=2pt]0}, olive, edge label'=$\rule{0pt}{10pt}$] (v5),
};
\end{feynman}
\end{tikzpicture}
\end{eqnarray*}
\vskip-.5cm
\caption{One-loop diagrams that contribute to the 4-point function of two supercurrents and two R-currents.}
\label{QQJJ-4pt-diags}
\end{figure}

The Feynman diagrams that contribute to the 4-point function are shown in fig.~\ref{QQJJ-4pt-diags}. They give
\bal
&\lb\wt\cq^\m(p_1)\wt{\bar\cq}{}^\n(p_2)\wt\cj^\r(p_3)\wt\cj^\s(p_4)\rb=\frac{1}{72\a^8} \int \frac{d^{4}q}{(2\p)^4}G_1^{(4)}(q,q',q'',q''')(q''+q''')^\r(q+q'')^\s\O_1^\m(-q)\slashed q'\O_2^\n(q'')\NO\\
&+\frac{m_1^2}{36\a^8} \int \frac{d^{4}q}{(2\p)^4}G_2^{(4)}(q,q',q'',q''')(q''+q''')^\r(q+q'')^\s\big(\g^\m\O_2^\n(q'')-\O_1^\m(-q)\g^\n+\g^\m\slashed q'\g^\n\big)\NO\\
&-\frac{1}{144\a^8} \int \frac{d^{4}q}{(2\p)^4}G_1^{(4)}(q,q',q'',q''')\O_1^\m(q')\slashed q\g^\s\slashed q'''\g^\r\slashed q''\O_2^\n(-q')\NO\\
&-\frac{m_1^2}{72\a^8} \int \frac{d^{4}q}{(2\p)^4}G_2^{(4)}(q,q',q'',q''')\Big(\g^\m\slashed q\g^\s\slashed q'''\g^\r\slashed q''\g^\n+\O_1^\m(q')(\slashed q\g^\s\g^\r+\g^\s\slashed q'''\g^\r+\g^\s\g^\r\slashed q'')\O_2^\n(-q')\NO\\
&\hspace{.7cm}+\O_1^\m(q')(\slashed q\g^\s\slashed q'''\g^\r+\slashed q\g^\s\g^\r\slashed q''+\g^\s\slashed q'''\g^\r\slashed q'')\g^\n-\g^\m(\slashed q\g^\s\slashed q'''\g^\r+\slashed q\g^\s\g^\r\slashed q''+\g^\s\slashed q'''\g^\r\slashed q'')\O_2^\n(-q')\Big)\NO\\
&-\frac{m_1^4}{36\a^8} \int \frac{d^{4}q}{(2\p)^4}\big(P_{\vf_2}(q)P_{\vf_2}(q')P_{\vf_2}(q'')P_{\vf_2}(q''')-P_{\vf_1}(q)P_{\vf_1}(q')P_{\vf_1}(q'')P_{\vf_1}(q''')\big)\times\NO\\
&\hspace{.7cm}\times\Big(\Om_1^\m(q')\g^\s\g^\r\g^\n-\g^\m\g^\s\g^\r\O_2^\n(-q')+\g^\m(\slashed q\g^\s\g^\r+\g^\s\slashed q'''\g^\r+\g^\s\g^\r\slashed q'')\g^\n\Big)\NO\\
&-\frac{1}{144\a^8} \int \frac{d^{4}k}{(2\p)^4}G_1^{(4)}(k,k',k'',k''')\Big((k+k''')^\s\O_1^\m(-k)\slashed k'\g^\r\slashed k''\O_2^\n(k''')-(k'+k'')^\r\O_1^\m(k')\slashed k\g^\s\slashed k'''\O_2^\n(-k'')\Big)\NO\\
&-\frac{m_1^2}{72\a^8} \int \frac{d^{4}k}{(2\p)^4}G_2^{(4)}(k,k',k'',k''')\times\NO\\
&\Big[(k+k''')^\s\big(\O_1^\m(-k)\g^\r\O_2^\n(k''')+\g^\m\slashed k'\g^\r\slashed k''\g^\n+\O_1^\m(-k)(\g^\r\slashed k''-\slashed k'\g^\r)\g^\n+\g^\m(\g^\r\slashed k''-\slashed k'\g^\r)\O_2^\n(k''')\big)\NO\\
&-(k'+k'')^\r\big(\O_1^\m(k')\g^\s\O_2^\n(-k'')+\g^\m\slashed k\g^\s\slashed k'''\g^\n-\O_1^\m(k')(\g^\s\slashed k'''-\slashed k\g^\s)\g^\n-\g^\m(\g^\s\slashed k'''-\slashed k\g^\s)\O_2^\n(-k'')\big)\Big]\NO\\
&+\frac{m_1^4}{36\a^8} \int \frac{d^{4}k}{(2\p)^4}\Big(P_{\vf_2}(k)P_{\vf_2}(k')P_{\vf_2}(k'')P_{\vf_2}(k''')-\frac12P_{\vf_1}(k)P_{\vf_1}(k')P_{\vf_1}(k'')P_{\vf_1}(k''')\Big)\times\NO\\
&\hspace{.7cm}\times\big((k+k''')^\s\g^\m\g^\r\g^\n-(k'+k'')^\r\g^\m\g^\s\g^\n\big)+\r\leftrightarrow \s,\; p_2\leftrightarrow p_3,
\eal
where $q'=q+p_1$, $q''=q+p_1+p_2$, $q'''=q+p_1+p_2+p_3$, $k'=k+p_1$, $k''=k+p_1+p_3$, and $k'''=k+p_1+p_2+p_3$. Simple power counting shows that all UV divergences in this expression as properly regulated. This completes the proof that the PV fields regulate all correlation functions relevant for the analysis of the 4-point function supersymmetry Ward identity.  

\section{Contact terms from correlators with insertions of $\cb_W$, $\cb_R$ and $\cb_S$}
\label{sec:local-correlators}

In this appendix we evaluate all 1-loop correlators that involve the symmetry breaking operators $\cb_W$, $\cb_R$ and $\cb_S$ in the FZ multiplet Ward identities \eqref{S-susy-WIDs-2pt-FZ}-\eqref{S-susy-WIDs-4pt-FZ}, \eqref{R-symmetry-WIDs-JJ-FZ}-\eqref{R-symmetry-WIDs-QQJJ-FZ} and \eqref{OM-susy-WID-3pt-simple}-\eqref{OM-susy-WID-4pt-simple}. Since these operators depend on the PV fields only, all such correlators are pure contact terms once the PV masses are sent to infinity. It this appendix we set $\a=1/\sqrt{2}$ in the propagators \eqref{WZ-propagators}, \eqref{PV-propagators} and, as in appendix \ref{sec:divergences}, we drop the hat $\hskip2pt\hat\cdot\hskip2pt$ to simplify the notation. In particular, $\cb_W$, $\cb_R$ and $\cb_S$ here refer to the flat space operators \eqref{PV-non-conservation-laws}, but we will utilize the more general form in the presence of supergravity fields given in \eqref{Bs} to obtain the relevant seagull terms.

\begin{flushleft}
\noindent\rule{\textwidth}{0.8pt}
\raisebox{-0.1cm}{$\<\cb_R\wt\cj\>$}
\noindent\rule{\textwidth}{0.8pt}
\end{flushleft}

This 2-point function enters in the Ward identity \eqref{R-symmetry-WIDs-JJ-FZ} encoding the breaking of R-symmetry at the level of the R-current 2-point function. It can be easily computed directly, but this is not necessary since we have already evaluated both the regulated and renormalized R-current 2-point function, respectively in \eqref{JJ} and \eqref{JJren}. Using the Ward identity \eqref{R-symmetry-WIDs-JJ-FZ}, we read off
\be\label{BR-J-reg}
\<\cb_R(p)\wt\cj^\n(-p)\>=ip_\m\<\wt\cj^\m(p)\wt\cj^\n(-p)\>=\frac{2i\p^2}{9(2\p)^4}\Big(4m_1^2\log2-\frac{1}{3}p^2\Big)p^\n,
\ee
and
\bbxd
\vskip.35cm
\be\label{BR-J-ren}
\<\cb_R(p)\wt\cj^\n(-p)\>_{\rm ren}=ip_\m\<\wt\cj^\m(p)\wt\cj^\n(-p)\>_{\rm ren}=-\frac{2i\p^2}{27(2\p)^4}p^2 p^\n.
\ee
\ebxd
This 2-point function reflects the explicit breaking of R-symmetry in the FZ multiplet.

\begin{flushleft}
\noindent\rule{\textwidth}{0.8pt}
\raisebox{-0.1cm}{$\<\cb_S\wt{\bar\cq}{}\,\>$}
\noindent\rule{\textwidth}{0.8pt}
\end{flushleft}

This 2-point function appears in the Ward identity \eqref{S-susy-WIDs-2pt-FZ}, which reflects the S-supersymmetry breaking in the 2-point function of the supercurrent. Since we have already determined both the  regulated and renormalized form of the supercurrent 2-point function respectively in \eqref{QQ-seagull}-\eqref{QQ} and \eqref{QQren}, we immediately deduce that
\be\label{BS-Q-reg}
\<\cb_S(p)\wt{\bar\cq}{}^\n(-p)\>=\g_\m\<\wt\cq^\m(p)\wt{\bar\cq}{}^\n(-p)\>=\frac{i\p^2}{3(2\p)^4}\Big(m_1^2\log2-\frac{1}{12}p^2\Big)i\e^{\m\n\r\s}\g_{\m\s}\g^5p_\r,
\ee
and
\bbxd
\vskip.35cm
\be\label{BS-Q-ren}
\<\cb_S(p)\wt{\bar\cq}{}^\n(-p)\>_{\rm ren}=\g_\m\<\wt\cq^\m(p)\wt{\bar\cq}{}^\n(-p)\>_{\rm ren}=-\frac{i\p^2}{36(2\p)^4}i\e^{\m\n\r\s}\g_{\m\s}\g^5p_\r p^2.
\ee
\ebxd
This 2-point function reflects the explicit breaking of S-supersymmetry in the FZ multiplet.

\begin{flushleft}
\noindent\rule{\textwidth}{0.8pt}
\raisebox{-0.1cm}{$\<\cb_R\wt\cj\wt\cj\>$}
\noindent\rule{\textwidth}{0.8pt}
\end{flushleft}

It is straightforward to see from \eqref{Bs} and \eqref{operator-derivatives} that the functional derivative and operator insertion definitions of this 3-point function are related as
\be
\<\cb_R(p_1)\wt\cj^\m(p_2)\wt\cj^\n(p_3)\>=\lb\cb_R(p_1)\wt\cj^\m(p_2)\wt\cj^\n(p_3)\rb-\frac{8}{9}\h^{\m\n}\lb\cb_R(p_1){\wt s}_{(1|0)}(-p_1)\rb,
\ee
where the seagull operator ${\wt s}_{(1|0)}$ is defined in \eqref{seagull-operators}. Since $\cb_R$ contains spinor fields only and ${\wt s}_{(1|0)}$ only scalars, the 2-point function $\lb\cb_R{\wt s}_{(1|0)}\rb$ vanishes identically.   

The Feynman diagram contribution to this 3-point function takes the form
\bal
\lb \cb_R(p_3) \wt{\cj}^\l(p_4)\wt{\cj}^\s(p_1)\rb=&\; {i^2\times}\frac{i}{27}\int \frac{d^4q}{(2\p)^4}\Big(m_2\tr\big(\g^\s\g^5 P_{\l_2}(q)\g^\l\g^5P_{\l_2}(q+p_4)\g^5P_{\l_2}(q-p_1)\big)\NO\\
&+m_2\tr\big(\g^\s\g^5 P_{\l_2}(q)\g^5P_{\l_2}(q+p_3)\g^\l\g^5P_{\l_2}(q-p_1)\big)\NO\\
&-2m_1\tr\big(\g^\s\g^5 P_{\l_1}(q)\g^\l\g^5P_{\l_1}(q+p_4)\g^5P_{\l_1}(q-p_1)\big)\NO\\
&-2m_1\tr\big(\g^\s\g^5P_{\l_1}(q)\g^5P_{\l_1}(q+p_3)\g^\l\g^5P_{\l_1}(q-p_1)\big)\Big).
\eal
Using the trace identities \eqref{d=4-gamma-traces} and introducing suitable Feynman parameters, this gives
\bal
&\lb \cb_R(p_3) \wt{\cj}^\l(p_4)\wt{\cj}^\s(p_1)\rb={i^2\times}\frac{4m_1^2}{27}\frac{2i\p^2}{(2\p)^4}\int d\ell_E\ell_E^3\int_{0}^{1} dx\,dy\, dz\,\d(x+y+z-1)\times\NO\\
&\bigg[\bigg(\frac{1}{(\ell_E^2+\D_2(p_4,p_1))^3}-\frac{1}{(\ell_E^2+\D_1(p_4,p_1))^3}\bigg)
\tr(\g^\s\g^\a\g^\l\g^\b\g^5)(2p_4\,y-2p_1\,z+p_4)_\a p_{3\b}\NO\\&
+\bigg(\frac{1}{(\ell_E^2+\D_2(p_3,p_1))^3}-\frac{1}{(\ell_E^2+\D_1(p_3,p_1))^3}\bigg)\tr(\g^\s\g^\a\g^\b\g^\l\g^5)(2p_3\,y-2p_1\,z-p_1)_\a p_{3\b}\bigg],
\eal
where $\D_i(p,q)\equiv m_i^2+p^2\,y+q^2\,z-(p\, y-q\, z)^2$ with $i=1,2$. The integration over $\ell_E$ gives
\bal
& \lb \cb_R(p_3)\wt{\cj}^\l(p_4)\wt{\cj}^\s(p_1)\rb={i^2\times}\frac{m_1^2}{27}\frac{2i\p^2}{(2\p)^4}\int_{0}^{1} dx\,dy\, dz\,\d(x+y+z-1)\times\NO\\
&\Big[\Big(\frac{1}{\D_2(p_4,p_1)}-\frac{1}{\D_1(p_4,p_1)}\Big)
\tr(\g^\s\g^\a\g^\l\g^\b\g^5)(2p_4\,y-2p_1\,z+p_4)_\a p_{3\b}\NO\\&
+\Big(\frac{1}{\D_2(p_3,p_1)}-\frac{1}{\D_1(p_3,p_1)}\Big)
\tr(\g^\s\g^\a\g^\b\g^\l\g^5)(2p_3\,y-2p_1\,z-p_1)_\a p_{3\b}\Big].
\eal

This expression has a finite limit as the PV mass $m_1$ is sent to infinity. Namely,  
\bal
&\lb \cb_R(p_3) \wt{\cj}^\l(p_4)\wt{\cj}^\s(p_1)\rb\NO\\
&=-i^2\times\frac{1}{54}\frac{2i\p^2}{(2\p)^4}\int dy dz\,
\tr(\g^\s\g^\a\g^\l\g^\b\g^5) p_{3\b}(2p_4\,y-2p_1\,z+p_4-2p_3\,y+2p_1\,z+p_1)_\a\NO\\&
=-{i^2\times}\frac{2}{54}\frac{2i\p^2}{(2\p)^4}\int dy dz\,
\tr(\g^\s\g^\a\g^\l\g^\b\g^5) p_{3\b}p_{4\a}\,y=-{i^2\times}\frac{1}{324\p^2}\e^{\s\l \b \a} p_{3\b}p_{4\a}.
\eal
Hence, this 3-point function is equal to the finite contact term 
\bbxd
\vskip.0cm
\be\label{BR-JJ}
\<\cb_R(p_1)\wt\cj^\m(p_2)\wt\cj^\n(p_3)\>=\<\cb_R(p_1)\wt\cj^\m(p_2)\wt\cj^\n(p_3)\>_{\rm ren}=\frac{1}{648\p^2}\e^{\m\n\r\s}p_{1\r}(p_{3\s}-p_{2\s}).
\ee
\ebxd
Using the values of the anomaly coefficients for the WZ model \eqref{WZ-anomaly-coefficients},
this 3-point function matches the R-symmetry anomaly in the conformal multiplet 3-point function of R-currents given in \eqref{R-symmetry-WID-axial}.

\begin{flushleft}
\noindent\rule{\textwidth}{0.8pt}
\raisebox{-0.1cm}{$\<\cb_S\wt{\bar\cq}\wt\cj\>$}
\noindent\rule{\textwidth}{0.8pt}
\end{flushleft}

From \eqref{Bs} and \eqref{operator-derivatives} again follows that
\be
\<\cb_S(p_1)\wt{\bar\cq}{}^\m(p_2)\wt{\cj}^\n(p_3)\>=\lb\cb_S(p_1)\wt{\bar\cq}{}^\m(p_2)\wt{\cj}^\n(p_3)\rb-\frac{\sqrt{2}}{3}\h^{\m\n}\lb\cb_S(p_1)\wt{\bar s}_{\(3|\frac12\)}(-p_1)\rb,
\ee
where the seagull operator ${\wt s}_{\(3|\frac12\)}$ is given in \eqref{seagull-operators}. The 2-point function involving the seagull term can be shown to be proportional to the 2-point function $\<\wt\co_M\wt\co_{M^*}\>$. Using \eqref{OOreg}, we determine 
\bbxd
\be\label{BS-S3-1/2}
\hskip-.45cm\lb \cb_S(p_1)\wt{\bar{s}}_{(3|\frac{1}{2})}(-p_1)\rb=2\sqrt{2}\,i\g^5\<\wt\co_M(p)\wt\co_{M^*}(-p)\>=-\frac{\sqrt{2}}{8\p^2}i\g^5\Big(m_1^2\log 2-\frac{p_1^2}{12}+\co(m_1^{-2})\Big).
\ee
\ebxd

The Feynman diagram contribution to the 3-point function is  
\bal
&\lb \cb_S(p_1)\wt{\bar\cq}{}^\n(p_2)\wt{\cj}^\k(p_3)\rb=\NO\\
&{i^2\times}\frac{i}{3}\int \frac{d^4q}{(2\p)^4}\Big(m_2P_{\vf_2}(q)P_LP_{\l_2}(p_1-q)\g^\k\g^5P_{\l_2}(-q-p_2)\big(i\O_2^\n(-q)-m_2\g^{\n}\big)P_L\NO\\
&+2im_2(2q+p_3)^\k P_{\vf_2}(q)P_{\vf_2}(q+p_3)P_L\g^5P_{\l_2}(p_1-q)\big(i\O_2^\n(-q-p_3)-m_2\g^{\n}\big)P_L\NO\\&
-2m_1P_{\vf_1}(q)P_LP_{\l_1}(p_1-q)\g^\k\g^5P_{\l_1}(-q-p_2)\big(i\O_2^\n(-q)-m_1\g^{\n}\big)P_L\\
&-4im_1(2q+p_3)^\k P_{\vf_1}(q)P_{\vf_1}(q+p_3)P_L\g^5P_{\l_1}(p_1-q)\big(i\O_2^\n(-q-p_3)-m_1\g^{\n}\big)P_L+P_L\leftrightarrow P_R\Big),\NO
\eal
where $\O_{1,2}^\m(q)$ were defined in \eqref{Omega}. Introducing Feynman parameters this gives
\bal
&\lb\cb_S(p_1)\wt{\bar\cq}{}^\n(p_2)\wt{\cj}^\k(p_3)\rb=\frac{2i\p^2}{(2\p)^4}\frac{4m_1^2}{3}\int d\ell_E \int_{0}^{1}dx\, dy\, dz\,\d(x+y+z-1)\times\NO\\
&\bigg[\bigg(\frac{\ell_E^3}{(\ell_E^2+\D_2(p_2,p_1))^3}-\frac{\ell_E^3}{(\ell_E^2+\D_1(p_2,p_1))^3}\bigg)\bigg(\frac14\ell_E^2\big(\g^\a\g^\k\g^5\g_\a\g^\n+\g^\a\g^\k\g^5\g^\n\g_\a+\g^\k\g^5\g^\a\g^\n\g_\a\big)\NO\\
&+(\slashed a+\slashed p_1)\g^\k\g^5\big((\slashed a-\slashed p_2)\g^\n+\g^\n\slashed a\big)-\frac23\big((\slashed a+\slashed p_1)\g^\k\g^5+\g^\k\g^5(\slashed a-\slashed p_2)\big)\g^{\n\r} p_{2\r}+\g^\k\g^5(\slashed a-\slashed p_2)\g^\n\slashed a \bigg)\NO\\
&+m_1^2\bigg(\frac{2\ell_E^3}{(\ell_E^2+\D_2(p_2,p_1))^3}-\frac{\ell_E^3}{(\ell_E^2+\D_1(p_2,p_1))^3}\bigg)\g^5\g^\k\g^\n+2\bigg(\frac{\ell_E^3}{(\ell_E^2+\D_2(p_3,p_1))^3}-\frac{\ell_E^3}{(\ell_E^2+\D_1(p_3,p_1))^3}\bigg)\times\NO\\
&\bigg(\ell_E^2\g^5\h^{\n\k}+(2b-p_3)^\k\Big(\g^5(\slashed p_1+\slashed b)\g^\n+\g^5\g^\n(\slashed b-\slashed p_3)-\frac{2}{3} p_{2\r}\g^{\n\r}\g^5\Big)\bigg)\bigg],
\eal
where again $\D_i(p,q)\equiv m_i^2+p^2\,y+q^2\,z-(p\, y-q\, z)^2$ with $i=1,2$, $a\equiv p_2\,y -p_1\,z$ and $b\equiv p_3\,y -p_1\,z$. Dropping terms that diverge as the PV mass is sent to infinity, this gives
\bbxd
\vskip.0cm
\begin{align}
&\lb {\cal B}_S(p_1)\wt{\bar{{\cal Q }}}{}^\m(p_2)\wt{{\cal J }}^\n(p_3)\rb_{\text{ren}}=\frac{2i\pi^2}{432(2\pi)^4}\Big(i\g_{\k\l}\big(2\h^{\m\n}\e^{\k\l\r\s}p_{1\r}p_{2\s}-\e^{\k\l\m\r}p_{1\r}p_{2}^\n+\e^{\k\l\m\r}p_{2\r}p_{1}^\n\NO\\&
+3\e^{\k\l\n\r}p_{1\r}p_{2}^\m+\e^{\k\l\n\r}p_{2\r}p_{1}^\m+4\e^{\k\l\m\n}p_{1}\cdot p_{2}+4\e^{\k\l\m\n}p_{2}^2-8\e^{\k\l\m\r}p_{2\r}p_{2}^\n+4\e^{\k\l\n\r}p_{2\r}p_{2}^\m\big)\NO\\&
-8i\g^\m{}_{\k}\e^{\n\k\r\s}p_{1\r}p_{2\s}+2i{\g_\k{}^{\s}}\e^{\m\n\k\r}p_{1\r}p_{2\s}+6i{\g_\k{}^{\s}}\e^{\m\n\k\r}p_{2\r}p_{1\s}-4i\e^{\m\n\r\s}p_{1\r}p_{2\s}\NO\\&
+\g^5\big(20\h^{\m\n}p_1\cdot p_2+36\h^{\m\n}p_1^2+28\h^{\m\n}p_2^2+12p_2^\m p_1^\n+4p_1^\m p_2^\n+24p_1^\m p_1^\n-16p_2^\m p_2^\n\big)\Big).
\end{align}
\ebxd

\begin{flushleft}
\noindent\rule{\textwidth}{0.8pt}
\raisebox{-0.1cm}{$\<\cb_R\wt\cq\wt{\bar\cq}{}\,\>$}
\noindent\rule{\textwidth}{0.8pt}
\end{flushleft}

Using \eqref{Bs} and \eqref{operator-derivatives} we determine that
\bal
&\<\wt{{\cal Q}}^{\m}(p_1)\wt{{\cal\bar{Q}}}{}^{\n}(p_2)\cb_R(p_3)\>=\lb\wt{{\cal Q}}^{\m}(p_1)\wt{{\cal\bar{Q}}}{}^{\n}(p_2)\cb_R(p_3)\rb\NO\\
&-\frac{i}{3}\g^\m\g^5\lb\cb_S(-p_2)\wt{{\cal\bar{Q}}}{}^{\n}(p_2)\rb-\frac{i}{3}\lb\wt\cq^\m(p_1)\bar\cb_S(-p_1)\rb\g^\n\g^5\NO\\
&+\frac38\e^{\m\n\r\s}\lb\wt\cj_\r(-p_3)\cb_R(p_3)\rb\g_\s+\frac18\lb\wt s_{(2|1)}^{\r}(-p_3)\cb_R(p_3)\rb\g^5\g_\r\h^{\m\n},
\eal
where the seagull operator $s_{(2|1)}$ is given in \eqref{seagull-operators}. We have already computed the 2-point functions $\lb\cb_R\wt\cj\rb$ and $\lb\cb_S\wt{{\cal\bar{Q}}}\rb$ respectively in \eqref{BR-J-ren} and \eqref{BS-Q-ren}. Moreover, $s_{(2|1)}$ is proportional to the fermionic part of the R-current and $\cb_R$ contains only fermions so that
\be
\lb\wt s_{(2|1)}^{\r}(-p_3)\cb_R(p_3)\rb=-6i\lb\wt \cj^{\r}(-p_3)\cb_R(p_3)\rb.
\ee  

The Feynman diagram contribution to the 3-point function is 
\bal
&\lb \wt{{\cal Q}}^{\m}(p_1)\wt{{\cal\bar{Q}}}{}^{\n}(p_2)\cb_R(p_3)\rb\\
&=i^2\times\frac{i}{3}\int\frac{d^{4}q}{(2\p)^{4}}\Big(m_2P_R\big(i\O_1^\m(q)+m_2\g^{\m}\big)P_{\l_2}(p_1-q)\g^5P_{\l_2}(-p_{2}-q)\big(i\O_2^\n(-q)-m_2\g^{\n}\big)P_LP_{\vf_2}(q)\NO\\
&-2m_1P_R\big(i\O_1^\m(q)+m_1\g^{\m}\big)P_{\l_1}(p_1-q)\g^5P_{\l_1}(-p_{2}-q) \big(i\O_2^\n(-q)-m_1\g^{\n}\big)P_LP_{\vf_1}(q)
+P_L\leftrightarrow P_R\Big),\NO
\eal
where $\O_{1,2}^\m(q)$ were defined in \eqref{Omega}. Introducing Feynman parameters this becomes
\bal
&\lb \wt{{\cal Q}}^{\m}(p_1)\wt{{\cal\bar{Q}}}{}^{\n}(p_2)\cb_R(p_3)\rb={i^2\times}\frac{2i\p^2}{(2\p)^{4}}\frac{im_1^2}{3}\g^5\int d\ell_E\int_{0}^{1} dx\,dy\, dz\,\d(x+y+z-1)\times\NO\\
&\bigg[-2\bigg(\frac{\ell_E^3}{(\ell_E^2+\D_2(p_2))^2}-\frac{\ell_E^3}{(\ell_E^2+\D_1(p_2))^2}\bigg)\Big(\slashed{c}\g^\m\g^\n-\g^\m\g^\n\slashed{c} +\g^\m\g^{\n\r}\frac{2p_{2\r}}{3}-\g^{\m\s}\g^\n\frac{2p_{1\s}}{3}\Big)\NO\\
&+\bigg(\frac{\ell_E^5}{(\ell_E^2+\D_2(p_2,p_1))^3}-\frac{\ell_E^5}{(\ell_E^2+\D_1(p_2,p_1))^3}\bigg)\big(\g^\x\g^\m\g^\k\g^\n\g_\x+\g^\m\g^\x\g^\k\g^\n\g_\x-\g^\x\g^\m\g_\x\g^\k\g^\n\big)p_{3\k}\NO\\
&+4\bigg(\frac{\ell_E^3}{(\ell_E^2+\D_2(p_2,p_1))^3}-\frac{\ell_E^3}{(\ell_E^2+\D_1(p_2,p_1))^3}\bigg)\bigg(-\slashed{a}\g^\m(\slashed{p}_1+\slashed{a})\slashed{p}_3\g^\n+\slashed{a}\g^\m\slashed{p}_3\g^\n\slashed{a}-\slashed{a}\g^\m\slashed{p}_3\g^{\n\r}\frac{2p_{2\r}}{3}\NO\\
&+\g^\m(\slashed{p}_1+\slashed{a})\slashed{p}_3\Big(\g^\n\slashed{a}-\g^{\n\r}\frac{2p_{2\r}}{3}\Big)+\g^{\m\s}(\slashed{p}_1+\slashed{a})\slashed{p}_3\g^\n\frac{2p_{1\s}}{3}-\g^{\m\s}\slashed{p}_3\g^\n\slashed{a}\frac{2p_{1\s}}{3}+\g^{\m\s}\slashed{p}_3\g^{\n\r}\frac{4p_{1\s}\,p_{2\r}}{9}\bigg)\NO\\&
+4m_1^2\g^\m\slashed{p}_3\g^\n\left(\frac{2\ell_E^3}{(\ell_E^2+\D_2(p_2,p_1))^3}-\frac{\ell_E^3}{(\ell_E^2+\D_1(p_2,p_1))^3}\right)\bigg],
\eal
where for $i=1,2$, we define $\D_i(p)\equiv m_i^2+p^2(1-y)y$, $\D_i(p,q)\equiv m_i^2+p^2\,y+q^2\,z-(p\, y-q\, z)^2$, $a\equiv p_2\,y -p_1\,z$ and $c=p_2 \,y\,$. Evaluating these expressions we determine 
\bbxd
\vskip.0cm
\begin{align}\label{QQ-BR-ren}
&\lb \wt{{\cal Q}}^{\m}(p_1)\wt{{\cal\bar{Q}}}{}^{\n}(p_2){\cal B}_R(p_3)\rb_{\text{ren}}=\frac{2i\pi^2}{108(2\pi)^4}\Big(\g_\k\big(3\e^{\m\k\r\s}p_{1\r}p_{2\s}p_1^\n-\e^{\n\k\r\s}p_{1\r}p_{2\s}p_1^\m+2\e^{\n\k\r\s}p_{1\r}p_{2\s}p_2^\m\NO\\&
+\e^{\m\n\k\r}p_{1\r}p_{2}^2+2\e^{\m\n\k\r}p_{2\r}p_{1}^2-\e^{\m\n\k\r}p_{1\r}\,p_{2}\cdot p_1+\e^{\m\n\k\r}p_{1\r}p_{1}^2+\e^{\m\n\k\r}p_{2\r}p_{2}^2\big)\NO\\&
+\g^5\slashed{p}_1\big(3i\h^{\m\n}p_1\cdot p_2+2i\h^{\m\n}p_1^2+2i\h^{\m\n}p_2^2+ip_2^\m p_1^\n-ip_1^\m p_2^\n-2ip_1^\m p_1^\n\big)\NO\\&
+\g^5\slashed{p}_2\big(3i\h^{\m\n}p_1\cdot p_2+2i\h^{\m\n}p_1^2+2i\h^{\m\n}p_2^2+ip_2^\m p_1^\n-ip_1^\m p_2^\n-2ip_2^\m p_2^\n\big)-\slashed{p}_1\e^{\m\n\r\s}p_{1\r}p_{2\s}\\&
+\g^5\g^\m\big(ip_1^2 p_2^\n+ip_2^2 p_1^\n+2ip_1\cdot p_2 p_1^\n+3ip_1^2 p_1^\n\big)
+\g^5\g^\n\big(ip_2^2 p_1^\m+ip_1^2 p_2^\m+2ip_2\cdot p_1 p_2^\m+3ip_2^2 p_2^\m\big)\Big).\NO
\end{align}
\ebxd

\begin{flushleft}
\noindent\rule{\textwidth}{0.8pt}
\raisebox{-0.1cm}{$\<\cb_S\wt{\bar\cq}{}\wt\cj\wt\cj\>$}
\noindent\rule{\textwidth}{0.8pt}
\end{flushleft}

Equations \eqref{Bs} and \eqref{operator-derivatives} imply that
\bal
&\<\cb_S(p_1)\wt{\bar\cq}{}^\n(p_2)\wt{\cj}^\k(p_3)\wt{\cj}^\l(p_4)\>=\lb\cb_S(p_1)\wt{\bar\cq}{}^\n(p_2)\wt{\cj}^\k(p_3)\wt{\cj}^\l(p_4)\rb\NO\\
&-\frac{\sqrt{2}}{3}\h^{\n\k}\lb\cb_S(p_1){\wt{\bar s}}_{\(3|\frac12\)}(-p_1-p_4)\wt{\cj}^\l(p_4)\rb-\frac{\sqrt{2}}{3}\h^{\n\l}\lb\cb_S(p_1){\wt{\bar s}}_{\(3|\frac12\)}(-p_1-p_3)\wt{\cj}^\k(p_3)\rb\NO\\
&-\frac{8}{9}\h^{\k\l}\lb\cb_S(p_1)\wt{\bar\cq}{}^\n(p_2){\wt s}_{(1|0)}(-p_1-p_2)\rb,
\eal
where the seagull operators ${\wt{\bar s}}_{\(3|\frac12\)}$ and ${\wt s}_{(1|0)}$ are given in \eqref{seagull-operators}.
 
Let us first consider the two 3-point functions involving seagull operators. The first is given by
\bal
&\frac{\sqrt{2}}{3}\lb \cb_S(p_1)\wt{\bar{s }}_{(3|\frac{1}{2})}(-p_1-p_3)\wt{\cj}^\k(p_3)\rb=\NO\\
&-i^2\times\frac{2}{9}\int \frac{d^4q}{(2\p)^4}\bigg[m_2\Big(P_{\vf_2}(q)P_LP_{\l_2}(p_1-q)\g^\k\g^5P_{\l_2}(p_1+p_3-q)P_L\g^5\NO\\
&+2i(2q+p_3)^\k P_{\vf_2}(q)P_{\vf_2}(q+p_3)P_L\g^5P_{\l_2}(p_1-q)P_L\g^5\Big)\NO\\
&-2m_1\Big(P_{\vf_1}(q)P_LP_{\l_1}(p_1-q)\g^\k\g^5P_{\l_1}(p_1+p_3-q)P_L\g^5\NO\\
&+2i(2q+p_3)^\k P_{\vf_1}(q)P_{\vf_1}(q+p_3)P_L\g^5P_{\l_1}(p_1-q)P_L\g^5\Big)+P_L\leftrightarrow P_R\bigg],
\eal
or, introducing Feynman parameters, 
\bal
&\frac{\sqrt{2}}{3}\lb \cb_S(p_1)\wt{\bar{s }}_{(3|\frac{1}{2})}(-p_1-p_3)\wt{\cj}^\k(p_3)\rb=-i^2\times\frac{2i\p^2}{(2\p)^4}\frac{8m_1^2}{9}\int d\ell_E\int_{0}^{1} dx\,dy\, dz\,\d(x+y+z-1)\times \NO\\
&\bigg[ \bigg(\frac{\ell_E^3}{(\ell_E^2+	\D_2(p_{1},-p_1-p_3))^2}-\frac{\ell_E^3}{(\ell_E^2+\D_1(p_{1},-p_1-p_3))^2}\bigg)\big((\slashed p_1+\slashed d)\g^\k-\g^\k(\slashed d+\slashed p_1+\slashed p_3)\big)\NO\\&
-2\bigg(\frac{\ell_E^3}{(\ell_E^2+\D_2(p_{1},p_{3}))^2}-\frac{\ell_E^3}{(\ell_E^2+\D_1(p_{1},p_{3}))^2}\bigg)(2yp_{1}+(1-2z)p_3)^\k\bigg]. 
\eal
As above, $\D_j(p,q)\equiv m_j^2+p^2\,y+q^2\,z-(p\, y-q\, z)^2$, and we have defined $d=-(p_1+p_3) z-p_1 y$.
We find that
\bbxd
\vskip.0cm
\begin{align}
\lb {\cal B}_S(p_1)\wt{\bar{ s}}_{(3|\frac{1}{2})}(-p_1-p_3)\wt{{\cal J}}^\k (p_3)\rb_{\text{ren}}=\frac{2i\pi^2}{36(2\pi)^4}\sqrt{2}\big(4p_1^\k+5p_3^\k+2\g^{\k\r}p_{1\r}+\g^{\k\r}p_{3\r}\big).
\end{align}
\ebxd
The second 3-point function with a seagull operator insertion takes the form 
\bal
&\lb \wt{{ s}}_{(1|0)}(-p_1-p_2)\cb_S(p_1)\wt{{\cal\bar{Q}}}{}^\n(p_2)\rb=\NO\\&
i^2\times\int \frac{d^4q}{(2\p)^4}\,\Big[m_2 P_{\vf_2}(q)P_{\vf_2}(q-p_1-p_2)P_LP_{\l_2}(p_1-q)\big(i\O_2^\n(p_1+p_2-q)-m_2\g^{\n}\big)P_L\NO\\
&-2m_1 P_{\vf_1}(q)P_{\vf_1}(q-p_1-p_2)P_LP_{\l_1}(p_1-q)\big(i\O_2^\n(p_1+p_2-q)-m_1\g^{\n}\big)P_L+P_L\leftrightarrow P_R\Big],
\eal
which evaluates to 
\bal
&\lb \wt{{ s}}_{(1|0)}(-p_1-p_2)\cb_S(p_1)\wt{{\cal\bar{Q}}}{}^\n(p_2)\rb=-i^2\times\frac{2i\p^2}{(2\p)^4} 4m_1^2\int d\ell_E\int_{0}^{1}dx\,dy\, dz\,\d(x+y+z-1)\times\\
&\bigg(\frac{\ell_E^3}{(\ell_E^2+\D_2(p_{1},-p_1-p_2))^2}-\frac{\ell_E^3}{(\ell_E^2+\D_1(p_{1},-p_1-p_2))^2}\bigg)\Big((\slashed p_1+\slashed e)\g^\n+\g^\n(\slashed e+\slashed p_1+\slashed p_2)-\g^{\n\r}\frac{2p_{2\r}}{3}\Big),\NO
\eal
where we defined $e=-(p_1+p_2)z-p_1 y$.
We get
\bbxd
\vskip.0cm
\begin{align}
&\lb \wt{{ s}}_{(1|0)}(-p_1-p_2) {\cal B}_S(p_1)\wt{\bar{{\cal Q }}}{}^\n(p_2)\rb_{\text{ren}}=-\frac{2i\pi^2}{12(2\pi)^4}\big(2p_1^\n+p_2^\n+\g^{\n\r}p_{2\r}\big).
\end{align}
\ebxd
Turing to the Feynman diagram contributions to the 4-point function, we have
\bal
&\lb\cb_S(p_1)\wt{\bar\cq}{}^\n(p_2)\wt{\cj}^\k(p_3)\wt{\cj}^\l(p_4)\rb=
-i^3\times \frac{1}{9}\int \frac{d^4q}{(2\p)^4}\NO\\
&\bigg[m_2\Big(P_{\vf_2}(q)P_LP_{\l_2}(p_1-q)\g^\k\g^5P_{\l_2}(-q'')\g^\l \g^5 P_{\l_2}(-q-p_2)\big(i\O_2^\n(-q)-m_2\g^{\n}\big)P_L\NO\\&
-4(2q+p_3)^\k(2q+2p_3+p_4)^\l P_{\vf_2}(q)P_{\vf_2}(q+p_3)P_{\vf_2}(q')P_LP_{\l_2}(p_1-q)\big(i\O_2^\n(-q')-m_2\g^{\n}\big)P_L\NO\\
&+2i(2q+p_3)^\k P_{\vf_2}(q)P_{\vf_2}(q+p_3)\g^5P_RP_{\l_2}(p_1-q)\g^\l\g^5P_{\l_2}(-q''')\big(i\O_2^\n(-q-p_3)-m_2\g^{\n}\big)P_L\Big)\NO\\
&-2m_1\Big(P_{\vf_1}(q)P_LP_{\l_1}(p_1-q)\g^\k\g^5P_{\l_1}(-q'')\g^\l \g^5 P_{\l_1}(-q-p_2)\big(i\O_2^\n(-q)-m_1\g^{\n}\big)P_L\NO\\&
-4(2q+p_3)^\k(2q+2p_3+p_4)^\l P_{\vf_1}(q)P_{\vf_1}(q+p_3)P_{\vf_1}(q')P_LP_{\l_1}(p_1-q)\big(i\O_2^\n(-q')-m_1\g^{\n}\big)P_L\NO\\
&+2i(2q+p_3)^\k P_{\vf_1}(q)P_{\vf_1}(q+p_3)\g^5P_RP_{\l_1}(p_1-q)\g^\l\g^5P_{\l_1}(-q''')\big(i\O_2^\n(-q-p_3)-m_1\g^{\n}\big)P_L\Big)\NO\\
&+P_L\leftrightarrow P_R\bigg]+\Big(\begin{array}{ccc} p_{3} & \leftrightarrow & p_{4}\\ \k & \leftrightarrow & \l \end{array}\Big),
\eal
where $q'=q-p_1-p_2$, $q''=q-p_1-p_3$, $q'''=q-p_1-p_4$. This can be written in the form
\be
\lb\cb_S(p_1)\wt{\bar\cq}{}^\n(p_2)\wt{\cj}^\k(p_3)\wt{\cj}^\l(p_4)\rb= i^3\times i\big(Y_1^{\n\k\l}+ Y_2^{\n\k\l}+ Y_3^{\n\k\l}\big)
+\Big(\begin{array}{ccc} p_{3} & \leftrightarrow & p_{4}\\ \k & \leftrightarrow & \l \end{array}\Big),
\ee	
where $Y_1^{\n\k\l}$, $Y_2^{\n\k\l}$ and $Y_2^{\n\k\l}$ denote the loop integrals 
\bal
Y_1^{\n\k\l}\equiv&\;-\frac{2i\p^2}{(2\p)^4}\frac{8m_1^2}{3}\int_0^1dx\, dy\, dz\, dt\,\d(x+y+z+t-1) \times\NO\\
&\int d\ell_E \bigg(\frac{\ell_E^5}{(\ell_E^2+	\D_2(p_3+p_4,-p_{3},p_{1}))^4}-\frac{\ell_E^5}{(\ell_E^2+	\D_1(p_3+p_4,-p_{3},p_{1}))^4}\bigg)\NO\\
&
\bigg(\Big(2\h^{\k\l}(-a-p_1)^\b+\h^{\k \b}(-2a+2p_3+p_4)^\l+\h^{\l \b}(-2a+p_3)^\k\Big) \g_\b\g^\n\\
&
+\Big(2\h^{\k\l}(-a+p_3+p_4)^\b+\h^{\k \b}(-2a+2p_3+p_4)^\l+\h^{\l \b}(-2a+p_3)^\k\Big)\g^\n\g_\b\h^{\k\l}\g^{\n\r}\frac{2p_{2\r}}{3}\bigg),\NO
\eal
\bal
Y_2^{\n\k\l}\equiv&\;-\frac{2i\p^2}{(2\p)^4}\frac{8m_1^2}{3} \int_0^1 dx\, dy\, dz\, dt\,\d(x+y+z+t-1) \times\NO\\
&\int d\ell_E\bigg[\bigg(\frac{\ell_E^5}{(\ell_E^2+\D_2(p_{4},p_1+p_3,p_{1}))^4}
-\frac{\ell_E^5}{(\ell_E^2+\D_1(p_{4},p_1+p_3,p_{1}))^4}\bigg)\NO\\
&\bigg(\frac14\Big(2\h^{\l \a}(p_2+p_4-b)^\b+2\h^{\l \b}(-b-p_{1})^\a+\h^{\a \b}(-2b+p_{4})^\l\Big) \g_\a\g^\k\g_\b\g^\n\NO\\
&+\frac14\Big(2(-b+p_{4})^\b+2\h^{\l \b}(-b-p_{1})^\a+\h^{\a \b}(-2b+p_{4})^\l\Big) \g_\a\g^\k\g^\n\g_\b\NO\\
&+\h^{\l\a}\frac{p_{2\r}}{3}\g_\a\g^\k\g^{\n\r}-\h^{\l \a}\frac{p_{2\r}}{3}\g^\k\g_\a\g^{\n\r}\NO\\
&-\frac14\Big(2\h^{\l \a}(-b+p_{4})^\b+2\h^{\l \b}(p_2+p_4-b)^\a+\h^{\a \b}(-2b+p_{4})^\l\Big) \g^\k\g_\a\g^\n\g_\b\bigg)\\
&+m_1^2\left(\frac{2\ell_E^3}{(\ell_E^2+	\D_2(p_{4},p_1+p_3,p_{1}))^4}-\frac{\ell_E^3}{(\ell_E^2+	\D_1(p_{4},p_1+p_3,p_{1}))^4}\right)(-2b+p_4)^\l\g^\k\g^\n\bigg],\NO
\eal
\bal
Y_3^{\n\k\l}\equiv&\;\frac{2i\p^2}{(2\p)^4}\frac{4m_1^2}{3}\int_0^1dx\, dy\, dz\, dt\,\d(x+y+z+t-1) \times \NO\\
&\int d\ell_E\bigg[\frac{1}{4}\bigg(\frac{\ell_E^5}{(\ell_E^2+	\D_2(p_{2},p_1+p_3,p_{1}))^4}-\frac{\ell_E^5}{(\ell_E^2+	\D_1(p_{2},p_1+p_3,p_{1}))^4}\bigg)\NO\\
&\bigg(-\Big(\h^{\a \b}(-c+p_2)^\x+\h^{\b \x}(-c-p_1)^\a+\h^{\a \x}(p_2+p_4-c)^\b\Big)\g_\a\g^\k\g_\b\g^\l\g_\x\g^\n\NO\\
&-\Big(-\h^{\a \b}c^\x+\h^{\b \x}(-c-p_1)^\a+\h^{\a \x}(p_2+p_4-c)^\b\Big)\g_\a\g^\k\g_\b\g^\l\g^\n\g_\x\NO\\
&+\Big(-\h^{\a \b}c^\x+\h^{\b \x}(-c-p_1)^\a+\h^{\a \x}(-c+p_{2})^\b\Big)\g_\a\g^\k\g^\l\g_\b\g^\n\g_\x\NO\\
&-\Big(-\h^{\a \b}c^\x+\h^{\b \x}(p_2+p_4-c)^\a+\h^{\a \x}(-c+p_{2})^\b\Big)\g^\k\g_\a\g^\l\g_\b\g^\n\g_\x\NO\\
&-\h^{\a \b}\frac{2p_{2\r}}{3}\g_\a\g^\k\g_\b\g^\l\g^{\n\r}+\h^{\a \b}\frac{2p_{2\r}}{3}\g_\a\g^\k\g^\l\g_\b\g^{\n\r}-\h^{\a \b}\frac{2p_{2\r}}{3}\g^\k\g_\a\g^\l\g_\b\g^{\n\r}\bigg)\NO\\
&+m_1^2 \bigg(\frac{2\ell_E^3}{(\ell_E^2+	\D_2(p_{2},p_1+p_3,p_{1}))^4}-\frac{\ell_E^3}{(\ell_E^2+	\D_1(p_{2},p_1+p_3,p_{1}))^4}\bigg)\NO\\
&\Big((c+p_1)^\a\g_\a\g^\k\g^\l\g^\n+(p_2+p_4-c)^\a\g^\k\g_\a\g^\l\g^\n\NO\\
&+(c-p_2)^\a\g^\k\g^\l\g_\a\g^\n+c^\a\g^\k\g^\l\g^\n\g_\a-\frac{2p_{2\r}}{3}\g^\k\g^\l\g^{\n\r}\Big)\bigg].
\eal
In these expressions we have introduced the notation $\D_j(p,q,r)=m_j^2+p^2 y+q^2 z+r^2 t-(p y-qz-r t)^2$, $a=(p_3+p_4) y+p_{3}z-p_1 t$, $b=p_4 y-(p_1+p_3)z-p_1 t$, $c=p_2 y-(p_1+p_3)z-p_1 t$. Moreover, integrals that vanish in the large PV mass limit have not been included.

The 4-point function is given by 
\bbxd
\vskip.0cm
\begin{align}\label{BS-QJJ-ren}
&\lb {\cal B}_S(p_1)\wt{\bar{{\cal Q }}}{}^\n(p_2)\wt{{\cal J }}^\k(p_3)\wt{{\cal J }}^\l(p_4)\rb_{\text{ren}}=\frac{2i\pi^2}{324(2\pi)^4}\Big(22\g^{\n\r}\h^{\k\l}p_{1\r}+12\g^{\l\r}\h^{\k\n}p_{1\r}+12\g^{\k\r}\h^{\l\n}p_{1\r}\NO\\&
+25\g^{\n\r}\h^{\k\l}\left(p_{3\r}+p_{4\r}\right)+6\g^{\l\n}\left(p_1^\k +p_3^\k\right)+6\g^{\k\n}(p_1^\l +p_4^\l)+9\g^{\k\l}(p_3^\n -p_4^\n)\NO\\&
+3i\e^{\k\l\n\r}\g^5(p_{4\r}-p_{3\r})-36p_1^\n\h^{\k\l}+36p_1^\l\h^{\k\n}+36p_1^\k\h^{\n\l}+18p_3^\n\h^{\k\l}+12p_3^\l\h^{\k\n}\NO\\&
+36p_3^\k\h^{\n\l}+18p_4^\n\h^{\k\l}+12p_4^\k\h^{\l\n}+36p_4^\l\h^{\n\k}\Big).
\end{align}
\ebxd

\begin{flushleft}
\noindent\rule{\textwidth}{0.8pt}
\raisebox{-0.1cm}{$\<\cb_R\wt\cq\wt{\bar\cq}{}\wt\cj\>$}
\noindent\rule{\textwidth}{0.8pt}
\end{flushleft}

From \eqref{Bs} and \eqref{operator-derivatives} we once again determine that
\bal
&\<\wt{{\cal Q}}^{\m}(p_1)\wt{{\cal\bar{Q}}}{}^{\n}(p_2)\cb_R(p_3)\wt{\cj}^{\l}(p_4)\>=\lb\wt{ {\cal Q}}^{\m}(p_1)\wt{{\cal\bar{Q}}}{}^{\n}(p_2)\cb_R(p_3)\wt{\cj}^{\l}(p_4)\rb\NO\\
&-\frac{i}{3}\g^\m\g^5\lb\cb_S(-p_2-p_4)\wt{{\cal\bar{Q}}}{}^{\n}(p_2)\wt{\cj}^{\l}(p_4)\rb-\frac{i}{3}\lb\wt\cq^\m(p_1)\bar\cb_S(-p_1-p_4)\wt{\cj}^{\l}(p_4)\rb\g^\n\g^5\NO\\
&+\frac38\e^{\m\n\r\s}\lb\wt\cj_\r(-p_3-p_4)\cb_R(p_3)\wt{\cj}^{\l}(p_4)\rb\g_\s+\frac18\lb\wt s_{(2|1)}^{\r}(-p_3-p_4)\cb_R(p_3)\wt{\cj}^{\l}(p_4)\rb\g^5\g_\r\h^{\m\n}\NO\\
&-\frac{\sqrt{2}}{3}\h^{\m\l}\lb{\wt s}_{\(3|\frac12\)}(-p_2-p_3)\wt{{\cal\bar{Q}}}{}^{\n}(p_2)\cb_R(p_3)\rb-\frac{\sqrt{2}}{3}\h^{\n\l}\lb\wt{ {\cal Q}}^{\m}(p_1){\wt {\bar s}}_{\(3|\frac12\)}(-p_1-p_3)\cb_R(p_3)\rb\\
&+\frac{\sqrt{2}\,i}{9}\h^{\m\l}\lb{\wt s}_{\(3|\frac12\)}(-p_2-p_3)\cb_S(p_2+p_3)\rb\g^\n\g^5+\frac{\sqrt{2}\,i}{9}\h^{\n\l}\g^\m\g^5\lb\cb_S(p_1+p_3){\wt {\bar s}}_{\(3|\frac12\)}(-p_1-p_3)\rb,\NO
\eal
where the seagull operators $\wt s_{(2|1)}$ and ${\wt s}_{\(3|\frac12\)}$ are given in \eqref{seagull-operators}. We have already computed the correlators $\lb\cb_S\wt{\bar\cq}\wt\cj\rb$, $\lb\cb_R\wt\cj\wt\cj\rb$ and $\lb\cb_S{\wt {\bar s}}_{\(3|\frac12\)}\rb$ (see eq.~\eqref{BS-S3-1/2} for the 2-point function). Moreover, $\lb\wt s_{(2|1)}\cb_R\wt\cj\rb=-6i\lb\wt \cj\cb_R\wt\cj\rb$ since $\cb_R$ contains only fermions, $\wt s_{(2|1)}$ is proportional to the fermionic part of the R-current, and only fermions contribute to the 3-point function $\lb\cb_R\wt\cj\wt\cj\rb$. Besides the Feynman diagram contribution to the 4-point function, therefore, we only need to determine the 3-point function $\lb{\wt s}_{\(3|\frac12\)}\,\wt{{\cal\bar{Q}}}\cb_R\rb$. 

However, it is convenient to evaluate instead the linear combination of this 3-point function and the 2-point function $\lb\cb_S{\wt {\bar s}}_{\(3|\frac12\)}\rb$ that appears in the 4-point function. We have
\bal
&\frac{\sqrt{2}}{3}\lb\wt{ {\cal Q}}^\m(p_1)\wt{\bar{s }}_{(3|\frac{1}{2})}(-p_1-p_3)\cb_R(p_3)\rb=\NO\\
&i^2\times{\frac{-2}{9}}\int \frac{d^4q}{(2\p)^4}\bigg(m_2 P_{\vf_2}(q)P_R\big(i\O_1^\m(q)+m_2\g^{\m}\big)P_{\l_2}(p_1-q)\g^5P_{\l_2}(p_1+p_3-q)P_L\g^5\NO\\
&-2m_1 P_{\vf_1}(q)P_R\big(i\O_1^\m(q)+m_1\g^{\m}\big)P_{\l_1}(p_1-q)\g^5P_{\l_1}(p_1+p_3-q)P_L\g^5\bigg)+P_L\leftrightarrow P_R,
\eal
as well as,
\bal
&{-\frac{\sqrt{2}}{9}}\g^\m\g^5\lb \cb_S(p_1+p_3)\wt{\bar{s }}_{(3|\frac{1}{2})}(-p_1-p_3)\rb=\NO\\
&i\times {\frac{-2}{9}}i\g^\m\int \frac{d^4q}{(2\p)^4}\Big(m_2^2 P_{\vf_2}(q)P_{\vf_2}(p_1+p_3-q)-2m_1^2 P_{\vf_1}(q)P_{\vf_1}(p_1+p_3-q)\Big).
\eal
Introducing Feynman parameters these give 
\bal
&\frac{\sqrt{2}}{3}\lb\wt{{\cal Q}}^\m(p_1)\wt{\bar{s }}_{(3|\frac{1}{2})}(-p_1-p_3)\cb_R(p_3)\rb{-\frac{\sqrt{2}\,i}{9}}\g^\m\g^5\lb \cb_S(p_1+p_3)\wt{\bar{s }}_{(3|\frac{1}{2})}(-p_1-p_3)\rb=\NO\\&
=i^2\times\frac{2i\p^2}{(2\p)^4}\frac{8im_1^2}{9}\int d\ell_E\int_0^1 dx\,dy\, dz\,\d(x+y+z-1)\times\NO\\
&\times\bigg(\frac{\ell_E^3}{(\ell_E^2+\D_2(p_2+p_4,p_1))^3}-\frac{\ell_E^3}{(\ell_E^2+\D_1(p_2+p_4,p_1))^3}\bigg)\NO\\
&\times \Big(-\big((\slashed p_2+\slashed p_4)\,y-\slashed p_1\,z\big)\g^\m\slashed{p}_3 -\g^\m\big((\slashed p_2+\slashed p_4)\,y+(1-z)\slashed p_1\big)\slashed{p}_3+\frac{2}{3}\g^{\m\s}\slashed{p}_3 p_{1\s}\Big),
\eal
where, as above, $\D_j(p,q)\equiv m_j^2+p^2\,y+q^2\,z-(p\, y-q\, z)^2$. Integrating over $\ell_E$ and taking the limit $m_1\to\infty$ we get
\bal
&\frac{\sqrt{2}}{3}\lb\wt{ {\cal Q}}^\m(p_1)\wt{\bar{s }}_{(3|\frac{1}{2})}(-p_{1}-p_3)\cb_R(p_3)\rb{-\frac{\sqrt{2}\,i}{9}}\g^\m\g^5\lb \cb_S(p_{13})\wt{\bar{s }}_{(3|\frac{1}{2})}(-p_{1}-p_3)\rb=\NO\\
&-i^2\times \frac{2i\p^2}{(2\p)^4}\frac{i}{9}\int_0^1 dx\,dy\, dz\,\d(x+y+z-1)\times\NO\\
&\Big(-\big((\slashed p_2+\slashed p_4)\,y-\slashed p_1\,z\big)\g^\m\slashed{p}_3 -\g^\m\big((\slashed p_2+\slashed p_4)\,y+(1-z)\slashed p_1\big)\slashed{p}_3+\frac{2}{3}\g^{\m\s}\slashed{p}_3 p_{1\s}\Big).
\eal
\bbxd
\vskip.0cm
\begin{align}
&\frac{\sqrt{2}}{3}\lb\wt{ {\cal Q}}^\m(p_1)\wt{\bar{s }}_{(3|\frac{1}{2})}(-p_{1}-p_3){\cal B}_R(p_3)\rb_{\text{ren}}-\frac{\sqrt{2}i}{9}\g^\m\g^5\lb {\cal B}_S(p_{1}+p_3)\wt{\bar{s }}_{(3|\frac{1}{2})}(-p_{1}-p_3)\rb_{\text{ren}}\NO\\&
=\frac{2i\pi^2}{54(2\pi)^4}\Big(\g_\n\g^5\e^{\m\n\r\s}p_{1\r}p_{3\s}-i\g^\m p_1\cdot p_3+i\slashed{p}_1 p_3^\m+ip_1^\m\slashed{p}_3+2ip_3^\m\slashed{p}_3\Big).
\end{align}
\ebxd
	
The Feynman diagram contribution to the 4-point function takes the form	
\bal
&\lb\wt{ {\cal Q}}^{\m}(p_1)\wt{{\cal\bar{Q}}}{}^{\n}(p_2)\cb_R(p_3)\wt{\cj}^{\l}(p_4)\rb=\NO\\&
i^3\times \frac{i^2}{9}\int \frac{d^4q}{(2\p)^4}\times\NO\\
&\Big(m_2P_{\vf_2}(q)P_L\big(i\O_1^\m(q)+m_2\g^{\m}\big)P_{\l_2}(p_1-q)\g^5P_{\l_2}(p_1+p_3-q)\g^\l \g^5 P_{\l_2}(-q-p_2)\big(i\O_2^\n(-q)-m_2\g^{\n}\big)P_L\NO\\
&-2m_1P_{\vf_1}(q)P_L\big(i\O_1^\m(q)+m_1\g^{\m}\big)P_{\l_1}(p_1-q)\g^5P_{\l_1}(p_1+p_3-q)\g^\l \g^5 P_{\l_1}(-q-p_2)\big(i\O_2^\n(-q)-m_1\g^{\n}\big)P_L\Big)\NO\\
&+i^3\times \frac{i^2}{9}\int \frac{d^4q}{(2\p)^4}\times \NO\\
&\Big(m_2P_{\vf_2}(q)P_L\big(i\O_1^\m(q)+m_2\g^{\m}\big)P_{\l_2}(p_1-q)\g^\l\g^5P_{\l_2}(p_1+p_4-q) \g^5 P_{\l_2}(-q-p_2)\big(i\O_2^\n(-q)-m_2\g^{\n}\big)P_L\NO\\
&-2m_1P_{\vf_1}(q)P_L\big(i\O_1^\m(q)+m_1\g^{\m}\big)P_{\l_1}(p_1-q)\g^\l\g^5P_{\l_1}(p_1+p_4-q) \g^5 P_{\l_1}(-q-p_2)\big(i\O_2^\n(-q)-m_1\g^{\n}\big)P_L\Big)\NO\\
&+i^3\times \frac{2i}{9}\int \frac{d^4q}{(2\p)^4}(2q+p_4)^\l \times\NO\\
&\Big(m_2P_{\vf_2}(q)P_{\vf_2}(q+p_4)\g^5P_R\big(i\O_1^\m(q)+m_2\g^{\m}\big)P_{\l_2}(p_1-q)\g^5P_{\l_2}(p_1+p_3-q)\big(i\O_2^\n(-q-p_4)-m_2\g^{\n}\big)P_L\NO\\
&-2m_1P_{\vf_1}(q)P_{\vf_1}(q+p_4)\g^5P_R\big(i\O_1^\m(q)+m_1\g^{\m}\big)P_{\l_1}(p_1-q)\g^5P_{\l_1}(p_1+p_3-q)\big(i\O_2^\n(-q-p_4)-m_1\g^{\n}\big)P_L\Big)\NO\\
&+P_L\leftrightarrow P_R.
\eal
Notice that the first and second integrals in this expression are related by charge conjugation (i.e. $ C(...)^TC^{-1}$) and the interchange $\m\leftrightarrow\n$, $p_1\leftrightarrow p_2$. Introducing Feynman parameters, the sum of all three integrals can be written in the form
\be
\lb\wt{ {\cal Q}}^{\m}(p_1)\wt{{\cal\bar{Q}}}{}^{\n}(p_2)\cb_R(p_3)\wt{\cj}^{\l}(p_4)\rb=i^3\times i\Big( (\Xi_1^{\m\n\l}+\text{Ch.C.})+(\Xi_2^{\m\n\l}+\text{Ch.C.})+\Theta_1^{\m\n\l}+\Theta_2^{\m\n\l}\Big),	
\ee	
where
\bal
&\Th_1^{\m\n\l}\equiv
\frac{2i\p^2}{(2\p)^4}\frac{-4im_1^2}{9}\int_0^1 dx\,dy\, dz\,\d(x+y+z-1)\times \NO\\
&\int
d\ell_E\bigg[(\g^\l\g^\m\g^\n-\g^\m\g^\n\g^\l)\bigg(\frac{\ell_E^5}{(\ell_E^2+\D_2(p_{4},p_1+p_3))^3}-\frac{\ell_E^5}{(\ell_E^2+\D_1(p_{4},p_1+p_3))^3}\bigg)\NO\\
&+2(2w-p_4)^\l\bigg(\frac{\ell_E^3}{(\ell_E^2+\D_2(p_{4},p_1+p_3))^3}-\frac{\ell_E^3}{(\ell_E^2+\D_1(p_{4},p_1+p_3))^3}\bigg)\times\NO\\
&\times\Big(\g^\a\g^\m\g^\n w_\a-\g^\m\g^\n\g^\a(w-p_4)_\a+\g^\m\g^{\n\r}\frac{2p_{2\r}}{3}-\g^{\m\s}\g^\n\frac{2p_{1\s}}{3}\Big)\bigg],
\eal
\bal
&\Th_2^{\m\n\l}\equiv\frac{2i\p^2}{(2\p)^4}\frac{-24im_1^2}{9}\int_0^1 dx\,dy\, dz\,dt\,\d(x+y+z+t-1)\times\NO\\
&\int d\ell_E\bigg[m_1^2\bigg(\frac{2\ell_E^3}{(\ell_E^2+\D_2(p_{4},p_{1},p_1+p_3))^4}-\frac{\ell_E^3}{(\ell_E^2+\D_1(p_{4},p_{1},p_1+p_3))^4}\bigg)\g^\m\slashed{p}_3\g^\n (-2e+p_4)^\l\NO\\
&+\bigg(\frac{\ell_E^5}{(\ell_E^2+\D_2(p_{4},p_{1},p_1+p_3))^4}-\frac{\ell_E^5}{(\ell_E^2+\D_1(p_{4},p_{1},p_1+p_3))^4}\bigg)\times\NO\\
&\bigg(-\frac14\g_\b\g^\m\g_\a\slashed{p}_3\g^\n\big(-2\h^{\l \a}e^\b-2\h^{\l \b}(e+p_1)^\a+\h^{\a \b}(-2e+p_4)^\l\big)\NO\\
&+\frac14\g_\a\g^\m\slashed{p}_3\g^\n\g_\b\big(-2\h^{\l \a}(e-p_4)^\b-2\h^{\l \b}e^\a+\h^{\a \b}(-2e+p_4)^\l\big)\NO\\
&{+}\frac14\g^\m\g_\a\slashed{p}_3\g^\n\g_\b\big(-2\h^{\l \a}(e-p_4)^\b-2\h^{\l \b}(e+p_1)^\a+\h^{\a \b}(-2e+p_4)^\l\big)\NO\\
&+\g^\l\g^\m\slashed{p}_3\g^{\n\r}\frac{p_{2\r}}{3}+\g^\m\g^\l\slashed{p}_3\g^{\n\r}\frac{p_{2\r}}{3}-\g^{\m\s}\g^\l\slashed{p}_3\g^\n\frac{p_{1\s}}{3}+\g^{\m\s}\slashed{p}_3\g^\n\g^\l\frac{p_{1\s}}{3}\bigg)\bigg],
\eal
\bal
&\Xi_1^{\m\n\l}\equiv
\frac{2i\p^2}{(2\p)^4}\frac{-4im_1^2}{9}\int_0^1 dx\,dy\, dz\,\d(x+y+z-1)\times \NO\\
&\int d\ell_E\bigg[\bigg(\frac{\ell_E^5}{(\ell_E^2+\D_2(p_{2},p_1+p_3))^3}-\frac{\ell_E^5}{(\ell_E^2+\D_1(p_{2},p_1+p_3))^3}\bigg)(\g^\l\h^{\m\n}-\g^\m\h^{\l\n})\NO\\
&+\bigg(\frac{\ell_E^3}{(\ell_E^2+\D_2(p_{2},p_1+p_3))^3}-\frac{\ell_E^3}{(\ell_E^2+\D_1(p_{2},p_1+p_3))^3}\bigg)\times\NO\\
&\Big(\slashed{n}\g^\m\g^\l(\slashed{n}-\slashed{p}_2)\g^\n+\slashed{n}\g^\m\g^\l\g^\n\slashed{n}-\slashed{n}\g^\m\g^\l\g^{\n\r}\frac{2p_{2\r}}{3}+\g^\m\g^\l(\slashed{n}-\slashed{p}_2)\g^\n\slashed{n}\NO\\
&+\g^\m\g^\l(-\slashed{n}+\slashed{p}_2)\g^{\n\r}\frac{2p_{2\r}}{3}+\g^{\m\s}\g^\l(-\slashed{n}+\slashed{p}_2)\g^\n\frac{2p_{1\s}}{3}-\g^{\m\s}\g^\l\g^\n\slashed{n}\frac{2p_{1\s}}{3}+\g^{\m\s}\g^\l\g^{\n\r}\frac{4p_{2\r}p_{1\s}}{9}\Big)\NO\\
&-{m_1^2}\int d\ell_E\bigg(\frac{2\ell_E^3}{(\ell_E^2+\D_2(p_{2},p_1+p_3))^3}-\frac{\ell_E^3}{(\ell_E^2+\D_1(p_{2},p_1+p_3))^3}\bigg)\g^\m\g^\l\g^\n\bigg],
\eal
\bal
&	\Xi_2^{\m\n\l}=
\frac{2i\p^2}{(2\p)^4}\frac{im_1^2}{3}\int_0^1 dx\,dy\, dz\, dt\d(x+y+z+t-1)\times\NO\\
&\int d\ell_E\bigg[4m_1^2\bigg(\frac{2\ell_E^3}{(\ell_E^2+\D_2(p_{2},p_{1},p_1+p_3))^4}-\frac{\ell_E^3}{(\ell_E^2+\D_1(p_{2},p_{1},p_1+p_3))^4}\bigg)\NO\\
&\Big(\big(\slashed{u}\g^\m+\g^\m(\slashed{p}_1+\slashed{u})\big)\slashed{p}_3\g^\l\g^\n+\g^\m\slashed{p}_3\g^\l\big((\slashed{u}-\slashed{p}_2)\g^\n+\g^\n\slashed{u}\big)-\g^\m\slashed{p}_3\g^\l\g^{\n\r}\frac{2p_{2\r}}{3}-\g^{\m\s}\slashed{p}_3\g^\l\g^\n\frac{2p_{1\s}}{3}\Big)\NO\\
&+\bigg(\frac{\ell_E^5}{(\ell_E^2+\D_2(p_{2},p_{1},p_1+p_3))^4}-\frac{\ell_E^5}{(\ell_E^2+\D_1(p_{2},p_{1},p_1+p_3))^4}\bigg)\times\NO\\
&\bigg(\g^\a\g^\m\g_\a\slashed{p}_3\g^\l(-\slashed{u}+\slashed{p}_2)\g^\n+\g^\a\g^\m(-\slashed{u}-\slashed{p}_1)\slashed{p}_3\g^\l\g_\a\g^\n-\slashed{u}\g^\m\g^\a\slashed{p}_3\g^\l\g_\a\g^\n-\g^\a\g^\m\g_\a\slashed{p}_3\g^\l\g^\n\slashed{u}\NO\\
&
+\g^\a\g^\m(-\slashed{u}-\slashed{p}_1)\slashed{p}_3\g^\l\g^\n\g_\a-\slashed{u}\g^\m\g_\a\slashed{p}_3\g^\l\g^\n\g^\a+\g^\a\g^\m\g_\a\slashed{p}_3\g^\l\g^{\n\r}\frac{2p_{2\r}}{3}-\g^\a\g^\m\slashed{p}_3\g^\l\g_\a\g^\n\slashed{u}\NO\\
&
+\g^\a\g^\m\slashed{p}_3\g^\l(-\slashed{u}+\slashed{p}_2)\g^\n\g_\a-\slashed{u}\g^\m\slashed{p}_3\g^\l\g^\a\g^\n\g_\a+\g^\a\g^\m\slashed{p}_3\g^\l\g_\a\g^{\n\r}\frac{2p_{2\r}}{3}-\g^\m\g^\a\slashed{p}_3\g^\l\g_\a\g^\n\slashed{u}\NO\\
&
+\g^\m\g^\a\slashed{p}_3\g^\l\g_\a\g^{\n\r}\frac{2p_{2\r}}{3}+\g^{\m\s}\g^\a\slashed{p}_3\g^\l\g_\a\g^\n\frac{2p_{1\s}}{3}+\g^\m\g^\a\slashed{p}_3\g^\l(-\slashed{u}+\slashed{p}_2)\g^\n\g_\a\NO\\
&+\g^\m(-\slashed{u}-\slashed{p}_1)\slashed{p}_3\g^\l\g^\a\g^\n\g_\a+
\g^{\m\s}\g^\a\slashed{p}_3\g^\l\g^\n\g_\a\frac{2p_{1\s}}{3}+\g^{\m\s}\slashed{p}_3\g^\l\g^\a\g^\n\g_\a\frac{2p_{1\s}}{3}\bigg)\bigg].
\eal
In all these expressions $\D_j(p,q)\equiv m_j^2+p^2\,y+q^2\,z-(p\, y-q\, z)^2$, $\D_j(p,q,r)=m_j^2+p^2 y+q^2 z+r^2 t-(p y-qz-r t)^2$, $w=p_4 y-p_{13}z$, $n=p_2 y-p_{13}z$, $e=p_4 y-p_{13}t-p_1 z$, $u=p_2 y-p_{13}t-p_1 z$. Moreover, as above, we have dropped integrals that vanish in the large PV mass limit, e.g.  
\be
m_1^2\int d\ell_E \bigg(\frac{ \ell_E^3}{(\ell_E^2+\D_2(p_{4},p_{1},p_1+p_3))^4}-\frac{\ell_E^3}{(\ell_E^2+\D_1(p_{4},p_{1},p_1+p_3))^4}\bigg).
\ee

Evaluating the above expressions we find that the 4-point function takes the form 
\bbxd
\vskip.2cm
\begin{align}\label{QQBR-J-ren}
&\lb\wt{ {\cal Q}}^{\m}(p_1)\wt{{\cal\bar{Q}}}{}^{\n}(p_2){\cal B}_R(p_3)\wt{{\cal J}}^{\l}(p_4)\rb_{\text{ren}}\NO\\&=\frac{2i\pi^2}{324(2\pi)^4}\bigg[i\slashed{p}_3\big(9(p_1^\m\h^{\l\n}-p_2^\n\h^{\l\m})+18(p_3^\m\h^{\l\n}-p_3^\n\h^{\l\m})	+4(p_2^\m\h^{\l\n}-p_1^\n\h^{\l\m})+4\h^{\n\m}(p_1^\l-p_2^\l)\big)\NO\\&
+i\slashed{p}_2\big(3p_1^\n\h^{\l\m}+2p_2^\n\h^{\l\m}-3p_3^\n\h^{\l\m}
-p_1^\m\h^{\l\n}-2p_3^\m\h^{\l\n}-4p_2^\l\h^{\m\n}-3p_3^\l\h^{\n\m}\big)\NO\\&
-i\slashed{p}_1\big(3p_2^\m\h^{\l\n}+2p_1^\m\h^{\l\n}-3p_3^\m\h^{\l\n}-p_2^\n\h^{\l\m}-2p_3^\n\h^{\l\m}-4p_1^\l\h^{\m\n}-3p_3^\l\h^{\n\m}\big)\NO\\&
+2i\g^\l\big(p_1^\m p_1^\n+5p_3^\m p_1^\n-p_2^\m p_2^\n+3p_3^\m p_2^\n-3p_1^\m p_3^\n-5p_2^\m p_3^\n+(p_2-p_1)\cdot(p_2+p_1+2p_3)\h^{\m\n}\big)\NO\\&
+i\g^\n\big(4p_1^\m p_1^\l-2p_2^\l p_1^\m+3p_1^\m p_3^\l-2p_2^\m p_2^\l+3p_2^\m p_3^\l+2p_3^\m p_1^\l-4p_2^\l p_3^\m \NO\\&
-7p_1^2 \h^{\m\l}-9p_2\cdot p_1\h^{\m\l}-8p_3\cdot p_1\h^{\m\l}-15p_2\cdot p_3\h^{\m\l}-11p_2^2 \h^{\m\l}-9p_3^2 \h^{\m\l}\big)\NO\\&
-i\g^\m\big(4p_2^\n p_2^\l-2p_1^\l p_2^\n+3p_2^\n p_3^\l-2p_1^\n p_1^\l+3p_1^\n p_3^\l+2p_3^\n p_2^\l-4p_1^\l p_3^\n \NO\\&
-7p_2^2 \h^{\n\l}-9p_1\cdot p_2\h^{\n\l}-8p_3\cdot p_2\h^{\n\l}-15p_1\cdot p_3\h^{\n\l}-11p_1^2 \h^{\n\l}-9p_3^2 \h^{\n\l}\big)\NO\\&
+\frac12\e^{\l\m\r\s}\g^\n\g^5(4p_{2\r}-11p_{1\r})p_{3\s}+\frac12\e^{\l\n\r\s}\g^\m\g^5(24p_{1\r}-7p_{2\r})p_{3\s}\NO\\&
+\frac14\e^{\m\n\r\s}\g^\l\g^5(13p_{2\r}-5p_{1\r})p_{3\s}+\frac14\e^{\l\m\n\r}\slashed{p}_3\g^5\big(5p_{2\r}-13p_{1\r}\big)+2\e^{\l\m\n\r}(2\slashed{p}_1-\slashed{p}_2)\g^5p_{3\r}\NO\\&
+\frac14\g_\k\g^5\Big(8\e^{\l\m\n\k}p_1^2+\e^{\l\m\n\k}p_1\cdot p_3-8\e^{\l\m\n\k}p_2^2-\e^{\l\m\n\k}p_2\cdot p_3-22\h^{\m\n}\e^{\l\k\r\s}p_{1\r}p_{3\s}\NO\\&
+10\h^{\m\n}\e^{\l\k\r\s}p_{2\r}p_{3\s}+7\h^{\l\n}\e^{\m\k\r\s}p_{1\r}p_{3\s}+3\h^{\l\n}\e^{\m\k\r\s}p_{2\r}p_{3\s}+51\h^{\l\m}\e^{\n\k\r\s}p_{1\r}p_{3\s}\NO\\&
+23\h^{\l\m}\e^{\n\k\r\s}p_{2\r}p_{3\s}-\e^{\l\m\k\r}p_{1\r}p_2^\n-17\e^{\l\m\k\r}p_{1\r}p_3^\n-3\e^{\l\m\k\r}p_{2\r}p_1^\n-8\e^{\l\m\k\r}p_{2\r}p_2^\n \NO\\&
-\e^{\l\m\k\r}p_{2\r}p_3^\n-14\e^{\l\m\k\r}p_{3\r}p_1^\n-8\e^{\l\n\k\r}p_{1\r}p_1^\m -3\e^{\l\n\k\r}p_{1\r}p_2^\m +31\e^{\l\n\k\r}p_{1\r}p_3^\m  \NO\\&
-4\e^{\l\n\k\r}p_{2\r}p_1^\m-17\e^{\l\n\k\r}p_{2\r}p_3^\m -32\e^{\l\n\k\r}p_{3\r}p_1^\m -14\e^{\l\n\k\r}p_{3\r}p_2^\m +16\e^{\m\n\k\r}p_{1\r}p_1^\l\NO\\&
+8\e^{\m\n\k\r}p_{1\r}p_3^\l-16\e^{\m\n\k\r}p_{2\r}p_2^\l+8\e^{\m\n\k\r}p_{2\r}p_3^\l-19\e^{\m\n\k\r}p_{3\r}p_1^\l+3\e^{\m\n\k\r}p_{3\r}p_2^\l\Big)\bigg].
\end{align}
\ebxd


\bibliographystyle{JHEP}
\bibliography{susy-anomalies,CSanomalies}

\end{document}